\documentclass[superscriptaddress,longbibliography,
reprint,
 amsmath,amssymb,
 aps,
pra,
]{revtex4-1}

\pdfoutput=1 
\usepackage[compatibility=false]{caption}
\usepackage{feynmp}
\usepackage{simpler-wick}
\usepackage{graphicx}
\usepackage{dcolumn}
\usepackage{bm}
\usepackage{subfig}

\usepackage[
hmargin=0.6in,
vmargin=0.8in,
]{geometry}

\usepackage[colorlinks=true,linkcolor=blue,citecolor=blue,urlcolor=blue]{hyperref}
\usepackage{braket}

\newcommand{\MP}{\mathrm{MP}}

\usepackage{bbm}

\newcommand{\h}{\mathrm{h}}
\newcommand{\p}{\mathrm{p}}
\newcommand{\CC}{\mathrm{CC}}

\newcommand{\IP}{\mathrm{IP}}

\newcommand{\B}{\mathrm{B}}

\newcommand{\pt}{\mathrm{pt}}
\newcommand{\sco}{\mathrm{sc}}

\newcommand{\dRPA}{\mathrm{RPA}}
\newcommand{\RPA}{\mathrm{RPA}}
\newcommand{\rCCD}{\mathrm{rCCD}}

\newcommand{\rf}{\mathrm{ref}}
\newcommand{\CCSD}{\mathrm{CCSD}}

\newcommand{\F}{\mathrm{F}}
\newcommand{\Ba}{\mathrm{B}}

\begin{document}

\title{Non-hermitian Green's function theory with $N$-body interactions: the coupled-cluster similarity transformation}

\author{Christopher J. N. Coveney}
\email{christopher.coveney@physics.ox.ac.uk}
\affiliation{ Department of Physics, University of Oxford, Oxford OX1 3PJ, United Kingdom}
\author{David P. Tew}%
\email{david.tew@chem.ox.ac.uk}
\affiliation{Physical and Theoretical Chemistry Laboratory, University of Oxford, South Parks Road, United Kingdom}

\date{\today}

\begin{abstract}
            We present the diagrammatic theory of the irreducible self-energy and Bethe-Salpeter kernel that naturally arises within the Green's function formalism for a general $N$-body non-hermitian interaction. In this work, we focus specifically on the coupled-cluster self-energy generated by the similarity transformation of the electronic structure Hamiltonian. The main result of this work is a novel Green's function formalism for non-hermitian many-body quantum systems. We develop the biorthogonal quantum theory to construct dynamical correlation functions where the time-dependence of operators is governed by a non-hermitian Hamiltonian. We extend the Gell-Mann and Low theorem to include non-hermitian interactions and to generate perturbative expansions of many-body Green's functions. We introduce the single-particle coupled-cluster Green's function and derive the perturbative diagrammatic expansion for the non-hermitian coupled-cluster self-energy in terms of the `non-interacting' reference Green's function, $\tilde{\Sigma}[G_0]$. From the exact equation-of-motion of the single-particle coupled-cluster Green's function, we derive the self-consistent renormalized coupled-cluster self-energy, $\tilde{\Sigma}[\tilde{G}]$, and demonstrate its relationship to the perturbative expansion of the self-energy, $\tilde{\Sigma}[G_0]$. Subsequently, we show that the usual electronic self-energy can be recovered from the coupled-cluster self-energy by neglecting the effects of the similarity transformation. We show how the coupled-cluster ground state energy can be obtained from the coupled-cluster self-energy and provide an overview of the relationship between approximations for the coupled-cluster self-energy, ionization potential/electron affinity equation-of-motion coupled-cluster theory (IP/EA-EOM-CC) and the $G_0W_0$ approximation. As a result, we introduce the CC-$G_0W_0$ self-energy by leveraging the connections between Green's function and coupled-cluster theory. Finally, we derive the diagrammatic expansion of the coupled-cluster Bethe-Salpeter kernel. 
\end{abstract}

\maketitle

\begin{fmffile}{diagram}
\section{Introduction}

The Green's function formalism is one of the major pillars in the \emph{ab initio} theory of many interacting particles~\cite{Quantum,mahan2000many,stefanucci2013nonequilibrium}. Green's function theory is based on the analysis of $N$-point dynamical correlation functions, with the single-particle Green's function occupying the central position. A significant conceptual and computational advantage of Green's functions is that they provide a simultaneous description of ground and excited state many-body correlation. Knowledge of the single-particle Green's function gives access to the exact addition and removal energies, ground state expectation values of any single-particle observable and, for two-body interactions, the exact ground state energy of the system. Similarly, knowledge of the $N$-body Green's function yields ground and excited state properties of $N$-body observables. 

Direct computation of the single-particle Green's function scales exponentially with the number of interacting particles, requiring knowledge of exact eigenstates of the full many-body Hamiltonian containing different numbers of particles. However, the single-particle Green's function can also be constructed from the Dyson equation. The Dyson equation relates a known, reference Green's function to the exact, interacting Green's function through the irreducible self-energy. The irreducible self-energy contains all the many-body interactions a single-particle experiences as it propagates through the system. The self-energy itself depends on the exact Green's function such that the Green's function formalism constitutes a self-consistent theory~\cite{hedin1965new}. The self-energy can be expressed either as a perturbative or self-consistent Feynman diagrammatic expansion. For systems of particles governed by a two-body interaction, theories of the self-energy have been well established and extensively explored in nuclear and condensed matter systems~\cite{barbieri2017self,aryasetiawan1998gw,hybertsen1986electron,rohlfing2000electron,stefanucci2013nonequilibrium,romaniello2009self,romaniello2012beyond,hirata2017one}. Only recently, in the context of nuclear theory, has the Green's function formalism been extended to include three-body interactions~\cite{carbone2013self,raimondi2018algebraic}. Until this work, the development of the Green's function formalism in the context of non-hermitian $N$-body interactions has been entirely unexplored.

The discovery that non-hermitian parity-time ($\mathcal{PT}$) symmetric Hamiltonians can possess real spectra~\cite{bender1998real} has lead to new active areas of theoretical and experimental physics research~\cite{bender2007making,bender2024pt,ashida2020non,meden2023mathcal,bergholtz2021exceptional}. There has been increasing interest in exploring non-hermitian Hamiltonians as they can describe experimentally verifiable and novel properties of physical systems~\cite{bender2024pt}. For example, non-hermitian Hamiltonians play a role in the theory of open quantum systems~\cite{barontini2013controlling,stefanucci2013nonequilibrium,meden2023mathcal}, dissipative Lindbladian dynamics~\cite{shibata2019dissipative,stefanucci2024kadanoff} as well topological quantum phase transitions~\cite{vsindelka2017excited,wang2019simulating,wang2019observation,zhang2025nonadiabatic}. Experimental work on probing the spectra and eigenstates of non-hermitian Hamiltonians has also recently emerged in the context of optics and photonics~\cite{el2018non,miri2019exceptional}, magnetic systems~\cite{hurst2022non}, exciton-polaritons~\cite{liang2025twist} as well as ultracold atoms~\cite{liang2022dynamic,zhao2025two}. The continued experimental and theoretical exploration of non-hermitian many-body quantum systems further motivates the development of a new Green’s function formalism capable of describing their spectral properties.

Another central pillar in the \emph{ab initio} theory of many interacting particles is coupled-cluster (CC) theory~\cite{shavitt2009many,helgaker2013molecular}. Coupled-cluster theory is directly concerned with obtaining ground state properties of a system of many interacting particles. It is based on a size-extensive reformulation of the Configuration Interaction  that leads to hierarchical and systematically controllable sets of equations that are, in principle, exact. CC theory is formulated in terms of the non-hermitian similarity transformed Hamiltonian. The CC similarity transformation is determined by the ground state CC amplitude equations. In general, when the CC transformation is not truncated, the similarity transformed Hamiltonian is an $N$-body non-hermitian operator that possesses different left and right eigenstates. Excited state properties are then usually obtained from Configuration Interaction expansions of the CC similarity transformed Hamiltonian, commonly referred to as equation-of-motion coupled-cluster (EOM-CC) theory.

Previous attempts at combining coupled-cluster and Green's function methods have been largely centered around direct construction of the electronic single-particle Green's function using the eigenstates and eigenvalues resulting from ground state and ionization potential/electron affinity equation-of-motion coupled-cluster theory (IP/EA-EOM-CC)~\cite{nooijen1992coupled,nooijen1993coupled,peng2016coupled,peng2018green,shee2019coupled}. In our work, we refer to the approach first presented in Refs~\cite{nooijen1992coupled} and~\cite{nooijen1993coupled} as the coupled-cluster representation of the electronic Green's function. The approach outlined in Refs~\cite{nooijen1992coupled,nooijen1993coupled} has led to many practical advantages over conventional many-body Green's function techniques for computing spectral functions and charged excitation energies~\cite{shee2019coupled,peng2021coupled,shee2022exploring,backhouse2022constructing}. However, these approaches to combining coupled-cluster with Green's function theory do not connect to a diagrammatic construction of the irreducible coupled-cluster self-energy~\cite{coveney2023coupled}. As a result, the connections to the coupled-cluster representation of the electronic two-particle Green's function remain largely unexplored.

Recently, there has been increasing interest in understanding and exploring the connections between coupled-cluster and Green's function theories based on the approximate $G_0W_0$ self-energy~\cite{quintero2022connections,monino2023connections,tolle2023exact,rishi2020route,lange2018relation}. It has also been demonstrated that the Random Phase (RPA) and static $GW$-BSE approximations can be recast as approximate forms of CCD theory where the doubles amplitudes are solved for by keeping only the `ring' diagrams (rCCD)~\cite{scuseria2008ground,quintero2022connections}. In the case of the static $GW$-BSE approximation, the associated rCCD equations contain screened interaction intermediates as well as bare Coulomb vertices~\cite{quintero2022connections}. These established connections have also been useful in the development of analytic nuclear gradients for the static through its connection to \emph{unitary} coupled-cluster theory~\cite{tolle2023exact,tolle2025fully_an,tolle2025fully}. This work has been crucial in furthering the conceptual understanding of the nature of the many-body self-energy and Bethe-Salpeter kernel in the context of wavefunction based approaches. However, Green's function theory is centered about \emph{universal} and \emph{exact} relationships between the exact single-particle Green's function, irreducible self-energy functional $\Sigma[G]$, and its higher order functional derivatives~\cite{stefanucci2013nonequilibrium}. To fully understand the relationship between the functional-diagrammatic framework of Green's function theory and the coupled-cluster similarity transformation, it is necessary to develop new methods in order to construct the self-energy generated by a higher than two-body non-hermitian interaction Hamiltonian. This leads us to introduce a formally exact formalism that is related to but distinct from prior work on connecting approximate Green's function and coupled-cluster theories.

In Ref.~\cite{coveney2023coupled}, the authors of this work demonstrated the existence of the irreducible coupled-cluster self-energy and Bethe-Salpeter kernel within the Breuckner formulation of coupled-cluster (BCC) theory. In BCC theory the singles excitation amplitudes are eliminated by construction.  This was achieved by taking successive functional derivatives of the BCC Lagrangian with respect to the non-interacting reference Green's function. This approach was motivated by the fact that calculation of the ground state energy at any level of approximation for the coupled-cluster representation of the electronic Green's function~\cite{nooijen1992coupled,nooijen1993coupled} results in a different ground state correlation energy from that obtained at the same level of approximation of the ground state coupled-cluster equations~\cite{nooijen1993coupled,lange2018relation}. However, the underlying general non-hermitian Green's function formalism was not presented and the approach taken was based on the assumption of the diagrammatic method for non-hermitian systems. 

In Ref.~\cite{coveney2025uncovering}, Coveney also demonstrated the emergence of the complete CCD amplitude equations starting from the electronic self-energy. This work uncovered the set of electronic self-energy diagrams relevant to ground state CCD theory and subsequently how the IP/EA-EOM-CCD approximation can be recovered. The relationships presented in Ref.~\cite{coveney2025uncovering} are related to but distinct from the diagrammatic coupled-cluster self-energy of the similarity transformed Hamiltonian. This is because the relevant analysis in Ref.~\cite{coveney2025uncovering} decouples the occupied and virtual sectors of the electronic self-energy leading to a `non-Dyson approximation'. Therefore, the self-energy that gives rise to the CCD amplitude equations cannot be derived from a functional-diagrammatic expansion with respect to the single-particle coupled-cluster Green's function.

In this work, we generalize the conventional formulation of Green's function theory, based on hermitian two-body (and recently three-body~\cite{carbone2013self}) interactions, to include up to $N$-body non-hermitian interactions. We present the diagrammatic theory of the self-energy generated by a non-hermitian $N$-body interaction, focusing specifically on the coupled-cluster similarity transformed Hamiltonian. Our non-hermitian Green's function formalism provides a rigorous formulation and extension of the results presented in Ref.~\cite{coveney2023coupled} to coupled-cluster theory formulated about an arbitrary reference state. This formalism unifies and combines the techniques of CC and Green's function theory, thereby furthering the conceptual understanding of ground and excited state many-body correlation in the context of condensed matter and nuclear physics. Our approach is based on taking the coupled-cluster similarity transformed Hamiltonian as the fundamental interaction Hamiltonian. This change of perspective unveils the derivation of the resulting theory of the self-energy and single-particle Green's function that naturally arise within the context of coupled-cluster theory.

The diagrammatic content of the irreducible self-energy depends on whether it is  expanded with respect to the non-interacting or exact Green's function. Expansion of the electronic self-energy with respect to the exact Green's function gives rise to self-consistent renormalization and leads to the GF2 and $GW$ approximations~\cite{phillips2014communication,neuhauser2017stochastic,kananenka2016efficient,reining2018gw,backhouse2020efficient,backhouse2020wave,pokhilko2021evaluation,hedin1965new,scott2023moment}. In the case of a two-body interaction, the interaction lines/vertices appearing in irreducible self-energy diagrams are `bare', exhibiting no dependence on the form of the Green's function itself. While it is true that Hedin's equations introduce the screened interaction $W[G]$, which is also a functional of the single-particle Green's function, this is a choice that is not fundamentally necessary to construct the self-consistent self-energy expansion. However, in the presence of higher-body interactions, the interaction lines/vertices appearing in the self-energy expansion are fundamentally dependent on which Green's function the self-energy is expanded with respect to. For example, if the self-energy is expanded with respect to the non-interacting Green's function, $G_0$, it is composed of one-particle irreducible diagrams that are constructed from effective interactions that arise when normal-ordering the Hamiltonian with respect to the reference state. However, to construct the self-energy as a functional of the exact Green's function requires effective interactions that arise when normal-ordering the Hamiltonian with respect to the exact ground state. It is a fundamental consequence of the Green's function formalism that interaction vertices of the self-energy are dressed by the Green's function. However, this consideration has been entirely unexplored in the context of non-hermitian interactions and coupled-cluster theory. This analysis is not only crucial to construct the correct renormalized self-energy functional but also to generate the correct diagrams of the Bethe-Salpeter kernel which arise by taking the functional derivative of the renormalized self-energy with respect to the exact Green's function.

This paper is organized as follows. In Section~\ref{sec:key_results}, we present a brief summary of the key results presented in this paper. In Section~\ref{sec:background}, we provide an overview of the conventional Green's function formalism. Here, we touch on the fundamental principles of Green's function theory and considerations that must be made when constructing the irreducible self-energy. In Section~\ref{sec:bio_qm}, we review and extend the biorthogonal quantum theory in order to define the pictures of quantum mechanics in the presence of a non-hermitian Hamiltonian. Here, we define the biorthogonal Schrodinger, Heisenberg and Interaction pictures, extending the Gell-Mann and Low theorem to include non-hermitian interactions. In Section~\ref{sec:non_fund}, we define the single-particle Green's function and Dyson equation for a system governed by an $N$-body non-hermitian Hamiltonian. We demonstrate the generalization of effective interactions required to construct the non-hermitian irreducible self-energy as a functional of the exact Green's function. In Section~\ref{sec:cc_overview}, we provide an overview of coupled-cluster theory and the similarity transformed Hamiltonian. This leads us to introduce a novel normal-ordering of the Hamiltonian required for the self-consistent theory of the coupled-cluster self-energy. In Section~\ref{subsec:cc_rep}, we discuss the coupled-cluster representation of the electronic Green's function which leads us to our novel approach to the genuine single-particle coupled-cluster Green's function. This change of perspective allows us to define a functional-diagrammatic formalism that unifies the Green's function formalism and coupled-cluster theory. In Section~\ref{sec:ex_sp_ccgf}, we derive the perturbative expansion of the single-particle coupled-cluster Green's function, thereby defining the irreducible coupled-cluster self-energy. In Section~\ref{sec:pert}, we derive the diagrammatic perturbation expansion of the irreducible coupled-cluster self-energy, demonstrating its relationship to the conventional electronic self-energy. In Section~\ref{sec:cc_se}, we use the biorthogonal theory developed in Section~\ref{sec:bio_qm} to construct the non-perturbative renormalized coupled-cluster self-energy from the exact equation-of-motion of the single-particle coupled-cluster Green's function. Here, we describe how the perturbative expansion is recovered from the self-consistent renormalization procedure. In Section~\ref{sec:gs_energy}, we demonstrate how the exact coupled-cluster ground state energy can be obtained from the coupled-cluster self-energy. In Section~\ref{sec:adc}, we use the spectral representation of the coupled-cluster self-energy to derive the associated coupled-cluster Dyson supermatrix. Here, we also introduce several different approximations for the coupled-cluster self-energy that may be readily implemented. In Section~\ref{sec:comp}, we demonstrate the relationship between the coupled-cluster self-energy, IP/EA-EOM-CCSD and the $G_0W_0$ approximation. This motivates us to introduce CC-$G_0W_0$ theory that combines aspects of Green's function theory and Hedin's equations with coupled-cluster theory. Finally, in Section~\ref{sec:BSE}, we derive the relationship between the coupled-cluster self-energy and the Bethe-Salpeter kernel. We present our conclusions and outlook for future work in Section~\ref{sec:conclusions}.

\begin{table*}[ht]
\centering
\renewcommand{\arraystretch}{3.0}
\begin{tabular}{>{\centering\arraybackslash}m{3cm}|>{\centering\arraybackslash}m{7cm}|>{\centering\arraybackslash}m{8cm}}
\hline
\hline
\textbf{Self-energy} & \textbf{Electronic:} $\Sigma[G]$ & \textbf{Coupled-cluster (This work):} $\tilde{\Sigma}[\tilde{G}]$ \\
\hline
Static & 
$\begin{gathered}
\begin{fmfgraph*}(35,35)
    \fmfset{arrow_len}{3mm}
    \fmfleft{i1}
    \fmfright{o1}
    \fmf{wiggly}{o1,i1}
    \fmf{double_arrow}{o1,o1}
    \fmfdot{i1,o1}
\end{fmfgraph*}
\end{gathered}\hspace{12.5mm}-\hspace{5mm}
\begin{gathered}
\begin{fmfgraph*}(35,35)
    \fmfset{arrow_len}{3mm}
    \fmfleft{i1}
    \fmfright{o1}
    \fmf{wiggly}{o1,i1}
    \fmf{fermion}{o1,o1}
    \fmfdot{i1,o1}
\end{fmfgraph*}
\end{gathered}$
& $\begin{gathered}
\begin{fmfgraph*}(40,40)
    \fmfset{arrow_len}{3mm}
    \fmfleft{i1}
    \fmfright{o1}
    \fmf{zigzag}{o1,i1}
    \fmfv{decor.shape=cross,decor.filled=full, decor.size=1.5thic}{o1}
    \fmfdot{i1}
\end{fmfgraph*}
\end{gathered}$ \\ 
\hline
\vspace{5mm} 
Second-order \vspace{5mm} & $\begin{gathered}\begin{fmfgraph*}(40,40)
    \fmfcurved
    \fmfset{arrow_len}{3mm}
    \fmfleft{i1,i2}
    \fmflabel{}{i1}
    \fmflabel{}{i2}
    \fmfright{o1,o2}
    \fmflabel{}{o1}
    \fmflabel{}{o2}
    \fmf{double_arrow}{i1,i2}
    \fmf{wiggly}{o1,i1}
    \fmf{double_arrow,left=0.2,tension=0}{o1,o2}
    \fmf{wiggly}{o2,i2}
    \fmf{double_arrow,left=0.2,tension=0}{o2,o1}
    \fmfdot{o1,o2,i1,i2}
\end{fmfgraph*}
\end{gathered}$ \vspace{0.5mm} & $\begin{gathered}\begin{fmfgraph*}(40,40)
    \fmfcurved
    \fmfset{arrow_len}{3mm}
    \fmfleft{i1,i2}
    \fmflabel{}{i1}
    \fmflabel{}{i2}
    \fmfright{o1,o2}
    \fmflabel{}{o1}
    \fmflabel{}{o2}
    \fmf{double_arrow}{i1,i2}
    \fmf{dbl_zigzag}{o1,i1}
    \fmf{double_arrow,left=0.2,tension=0}{o1,o2}
    \fmf{dbl_zigzag}{o2,i2}
    \fmf{double_arrow,left=0.2,tension=0}{o2,o1}
    \fmfdot{o1,o2,i1,i2}
\end{fmfgraph*}
\end{gathered}$ \vspace{0.5mm}
\\ 
\hline
\vspace{7.5mm}
Third-order (2p1h/2h1p) \vspace{5mm} & $ \begin{gathered}
    \begin{fmfgraph*}(30,50)
    \fmfcurved
    \fmfset{arrow_len}{3mm}
    \fmfleft{i1,i2,i3}
    \fmflabel{}{i1}
    \fmflabel{}{i2}
    \fmfright{o1,o2,o3}
    \fmflabel{}{o1}
    \fmflabel{}{o2}
    \fmf{wiggly}{i2,o2}
    \fmf{wiggly}{i3,o3}
    \fmf{wiggly}{i1,o1}
    \fmf{double_arrow}{i1,i2}
    \fmf{double_arrow}{i2,i3}
    \fmf{double_arrow}{o1,o2}
    \fmf{double_arrow}{o2,o3}
    \fmf{double_arrow,left=0.3}{o3,o1}
    \fmfforce{(0.0w,0.0h)}{i1}
    \fmfforce{(1.0w,0.0h)}{o1}
    \fmfforce{(0.0w,1.0h)}{i3}
    \fmfforce{(1.0w,1.0h)}{o3}
    \fmfdot{i1,i2,i3}
    \fmfdot{o1,o2,o3}
\end{fmfgraph*}
\end{gathered}
\hspace{7.5mm}+\hspace{5mm}
\begin{gathered}
\begin{fmfgraph*}(30,50)
    \fmfcurved
    \fmfset{arrow_len}{3mm}
    \fmfleft{i1,i2,i3}
    \fmflabel{}{i1}
    \fmflabel{}{i2}
    \fmfright{o1,o2,o3}
    \fmflabel{}{o1}
    \fmflabel{}{o2}
    \fmf{wiggly}{i1,v1}
    \fmf{wiggly}{v1,o1}
    \fmf{wiggly}{v2,o2}
    \fmf{wiggly}{i3,v3}
    \fmf{double_arrow}{i1,i3}
    \fmf{double_arrow,left=0.3}{o1,o2}
    \fmf{double_arrow,left=0.3}{o2,o1}
    \fmf{double_arrow,left=0.3}{v2,v3}
    \fmf{double_arrow,left=0.3}{v3,v2}
    \fmfforce{(0.0w,0.0h)}{i1}
    \fmfforce{(1.0w,0.0h)}{o1}
    \fmfforce{(0.5w,0.5h)}{v2}
    \fmfforce{(0.5w,0.0h)}{v1}
    \fmfforce{(0.5w,1.0h)}{v3}
    \fmfforce{(0.0w,1.0h)}{i3}
    \fmfforce{(1.0w,1.0h)}{o3}
    \fmfdot{v2,v3}
    \fmfdot{i1,i3}
    \fmfdot{o1,o2}
\end{fmfgraph*}
\end{gathered}$   \vspace{0.5mm} & $\begin{gathered}
    \begin{fmfgraph*}(30,50)
    \fmfcurved
    \fmfset{arrow_len}{3mm}
    \fmfleft{i1,i2,i3}
    \fmflabel{}{i1}
    \fmflabel{}{i2}
    \fmfright{o1,o2,o3}
    \fmflabel{}{o1}
    \fmflabel{}{o2}
    \fmf{dbl_zigzag}{i2,o2}
    \fmf{dbl_zigzag}{i3,o3}
    \fmf{dbl_zigzag}{i1,o1}
    \fmf{double_arrow}{i1,i2}
    \fmf{double_arrow}{i2,i3}
    \fmf{double_arrow}{o1,o2}
    \fmf{double_arrow}{o2,o3}
    \fmf{double_arrow,left=0.3}{o3,o1}
    \fmfforce{(0.0w,0.0h)}{i1}
    \fmfforce{(1.0w,0.0h)}{o1}
    \fmfforce{(0.0w,1.0h)}{i3}
    \fmfforce{(1.0w,1.0h)}{o3}
    \fmfdot{i1,i2,i3}
    \fmfdot{o1,o2,o3}
\end{fmfgraph*}
\end{gathered}
\hspace{5mm}+\hspace{5mm}
\begin{gathered}
\begin{fmfgraph*}(30,50)
    \fmfcurved
    \fmfset{arrow_len}{3mm}
    \fmfleft{i1,i2,i3}
    \fmflabel{}{i1}
    \fmflabel{}{i2}
    \fmfright{o1,o2,o3}
    \fmflabel{}{o1}
    \fmflabel{}{o2}
    \fmf{dbl_zigzag}{i1,v1}
    \fmf{dbl_zigzag}{v1,o1}
    \fmf{dbl_zigzag}{v2,o2}
    \fmf{dbl_zigzag}{i3,v3}
    \fmf{double_arrow}{i1,i3}
    \fmf{double_arrow,left=0.3}{o1,o2}
    \fmf{double_arrow,left=0.3}{o2,o1}
    \fmf{double_arrow,left=0.3}{v2,v3}
    \fmf{double_arrow,left=0.3}{v3,v2}
    \fmfforce{(0.0w,0.0h)}{i1}
    \fmfforce{(1.0w,0.0h)}{o1}
    \fmfforce{(0.5w,0.5h)}{v2}
    \fmfforce{(0.5w,0.0h)}{v1}
    \fmfforce{(0.5w,1.0h)}{v3}
    \fmfforce{(0.0w,1.0h)}{i3}
    \fmfforce{(1.0w,1.0h)}{o3}
    \fmfdot{v2,v3}
    \fmfdot{i1,i3}
    \fmfdot{o1,o2}
\end{fmfgraph*}
\end{gathered}\hspace{5mm}+\hspace{5mm}
\begin{gathered}
    \begin{fmfgraph*}(30,50)
    \fmfcurved
    \fmfset{arrow_len}{3mm}
    \fmfleft{i1,i2,i3}
    \fmflabel{}{i1}
    \fmflabel{}{i2}
    \fmfright{o1,o2,o3}
    \fmflabel{}{o1}
    \fmflabel{}{o2}
    \fmf{dbl_zigzag}{i1,v1}
    \fmf{dbl_dashes}{i2,v2}
    \fmf{dbl_dashes}{v2,o2}
    \fmf{dbl_zigzag}{i3,v3}
    \fmf{double_arrow}{i1,i2}
    \fmf{double_arrow}{v1,v2}
    \fmf{double_arrow}{o2,v1}
    \fmf{double_arrow}{v2,v3}
    \fmf{double_arrow}{v3,o2}
    \fmf{double_arrow}{i2,i3}
    \fmfforce{(0.0w,0.0h)}{i1}
    \fmfforce{(1.0w,0.0h)}{o1}
    \fmfforce{(0.5w,0.5h)}{v2}
    \fmfforce{(0.5w,0.0h)}{v1}
    \fmfforce{(0.5w,1.0h)}{v3}
    \fmfforce{(0.0w,1.0h)}{i3}
    \fmfforce{(1.0w,1.0h)}{o3}
    \fmfdotn{v}{3}
    \fmfdot{i1,i2,i3}
    \fmfdot{o2}
\end{fmfgraph*}
\end{gathered}$  \vspace{0.5mm} \\
\hline
\hline
\end{tabular}
\caption{Summary of the renormalized electronic and coupled-cluster self-energy diagrams up to third-order. The third-order coupled-cluster self-energy diagrams presented here are restricted to those that display 2-particle,1-hole/2-hole,1-particle excitation character. The complete set of third-order diagrams can be found in Section~\ref{sec:cc_se}.}
\label{tab:cc_se}
\end{table*}

Throughout this work, indices and wavevectors $i\mathbf{k}_i,j\mathbf{k}_j,\cdots$ denote occupied (valence band) spin-orbitals and $a\mathbf{k}_a,b\mathbf{k}_b,\cdots$ will denote virtuals (conduction band). A mixture of latin and greek indices $p\mathbf{k}_p,q\mathbf{k}_q, \alpha\mathbf{k}_{\alpha},\cdots$ is used to denote general spin-orbitals. The wavevector dependence will be omitted for clarity. We will refer to the Green's function and self-energy for the two-body electronic Hamiltonian as the electronic Green's function and the electronic self-energy. The electronic Green's function will be denoted by $G$ and the electronic self-energy will be denoted by $\Sigma$. The Green's function and self-energy for the $N$-body similarity transformed Hamiltonian will be referred to as the single-particle coupled-cluster Green's function and coupled-cluster self-energy. The single-particle coupled-cluster Green's function will be denoted by $\tilde{G}$ and the coupled-cluster self-energy will be denoted by $\tilde{\Sigma}$.

\section{Summary of Key Results}~\label{sec:key_results}

Our goal is to present the diagrammatic theory of the irreducible self-energy and Bethe-Salpter kernel for an $N$-body non-hermitian Hamiltonian. We then directly apply the general formalism presented in this work to the coupled-cluster similarity transformation of the electronic structure Hamiltonian.

For the two-body electronic structure Hamiltonian, at equilibrium and at zero temperature, the single-particle electronic Green's function is defined (in the Heisenberg picture) as
\begin{gather*}
    \begin{split}
        iG_{pq}(t_1,t_2) = \braket{\Psi_0|\mathcal{T}\left\{a_{p}(t_1)a^\dag_{q}(t_2)\right\}|\Psi_0} . 
    \end{split}
\end{gather*}
From its \emph{exact} equation-of-motion, it can be demonstrated that the electronic Green's function obeys the Dyson equation:
\begin{gather}
    \begin{split}~\label{eq:dyson_eq}
        G = G_0 + G_0\Sigma[G]G \ , 
    \end{split}
\end{gather}
where $G_0$ is a reference Green's function which we take within the Hartree--Fock (HF) approximation. $\Sigma[G]$ is the diagrammatic irreducible electronic self-energy expressed as a functional of the exact electronic single-particle Green's function. We have chosen to suppress the time/frequency dependence of Eq.~\ref{eq:dyson_eq} for notational brevity. 

The electronic self-energy, $\Sigma$, is defined diagrammatically as the set of all \emph{one-particle irreducible} diagrams. These diagrams are constructed by the contraction of Green's function lines with the interactions of the underlying many-body Hamiltonian (which in this case is the electronic structure Hamiltonian). The diagrammatic construction of the electronic self-energy yields a formally exact eigenvalue problem for the poles and residues of the Fourier transform of the single-particle electronic Green's function. The poles correspond to the exact addition and removal energies of the system.  

Taking the functional derivative of the electronic self-energy with respect to the exact single-particle electronic Green's function yields the kernel of the electronic Bethe-Salpeter equation (BSE):
\begin{gather}
    \begin{split}~\label{eq:el_bse}
        \Xi[G] = i\frac{\delta\Sigma[G]}{\delta G} \ .
    \end{split}
\end{gather}
Knowledge of the BSE kernel allows for the calculation of the two-particle Green's function and therefore two-particle excitation energies. It is central to understand that the Green's function formalism rests on the functional-diagrammatic relationships contained in Eqs~\ref{eq:dyson_eq} and~\ref{eq:el_bse}. 

\begin{table*}[ht]
\centering
\renewcommand{\arraystretch}{3.0}
\begin{tabular}{>{\centering\arraybackslash}m{3cm}|>{\centering\arraybackslash}m{7.5cm}|>{\centering\arraybackslash}m{7.5cm}}
\hline
\hline
\textbf{GF formalism} & \textbf{Electronic} & \textbf{Coupled-cluster (This work)}  \\
\hline
Energy & 
$E_c = \frac{1}{2}\lim_{\eta\to0}\sum_{pq}\int^{\infty}_{-\infty}\frac{d\omega}{2\pi i} e^{i\eta\omega}\Sigma^{c}_{pq}(\omega)G_{qp}(\omega)$ & $E^{\CC}_c = \frac{1}{2}\sum_{i} \left(\tilde{\Sigma}^{\infty(0)}_{ii}+\sum_{a}h_{ia}t^{a}_{i}\right)$ 
\\
\hline
\vspace{0.5mm} Dyson Supermatrix \vspace{0.25mm} & \vspace{0.5mm}
\(
\mathbf{D}^{\text{el}} =
\left(
\renewcommand{\arraystretch}{1.2}
\begin{array}{ccc}
\mathbf{f}+\mathbf{\Sigma}^{\infty} & \mathbf{U}^\dag & \mathbf{V} \\
\mathbf{U} & \mathbf{K}^{>}+\mathbf{C}^{>} & \mathbf{0} \\
\mathbf{V}^\dag & \mathbf{0} & \mathbf{K}^{<}+\mathbf{C}^{<}
\end{array}
\right)
\) \vspace{0.25mm}
& \vspace{0.5mm}
$\mathbf{\tilde{D}}^{\CC} = \left(\renewcommand{\arraystretch}{1.2}\begin{array}{ccc}
            \mathbf{f}+\mathbf{\tilde{\Sigma}}^{\infty} & \mathbf{\tilde{U}} & \mathbf{\bar{V}} \\
            \mathbf{\bar{U}} & \mathbf{\bar{K}}^{>}+\mathbf{\bar{C}}^{>} & \mathbf{0} \\
             \mathbf{\tilde{V}} & \mathbf{0} & \mathbf{\bar{K}}^{<}+\mathbf{\bar{C}}^{<}
        \end{array}\right)$ \vspace{0.25mm}
\\
\hline
\hline
\end{tabular}
\caption{Summary of quantities obtained from the Green's function (GF) formalism for the electronic structure and coupled-cluster similarity-transformed Hamiltonians.}
\label{tab:quantities}
\end{table*}

In this work, we introduce the inherently non-hermitian single-particle Green's function, irreducible self-energy and BSE kernel for an $N$-body non-hermitian Hamiltonian. Taking the coupled-cluster similarity-transformed Hamiltonian $\bar{H}$ as the $N$-body non-hermitian Hamiltonian of interest, we define the single-particle coupled-cluster Green's function (in the biorthogonal Heisenberg picture defined in Section~\ref{sec:bio_qm}) as 
\begin{equation*}
    i\tilde{G}_{pq}(t_1,t_2) = \braket{\tilde{\Psi}_0|\mathcal{T}\left\{a_{p}(t_1)a^\dag_{q}(t_2)\right\}|\Phi_0} \ ,
\end{equation*}
where the time-dependence of the operators is governed by $\bar{H}$. Here, $\ket{\Phi_0}$ and $\ket{\tilde{\Psi}_0}$ are the right and left eigenstates of $\bar{H}$, respectively. To the best our knowledge, this propagator has not been previously defined and, as discussed in Section~\ref{subsec:cc_rep}, is distinct from the formalism introduced in Refs~\cite{nooijen1992coupled,nooijen1993coupled}.  As demonstrated in Section~\ref{sec:ex_sp_ccgf}, the propagator $\tilde{G}$ gives rise to the formally exact coupled-cluster Dyson equation. This allows us to generate a functional-diagrammatic Green's function formalism within coupled-cluster theory.

From the equation-of-motion for $\tilde{G}$, we show that $\tilde{G}$ obeys the \emph{exact} Dyson equation:
\begin{gather*}
    \begin{split}
        \tilde{G} = G_0 + G_0\tilde{\Sigma}[\tilde{G}]\tilde{G} \ , 
    \end{split}
\end{gather*}
where $\tilde{\Sigma}[\tilde{G}]$ is the diagrammatic irreducible coupled-cluster self-energy expressed as a functional of the exact single-particle coupled-cluster  Green's function. Importantly, $G_0$ is the same reference Green's function appearing in Eq.~\ref{eq:dyson_eq} (the electronic Dyson equation) and we have again suppressed the time/frequency dependence for notational brevity. The poles of the Fourier transform of $\tilde{G}$ are the \emph{exact} addition and removal energies of the system.

In this work we demonstrate that the irreducible coupled-cluster self-energy, $\tilde{\Sigma}$, is defined diagrammatically as the set of all \emph{one-particle irreducible} diagrams resulting from the contraction of the single-particle coupled-cluster Green's function lines with the interactions of the underlying similarity transformed Hamiltonian. In Section~\ref{sec:adc}, we demonstrate that the construction of the coupled-cluster self-energy yields a formally exact eigenvalue problem for the poles and residues of the single-particle coupled-cluster Green's function, $\tilde{G}$.   

The functional derivative of the coupled-cluster self-energy with respect to the exact single-particle CC Green's function yields the coupled-cluster BSE kernel:
\begin{equation*}
    \tilde{\Xi} = i \frac{\delta\tilde{\Sigma}[\tilde{G}]}{\delta\tilde{G}} \ .
\end{equation*} 
Knowledge of the coupled-cluster BSE kernel allows for the calculation of the two-particle coupled-cluster Green's function. The Fourier transform of the exact two-particle CC Green's function contains poles at the \emph{exact} two-particle excitation energies of the system.

In order to demonstrate that our formalism is diagrammatically consistent, we also derive the perturbative expansion of the coupled-cluster self-energy with respect to the reference Green's function, $G_0$. This demonstrates the importance of the emergence of effective interactions that are naturally generated by Green's function theory with three-body or higher interactions. To summarize our key findings, we present the corresponding standard equations of Green's function theory, and their exact generalization to the non-hermitian CC $N$-body Hamiltonian.

In Table~\ref{tab:cc_se}, we depict the diagrammatic content of the self-consistent renormalized self-energy of the electronic and coupled-cluster similarity transformed Hamiltonians, respectively. For the diagrams depicted in the second column of Table~\ref{tab:cc_se}, the double arrows represent the exact electronic Green's function $G$, the single-arrows represent the reference Green's function $G_0$, and the interaction line is the antisymmetrized Coulomb interaction. For the diagrams depicted in the third column, the double arrows represent the exact single-particle coupled-cluster Green's function $\tilde{G}$ and the interaction lines are now the one-body, two-body and three-body coupled-cluster effective interactions, which are all functionals of $\tilde{G}$.

The static component of the standard hermitian electronic self-energy is given by the diagrams contained in the second column--second row of Table~\ref{tab:cc_se}. 
The two diagrams evaluate to~\cite{schirmer2018many,stefanucci2013nonequilibrium}
\begin{gather*}
\begin{split}
\Sigma^{\infty}_{pq}[G]&=-i\sum_{rs}\braket{pr||qs}\Big[G_{sr}(t-t^+)-G^{0}_{sr}(t-t^+)\Big]\\
&=\sum_{rs}\braket{pr||qs}(\gamma_{rs}-\gamma^{\rf}_{rs}) \ ,
\end{split}
\end{gather*}
where $\gamma_{rs}/\gamma^{\rf}_{rs}$ are the exact/reference electronic one-body reduced density matrices. Importantly, the two-body coulomb interaction is not a functional of the single-particle electronic Green's function~\cite{coveney2023coupled}. 

\begin{table*}[ht]
\centering
\renewcommand{\arraystretch}{3.0}
\begin{tabular}{>{\centering\arraybackslash}m{3cm}|>{\centering\arraybackslash}m{7.5cm}|>{\centering\arraybackslash}m{7.5cm}}
\hline
\hline
\textbf{BSE kernel} & \textbf{Electronic:} $\Xi[G]=i\frac{\delta\Sigma[G]}{\delta G}$ & \textbf{Coupled-cluster (This work):} $\tilde{\Xi}[\tilde{G}]=i\frac{\delta\tilde{\Sigma}[\tilde{G}]}{\delta \tilde{G}}$ \\
\hline
Static & 
$\begin{gathered}
\begin{fmfgraph*}(40,40)
    \fmfset{arrow_len}{3mm}
    \fmfleft{i1}
    \fmfright{o1}
    \fmf{wiggly}{o1,i1}
    \fmfdot{i1,o1}
\end{fmfgraph*}
\end{gathered}$
& 
$\begin{gathered}
\begin{fmfgraph*}(40,40)
    \fmfset{arrow_len}{3mm}
    \fmfleft{i1}
    \fmfright{o1}
    \fmf{dbl_zigzag}{o1,i1}
    \fmfdot{i1,o1}
\end{fmfgraph*}
\end{gathered}$
\\
\hline
\vspace{5mm}
Second-order (Two-body only) \vspace{5mm} &
$\begin{gathered}
\begin{fmfgraph*}(40,40)
    \fmfcurved
    \fmfset{arrow_len}{3mm}
    \fmfleft{i1,i2} \fmfright{o1,o2}
    \fmf{double_arrow}{i1,i2}
    \fmf{wiggly}{o1,i1} \fmf{wiggly}{o2,i2}
    \fmf{double_arrow,tension=0}{o2,o1}
    \fmfdot{o1,o2,i1,i2}
\end{fmfgraph*}
\end{gathered}
\hspace{2.5mm}+\hspace{2.5mm}
\begin{gathered}
\begin{fmfgraph*}(40,40)
    \fmfcurved
    \fmfset{arrow_len}{3mm}
    \fmfleft{i1,i2} \fmfright{o1,o2}
    \fmf{double_arrow}{i1,i2}
    \fmf{wiggly}{o1,i1} \fmf{wiggly}{o2,i2}
    \fmf{double_arrow,tension=0}{o1,o2}
    \fmfdot{o1,o2,i1,i2}
\end{fmfgraph*}
\end{gathered}
\hspace{2.5mm}+\hspace{2.5mm}
\begin{gathered}
\begin{fmfgraph*}(40,40)
    \fmfcurved
    \fmfset{arrow_len}{3mm}
    \fmfleft{i1,i2} \fmfright{o1,o2}
    \fmf{wiggly}{o1,i1}
    \fmf{double_arrow,left=0.2,tension=0}{o1,o2}
    \fmf{wiggly}{o2,i2}
    \fmf{double_arrow,left=0.2,tension=0}{o2,o1}
    \fmfdot{o1,o2,i1,i2}
\end{fmfgraph*}
\end{gathered}$ \vspace{2.5mm}
&
$\begin{gathered}
\begin{fmfgraph*}(40,40)
    \fmfcurved
    \fmfset{arrow_len}{3mm}
    \fmfleft{i1,i2}
    \fmflabel{}{i1}
    \fmflabel{}{i2}
    \fmfright{o1,o2}
    \fmflabel{}{o1}
    \fmflabel{}{o2}
    \fmf{double_arrow}{i1,i2}
    \fmf{dbl_zigzag}{o1,i1}
    \fmf{dbl_zigzag}{o2,i2}
    \fmf{double_arrow,tension=0}{o2,o1}
    \fmfdot{o1,o2,i1,i2}
\end{fmfgraph*}
\end{gathered} \hspace{2.5mm}+\hspace{2.5mm} 
\begin{gathered}
\begin{fmfgraph*}(40,40)
    \fmfcurved
    \fmfset{arrow_len}{3mm}
    \fmfleft{i1,i2}
    \fmflabel{}{i1}
    \fmflabel{}{i2}
    \fmfright{o1,o2}
    \fmflabel{}{o1}
    \fmflabel{}{o2}
    \fmf{double_arrow}{i1,i2}
    \fmf{dbl_zigzag}{o1,i1}
    \fmf{dbl_zigzag}{o2,i2}
    \fmf{double_arrow,tension=0}{o1,o2}
    \fmfdot{o1,o2,i1,i2}
\end{fmfgraph*}
\end{gathered} \hspace{3.5mm}+\hspace{2.5mm}
\begin{gathered}
\begin{fmfgraph*}(40,40)
    \fmfcurved
    \fmfset{arrow_len}{3mm}
    \fmfleft{i1,i2}
    \fmflabel{}{i1}
    \fmflabel{}{i2}
    \fmfright{o1,o2}
    \fmflabel{}{o1}
    \fmflabel{}{o2}
    \fmf{dbl_zigzag}{o1,i1}
    \fmf{double_arrow,left=0.2,tension=0}{o1,o2}
    \fmf{dbl_zigzag}{o2,i2}
    \fmf{double_arrow,left=0.2,tension=0}{o2,o1}
    \fmfdot{o1,o2,i1,i2}
\end{fmfgraph*}
\end{gathered}$ \vspace{2.5mm} \\
\hline
\hline
\end{tabular}
\caption{Summary of the renormalized electronic and coupled-cluster BSE kernel diagrams up to second-order. The second-order diagrams of the CC-BSE kernel depicted here contain only those with two-body effective interactions. The complete set of second-order CC-BSE kernel diagrams can be found in Section~\ref{sec:BSE}.}
\label{tab:bse}
\end{table*}
The exact static component of the non-hermitian coupled-cluster self-energy is depicted in the diagram of the third column--second row of Table~\ref{tab:cc_se} and evaluates to 
\begin{gather*}
\begin{split}
\tilde{\Sigma}^{\infty}_{pq}[\tilde{G}] &=\hspace{5mm} 
\begin{gathered}
\begin{fmfgraph*}(40,40)
    \fmfset{arrow_len}{3mm}
    \fmfleft{i1}
    \fmfright{o1}
    \fmf{zigzag}{o1,i1}
    \fmfv{decor.shape=cross,decor.filled=full, decor.size=1.5thic}{o1}
    \fmfdot{i1}
    \fmflabel{$\substack{p\\q}$}{i1}
\end{fmfgraph*}
\end{gathered}\hspace{2.5mm}=\tilde{F}_{pq}[\tilde{G}]-f_{pq} \ ,
\end{split}
\end{gather*}
where $\tilde{F}_{pq}$ is the irreducible one-body effective interaction arising from the coupled-cluster similarity transformation, which is explicitly a functional of the single-particle coupled-cluster Green's function, $\tilde{G}$. We subsequently demonstrate that this diagram reduces to the static electronic self-energy when the similarity transformation of the electronic structure Hamiltonian, $\bar{H}=e^{-T}He^{T}$, is removed ($T=0$). This term was first derived in Ref.~\cite{coveney2023coupled} within BCC theory by taking the functional derivative of the Lagrangian with respect to the non-interacting reference Green's function.

The standard second-order electronic self-energy is given by the diagram in the second column--third row of Table~\ref{tab:cc_se}. In the case of coupled-cluster theory, we find the second-order contribution in the third column--third row of Table~\ref{tab:cc_se}. In the CC self-energy diagram, the interaction is now the two-body coupled-cluster effective interaction $\tilde{\Xi}_{pq,rs}[\tilde{G}]$, which is also a functional of $\tilde{G}$. This diagram is derived in Section~\ref{sec:cc_se} and again reduces to the second-order electronic self-energy when the similarity transformation of the electronic structure Hamiltonian is neglected. Three-body and higher diagrams do not contribute at second-order as they evaluate to zero due to the coupled-cluster amplitude equations.

The standard third-order diagrams of the electronic self-energy are given in the second column--final row of Table~\ref{tab:cc_se}. For the coupled-cluster self-energy, third column--final row of Table~\ref{tab:cc_se}, we see the appearance of many more diagrams at third-order which demonstrate the presence of two and three-body effective interactions (introduced by the coupled-cluster similarity transformation) which are themselves also functionals of $\tilde{G}$. These terms are defined in Section~\ref{sec:cc_se}, where the three-body interaction, $\tilde{\chi}_{pqr,stu}[\tilde{G}]$, is also defined.

The first two diagrams of the third-order CC self-energy are topologically identical to the electronic case, containing correlation effects only due to the two-body effective interaction. However, we also have a special two-particle,one-hole/two-hole,one-particle diagram containing the three-body effective interaction. The complete set of third-order CC self-energy diagrams is presented in Section~\ref{sec:cc_se}.

In Table~\ref{tab:quantities}, we collect the standard equations of the Green's function formalism for the electronic and coupled-cluster similarity-transformed Hamiltonians, respectively. Within Green's function theory, the ground state electronic correlation energy is obtained from the electronic Green's function and self-energy through the Galitskii-Migdal formula~\cite{galitskii1958application}. This formula is exact because the electronic structure Hamiltonian is a two-body operator. In coupled-cluster theory, the ground state correlation energy is found directly from the first-order contribution to the static self-energy
as derived in Section~\ref{sec:gs_energy}, and is equivalent to the Galitskii-Migdal formula for the electronic correlation energy for a two-body interaction. 

From the spectral representation of the exact coupled-cluster self-energy, in Section~\ref{sec:adc} we derive the coupled-cluster Dyson supermatrix~\cite{coveney2023coupled}. The CC Dyson supermatrix can be found in column three--row three of Table~\ref{tab:quantities}. 
The upper left block, $\mathbf{f}+\mathbf{\tilde{\Sigma}}^{\infty}$, is defined over the full set of occupied and virtual spin-orbitals and the off-diagonal `coupling matrices' $\mathbf{\tilde{U}}/\mathbf{\bar{U}} $, $\mathbf{\bar{V}}/\mathbf{\tilde{V}}$ couple the single-particle states to many-particle excited state configurations. The $\mathbf{\bar{K}}^{\lessgtr}+\mathbf{\bar{C}}^{\lessgtr}$ matrices contain the interactions between the different excited state configurations mediated by the similarity-transformed Hamiltonian. The CC Dyson supermatrix is to be contrasted with the electronic Dyson supermatrix $\mathbf{D}^{\text{el}}$, written in column two--row three of Table~\ref{tab:quantities}, which is manifestly hermitian. The eigenvalues of the exact electronic and exact coupled-cluster Dyson supermatrices both yield the exact ionization and electron affinities of the system. Using the CC Dyson supermatrix representation, in Section~\ref{sec:adc} we also derive different approximations for the coupled-cluster self-energy, restricted to the space of two-particle--one-hole (two-hole--one-particle) excited state configurations, that preserve the correct analytic pole-structure of the Green's function and self-energy.

We arrive at CC-$G_0W_0$ theory by employing the ground-state ring CCD (rCCD) approximation~\cite{scuseria2008ground} while truncating the interaction matrices to consist of 2p1h/2h1p excitations:
    \begin{gather*}
    \begin{split}
        &\mathbf{\tilde{D}}^{\text{CC-$G_0W_0$}} = \left(\begin{array}{ccc}
            f_{pq} + \tilde{\Sigma}^{\infty{\rCCD}}_{pq} & \chi^{\rCCD}_{pi,ab} &\chi^{\rCCD}_{pa,ij} \\
            \chi^{\rCCD}_{ab,qi} & \mathbf{\bar{\Lambda}}^{G_0W_0>}_{iab,jcd} & \mathbf{0} \\
             \chi^{\rCCD}_{ij,qa}& \mathbf{0} & \mathbf{\bar{\Lambda}}^{G_0W_0<}_{ija,klb}
        \end{array}\right) \ .
    \end{split}
\end{gather*}
Here $\tilde{\Sigma}^{\infty{\rCCD}}_{pq}$ is the static component of the coupled-cluster self-energy evaluated in the rCCD approximation and $\chi^{\rCCD}_{pq,rs}$ is the rCCD two-body interaction that arises when normal-ordering the similarity transformed Hamiltonian with respect to the reference determinant. The interaction matrices are given by $\mathbf{\bar{\Lambda}}^{G_0W_0>}_{iab,jcd}=(\epsilon_{a}\delta_{bd}\delta_{ij}+H^{\dRPA}_{ib,jd})\delta_{ac}$ and $\mathbf{\bar{\Lambda}}^{G_0W_0>}_{ija,klb}=(\epsilon_{i}\delta_{jl}\delta_{ab}-H^{\dRPA}_{ja,lb})\delta_{ik}$, respectively, where $H^{\dRPA}$ is the RPA Hamiltonian. All matrix elements are defined in Section~\ref{sec:adc}, where it is also shown that the $\mathbf{\tilde{D}}^{\text{CC-$G_0W_0$}}$ supermatrix provides an infinite-order summation of coupled-cluster self-energy bubble diagrams. 

In Table~\ref{tab:bse}, we present the diagrammatic content of the electronic and coupled-cluster BSE kernels, respectively. The diagrammatic content of the electronic BSE kernel is given in the second column of Table~\ref{tab:bse}. 
The coupled-cluster BSE kernel (up to second-order) is also given in the third column of Table~\ref{tab:bse}. Within coupled-cluster theory, derivation of the BSE kernel is significantly more complicated than for the electronic case as the interaction lines are also functionals of $\tilde{G}$. Therefore, one must take functional derivatives of both interaction and Green's function lines of the self-energy to obtain the correct expression for the BSE kernel. The full diagrammatic content of the CC BSE kernel at second-order, including three- and higher-body effects, is given in Section~\ref{sec:BSE}. 

\section{Review of Conventional Green's function Theory}~\label{sec:background}

In this section, we present an overview of the conventional formulation of many-body quantum field theory where the Hamiltonian is hermitian. This is necessary in order to provide a complete description of the forthcoming non-hermitian theory where the time-dependence of states and operators must be carefully addressed.

\subsection{The pictures of Quantum Mechanics} 
Let us assume that we have a hermitian Hamiltonian $H$, that contains no explicit time-dependence. The Schrödinger equation is therefore
\begin{equation}
    i\frac{\partial}{\partial t}\ket{\Psi_S(t)} = H\ket{\Psi_S(t)} \ .
\end{equation}
Given an initial state $\ket{\Psi_{S}(0)}$, we write the formal solution in terms of the time-evolution operator as  
\begin{equation}
    \ket{\Psi_S(t)} = \exp(-iHt)\ket{\Psi_S(0)} \ .
\end{equation}
In the Schrödinger representation of quantum mechanics, all operators are time-independent and the time-dependence enters via the state vector, $\ket{\Psi_S(t)}$. Alternatively, we can use the Heisenberg picture where we instead define time-dependent operators and stationary state vectors. This formulation corresponds more intuitively to the classical picture where observables such as position and momentum vary with time. The state-vector in the Heisenberg representation is time-independent and given by $\ket{\Psi_{H}} = \ket{\Psi_S(0)}$. Heisenberg operators are related to Schrödinger operators via~\cite{Quantum,mahan2000many} 
\begin{equation}
    O_H(t) = \exp(iHt)O_S\exp(-iHt) \ .
\end{equation}
Finally, we have the Interaction picture where we assume that the Hamiltonian may be written as a sum of two terms $H = H_0 + H_1$, where $H_0$ is the reference Hamiltonian. The interaction state vector is written as 
\begin{equation}
    \ket{\Psi_I(t)} = \exp(iH_0t)\ket{\Psi_S(t)} \ .
\end{equation}
This time-dependent unitary transformation results in the Interaction state vector obeying the Schrödinger equation
\begin{equation}
    i\frac{\partial }{\partial t}\ket{\Psi_I(t)} = H_1(t)\ket{\Psi_{I}(t)}
\end{equation}
where $H_1(t) = \exp(iH_0t)O_S\exp(-iH_0t)$. The formal solution of the Interaction state vector Schrödinger equation is given by~\cite{Quantum}
\begin{gather}~\label{eq:uni_time_ev}
\begin{split}
    \ket{\Psi_I(t)} &= U(t,t_0)\ket{\Psi_I(t_0)} \\
    &= \mathcal{T}\left\{\exp\left(-i\int^{t}_{t_0}dt'\ H_1(t')\right)\right\}\ket{\Psi_I(t_0)} 
\end{split}
\end{gather}
where $\mathcal{T}$ represents the time-ordering operator. Clearly we have $\ket{\Psi_H} = \ket{\Psi_I(0)}$, with the relationship between Heisenberg and Interaction picture operators as 
\begin{equation}
    O_H(t) = U(0,t)O_I(t)U(t,0) \ ,
\end{equation}
where we have used the unitarity of the time-evolution operator. All three pictures coincide at $t=0$. 

\subsection{The ground state in quantum field theory: Gell-Mann and Low theorem }

 In quantum field theory, where the particle number is no longer a `good quantum number', the Gell-Mann and Low (GML) theorem provides the central key to find the ground state and construct propagators perturbatively. In order to derive a consistent Green's function theory for non-hermitian interactions we must understand the implicit assumptions contained in the GML theorem for hermitian systems. 
 
 To derive the GML theorem requires the introduction of the explicitly time-dependent adiabatic `switching-on' Hamiltonian as
\begin{equation}
    H(t) = H_0 + e^{-\eta|t|}H_1 , 
\end{equation}
where $\eta$ is a positive infinitesimal. At large times, this Hamiltonian reduces to the reference and at $t=0$ it corresponds to the full Hamiltonian of our system. Using the results on the relationships between state vectors in the pictures of quantum mechanics, we immediately write the Heisenberg state vector as
\begin{equation}~\label{eq:GM+low}
    \ket{\Psi_H} = U_{\eta}(0,t_0)\ket{\Psi_{I}(t_0)} \ ,
\end{equation}
where $U_{\eta}$ is given by replacing $H_1(t')$ in Eq.~\ref{eq:uni_time_ev} with $e^{-\eta|t'|}H_1(t')$.
Letting the time $t_0$ approach $-\infty$, while taking the physical limit as $\eta\to0$ of Eq.~\ref{eq:GM+low}, leads immediately to the GML theorem on the ground state of a quantum field theory~\cite{Quantum,gell1954quantum}. The theorem states that if the quantity 
\begin{equation}~\label{eq:GM_low}
    \frac{\ket{\Psi_0}}{\braket{\Phi_0|\Psi_0}} = \lim_{\eta\to0} \frac{U_{\eta}(0,\pm\infty)\ket{\Phi_0}}{\braket{\Phi_0|U_{\eta}(0,\pm\infty)|\Phi_0}}
\end{equation}
converges, then it is an eigenstate of the full Hamiltonian such that 
\begin{equation}
    H\frac{\ket{\Psi_0}}{\braket{\Phi_0|\Psi_0}} = E\frac{\ket{\Psi_0}}{\braket{\Phi_0|\Psi_0}} \ .
\end{equation}
In this way an eigenstate of the full Hamiltonian is generated adiabatically from $\ket{\Phi_0}$ as the interaction is switched on. If $\ket{\Phi_0}$ is the ground state of the reference Hamiltonian, then the eigenstate generated from this procedure is usually the true ground state. However, this is not always the case. Importantly, the limits of the numerator and denominator as $\eta\to0$ in Eq.~\ref{eq:GM_low} do not exist separately~\cite{gell1954quantum}. The theorem only asserts that if the limit of their ratio exists, then the eigenstate is well defined and is an eigenstate of the full Hamiltonian.

\subsection{The single-particle Green's function and irreducible self-energy}
At equilibrium and at zero temperature, the single-particle Green's function is defined as 
\begin{equation}
    iG_{pq}(t_1-t_2) = \frac{\braket{\Psi_0|\mathcal{T}\left\{a_{p}(t_1)a^\dag_{q}(t_2)\right\}|\Psi_0}}{\braket{\Psi_0|\Psi_0}} ,
\end{equation}
where $a^\dag_p(t)/a_p(t)$ create/annihilate electrons in spin-orbital $\ket{\phi_p}$ in the Heisenberg representation, $\mathcal{T}$ is the time-ordering operator and $\ket{\Psi_0}$ is the exact ground state. In general, the single-particle Green's function contains information on the following observables:
\begin{enumerate}
    \item The exact single-particle excitation spectrum.
    \item The ground state expectation value of any \emph{single-particle} operator. 
    \item The exact ground-state energy (in the case of a two-body interaction).  
\end{enumerate}
The single-particle Green's function is related to the one-body reduced density matrix by 
\begin{gather}
    \begin{split}~\label{eq:gf_rdm}
        \gamma_{pq} = -iG_{qp}(t_1-t_1^+) = \braket{\Psi_0|a^\dag_pa_q|\Psi_0} \ ,
    \end{split}
\end{gather}
where the notation $t_1^+=\lim_{\eta\to0^+}t_1+\eta$ indicates evaluation of time argument from above. 

Using the GML theorem (Eq.~\ref{eq:GM_low}), we construct the perturbative expansion for the Green's function as
    \begin{gather}~\label{eq:GF_PT}
\begin{split}
     iG_{pq}(t_1-t_2) =\lim_{\eta\to0}\braket{\Phi_0|\mathcal{T}\left\{S_{\eta}a_{p}(t_1)a^\dag_q(t_2)\right\}|\Phi_0}_{\mathcal{C}}
\end{split}
\end{gather}
where $S_{\eta}=\mathcal{T}\left\{\exp\left(-i\int^{\infty}_{-\infty}dt e^{-\eta|t|}H_1(t)\right)\right\}$ and the subscript $\mathcal{C}$ corresponds to retaining only the connected diagrams in the expansion using Goldstone's theorem~\cite{Quantum}. 

The number of diagrams (terms) generated by Eq.~\ref{eq:GF_PT} at each order is reduced by collecting all \emph{one-particle irreducible} (1PI) diagrams in the expansion. 1PI diagrams are diagrams that cannot be divided into different sub-diagrams by cutting a single propagator line and their complete set \emph{defines} the irreducible self-energy kernel, $\Sigma_{pq}(\omega)$. All reducible diagrams are then generated by the geometric Dyson series
\begin{equation}~\label{eq:dyson}
    G_{pq}(\omega) = G^0_{pq}(\omega) + \sum_{rs}G^{0}_{pr}(\omega)\Sigma_{rs}(\omega) G_{sq}(\omega) \ .
\end{equation}
Here, $G^{0}_{pq}(\omega)$ is the `non-interacting' propagator that results from independent-particle propagation corresponding to time evolution under the action of the one-body reference operator $H_0$. The Dyson equation allows us to move away from direct Green's function perturbation theory and instead focus on the reduced set of diagrams that constitute the self-energy. In general, the self-energy is composed of a static component and a dynamical contribution
\begin{gather}
    \begin{split}
        \Sigma_{pq}(\omega) = \Sigma^{\infty}_{pq} + \Sigma^{\F}_{pq}(\omega) + \Sigma^{\Ba}_{pq}(\omega)
    \end{split}
\end{gather}
where $\Sigma^{\infty}_{pq}$ is the static component and $\Sigma^{\F}_{pq}(\omega)/\Sigma^{\Ba}_{pq}(\omega)$ are the forward/backward time dynamical contributions. In this work, we choose to define the static component of the self-energy with respect to the Fock operator as we define the non-interacting Green's function, $G_0$, within the Hartree--Fock approximation. 

The set of diagrams that constitute the irreducible self-energy may be expressed with respect to the reference Green's function $G_0$, or the exact Green's function, $G$. If the self-energy is expanded in terms of the exact Green's function, this restricts the self-energy diagrams to be only those which are 1PI, skeleton and interaction-irreducible. Skeleton diagrams are self-energy diagrams that cannot be separated into two pieces by cutting a single fermion line twice at two different points~\cite{stefanucci2013nonequilibrium}. Interaction-irreducible diagrams are self-energy diagrams that are constructed from the effective interactions that arise when normal-ordering the Hamiltonian with respect to the exact ground state~\cite{carbone2013self,raimondi2018algebraic}. Expressing the self-energy with respect to the exact Green's function $\Sigma[G]$, is referred to as self-consistent renormalization and considerably reduces the number of diagrams constituting the self-energy expansion. It is also crucial to construct the functional $\Sigma[G]$ to correctly derive the BSE kernel. For example, the effective interactions corresponding to the two-body Hamiltonian 
\begin{gather}
    \begin{split}
        H = \sum_{pq} h_{pq} a^\dag_pa_q + \frac{1}{(2!)^2}\sum_{pq,rs} v_{pq,rs} a^\dag_{p}a^\dag_qa_sa_r 
    \end{split}
\end{gather}
are given by normal-ordering $H$ with respect to the exact ground state $\ket{\Psi_0}$. This gives  
\begin{gather}
    \begin{split}
        H = E^{N}_0 +  \sum_{pq} F_{pq} \{a^\dag_pa_q\} + \frac{1}{(2!)^2}\sum_{pq,rs} v_{pq,rs} \{a^\dag_{p}a^\dag_qa_sa_r\}
    \end{split}
\end{gather}
where the effective interactions that are used to construct interaction-irreducible diagrams are defined by 
    \begin{gather}
    \begin{split}
        F_{pq} &= f_{pq}+\sum_{rs}v_{pr,qs}\Big(\gamma_{rs}-\gamma^{\rf}_{rs}\Big)\\
        &= f_{pq}+\Sigma^{\infty}_{pq} , 
    \end{split}
    \end{gather}
    where $f_{pq}=h_{pq} + \sum_{rs}v_{pr,qs}\gamma^{\rf}_{rs}$ is the Fock operator, $\gamma_{rs} = \braket{\Psi^N_0|a^\dag_{r}a_{s}|\Psi^N_0}$ is the exact one-particle reduced density matrix and $\gamma^{\rf}_{rs} = \braket{\Phi_0|a^\dag_{r}a_{s}|\Phi_0}$ is the reference one-particle reduced density matrix. 
Here $v_{pq,rs}$ is the antisymetrized two-body interaction and $E^{N}_0 = \braket{\Psi^N_0|H|\Psi^N_0}$ is the exact ground state energy. The notation $\{\cdots\}$ indicates a string of operators that is normal-ordered with respect to the exact ground state. 

By the relationship between the one-particle reduced density matrix and the Green's function (Eq.~\ref{eq:gf_rdm}), we see that the contribution $\Sigma^{\infty}_{pq}$ to the one-body effective interaction is the exact static component of the self-energy
\begin{gather}
    \begin{split}~\label{eq:stat_se}
        \Sigma^{\infty}_{pq} &= \sum_{rs}v_{pr,qs}\Big(\gamma_{rs}-\gamma^{\rf}_{rs}\Big)
    \end{split}
\end{gather}
and depends self-consistently on the exact Green's function~\cite{hedin1965new,stefanucci2013nonequilibrium}.

In the case of a two-body interaction Hamiltonian, the effective two-body interaction is the same as the `bare' two-body interaction. This is because there are no higher-body terms present that may dress this interaction. Therefore, the irreducible self-energy may be expressed in terms of the 1PI, skeleton and interaction-irreducible diagrams constructed from the bare two-body interaction using exact propagators. Therefore, we are usually not concerned with the effective interactions that arise when normal-ordering $H$, as the self-consistent expansion for the self-energy is unchanged when using the `bare' two-body interaction.

However, generating the effective interactions for a three-body Hamiltonian is more involved. Consider the three-body Hamiltonian
\begin{gather}
    \begin{split}
        H &= \sum_{pq} h_{pq} a^\dag_pa_q + \frac{1}{(2!)^2}\sum_{pq,rs} v_{pq,rs} a^\dag_pa^\dag_qa_sa_r \\
        &+\frac{1}{(3!)^2}\sum_{pqr,stu} W_{pqr,stu} a^\dag_pa^\dag_qa^\dag_ra_ua_ta_s \ .
    \end{split}
\end{gather}
Normal-ordering this Hamiltonian with respect to the exact ground state gives 
\begin{gather}
    \begin{split}
        H &= E^{N}_0+ \sum_{pq} F_{pq} \{a^\dag_pa_q\} + \frac{1}{(2!)^2}\sum_{pq,rs} V_{pq,rs} \{a^\dag_pa^\dag_qa_sa_r\} \\
        &+\frac{1}{(3!)^2}\sum_{pqr,stu} W_{pqr,stu} \{a^\dag_pa^\dag_qa^\dag_ra_ua_ta_s\} 
    \end{split}
\end{gather}
where the effective interactions are now 
\begin{subequations}
    \begin{align}
    \begin{split}
            F_{pq} 
            &= f_{pq} +\sum_{rs}v_{pr,qs}\Big(\gamma_{rs}-\gamma^{\rf}_{rs}\Big)+\frac{1}{(2!)^2}\sum_{rs,tu}W_{prs,qtu}\Gamma_{rs,tu}\\
            &= f_{pq}+\Sigma^{\infty}_{pq}
    \end{split}
    \end{align}
    with
    \begin{align}
    \begin{split}~\label{eq:eff_2p}
        V_{pq,rs} &= v_{pq,rs} + \sum_{tu} W_{pqt,rsu}\gamma_{tu} \ .
    \end{split}
    \end{align}
\end{subequations}
where $W_{pqr,stu}$ is the anti-symmetrized three-body interaction and $\Gamma_{pq,rs} = \braket{\Psi^N_0|a^\dag_{p}a^\dag_{q}a_{s}a_{r}|\Psi^N_0}$ is the exact two-particle reduced density matrix. Here, we clearly see that generating the self-consistent renormalized 1PI, skeleton and interaction-irreducible diagrams for $\Sigma[G]$ corresponding to a three-body Hamiltonian is more involved~\cite{carbone2013self}. The exact static contribution to the self-energy: 
\begin{gather}
    \begin{split}
        \Sigma^{\infty}_{pq} = \sum_{rs}v_{pr,qs}\Big(\gamma_{rs}-\gamma^{\rf}_{rs}\Big)+\frac{1}{(2!)^2}\sum_{rs,tu}W_{prs,qtu}\Gamma_{rs,tu} \ ,
    \end{split}
\end{gather}
now involves the two-particle reduced density matrix contracted with the three-body interaction. 

It is central to note that the effective two-body interaction given in Eq.~\ref{eq:eff_2p} is significantly modified and is now a functional of the exact single-particle Green's function, $V[G]$. This is due to the contraction of the bare three-body interaction with the exact one-particle reduced density matrix. This fact creates a fundamental distinction between Green's function theories of two-body and three-body (and higher) interactions. Therefore, the self-consistent self-energy functional $\Sigma[G]$, is constructed only from skeleton diagrams using the effective interactions, $V$ and $W$, with exact propagators~\cite{carbone2013self,raimondi2018algebraic}. In the case of a three-body interaction, this greatly reduces the number of terms in the diagrammatic expansion and this procedure is absolutely necessary when expressing the self-energy as a functional of the exact Green's function. When there are three-body interactions present, the ground state energy is no longer directly obtainable from knowledge only of the single-particle Green's function and self-energy~\cite{carbone2013self}. 

This concludes our summary of the standard results that are pertinent to the similarity transformed formulation.

\section{Biorthogonal Quantum Theory}~\label{sec:bio_qm}

Within non-hermitian quantum theory, time evolution of states and operators must be formulated carefully. This is due to the fact that the left and right states can be related to each other by a non-trivial metric and therefore evolve under different Hamiltonians. In this section, we will review aspects of biorthogonal quantum theory, as well as the biorthogonal Schrödinger and Heisenberg pictures. To the best of our knowledge, for the first time we present the biorthogonal Interaction picture and provide an extension of the GML theorem to include non-hermitian interactions. This analysis is required to define a consistent Green's function theory that generates a diagrammatic theory of the self-energy and Bethe-Salpeter kernel. Finally, the biorthogonal Heisenberg picture is central to the self-consistent Green's function theory presented in Section~\ref{sec:cc_se}. 

Suppose we have a diagonalizable non-hermitian Hamiltonian $\Bar{H}$ with eigenstates $\{\ket{\Psi_n}\}$ and eigenvalues $\{E_n\}$ such that
\begin{equation}
    \Bar{H}\ket{\Psi_n} = E_{n}\ket{\Psi_n} \hspace{2.5mm}\Rightarrow\hspace{2.5mm} \bra{\Psi_n}\Bar{H}^\dag = \bra{\Psi_n}E^*_n \ .
\end{equation}
The eigenstates of the hermitian adjoint $\bar{H}^\dag$ are
\begin{equation}
    \Bar{H}^\dag\ket{\Phi_n} = \mathcal{E}_{n}\ket{\Phi_n} \hspace{2.5mm}\Rightarrow\hspace{2.5mm} \bra{\Phi_n}\Bar{H} = \bra{\Phi_n}\mathcal{E}^*_n \ .
\end{equation}
The set $\{\ket{\Psi_n}\}$ is complete in the Hilbert space, with $\braket{\Phi_n|\Psi_m} = \delta_{nm}$, $E_{n} = \mathcal{E}_n^* \hspace{1.5mm}\forall \  n$. Completeness of $\{\ket{\Psi_n}\}$ implies the completeness of $\{\ket{\Phi_n}\}$~\cite{brody2013biorthogonal,southall2022comparing}. The eigenstates of $\Bar{H}$ are generally not orthogonal~\cite{brody2013biorthogonal}. The basis set $\{\ket{\Phi_m}\}^{N}_{m=1}$ is referred to as the biorthonormal basis associated with $\{\ket{\Psi_n}\}_{n=1}^{N}$~\cite{brody2013biorthogonal,brody2016consistency,southall2022comparing}. A general state 
\begin{equation}~\label{eq:reg}
    \ket{\Psi} = \sum_{n=1}^{N} C_n \ket{\Psi_n} , 
\end{equation}
has an \emph{associated state} and a dual state
\begin{equation}~\label{eq:dual}
    \ket{\Tilde{\Psi}} := \sum_{n=1}^{N} C_n \ket{\Phi_n}  \hspace{2.5mm}\Rightarrow \hspace{2.5mm} \bra{\Tilde{\Psi}} = \sum_{n=1}^{N} C^*_n \bra{\Phi_n} \ . 
\end{equation}
The state dual to $\ket{\Psi}$ is given by $\bra{\Tilde{\Psi}}$ and the associated state is given by its hermitian conjugate: $\ket{\Tilde{\Psi}}$. The biorthogonal inner product is defined as
\begin{equation}
    \braket{\Psi,\Psi}_{bo} = \braket{\Tilde{\Psi}|\Psi} = \sum_{n} |C_n|^2 \ .
\end{equation}
where $\braket{.,.}_{bo}$ denotes the biorthogonal inner product. Under the biorthogonal inner product, the set of states, $\{\ket{\Psi_n},\ket{\Phi_n}\}_{n=1}^N$ forms a complete biorthogonal system. A generic operator, $A$, can be represented in the biorthogonal basis as 
\begin{equation}
    A = \sum_{nm} a_{nm} \ket{\Psi_n}\bra{\Phi_m} \ ,
\end{equation}
with $a_{nm}=\braket{\Phi_n|A|\Psi_m}$. The expectation value of an observable in a pure state $\ket{\Psi}$ is defined as 
\begin{equation}
    \braket{A} = \frac{\braket{\Tilde{\Psi}|A|\Psi}}{\braket{\Tilde{\Psi}|\Psi}} \ ,
\end{equation}
and if the operator is biorthogonally hermitian with $a_{nm}=a^*_{mn}$, the expectation value of the observable will be real. Non-hermitian quantum mechanics can also be formulated in terms of conventional quantum theory where the Hilbert space is endowed with a nontrivial metric. This is known as pseudo-hermitian quantum theory~\cite{brody2013biorthogonal,pauli1943dirac,mostafazadeh2002pseudo}.

\subsection{Biorthogonal Schrödinger and Heinsenberg pictures}

The formalism of quantum field theory takes place in the Heisenberg picture~\cite{dirac1981principles}. Therefore, we must carefully define the relationship between the Schrödinger and Heisenberg pictures in order to define the correct extension of the Green's function formalism for non-hermitian interactions. 

The time-dependent non-hermitian Schrödinger equation is written as 
\begin{equation}
    i\frac{\partial}{\partial t}\ket{\Psi_S(t)} = \Bar{H}\ket{\Psi_S(t)}
\end{equation}
where $\Bar{H}$ is a time-independent, pseudo-hermitian Hamiltonian~\cite{mostafazadeh2002pseudo}. The formal solution is given by 
\begin{equation}
    \ket{\Psi_S(t)} = \bar{U}(t,0)\ket{\Psi_S(0)} \ ,
\end{equation}
where $\bar{U}(t,0)=\exp(-i\Bar{H}t)$. It is important to note that the time-evolution operator $\bar{U}$ is not unitary. The associated states necessarily obey the conjugate equation 
\begin{equation}
    i\frac{\partial}{\partial t}\ket{\Tilde{\Psi}_S(t)} = \Bar{H}^\dag\ket{\Tilde{\Psi}_S(t)}
\end{equation}
yielding the formal solution
\begin{equation}
    \ket{\Tilde{\Psi}_S(t)} = \tilde{U}^\dag(t,0)\ket{\Tilde{\Psi}_S(0)} , 
\end{equation}
where $\ket{\Tilde{\Psi}_S(0)}$ is the associated state of $\ket{\Psi_S(0)}$ and $\tilde{U}^\dag(t,0)=\exp(-i\Bar{H}^\dag t)$. This ensures that the norm of the state is preserved under time-evolution as $\braket{\Tilde{\Psi}_S(t)|\Psi_S(t)} = \braket{\Tilde{\Psi}_S(0)|\Psi_S(0)}$. Clearly, we see that \emph{associated states} evolve according to a different Hamiltonian ($\bar{H}^\dag$)~\cite{bender2002complex}. This also has important consequences in the theory of non-hermitian quantum electrodynamics~\cite{southall2022comparing}.

The Heisenberg states are given by the relationship: $\ket{\Psi_H} = \ket{\Psi_S(0)}$ and $\ket{\Tilde{\Psi}_H} = \ket{\Tilde{\Psi}_S(0)}$. Therefore, the general relationship between an operator in the Heisenberg and Schrödinger representations under the birothogonal inner product is given by 
\begin{gather}
\begin{split}~\label{eq:bio_heisenberg}
     O_H(t) &= \exp(i\Bar{H}t)O_S\exp(-i\Bar{H}t) \ .
\end{split}
\end{gather}
It should be noted that the same form has been obtained in the pseudo-hermitian theory with respect to the $\tau$-inner product, where $\tau$ is the metric~\cite{miao2016investigation}. Now that we have defined the biorthogonal Schrödinger and Heisenberg pictures, we introduce the biorthogonal Interaction picture.

\subsection{Biorthogonal Interaction picture}

Assume that a time-independent diagonalizable non-hermitian Hamiltonian $
\Bar{H}$ may be expressed as the sum of two non-hermitian terms $\Bar{H}_0$ and $\Bar{H}_1$ such that $\Bar{H} = \Bar{H}_0 + \Bar{H}_1$. Let us define the Interaction state vector in the Hilbert space as $\ket{\Psi_I(t)} = \exp(i\Bar{H}_0t)\ket{\Psi_S(t)}$ and  it's associated state as $\ket{\Tilde{\Psi}_I(t)} = \exp(i\Bar{H}^\dag_0t)\ket{\Tilde{\Psi}_S(t)}$. This transformation leads to the following Schrödinger equations for the Interaction state vector 
\begin{subequations}
    \begin{gather}
    \begin{split}~\label{eq:int_se}
        i\frac{\partial}{\partial t}\ket{\Psi_I(t)} = \Bar{H}_1(t)\ket{\Psi_I(t)}
    \end{split}
\end{gather}
and its associated state
\begin{gather}
    \begin{split}~\label{eq:int_se_1}
        i\frac{\partial}{\partial t}\ket{\Tilde{\Psi}_I(t)} = \Bar{H}^\dag_1(t)\ket{\Tilde{\Psi}_I(t)}
    \end{split}
\end{gather}
\end{subequations}
where $\Bar{H}_1(t)=\exp(i\Bar{H}_0t)\Bar{H}_1\exp(-i\Bar{H}_0t)$ and $\Bar{H}^\dag_1(t)$ is the hermitian adjoint. An arbitrary matrix element in the Schrödinger picture requires a general operator in the biorthogonal Interaction picture to be defined as 
\begin{equation}
    O_I(t)=\exp(i\Bar{H}_0t)O_S\exp(-i\Bar{H}_0t) \ .
\end{equation}
In order to find an expression for the time-evolution operator in the biorthogonal Interaction picture, from Eqs~\ref{eq:int_se} and~\ref{eq:int_se_1}, we immediately find that 
\begin{gather}
\begin{split}
    \ket{\Psi_I(t)} &= \mathcal{T}\left\{\exp\left(-i\int^{t}_{t_0} dt' \Bar{H}_1(t')\right)\right\}\ket{\Psi_I(t_0)} \\
    &= \bar{U}(t,t_0)\ket{\Psi_I(t_0)}
\end{split}
\end{gather}
and 
\begin{gather}
\begin{split}
        \ket{\Tilde{\Psi}_I(t)} &= \mathcal{T}\left\{\exp\left(-i\int^{t}_{t_0} dt' \Bar{H}^\dag_1(t')\right)\right\}\ket{\Tilde{\Psi}_I(t_0)} \\
        &= \Tilde{U}^\dag(t,t_0)\ket{\Tilde{\Psi}_I(t_0)} \ .
\end{split}
\end{gather}
 These relations ensure that the biorthogonal norm of the interaction state-vector is time-independent. As a result, the general relationship between the birothogonal Interaction and Heisenberg pictures under the biorthogonal inner product is given by
\begin{subequations}
\begin{align}
    \begin{split}
        \ket{\Psi_H} &= \bar{U}(0,t_0)\ket{\Psi_I(t_0)}
        \end{split}\\
        \begin{split}
        \ket{\Tilde{\Psi}_H} &= \Tilde{U}^\dag(0,t_0)\ket{\Tilde{\Psi}_I(t_0)}
        \end{split}\\
        \begin{split}
        O_H(t) &= \bar{U}(0,t)O_I(t)\bar{U}(t,0) \ .
    \end{split}
\end{align}
\end{subequations}

\subsection{Extension of the Gell-Mann and Low theorem for non-hermitian interactions}

In order to generate a diagrammatically consistent Green's function formalism for non-hermitian interactions, here we demonstrate that we can extend the GML theorem. 

Exact eigenstates of the full non-hermitian Hamiltonian can be generated adiabatically by introducing the time-dependent Hamiltonian
\begin{equation}
    \Bar{H} = \Bar{H}_0 + e^{-\eta|t|}\Bar{H}_1 \ ,
\end{equation}
where $\eta$ is a positive infinitesimal. Here, we have made the implicit assumption that $\bar{H}$ can be partitioned as the sum of two diagonalizable non-hermitian Hamiltonians $\bar{H}_0$ and $\bar{H}_1$. At very large times, the Hamiltonian reduces to $\Bar{H}_0$ and at $t=0$, the Hamiltonian is the full non-hermitian Hamiltonian. The general solution for the interaction and associated states is given by 
\begin{subequations}
\begin{align}
\begin{split}
    \ket{\Psi_I(t)} &= \bar{U}_{\eta}(t,t_0)\ket{\Psi_I(t_0)}
    \end{split}\\
    \begin{split}
    \ket{\tilde{\Psi}_I(t)} &= \tilde{U}^\dag_{\eta}(t,t_0)\ket{\tilde{\Psi}_I(t_0)}
\end{split}
\end{align}
\end{subequations}
where the time-evolution operator now takes the form 
\begin{equation}
    \bar{U}_{\eta}(t,t_0) = \mathcal{T}\left\{\exp\left(-i\int^{t}_{t_0} dt' e^{-\eta|t'|}\Bar{H}_1(t')\right)\right\} \ ,
\end{equation}
with the analogous form for $\tilde{U}^\dag_{\eta}(t,t_0)$. Using these relationships, we take the time limit $t_0\to-\infty$ and the physical limit as $\eta\to0$ by extending the GML theorem (see Appendix~\ref{app_gm}) such that if the limits of the quantities 
\begin{subequations}
\begin{gather}
\begin{split}~\label{eq:GM}
   \frac{\ket{\Psi_0}}{\braket{\tilde{\Phi}_0|\Psi_0}} \equiv \lim_{\eta\to0^+}\frac{\bar{U}_{\eta}(0,-\infty)\ket{\Phi_0}}{\braket{\Tilde{\Phi}_0|\bar{U}_{\eta}(0,-\infty)|\Phi_0}} 
\end{split}
\end{gather}
and 
\begin{gather}
\begin{split}~\label{eq:GM_1}
    \frac{\ket{\Tilde{\Psi}_0}}{\braket{\Phi_0|\Tilde{\Psi}_0}} \equiv \lim_{\eta\to0^+}\frac{\Tilde{U}^\dag_{\eta}(0,-\infty)\ket{\tilde{\Phi}_0}}{\braket{\Phi_0|\Tilde{U}^\dag_{\eta}(0,-\infty)|\Tilde{\Phi}_0}} 
\end{split}
\end{gather}
\end{subequations}
exist, then they are the `right' and `left' eigenstates of the full Hamiltonian $\Bar{H}$: 
\begin{equation}
    \Bar{H}\frac{\ket{\Psi_0}}{\braket{\tilde{\Phi}_0|\Psi_0}} = E \frac{\ket{\Psi_0}}{\braket{\tilde{\Phi}_0|\Psi_0}} \hspace{2.5mm} ;  \hspace{2.5mm} \Bar{H}^\dag\frac{\ket{\Tilde{\Psi}_0}}{\braket{\Phi_0|\Tilde{\Psi}_0}} = E \frac{\ket{\Tilde{\Psi}_0}}{\braket{\Phi_0|\Tilde{\Psi}_0}} \ .
\end{equation}
Importantly, the limit as $\eta\to0$ of both the numerators and denominators of Eqs~\ref{eq:GM} and~\ref{eq:GM_1} do not separately exist. The proof can be derived in a similar way to the original proof presented for the Hermitian case but can only be obtained by careful analysis of the role of time-dependence of operators for non-hermitian quantum systems. The theorem also applies equally to the state obtained from evolving from $t=+\infty$~\cite{Quantum,gell1954quantum}. Using our definition, the associated state is written as 
\begin{gather}
    \begin{split}
        \frac{\bra{\Tilde{\Psi}_0}}{\braket{\Tilde{\Psi}_0|\Phi_0}^*} &= \lim_{\eta\to0^+}\frac{\bra{\Tilde{\Phi}_0}\Tilde{U}_{\eta}(0,-\infty)}{\braket{\Tilde{\Phi}_0|\Tilde{U}_{\eta}(0,-\infty)|\Phi_0}^*}\\
        &=\lim_{\eta\to0^+}\frac{\bra{\Tilde{\Phi}_0}\bar{U}_{\eta}(-\infty,0)}{\braket{\Tilde{\Phi}_0|\bar{U}_{\eta}(-\infty,0)|\Phi_0}^*} \ .
    \end{split}
\end{gather}
Therefore, the state obtained from the adiabatic procedure is an eigenstate of the full non-hermitian Hamiltonian, which may be obtained from either evolving `forward' or `backward' in time.

By making use of our extended GML theorem, we can construct perturbative expressions for time-dependent correlation functions by evolving on the right from $-\infty$ and on the left from $+\infty$ as follows (making use of the linked diagram theorem~\cite{Quantum,shavitt2009many})
\begin{gather}
\begin{split}~\label{eq:ext_GM}
    &\mathcal{R}(t_1,t_2,\cdots,t_n) =\frac{\langle{\Tilde{\Psi}_0|\mathcal{T}\left\{O_H(t_1)O_H(t_2)\cdots O_H(t_n)\right\}|\Psi_0\rangle}}{\braket{\Tilde{\Psi}_0|\Psi_0}}\\
    &=\lim_{\eta\to0^+}\langle\Tilde{\Phi}_0|\mathcal{T}\left\{\bar{S}_{\eta}O_{I}(t_1)O_I(t_2)\cdots O_I(t_n)\right\}|\Phi_0\rangle_{_{\mathcal{C}}} \ ,
    \end{split}
\end{gather}
where the subscript $\mathcal{C}$ indicates that only connected diagrams contribute to the expansion. Here, $\bar{S}_{\eta}=\bar{U}_{\eta}(\infty,-\infty)$ is the non-unitary scattering operator. If we choose the partitioning such that the reference Hamiltonian is hermitian, \emph{i.e.} $H_0=H^\dag_0$, the results above hold and we have 
\begin{gather}~\label{eq:ext_GM_1}
\begin{split}
     &\mathcal{R}(t_1,t_2,\cdots,t_n) =\\
     &\lim_{\eta\to0^+}\langle\Phi_0|\mathcal{T}\left\{\bar{S}_{\eta}O_{I}(t_1)O_I(t_2)\cdots O_I(t_n)\right\}|\Phi_0\rangle_{_{\mathcal{C}}}  
     \end{split}
\end{gather}
where now the left and right states of the hermitian reference are simply related to each other via hermitian conjugation. 

\section{The single-particle Green's function for $N$-body non-hermitian Interactions}~\label{sec:non_fund}

In this section, we define the single-particle Green's function for the case of an $N$-body non-hermitian interaction Hamiltonian and discuss the construction of the non-hermitian irreducible self-energy that emerges. This analysis provides the foundation for the analysis of the diagrammatic coupled-cluster self-energy presented in Sections~\ref{sec:pert} and~\ref{sec:cc_se}. 

Suppose we have an $N$-body non-hermitian Hamiltonian 
\begin{gather}
    \begin{split}
        \bar{H} &= \sum_{pq} \bar{h}_{pq} a^\dag_pa_q + \frac{1}{(2!)^2}\sum_{pq,rs} \bar{v}_{pq,rs} a^\dag_pa^\dag_qa_sa_r \\
        &+\frac{1}{(3!)^2}\sum_{pqr,stu} \bar{w}_{pqr,stu} a^\dag_pa^\dag_qa^\dag_ra_ua_ta_s 
        \\
        &+  \frac{1}{(4!)^2}\sum_{\substack{pqrs\\
        tuvz}} \bar{u}_{pqrs,tuvz} a^\dag_pa^\dag_qa^\dag_ra^\dag_s a_za_va_ua_t \\
        &+ \frac{1}{(5!)^2}\sum_{\substack{pqrsn\\tuvzm}} \bar{\kappa}_{pqrsn,tuvzm} a^\dag_pa^\dag_qa^\dag_ra^\dag_s a^\dag_{n} a_m a_za_va_ua_t + \cdots
    \end{split}
\end{gather}
with right and left eigenstates given by $\bar{H}\ket{\Psi^N_0} = E^N_0\ket{\Psi^N_0}$ and $\bar{H}^\dag\ket{\tilde{\Psi}^N_0} = E^N_0\ket{\tilde{\Psi}^N_0} $. The biorthogonal single-particle Green's function of the non-hermitian theory is defined as the biorthogonal expectation value 
\begin{gather}
    \begin{split}
        i\tilde{G}_{pq}(t_1,t_2) = \braket{\tilde{\Psi}^N_0|\mathcal{T}\{a_{p}(t_1)a^\dag_{q}(t_2)\}|\Psi^N_0}
    \end{split}
\end{gather}
where the time-dependence of the field-operators is given by $a_{p}(t_1)=e^{i\bar{H}t_1}a_pe^{-i\bar{H}t_1}$ as defined in the biorthogonal Heisenberg picture. The notation $\tilde{G}$ is used to indicate that the Green's function is defined with respect to the biorthogonal inner product. 

By partitioning the non-hermitian Hamiltonian into an arbitrary reference operator $\bar{H}_0$, and an interaction term $\bar{H}_1 = \bar{H}-\bar{H}_0$, we can use the extended GML theorem (Eq.~\ref{eq:ext_GM}) to write the perturbative expansion for the Green's function as 
\begin{gather}
\begin{split}~\label{eq:non_gf}
    &i\Tilde{G}_{pq}(t_1,t_2) = \\
    &\langle\tilde{\Phi}_0|\mathcal{T}\left\{\exp\left(-i\int^{\infty}_{-\infty}dt\  \Bar{H}_1(t)\right)a_{p}(t_1)a^\dag_q(t_2)\right\}|\Phi_0\rangle_{_\mathcal{C}} ,
\end{split}
\end{gather}
where the time-dependence of the field operators is now defined in the biorthogonal Interaction picture: $a_{p}(t_1)=e^{i\bar{H}_0t}a_pe^{-i\bar{H}_0t}$. Here, $\ket{\Phi_0}$ and $\bra{\tilde{\Phi}_0}$ are the right and left ground eigenstates of the reference operator, $\bar{H}_0$. We have dropped the implicit limit as $\eta\to0$ in Eq.~\ref{eq:ext_GM} for notational simplicity. Eq.~\ref{eq:non_gf} immediately gives rise to the exact geometric Dyson series  
\begin{gather}
    \begin{split}
        \tilde{G}_{pq}(\omega) = \tilde{G}^0_{pq}(\omega) + \sum_{rs} \tilde{G}^0_{pr}(\omega)\tilde{\Sigma}_{rs}(\omega)\tilde{G}_{sq}(\omega) \ ,
    \end{split}
\end{gather}
where $\tilde{G}^0$ is the non-interacting Green's function of the reference. The non-hermitian self-energy can be exactly written as the sum of the static and forward-/backward-time dynamical contributions
\begin{gather}
    \begin{split}
        \tilde{\Sigma}_{pq}(\omega) = \tilde{\Sigma}^{\infty}_{pq} + \tilde{\Sigma}^{\F}_{pq}(\omega) + \tilde{\Sigma}^{\Ba}_{pq}(\omega) \ ,
    \end{split}
\end{gather}
and is a functional of the exact biorthogonal single-particle Green's function, $\tilde{\Sigma}[\tilde{G}]$. The 1PI, skeleton and interaction-irreducible diagrams that make up the self-energy functional $\tilde{\Sigma}[\tilde{G}]$ are obtained from effective interactions. These interactions arise when normal-ordering $\bar{H}$ with respect to the biorthogonal ground state expectation value.  In this case, the one-body term is given by 
\begin{subequations}
    \begin{gather}
        \begin{split}
        \tilde{F}_{pq} &= \bar{h}_{pq} +\sum_{rs}\bar{v}_{pr,qs}\tilde{\gamma}_{rs}+\frac{1}{(2!)^2}\sum_{rs,tu}\bar{w}_{prs,qtu}\tilde{\Gamma}_{rs,tu} \\
        &+ \frac{1}{(3!)^2}\sum_{rst,uvz}\bar{u}_{prst,quvz}\tilde{\Gamma}_{rst,uvz} + \cdots \\
        &=\tilde{f}_{pq}+\sum_{rs}\bar{v}_{pr,qs}\Big(\tilde{\gamma}_{rs}-\tilde{\gamma}^{\rf}_{rs}\Big)+\frac{1}{(2!)^2}\sum_{rs,tu}\bar{w}_{prs,qtu}\tilde{\Gamma}_{rs,tu} \\
        &+ \frac{1}{(3!)^2}\sum_{rst,uvz}\bar{u}_{prst,quvz}\tilde{\Gamma}_{rst,uvz} + \cdots \\
        &= \tilde{f}_{pq}+\tilde{\Sigma}^{\infty}_{pq}\ ,
        \end{split}
    \end{gather}
    where $\tilde{f}_{pq}$ is the non-hermitian Fock operator, the exact one-particle reduced density matrix is defined as $\tilde{\gamma}_{pq}=\braket{\tilde{\Psi}^N_0|a^\dag_pa_q|\Psi^N_0}$ and the reference one-particle reduced density matrix is $\tilde{\gamma}^{\rf}_{pq}=\braket{\tilde{\Phi}_0|a^\dag_pa_q|\Phi_0}$.
    The two-body effective interaction is obtained as 
    \begin{gather}
        \begin{split}
            \bar{V}_{pq,rs} &= \bar{v}_{pq,rs} + \sum_{tu} \bar{w}_{pqt,rsu}\tilde{\gamma}_{tu} \\
            &+ \frac{1}{(2!)^2} \sum_{tu,vz}\bar{u}_{pqtu,rsvz}\tilde{\Gamma}_{tu,vz} + \cdots
        \end{split}
    \end{gather}
    with the three-body effective interaction 
    \begin{gather}
        \begin{split}
            \bar{W}_{pqr,stu} &= \bar{w}_{pqr,stu} + \sum_{vx} \bar{u}_{pqrv,stux}\tilde{\gamma}_{vx} \\
            &+ \frac{1}{(2!)^2}\sum_{vw,lo}\bar{\kappa}_{pqrvw,stulo}\tilde{\Gamma}_{vw,lo}+\cdots
        \end{split}
    \end{gather}
     The four-body effective interaction is 
     \begin{gather}
        \begin{split}
            \bar{U}_{pqrs,tuvo} &= \bar{u}_{pqrs,tuvo} + \sum_{yx} \bar{\kappa}_{pqrsy,tuvox}\tilde{\gamma}_{yx} + \cdots
        \end{split}
    \end{gather}
\end{subequations}
This procedure is continued up to the $N$-body interaction. The two-particle biorthogonal reduced density matrix is defined as $\tilde{\Gamma}_{pq,rs}=\braket{\tilde{\Psi}^N_0|a^\dag_pa^\dag_qa_sa_r|\Psi^N_0}$ and so on for the higher-body equivalents. 

We immediately find the exact static component of the non-hermitian self-energy $\tilde{\Sigma}^\infty[\tilde{G}]$ as the remainder of the one-body effective interaction, $\bar{F}-\tilde{f}$:
\begin{gather}
    \begin{split}
        \tilde{\Sigma}^{\infty}_{pq} &= \sum_{rs}\bar{v}_{pr,qs}\Big(\tilde{\gamma}_{rs}-\gamma^{\rf}_{rs}\Big)+\frac{1}{(2!)^2}\sum_{rs,tu}\bar{w}_{prs,qtu}\tilde{\Gamma}_{rs,tu} \\
        &+ \frac{1}{(3!)^2}\sum_{rst,uvz}\bar{u}_{prst,quvz}\tilde{\Gamma}_{rst,uvz} + \cdots 
    \end{split}
\end{gather}
To construct the non-hermitian self-energy functional self-consistently, it must be composed of the effective interactions $\bar{V}$, $\bar{W}$, $\bar{U}$ and so on, which are required in order to generate interaction-irreducible self-energy diagrams. Therefore, it is these effective interactions that appear as vertices in the self-consistent expression of the self-energy functional, $\tilde{\Sigma}[\tilde{G}]$. Importantly, all effective interactions up to the $N$-body term are explicitly functionals of the single-particle Green's function. The perturbative expansion for the self-energy $\tilde{\Sigma}[\tilde{G}_0]$ is obtained by expanding both the propagators and effective interactions with respect to the reference Green's function, $\tilde{G}_0$. For the remainder of this work we will focus on generating the interaction-irreducible Feynman diagrams resulting from the coupled-cluster similarity transformed Hamiltonian. However, the analysis presented in this work is general and can be applied to any $N$-body hermitian or non-hermitian interaction Hamiltonian.

\section{Overview of Coupled-Cluster Theory}~\label{sec:cc_overview}

Within coupled-cluster theory, the following ansatz is made for the exact ground state wavefunction of a system of $N$ interacting particles 
\begin{equation}
    \ket{\Psi^{N}_0} = e^{T}\ket{\Phi_0} \ ,
\end{equation}
where $\ket{\Phi_0}$ is the reference determinant. The wave operator $T$, creates all excitations with respect to the reference determinant and is defined as 
\begin{gather}
\begin{split}
        T &= \sum_{ia} t^{a}_{i} a^\dag_aa_i + \frac{1}{(2!)^2} \sum_{ijab} t^{ab}_{ij} a^\dag_a a^\dag_b a_j a_i \\
        &+\frac{1}{(3!)^2}\sum_{ijkabc}t^{abc}_{ijk} a^\dag_a a^\dag_b a^\dag_c a_k a_j a_i + \cdots  
\end{split}
\end{gather}
where $\{t^{a}_{i},t^{ab}_{ij},t^{abc}_{ijk},\cdots \}$ are the cluster amplitudes. Using the CC ansatz, we write the non-hermitian CC eigenvalue problem for the right and left eigenstates of the similarity transformed Hamiltonian as
\begin{gather}
\begin{split}
    \Bar{H}\ket{\Phi_0} &= E^{\CC}_0\ket{\Phi_0} \hspace{2.5mm};\hspace{2.5mm}
    \bra{\Tilde{\Psi}_0}\Bar{H} = \bra{\Tilde{\Psi}_0} E^{\CC}_0 \ ,
\end{split}
\end{gather}
where 
\begin{equation}
    \Bar{H} = e^{-T}He^{T}    
\end{equation}
is the similarity transformed Hamiltonian and $H$ is the bare Hamiltonian. Here, we see that the reference $\ket{\Phi_0}$ is the right eigenstate of the similarity transformed Hamiltonian. The left eigenstate is 
given by $\bra{\tilde{\Psi}_0}=\bra{\Phi_0}(\mathbbm{1}+\Lambda)$, where $\Lambda$ is the de-excitation operator defined as
\begin{gather}
    \begin{split}
        \Lambda &= \sum_{ai} \lambda^{i}_{a}a^\dag_ia_a + \frac{1}{(2!)^2}\sum_{abij} \lambda^{ij}_{ab}a^\dag_i a^\dag_j a_b a_a \\
        &+\frac{1}{(3!)^2} \sum_{abcijk} \lambda^{ijk}_{abc} a^\dag_i a^\dag_j a^\dag_k a_c a_b a_a + \cdots 
    \end{split}
\end{gather}
with $\{\lambda^{i}_{a},\lambda^{ij}_{ab},\lambda^{ijk}_{abc},\cdots \}$ representing the de-excitation amplitudes. The excitation, de-excitation amplitudes and ground state energy are obtained from minimization of the coupled-cluster Lagrangian~\cite{helgaker2013molecular} 
\begin{gather}
    \begin{split}
        \mathcal{L}^{\CC}_0(\mathbf{t},\mathbf{\lambda}) = \braket{\tilde{\Psi}_0|\bar{H}|\Phi_0} = \braket{\Phi_0|\Bar{H}|\Phi_0} + \sum_{\mu}\lambda_{\mu}\braket{\Phi_{\mu}|\Bar{H}|\Phi_0} \ ,
    \end{split}
\end{gather}
where $\ket{\Phi_{\mu}}$ represents the manifold of excited Slater determinants given by $\ket{\Phi^a_i} = a^\dag_aa_i\ket{\Phi_0}$ etc. Minimization of the Lagrangian gives rise to the excitation and de-excitation amplitude equations
\begin{subequations}
    \begin{equation}~\label{eq:amps}
        \frac{\partial\mathcal{L}^{\CC}_0}{\partial\lambda_{\mu}} = \braket{\Phi_{\mu}|\Bar{H}|\Phi_0} = 0
    \end{equation}
    \begin{equation}
        \frac{\partial\mathcal{L}^{\CC}_0}{\partial t_{\mu}} = \braket{\Phi_{0}|\Bar{H}\tau_{\mu}|\Phi_0} +\sum_{\nu}\lambda_{\nu}\braket{\Phi_{\nu}|[\bar{H},\tau_{\mu}]|\Phi_0} = 0
    \end{equation}
\end{subequations}
where $\tau_{\mu}$ is the excitation operator associated with the cluster amplitude $t_{\mu}$. Once the excitation amplitudes are determined, the ground state energy is simply given by 
\begin{equation}
    E^{\CC}_0 = \braket{\tilde{\Psi}_0|\bar{H}|\Phi_0} = \braket{\Phi_0|\bar{H}|\Phi_0} \ .
\end{equation}
    When $T$ is not truncated, the ground state energy obtained from this procedure is formally exact. Importantly, the reference state $\ket{\Phi_0}$ can be constructed such that the singles amplitudes, $t^{a}_{i}$, are zero by construction. This approach is referred to as BCC theory and implies that the reference state has maximum overlap with the exact ground state. BCC theory uses spin-orbitals that are determined by unitary rotation using the singles amplitude equations~\cite{chiles1981electron,handy1989size,tew2016explicitly}. 
    
    In terms of the cluster amplitudes and general spin-orbitals, the coupled-cluster ground state energy corresponding to the electronic structure Hamiltonian is given by 
\begin{gather}
    \begin{split}
        E^{\CC}_0 = E^{\rf}_0 + \frac{1}{4}\sum_{ijab}\braket{ij||ab}(t^{ab}_{ij}+2t^{a}_it^b_j) + \sum_{ia}f_{ia}t^{a}_{i} , 
    \end{split}
\end{gather}
where $E^{\rf}_0$ is the energy of the reference and $\braket{pq||rs}$ is the anti-symmetrized Coulomb interaction. 

The similarity transformed Hamiltonian $\bar{H}$ possesses the same spectrum as the electronic structure Hamiltonian, $H$. As $\bar{H}$ is non-hermitian but contains a \emph{real} spectrum, it is necessarily pseudo-hermitian (see Appendix~\ref{app:psd}). This fact is not commonly expressed in the literature on coupled-cluster theory. Additionally, $\bar{H}$ is not $\mathcal{PT}$-symmetric, in contrast to most of the Hamiltonians studied in the literature on non-hermitian quantum systems~\cite{bender2002complex,mostafazadeh2002pseudo,bender2007making,ashida2020non,takasu2020pt,bergholtz2021exceptional,meden2023mathcal,zhang2025nonadiabatic}.

In IP/EA-EOM-CC theory, the charged excitation energies are obtained by diagonalization of the similarity transformed Hamiltonian expressed in the basis of all Slater determinants containing $(N\pm1)$-electrons. This is exactly equivalent to a Configuration Interaction expansion of $\bar{H}$. Focusing on the removal energies, the IP-EOM-CC eigenvalue problem can be written as 
\begin{equation}~\label{eq:eom_cc_operator}
        [\Bar{H},R^{\IP}_{\nu}]\ket{\Phi_0} = -\varepsilon^{N-1}_{\nu}R^{\IP}_{\nu}\ket{\Phi_0} , 
\end{equation}
where $R^{\IP}_{\nu}\ket{\Phi_0}$ is the right $(N-1)$-particle eigenstate of $\bar{H}$, $\ket{\Psi^{N-1}_{\nu}}$. In IP-EOM-CC theory, the operator creates all determinants consisting of $(N-1)$ electrons:
\begin{equation}
    R^{\IP}_{\nu} = \sum_{i} r_{i}(\nu)a_{i} + \sum_{i<j,a}r^{a}_{ij}(\nu)a^\dag_aa_ja_i + \cdots
\end{equation}
The IP-EOM-CC eigenvalue problem is written in matrix form in terms of the non-Hermitian supermatrix:
\begin{gather}
    \begin{split}~\label{eq:eom_cc}
       \mathbf{\bar{H}}^{\text{IP-EOM-CC}} = -\left(\begin{array}{ccc}
           \braket{\Phi_{i}|\bar{H}_N|\Phi_{j}}  & \braket{\Phi_{i}|\bar{H}_N|\Phi^{a}_{kl}} & \cdots \\
            \braket{\Phi^{b}_{mn}|\bar{H}_N|\Phi_{j}} & \braket{\Phi^{b}_{mn}|\bar{H}_N|\Phi^{a}_{kl}} & \cdots \\
             \vdots & \vdots & \ddots
        \end{array}\right)  \ ,
    \end{split}
\end{gather}
where $\bar{H}_N=\bar{H}-E^{\CC}_0$. An analogous supermatrix for the EA-EOM-CC eigenvalue problem can also be constructed. In IP/EA-EOM-CC theory, the ionization potential and electron affinity sectors are automatically decoupled as the formalism is based on Configuration Interaction expansions of $\bar{H}$. This approach to obtaining charged excitation spectra can be made to be formally exact, but is distinct from the diagrammatic Green's function formalism which does not decouple the ionization potential and electron affinity sectors. 

\subsection{The similarity transformed Hamiltonian}

The analysis of the effective interactions generated by the CC similarity transformed Hamiltonian will be crucial when constructing the perturbative and renormalized expansions of the coupled-cluster self-energy and Bethe-Salpeter kernel in Sections~\ref{sec:pert}, ~\ref{sec:cc_se} and ~\ref{sec:BSE}. 

The full similarity transformed Hamiltonian is an $N$-body operator written as
\begin{gather}
\begin{split}~\label{eq:sim_ham}
    \Bar{H} &= \bar{h} + \bar{V}+\bar{W} + \cdots\\
    &= \sum_{pq}\bar{h}_{pq}a^\dag_{p}a_q + \frac{1}{(2!)^2}\sum_{pq,rs} \bar{h}_{pq,rs}a^\dag_{p}a^\dag_qa_{s}a_r\\
    &+\frac{1}{(3!)^2}\sum_{pqr,stu}\bar{h}_{pqr,stu} a^\dag_{p}a^\dag_q a^\dag_{r}a_ua_ta_s + \cdots \ ,
\end{split}
\end{gather}
where $\{\bar{h}_{pq},\bar{h}_{pq,rs},\bar{h}_{pqr,stu},\cdots\}$ are the one-body, antisymmetrized two-body, three-body and so on matrix elements. Using the notation of Ref.~\cite{coveney2023coupled}, normal-ordering this Hamiltonian with respect to the reference determinant gives~\cite{shavitt2009many} 
\begin{gather}
\begin{split}
    \Bar{H} &= E^{\CC}_0 + \sum_{pq}F_{pq}\{a^\dag_{p}a_q\}_0 + \frac{1}{(2!)^2}\sum_{pq,rs} \chi_{pq,rs}\{a^\dag_{p}a^\dag_qa_{s}a_r\}_0\\
    &+\frac{1}{(3!)^2}\sum_{pqr,stu}\chi_{pqr,stu} \{a^\dag_{p}a^\dag_q a^\dag_{r}a_ua_ta_s\}_0 + \cdots 
\end{split}
\end{gather}
where the notation $\{\cdots\}_0$ indicates normal-ordering with respect to the reference, $\ket{\Phi_0}$. The set of effective interaction matrix elements $\{F_{pq},\chi_{pq,rs},\chi_{pqr,stu},\cdots\}$, up to the four-body effective interaction, can be found in Refs~\citenum{shavitt2009many} and~\citenum{gauss1995coupled}. This is the conventional normal-ordering of $\bar{H}$ usually presented in the context of coupled-cluster theory and is central to the perturbative expansion of the irreducible coupled-cluster self-energy and Bethe-Salpeter kernel with respect to the reference Green's function.

However, we can also normal-order the Hamiltonian with respect to the biorthogonal expectation value $\braket{\Phi_0|\Phi_0}_{bo} := \braket{\tilde{\Psi}_0|\Phi_0}$, as defined in Section~\ref{sec:bio_qm}, yielding
\begin{gather}
\begin{split}~\label{eq:norm_tot}
    \Bar{H} &= E^{\CC}_0 + \sum_{pq}\tilde{F}_{pq}\{a^\dag_{p}a_q\} + \frac{1}{(2!)^2}\sum_{pq,rs} \tilde{\Xi}_{pq,rs}\{a^\dag_{p}a^\dag_qa_{s}a_r\}\\
    &+\frac{1}{(3!)^2}\sum_{pqr,stu}\tilde{\chi}_{pqr,stu} \{a^\dag_{p}a^\dag_q a^\dag_{r}a_ua_ta_s\} + \cdots  
\end{split}
\end{gather}
to the best of our knowledge, this normal-ordering of $\bar{H}$ has not been reported in the literature before. It is this normal-ordering that will be central to the development of the self-consistent Green's function formalism within coupled-cluster theory.

The relationship between all matrix elements $\{\bar{h}_{pq},\bar{h}_{pq,rs},\bar{h}_{pqr,stu},\cdots\},\{F_{pq},\chi_{pq,rs},\chi_{pqr,stu},\cdots\}$ and
$\{\tilde{F}_{pq},\tilde{\Xi}_{pq,rs},\tilde{\chi}_{pqr,stu},\cdots\}$ is derived in Appendix~\ref{appendix_2}. The coupled-cluster amplitude equations are given by the effective interaction elements$\{F_{ai},\chi_{ab,ij},\chi_{abc,ijk},\cdots\}$ which all evaluate to zero at convergence~\cite{shavitt2009many}. Since these effective interaction elements vanish, the interaction elements $\{\tilde{F}_{ai},\tilde{\Xi}_{ab,ij},\tilde{\chi}_{abc,ijk},\cdots\}$ also evaluate to zero. All effective interaction matrix elements with four or more lines below the vertex, apart from the two-body terms $\chi_{ij,ab}$ and $\tilde{\Xi}_{ij,ab}$, vanish as a result of the coupled-cluster similarity transformation. Finally, it is useful to split the similarity transformed Hamiltonian into two terms
\begin{gather}
\begin{split}
    \bar{H} &= F + \bar{H}_1\\
    &= \sum_{pq} f_{pq} a^\dag_pa_q + \sum_{pq} (\bar{h}_{pq}-f_{pq}) a^\dag_pa_q \\
    &+ \frac{1}{(2!)^2}\sum_{pq,rs} \bar{h}_{pq,rs}a^\dag_{p}a^\dag_qa_{s}a_r\\
    &+\frac{1}{(3!)^2}\sum_{pqr,stu}\bar{h}_{pqr,stu} a^\dag_{p}a^\dag_q a^\dag_{r}a_ua_ta_s + \cdots  
\end{split}
\end{gather}
where $F$ is the Fock operator and $\bar{H}_1=\bar{H}-F$, is the $N$-body operator containing all the interactions and effects introduced by the similarity transformed Hamiltonian.

\section{The coupled-cluster representation of the electronic Green's function}~\label{subsec:cc_rep}

As stated in Section~\ref{sec:key_results}, the single-particle Green's function of the hermitian, two-body electronic structure Hamiltonian $H$, is defined as 
\begin{gather}
    \begin{split}~\label{eq:el_gf}
        iG_{pq}(t_1-t_2) = \braket{\Psi^N_0|\mathcal{T}\left\{a_{p}(t_1)a^\dag_{q}(t_2)\right\}|\Psi^N_0}
    \end{split}
\end{gather}
where $\ket{\Psi^N_0}$ is the exact $N$-electron ground state wavefunction and the time-dependence of the field operators is governed by the electronic structure Hamiltonian: $a_{p}(t)=e^{iHt}a_pe^{-iHt}$. 

Attempts at combining coupled-cluster theory with the Green's function formalism have largely focused on \emph{direct computation} of Eq.~\ref{eq:el_gf} for the electronic Green's function using coupled-cluster theory~\cite{nooijen1992coupled,nooijen1993coupled,peng2016coupled,peng2018green}. This approach leads to the coupled-cluster representation of the electronic Green's function. We use the biorthogonal theory developed in Section~\ref{sec:bio_qm} to obtain the time-domain expression for this propagator as
\begin{equation}~\label{eq:cc_rep_el}
    i\Bar{G}_{pq}(t_1-t_2) = \braket{\Tilde{\Psi}_0|\mathcal{T}\left\{\Bar{a}_{p}(t_1)\Bar{a}^\dag_{q}(t_2)\right\}|\Phi_0} \ ,
\end{equation}
where $\Bar{a}_{p}(t_1)=\exp(i\Bar{H}t)\Bar{a}_p\exp(-i\Bar{H}t)$, $\Bar{a}^\dag_{q}(t_1)=\exp(i\Bar{H}t)\Bar{a}^\dag_q\exp(-i\Bar{H}t)$ with $\bar{a}_p=e^{-T}a_pe^{T}$ and $\bar{a}^\dag_p=e^{-T}a^\dag_pe^{T}$, respectively. The notation $\bar{G}$ is chosen to distinguish the coupled-cluster representation of the electronic Green's function from the electronic Green's function, denoted by $G$. 

Performing the similarity transformation on the operators entering Eq.~\ref{eq:cc_rep_el} gives
\begin{gather}
\begin{split}
        \Bar{a}_p(t_1) &= a_p(t_1) + e^{i\Bar{H}t_1}[a_p,T]e^{-i\Bar{H}t_1} \\
        &= a_p(t_1) + [a_p,T](t_1) \ ,
\end{split}
\end{gather}
where $[a_p,T](t_1)=e^{i\Bar{H}t_1}[a_p,T]e^{-i\Bar{H}t_1}$. Inserting this relation into the time-ordered expectation value gives 
\begin{widetext}
    \begin{gather}
\begin{split}~\label{eq:many_gfs}
    i\Bar{G}_{pq}(t_1-t_2) &= \braket{\Tilde{\Psi}_0|\mathcal{T}\left\{a_{p}(t_1)a^\dag_{q}(t_2)\right\}|\Phi_0} + \braket{\Tilde{\Psi}_0|\mathcal{T}\left\{[a_p,T](t_1)a^\dag_{q}(t_2)\right\}|\Phi_0}+ \braket{\Tilde{\Psi}_0|\mathcal{T}\left\{a_{p}(t_1)[a^\dag_q,T](t_2)\right\}|\Phi_0} \\
    &+ \braket{\Tilde{\Psi}_0|\mathcal{T}\left\{[a_p,T](t_1)[a^\dag_q,T](t_2)\right\}|\Phi_0}  \ .
\end{split}
\end{gather}
\end{widetext}
Due to the commutator of the creation/annihilation operators with $T$, it is clear that the coupled-cluster representation of the electronic single-particle Green's function is in fact a collection of `many-particle' Green's functions of specified time-orderings, including up to the $N$-body Green's function. 

However, the occupied--virtual part of the coupled-cluster representation of the electronic Green's function is in fact a single-particle propagator:
\begin{gather}
\begin{split}
    i\Bar{G}_{ia}(t_1-t_2) &= \braket{\Tilde{\Psi}_0|\mathcal{T}\left\{\Bar{a}_{i}(t_1)\Bar{a}^\dag_{a}(t_2)\right\}|\Phi_0}\\
    &=\braket{\Tilde{\Psi}_0|\mathcal{T}\left\{a_{i}(t_1)a^\dag_{a}(t_2)\right\}|\Phi_0} \ .
\end{split}
\end{gather}
This is because $[a_i,T] = [a^\dag_a,T]=0$. Through Eq.~\ref{eq:many_gfs}, we see that the full coupled-cluster representation of the electronic Green's function is a complicated combination of many-body Green's functions. However, the many-body Green's functions are explicitly coupled to each other through their exact equations of motion (see Section~\ref{sec:cc_se}) and it is exactly this hierarchy of coupled many-body Green's functions that we want to decouple by introduction of the coupled-cluster self-energy. 

To obtain the spectral from of the coupled-cluster representation of the electronic Green's function, we take the Fourier transform of Eq.~\ref{eq:cc_rep_el}, resolving the identity of $(N\pm1)$-particle eigenstates of $\Bar{H}$, to give:
\begin{gather}
    \begin{split}~\label{eq:cc_leh}
        \Bar{G}_{pq}(\omega) &= \sum_{\mu}\frac{\braket{\Tilde{\Psi}_0|\Bar{a}_p|\bar{\Psi}^{N+1}_{\mu}}\braket{\Tilde{\Psi}^{N+1}_{\mu}|\Bar{a}^\dag_q|\Phi_0}}{\omega-\varepsilon^{N+1}_{\mu}+i\eta}\\
        &+\sum_{\nu}\frac{\braket{\Tilde{\Psi}_0|\Bar{a}^\dag_q|\bar{\Psi}^{N-1}_{\nu}}\braket{\Tilde{\Psi}^{N-1}_{\nu}|\Bar{a}_p|\Phi_0}}{\omega-\varepsilon^{N-1}_{\nu}-i\eta} \ ,
    \end{split}
\end{gather}
where $\varepsilon^{N+1}_{\mu} = E^{N+1}_{\mu}-E^{N}_0$ and $\varepsilon^{N-1}_{\nu}=E^{N}_0-E^{N-1}_{\nu}$ are the exact addition and removal energies of the system of interacting electrons, within a complete basis when $T$ is not truncated. This is the expression that was first presented by Nooijen and Snijders in Refs~\cite{nooijen1992coupled,nooijen1993coupled}.

In principle, knowledge of the exact coupled-cluster representation of the electronic Green's function gives the exact single-particle spectrum, coupled-cluster ground state energy as well as the non-hermitian one-particle reduced density matrix
\begin{gather}
    \begin{split}
        \bar{\gamma}_{pq} = -i\bar{G}_{qp}(t-t^+) = \braket{\tilde{\Psi}_0|\bar{a}^\dag_{p}\bar{a}_q|\Phi_0} \ .
    \end{split}
\end{gather}
However, it should be noted that calculation of the ground state energy at any level of approximation for the coupled-cluster representation of the electronic Green's function results in a different ground state energy from that obtained at the same level of approximation of the ground state coupled-cluster equations~\cite{nooijen1992coupled,nooijen1993coupled,lange2018relation}. 

Formally the Dyson equation for the coupled-cluster representation of the electronic Green's function is written as 
\begin{equation}~\label{eq:cc_rep}
    \Bar{G}_{pq}(\omega) = G^{0}_{pq}(\omega) + \sum_{rs}G^{0}_{pr}(\omega)\Bar{\Sigma}_{rs}(\omega)\Bar{G}_{sq}(\omega) \ .
\end{equation}
However, it is important to note that the `self-energy', $\Bar{\Sigma}$, is not the same object as the electronic self-energy, $\Sigma$. This is because $\Bar{\Sigma}$ is an object that connects $G_0$ to $\Bar{G}$ rather than $G$ of the hermitian electronic structure Hamiltonian. As a result the `self-energy' is non-hermitian. 

Due to the fact that the coupled-cluster representation of the electronic Green's function introduces a fundamental coupling to every $N$-particle Green's function, the `self-energy' $\bar{\Sigma}$ entering Eq.~\ref{eq:cc_rep} cannot be defined in the diagrammatic way as the set of all 1PI diagrams of the underlying similarity transformed Hamiltonian. Therefore, the coupled-cluster representation of the electronic Green's function, $\bar{G}$, does not concern the diagrammatic coupled-cluster self-energy and therefore does not fully connect Green's function and coupled-cluster theories.

\section{The single-particle coupled-cluster Green's function and the coupled-cluster self-energy}~\label{sec:ex_sp_ccgf}

In this work, we take the similarity transformed Hamiltonian as the fundamental interaction Hamiltonian and use the biorthogonal formulation for non-hermitian Hamiltonians developed in this paper to define the \emph{single-particle} coupled-cluster Green's function (SP-CCGF) as 
\begin{equation}~\label{eq:sp_cc_gf}
    i\Tilde{G}_{pq}(t_1-t_2) = \braket{\Tilde{\Psi}_0|\mathcal{T}\left\{a_{p}(t_1)a^\dag_{q}(t_2)\right\}|\Phi_0} \ ,
\end{equation}
where the time-dependence of the operators is governed by the similarity transformed Hamiltonian, $\Bar{H}$ via Eq.~\ref{eq:bio_heisenberg}. This propagator arises simply by application of the general non-hermitian theory presented in Section~\ref{sec:non_fund} to the coupled-cluster similarity transformed Hamiltonian. As we shall demonstrate, it is by defining this quantity that we can develop a consistent and formally exact diagrammatic theory of the coupled-cluster self-energy. Therefore, Eq.~\ref{eq:sp_cc_gf} represents the diagrammatic Green's function native to both coupled-cluster and Green's function formalisms. We note that $\tilde{G}_{ia}(\omega) = \Bar{G}_{ia}(\omega)$. 

This single-particle propagator (Eq.~\ref{eq:sp_cc_gf}) remains coupled to all higher-order many-body Green's functions through its exact equation-of-motion. It will subsequently be shown that $\tilde{G}$ also contains the exact charged excitation spectrum as well as the coupled-cluster ground-state correlation energy through its associated self-energy. In the following we refer to $G$ as the electronic Green's function, $\bar{G}$ as the coupled-cluster representation of the electronic Green's function and $\tilde{G}$ as the single-particle coupled-cluster Green's function.

\subsection{Lehmann representation of the single-particle coupled-cluster Green's function}

The diagrammatic \emph{single-particle} coupled-cluster Green's function is defined in Eq.~\ref{eq:sp_cc_gf}. By expanding the time-ordered product, resolving the identity of $(N\pm1)$-particle eigenstates of $\Bar{H}$ and taking the Fourier transform, we find the Lehmann representation of the SP-CCGF:
\begin{gather}
    \begin{split}~\label{eq:lehmann}
        \Tilde{G}_{pq}(\omega) &= \sum_{\mu}\frac{\braket{\Tilde{\Psi}_0|a_p|\bar{\Psi}^{N+1}_{\mu}}\braket{\Tilde{\Psi}^{N+1}_{\mu}|a^\dag_q|\Phi_0}}{\omega-\varepsilon^{N+1}_{\mu}+i\eta}\\
        &+\sum_{\nu}\frac{\braket{\Tilde{\Psi}_0|a^\dag_q|\bar{\Psi}^{N-1}_{\nu}}\braket{\Tilde{\Psi}^{N-1}_{\nu}|a_p|\Phi_0}}{\omega-\varepsilon^{N-1}_{\nu}-i\eta} \ .
    \end{split}
\end{gather}
We immediately see that the SP-CCGF contains the exact single-particle spectrum and we will subsequently show how the associated self-energy also contains the exact ground state energy. The only difference between the SP-CCGF and the coupled-cluster representation of the electronic Green's function (Eq.~\ref{eq:cc_leh}) arises in the form of the residues, which in Eq.~\ref{eq:cc_leh} are different due to the explicit contribution of the higher-order $N$-particle Green's functions.  We introduce the residues of the coupled-cluster propagator defined in Eq.~\ref{eq:lehmann} as
$\Bar{Y}^{\mu}_{p} = \braket{\tilde{\Psi}^{N+1}_{\mu}|a^\dag_{p}|\Phi_0}, \tilde{Y}^{\mu}_{p} = \braket{\tilde{\Psi}_{0}|a_{p}|\bar{\Psi}^{N+1}_{\mu}}, \Bar{X}^{\nu}_{p} = \braket{\tilde{\Psi}^{N-1}_{\nu}|a_{p}|\Phi_0}$ and $ \tilde{X}^{\nu}_{p} = \braket{\tilde{\Psi}_{0}|a^\dag_{p}|\bar{\Psi}^{N-1}_{\nu}}$, and re-write the single-particle coupled-cluster Green's function as 
\begin{gather}
    \begin{split}~\label{eq:exact_leh}
        \tilde{G}_{pq}(\omega) &= \sum_{\mu}\frac{\tilde{Y}^{\mu}_{p}\bar{Y}^{\mu}_{q}}{\omega-\varepsilon^{N+1}_{\mu}+i\eta}+\sum_{\nu}\frac{\Tilde{X}^{\nu}_{q}\bar{X}^{\nu}_{p}}{\omega-\varepsilon^{N-1}_{\nu}-i\eta} \ .
    \end{split}
\end{gather}
The four blocks of the SP-CCGF are given by:
\begin{subequations}
    \begin{align}
        \begin{split}
            \tilde{G}_{ij}(\omega) &= \sum_{\nu} \frac{\Tilde{X}^{\nu}_{j}\bar{X}^{\nu}_{i}}{\omega-\varepsilon^{N-1}_{\nu}-i\eta}
        \end{split}\\
        \begin{split}
            \tilde{G}_{ab}(\omega) &= \sum_{\mu}\frac{\tilde{Y}^{\mu}_{a}\bar{Y}^{\mu}_{b}}{\omega-\varepsilon^{N+1}_{\mu}+i\eta}
        \end{split}\\
    \begin{split}
        \tilde{G}_{ia}(\omega) &= \sum_{\mu}\frac{\tilde{Y}^{\mu}_{i}\bar{Y}^{\mu}_{a}}{\omega-\varepsilon^{N+1}_{\mu}+i\eta}+\sum_{\nu}\frac{\Tilde{X}^{\nu}_{a}\bar{X}^{\nu}_{i}}{\omega-\varepsilon^{N-1}_{\nu}-i\eta} 
    \end{split}\\
    \begin{split}
        \tilde{G}_{ai}(\omega) &= 0  \ .
    \end{split}
\end{align}
\end{subequations}
Note that $\bar{Y}^{\mu}_{i}=\bar{X}^{\nu}_{a}=0$. Therefore, we see that the occupied block contains only backward time contributions, the virtual block contains only forward time contributions, the occupied--virtual block contains both forward and backward time contributions and that the virtual--occupied block vanishes.

\subsection{The irreducible coupled-cluster self-energy}~\label{sub:spec_se}

The perturbative expression for the SP-CCGF is given by (Eq.~\ref{eq:ext_GM_1}) 
    \begin{gather}
\begin{split}~\label{eq:cc_gf}
    i&\Tilde{G}_{pq}(t_1-t_2) = \\
    &\langle\Phi_0|\mathcal{T}\left\{\exp\left(-i\int^{\infty}_{-\infty}dt\  \Bar{H}_1(t)\right)a_{p}(t_1)a^\dag_q(t_2)\right\}|\Phi_0\rangle_{_\mathcal{C}} ,
\end{split}
\end{gather}
where the operators are implicitly defined in the biorthogonal Heisenberg picture in the first equality and the biorthogonal Interaction picture in the perturbative expansion: $a_{p}(t)=e^{iFt}a_pe^{-iFt}$. Here, we have taken the reference state $\ket{\Phi_0}$ corresponding to the Fock operator $F$, and have dropped the implicit limit of $\eta\to0$. This procedure was outlined in Section~\ref{sec:bio_qm} and results in a diagrammatic perturbation theory for the Green's function. 

The set of diagrams generated by Eq.~\ref{eq:cc_gf} is reduced by collecting all 1PI diagrams and performing a geometric series summation through the coupled-cluster Dyson equation 
\begin{equation}~\label{eq:cc_dyson}
    \Tilde{G}_{pq}(\omega) = G^{0}_{pq}(\omega) + \sum_{rs}G^{0}_{pr}(\omega)\Tilde{\Sigma}_{rs}(\omega)\tilde{G}_{sq}(\omega) ,
\end{equation}
where $G^{0}_{pq}(\omega)$ is the reference Green's function of the Fock operator and $\Tilde{\Sigma}_{pq}(\omega)$ is the coupled-cluster self-energy. As $\tilde{G}_{ia}=\bar{G}_{ia}$, it should be noted that $\tilde{\Sigma}_{ia}=\bar{\Sigma}_{ia}$. 

From the Lehmann representation of the SP-CCGF (Eq.~\ref{eq:lehmann}), we immediately identify the spectral form of the coupled-cluster self-energy as
\begin{gather}
    \begin{split}~\label{eq:spec_se}
        \Tilde{\Sigma}_{pq}(\omega) &= \Tilde{\Sigma}^{\infty}_{pq} + \Tilde{\Sigma}^{\F}_{pq}(\omega)+\Tilde{\Sigma}^{\Ba}_{pq}(\omega) \\
        &= \Tilde{\Sigma}^{\infty}_{pq} + \sum_{JJ'} \tilde{U}_{pJ}\left[(\omega+i\eta)\mathbbm{1}-\mathbf{(\bar{K}}^{>}+\bar{\mathbf{C}}^{>})\right]^{-1}_{JJ'}\bar{U}_{J'q}\\
        &+ \sum_{AA'} \bar{V}_{pA}\left[(\omega-i\eta)\mathbbm{1}-(\bar{\mathbf{K}}^{<}+\bar{\mathbf{C}}^{<})\right]^{-1}_{AA'}\tilde{V}_{A'q} \ .
    \end{split}
\end{gather}
Here $\tilde{\Sigma}^{\infty}_{pq}$ is the exact static component of the coupled-cluster self-energy. The indices $JJ'/AA'$ label forward-time/backward-time multi-particle-hole Intermediate State Configurations (ISCs). ISCs are excited-state configurations that contain $(N\pm1)$-particles resulting from interactions a single-particle experiences as it propagates through the system. The matrices $\mathbf{\bar{K}}^{>}_{JJ'}+\mathbf{\bar{C}}^{>}_{JJ'}$ and $\mathbf{\bar{K}}^{<}_{AA'}+\mathbf{\bar{C}}^{<}_{AA'}$ represent the interactions between the different ISCs and the $\mathbf{\bar{K}}$-matrices are block-diagonal by definition ($\mathbf{\bar{K}}^{>}_{JJ'}=\mathbf{\bar{K}}^{>}_{JJ}\delta_{JJ'}$). The quantities $\tilde{U}_{pJ}/\bar{U}_{Jp}$ and $\bar{V}_{pJ}/\tilde{V}_{Ap}$ represent the coupling matrices that link initial and final single-particle states to the different ISCs. The spectral form of Eq.~\ref{eq:spec_se} was initially proposed in Ref.~\cite{coveney2023coupled} through the diagrammatic analysis obtained by the functional derivatives of the BCC Lagrangian. Here, we see that the spectral form of the self-energy arises directly from the structure of the SP-CCGF.

For the occupied-occupied block of the self-energy we have 
\begin{gather}
    \begin{split}~\label{eq:spec_occ}
        \tilde{\Sigma}_{ij}(\omega) = \tilde{\Sigma}^{\infty}_{ij} + \sum_{AA'}\bar{V}_{iA}\left[(\omega-i\eta)\mathbbm{1}-(\mathbf{\bar{K}}^{<}+\mathbf{\bar{C}}^{<})\right]^{-1}_{AA'}\tilde{V}_{A'j} .
    \end{split}
\end{gather}
The occupied block of the coupled-cluster self-energy only contains poles above the real axis and is a consequence of backward time propagation alone. The self-energy of the virtual-virtual block is 
\begin{gather}
    \begin{split}~\label{eq:spec_vir}
        \tilde{\Sigma}_{ab}(\omega) = \tilde{\Sigma}^{\infty}_{ab} + \sum_{JJ'}\tilde{U}_{aJ}\left[(\omega+i\eta)\mathbbm{1}-(\mathbf{\bar{K}}^{>}+\mathbf{\bar{C}}^{>})\right]^{-1}_{JJ'}\bar{U}_{J'b}
    \end{split}
\end{gather}
and contains poles only beneath the real axis resulting from forward time propagation only. The self-energy of the occupied-virtual block contains the combined forward and backward propagation and has poles both above and below the real axis as 
\begin{gather}
    \begin{split}~\label{eq:spec_ov}
        \tilde{\Sigma}_{ia}(\omega) &= \tilde{\Sigma}^{\infty}_{ia} + \sum_{JJ'}\tilde{U}_{iJ}\left[(\omega+i\eta)\mathbbm{1}-(\mathbf{\bar{K}}^{>}+\mathbf{\bar{C}}^{>})\right]^{-1}_{JJ'}\bar{U}_{J'a}\\
        &+\sum_{AA'}\bar{V}_{iA}\left[(\omega-i\eta)\mathbbm{1}-(\mathbf{\bar{K}}^{<}+\mathbf{\bar{C}}^{<})\right]^{-1}_{AA'}\tilde{V}_{A'a} .
    \end{split}
\end{gather}
As $\tilde{G}_{ai}(\omega)=0$, the self-energy of the virtual-occupied block also vanishes: $\tilde{\Sigma}_{ai}(\omega) = 0$. The vanishing virtual-occupied block of the self-energy is directly the result of the coupled-cluster amplitude equations (Eq.~\ref{eq:amps}).

An important limit is when the cluster operator vanishes, \emph{i.e.} $T=0$ ($\tilde{G}\to G$). In this case, the coupled-cluster Dyson equation (Eq.~\ref{eq:cc_dyson}) reduces to the electronic Dyson equation, Eq.~\ref{eq:dyson}, where $H_1$ is the bare two-body Coulomb interaction subtracted by the HF potential. The coupled-cluster self-energy then reduces to the two-body electronic self-energy.

\section{Diagrammatic perturbation expansion of the coupled-cluster self-energy}~\label{sec:pert}

The perturbative expansion for the irreducible coupled-cluster self-energy is formally defined by the set of all one-particle irreducible diagrams constructed from reference propagators contracted with interactions generated by normal-ordering the underlying many-body Hamiltonian with respect to the reference state. In this section, we derive the perturbative diagrammatic expansion of the coupled-cluster self-energy functional, $\tilde{\Sigma}[G_0]$, to third-order. 

The number of diagrams in the perturbative expansion for the coupled-cluster self-energy can be reduced by restricting the diagrammatic expansion to contain only interaction-irreducible diagrams~\cite{carbone2013self}. This results in a combined series that involves renormalization of the propagators and interaction vertices. Diagrams are referred to as interaction-reducible when they can be split into two diagrams by cutting an interaction vertex into two parts~\cite{carbone2013self,raimondi2018algebraic}. As long as only interaction-irreducible diagrams are included, where all propagators are dressed, the diagrammatic expansion for the coupled-cluster self-energy is equivalent to replacing the perturbation $\bar{H}_1$ in Eq.~\ref{eq:cc_gf} with the effective interaction 
\begin{gather}
    \begin{split}
        \Tilde{\bar{H}}_1 &= \sum_{pq} (\tilde{F}_{pq}-f_{pq}) a^\dag_pa_q+ \frac{1}{(2!)^2}\sum_{pq,rs} \tilde{\Xi}_{pq,rs}a^\dag_{p}a^\dag_qa_{s}a_r\\
        &+\frac{1}{(3!)^2}\sum_{\substack{pqr\\stu}}\tilde{\chi}_{pqr,stu} a^\dag_{p}a^\dag_q a^\dag_{r}a_ua_ta_s 
    + \cdots
    \end{split}
\end{gather}
obtained from the normal-ordering of the similarity transformed Hamiltonian with respect to the biorthogonal ground state expectation value (Eq.~\ref{eq:norm_tot}). Obtaining the perturbative expression for the coupled-cluster self-energy in terms of effective interactions resulting from normal-ordering with respect to the reference $\ket{\Phi_0}$, requires the relationships between the effective interactions, which are defined in Appendices~\ref{appendix_2} and~\ref{app_pert_int}. The effective interactions arising from normal-ordering with respect to the reference arise when expanding the self-energy with respect to the reference Green's function, $G_0$. Therefore, the self-energy expansion with respect to the non-interacting propagator is obtained by the expansion of interaction-irreducible interactions that enter 1PI, skeleton self-energy diagrams. 

In the following, we employ the antisymmetrized Hugenholtz-Shavitt-Bartlett diagram convention~\cite{hirata2024nonconvergence}. Diagrammatically, the effective interactions that arise when normal-ordering the Hamiltonian with respect to the reference state are represented by 
\begin{gather*}
    \begin{split}
        \begin{gathered}
            \begin{fmfgraph*}(40,40)
    \fmfcurved
    \fmfset{arrow_len}{3mm}
    \fmfleft{i1}
    \fmflabel{}{i1}
    \fmfright{o1}
    \fmf{dbl_wiggly}{i1,o1}
    \fmfdot{o1,i1}
    \fmflabel{$\substack{p\\r}$}{i1}
    \fmflabel{$\substack{q\\s}$}{o1}
\end{fmfgraph*}
        \end{gathered}\hspace{5mm}&= \chi_{pq,rs}\ ,\\
       \begin{gathered}
            \begin{fmfgraph*}(60,40)
    \fmfcurved
    \fmfset{arrow_len}{3mm}
    \fmfleft{i1}
    \fmflabel{}{i1}
    \fmfright{o1}
    \fmf{dashes}{i1,v1}
    \fmf{dashes}{v1,o1}
    \fmfdot{o1,i1,v1}
    \fmflabel{$\substack{p\\s}$}{i1}
    \fmflabel{$\substack{q\\ \\ t}$}{v1}
    \fmflabel{$\substack{r\\u}$}{o1}
\end{fmfgraph*}
\end{gathered} \hspace{5mm}&= \chi_{pqr,stu} \ ,\\
\begin{gathered}
            \begin{fmfgraph*}(60,40)
    \fmfcurved
    \fmfset{arrow_len}{3mm}
    \fmfleft{i1}
    \fmflabel{}{i1}
    \fmfright{o1}
    \fmf{dbl_dashes}{i1,v1}
    \fmf{dbl_dashes}{v1,v2}
    \fmf{dbl_dashes}{v2,o1}
    \fmfdot{o1,i1,v1,v2}
    \fmflabel{$\substack{p\\s}$}{i1}
    \fmflabel{$\substack{q\\ \\ t}$}{v1}
    \fmflabel{$\substack{r\\ \\ u}$}{v2}
    \fmflabel{$\substack{w\\v}$}{o1}
\end{fmfgraph*}
\end{gathered} \hspace{5mm}&= \chi_{pqrw,stuv} \ ,
    \end{split}
\end{gather*}
and so on for the higher-body terms. The effective interactions that arise from normal-ordering the Hamiltonian with respect to the biorthogonal expectation value are obtained by contracting the diagrams above with the corresponding $\lambda$-matrices as outlined in Appendix~\ref{appendix_2}. 

\subsection{First-order diagrams}

The first-order contribution in the perturbative expansion is the static term
\begin{gather}
\begin{split}\label{eq:cc_se_corr_energy}
     \Tilde{\Sigma}^{\infty(0)}_{pq}[G_0] = \hspace{5mm}\begin{gathered}
\begin{fmfgraph*}(40,40)
    \fmfset{arrow_len}{3mm}
    \fmfleft{i1,i2,i3}
    \fmfright{o1,o2,o3}
    \fmf{fermion,label=$q$,label.side=left}{i1,i2}
    \fmf{fermion,label=$p$,label.side=left}{i2,i3}
    \fmf{dbl_dashes}{i2,o2}
    \fmfforce{(0.0w,0.h)}{i1}
    \fmfforce{(0.0w,0.5h)}{i2}
    \fmfforce{(0.0w,1.0h)}{i3}
    \fmfdot{i2}
    \fmfv{decor.shape=cross,decor.filled=full, decor.size=1.5thic}{o2}
\end{fmfgraph*}
\end{gathered}\hspace{5mm}=\hspace{5mm} F_{pq}-f_{pq} \ .
\end{split}
\end{gather}
This term was derived in Ref.~\cite{coveney2023coupled} as the first contribution to the functional derivative of the BCC Lagrangian with respect to the non-interacting Green's function. However, Eq.~\ref{eq:cc_se_corr_energy} represents the generalization to coupled-cluster theory formulated about an arbitrary reference state.

Depending on the external time-orderings of the legs, this self-energy contains different sets of diagrams. For simplicity, we take the reference to be the HF state such that the reference Green's function is the HF Green's function. Focusing on the occupied-occupied contribution, we have 
    \begin{gather}~\label{eq:se_stat}
\begin{split}
     \Tilde{\Sigma}^{\infty(0)}_{ij} &= \hspace{5mm}\begin{gathered}
\begin{fmfgraph*}(40,40)
    \fmfset{arrow_len}{3mm}
    \fmfleft{i1,i2,i3}
    \fmfright{o1,o2,o3}
    \fmf{fermion,label=$j$,label.side=right}{i3,i2}
    \fmf{fermion,label=$i$,label.side=right}{i2,i1}
    \fmf{dbl_dashes}{i2,o2}
    \fmfforce{(0.0w,0.h)}{i1}
    \fmfforce{(0.0w,0.5h)}{i2}
    \fmfforce{(0.0w,1.0h)}{i3}
    \fmfdot{i2}
    \fmfv{decor.shape=cross,decor.filled=full, decor.size=1.5thic}{o2}
\end{fmfgraph*}
\end{gathered}\\
&=\hspace{5mm} 
\begin{gathered}
\begin{fmfgraph*}(40,40)
    \fmfset{arrow_len}{3mm}
    \fmfleft{i1,i2,i3}
    \fmfright{o1,o2,o3}
    \fmf{fermion,label=$j$,label.side=right}{i3,i2}
    \fmf{fermion,label=$i$,label.side=right}{i2,i1}
    \fmf{wiggly}{i2,o2}
    \fmf{fermion,left=0.3}{o2,o1}
    \fmf{fermion,left=0.3}{o1,o2}
    \fmfforce{(0.0w,0.h)}{i1}
    \fmfforce{(0.0w,0.5h)}{i2}
    \fmfforce{(0.0w,1.0h)}{i3}
    \fmfforce{(0.9w,0.5h)}{o2}
    \fmfdot{i2,o2}
    \fmf{plain}{v3,o1}
    \fmf{plain}{o1,v4}
    \fmfforce{(0.8w,0.0h)}{v3}
    \fmfforce{(1.0w,0.0h)}{v4}
\end{fmfgraph*}
\end{gathered}\hspace{2.5mm}+\hspace{5mm}
\begin{gathered}
\begin{fmfgraph*}(40,40)
\fmfset{arrow_len}{3mm}
    \fmfleft{i1,i2}
    \fmfright{o1,o2}
    \fmf{fermion}{i1,i2}
    \fmf{plain}{i1,o1}
    \fmf{wiggly}{i2,o2}
    \fmf{fermion,left=0.3}{o1,o2}
    \fmf{fermion,left=0.3}{o2,o1}
    \fmf{fermion,label=$j$,label.side=right}{v1,i1}
    \fmf{fermion,label=$i$,label.side=left}{i2,v2}
    \fmfforce{(0.0w,0.5h)}{v1}
    \fmfforce{(0.2w,0.0h)}{i1}
    \fmfforce{(0.2w,1.0h)}{i2}
    \fmfforce{(0.4w,0.5h)}{v2}
    \fmfdot{i2,o2}
\end{fmfgraph*}
\end{gathered}
\hspace{2.5mm}+\hspace{2.5mm}
\begin{gathered}
\begin{fmfgraph*}(40,40)
\fmfset{arrow_len}{3mm}
    \fmfleft{i1,i2}
    \fmfright{o1,o2}
    \fmf{fermion}{i1,i2}
    \fmf{wiggly}{i2,o2}
    \fmf{fermion,left=0.3}{o1,o2}
    \fmf{fermion,left=0.3}{o2,o1}
    \fmfdot{i2,o2}
    \fmf{plain}{v1,i1}
    \fmf{plain}{i1,v2}
    \fmf{plain}{v3,o1}
    \fmf{plain}{o1,v4}
    \fmf{fermion,label=$j$,label.side=right}{v5,i1}
    \fmf{fermion,label=$i$,label.side=left}{i2,v6}
    \fmfforce{(-0.05w,0.5h)}{v5}
    \fmfforce{(0.1w,0.0h)}{i1}
    \fmfforce{(0.1w,1.0h)}{i2}
    \fmfforce{(0.4w,0.5h)}{v6}
    \fmfforce{(0.00w,0.0h)}{v1}
    \fmfforce{(0.2w,0.0h)}{v2}
    \fmfforce{(0.8w,0.0h)}{v3}
    \fmfforce{(1.0w,0.0h)}{v4}
\end{fmfgraph*}
\end{gathered}
\end{split}
\end{gather}
where we have employed the standard Goldstone diagram notation of Ref.~\cite{shavitt2009many}. This set of Goldstone diagrams corresponds to the equation
\begin{gather}
\begin{split}~\label{eq:cc_stat_se}
    \Tilde{\Sigma}^{\infty(0)}_{ij} &= \frac{1}{2}\sum_{kab}\braket{ik||ab}\Big(t^{ab}_{jk}+2t^{a}_{j}t^{b}_{k}\Big) + \sum_{ka} \braket{ik||ja}t^{a}_{k} \ .
\end{split}
\end{gather}
Likewise, the first-order self-energy of the virtual-virtual states is given by
    \begin{gather*}
\begin{split}
     \Tilde{\Sigma}^{\infty(0)}_{ab} &= \hspace{5mm}\begin{gathered}
\begin{fmfgraph*}(40,40)
    \fmfset{arrow_len}{3mm}
    \fmfleft{i1,i2,i3}
    \fmfright{o1,o2,o3}
    \fmf{fermion,label=$a$,label.side=left}{i2,i3}
    \fmf{fermion,label=$b$,label.side=left}{i1,i2}
    \fmf{dbl_dashes}{i2,o2}
    \fmfforce{(0.0w,0.h)}{i1}
    \fmfforce{(0.0w,0.5h)}{i2}
    \fmfforce{(0.0w,1.0h)}{i3}
    \fmfdot{i2}
    \fmfv{decor.shape=cross,decor.filled=full, decor.size=1.5thic}{o2}
\end{fmfgraph*}
\end{gathered}\\
&=\hspace{5mm} 
\begin{gathered}
\begin{fmfgraph*}(40,40)
    \fmfset{arrow_len}{3mm}
    \fmfleft{i1,i2,i3}
    \fmfright{o1,o2,o3}
    \fmf{fermion,label=$a$,label.side=left}{i2,i3}
    \fmf{fermion,label=$b$,label.side=left}{i1,i2}
    \fmf{wiggly}{i2,o2}
    \fmf{fermion,left=0.3}{o2,o1}
    \fmf{fermion,left=0.3}{o1,o2}
    \fmfforce{(0.0w,0.h)}{i1}
    \fmfforce{(0.0w,0.5h)}{i2}
    \fmfforce{(0.0w,1.0h)}{i3}
    \fmfforce{(0.9w,0.5h)}{o2}
    \fmfdot{i2,o2}
    \fmf{plain}{v3,o1}
    \fmf{plain}{o1,v4}
    \fmfforce{(0.8w,0.0h)}{v3}
    \fmfforce{(1.0w,0.0h)}{v4}
\end{fmfgraph*}
\end{gathered}\hspace{2.5mm}+\hspace{5mm}
\begin{gathered}
\begin{fmfgraph*}(40,40)
\fmfset{arrow_len}{3mm}
    \fmfleft{i1,i2}
    \fmfright{o1,o2}
    \fmf{fermion}{i2,i1}
    \fmf{plain}{i1,o1}
    \fmf{wiggly}{i2,o2}
    \fmf{fermion,left=0.3}{o1,o2}
    \fmf{fermion,left=0.3}{o2,o1}
    \fmf{fermion,label=$a$,label.side=left}{i1,v1}
    \fmf{fermion,label=$b$,label.side=right}{v2,i2}
    \fmfforce{(0.0w,0.5h)}{v1}
    \fmfforce{(0.2w,0.0h)}{i1}
    \fmfforce{(0.2w,1.0h)}{i2}
    \fmfforce{(0.4w,0.5h)}{v2}
    \fmfdot{i2,o2}
\end{fmfgraph*}
\end{gathered}
\hspace{2.5mm}+\hspace{2.5mm}
\begin{gathered}
\begin{fmfgraph*}(40,40)
\fmfset{arrow_len}{3mm}
    \fmfleft{i1,i2}
    \fmfright{o1,o2}
    \fmf{fermion}{i2,i1}
    \fmf{wiggly}{i2,o2}
    \fmf{fermion,left=0.3}{o1,o2}
    \fmf{fermion,left=0.3}{o2,o1}
    \fmfdot{i2,o2}
    \fmf{plain}{v1,i1}
    \fmf{plain}{i1,v2}
    \fmf{plain}{v3,o1}
    \fmf{plain}{o1,v4}
    \fmf{fermion,label=$a$,label.side=left}{i1,v5}
    \fmf{fermion,label=$b$,label.side=right}{v6,i2}
    \fmfforce{(-0.05w,0.5h)}{v5}
    \fmfforce{(0.1w,0.0h)}{i1}
    \fmfforce{(0.1w,1.0h)}{i2}
    \fmfforce{(0.4w,0.5h)}{v6}
    \fmfforce{(0.0w,0.0h)}{v1}
    \fmfforce{(0.2w,0.0h)}{v2}
    \fmfforce{(0.8w,0.0h)}{v3}
    \fmfforce{(1.0w,0.0h)}{v4}
\end{fmfgraph*}
\end{gathered}
\end{split}
\end{gather*}
which is written as  
\begin{gather}
    \begin{split}
        \Tilde{\Sigma}^{\infty(0)}_{ab} &= -\frac{1}{2}\sum_{ijc}\braket{ij||bc}\Big(t^{ac}_{ij}+2t^{a}_{i}t^{c}_{j}\Big) + \sum_{kc} \braket{ak||bc}t^{c}_{k} \ .
    \end{split}
\end{gather}
The contribution to the occupied-virtual block is given by $\Tilde{\Sigma}^{\infty(0)}_{ai} = \braket{\Phi^{a}_{i}|\Bar{H}|\Phi_0} = 0$, which are the singles amplitude equations. The virtual--occupied self-energy contribution is given by 
\begin{gather*}
\begin{split}
    \tilde{\Sigma}^{\infty(0)}_{ia} = \hspace{5mm}\begin{gathered}
\begin{fmfgraph*}(40,40)
    \fmfset{arrow_len}{3mm}
    \fmfleft{i1,i2}
    \fmfright{o1,o2}
    \fmf{fermion,label=$i$,label.side=right}{i2,v1}
    \fmf{fermion,label=$a$,label.side=right}{i1,i2}
    \fmf{wiggly}{i2,o2}
    \fmf{fermion,left=0.3}{o2,o1}
    \fmf{fermion,left=0.3}{o1,o2}
    \fmfforce{(0.2w,0.0h)}{i1}
    \fmfforce{(0.0w,0.0h)}{v1}
    \fmfforce{(0.1w,0.5h)}{i2}
    \fmfforce{(0.9w,0.5h)}{o2}
    \fmfdot{i2,o2}
    \fmf{plain}{v3,o1}
    \fmf{plain}{o1,v4}
    \fmfforce{(0.8w,0.0h)}{v3}
    \fmfforce{(1.0w,0.0h)}{v4}
\end{fmfgraph*}
\end{gathered}\hspace{2.5mm}=  \sum_{kb}\braket{ik||ab}t^{b}_{k} \ .
\end{split}
\end{gather*}
These terms represent the most important 1PI contribution to the coupled-cluster self-energy~\cite{coveney2023coupled}. It should be noted that if we work with the Brueckner reference, the first-order contribution decouples the IP/EA sectors of the self-energy. From this analysis, we see the general feature of the coupled-cluster self-energy in that it combines Feynman diagrams in terms of propagators with Goldstone diagrams that represent the interaction vertices arising from the similarity transformed Hamiltonian. 

When the similarity transformation is not present ($T=0$), our expression reduces to the perturbative electronic self-energy expansion and hence $\tilde{\Sigma}^{T=0,\infty(0)}_{pq} = \Sigma^{\infty(0)}_{pq}=0$ as required.

\subsection{Second-order diagrams}

At second-order, both static and frequency-dependent terms arise.
The only explicitly frequency-dependent second-order perturbative contribution to the coupled-cluster self-energy is the diagram 
\begin{gather}
\begin{split}~\label{eq:cc_II_pt}
\tilde{\Sigma}^{(2)}[G_0]= 
\begin{gathered}
    \begin{fmfgraph*}(50,50)
    \fmfcurved
    \fmfset{arrow_len}{3mm}
    \fmfleft{i1,i2}
    \fmflabel{}{i1}
    \fmflabel{}{i2}
    \fmfright{o1,o2}
    \fmflabel{}{o1}
    \fmflabel{}{o2}
    \fmf{fermion}{i1,i2}
    \fmf{dbl_wiggly}{o1,i1}
    \fmf{fermion,left=0.3,tension=0}{o1,o2}
    \fmf{dbl_wiggly}{o2,i2}
    \fmf{fermion,left=0.3,tension=0}{o2,o1}
    \fmfdot{o1,o2,i1,i2}
\end{fmfgraph*}
\end{gathered}
\end{split}
\end{gather}
Fixing the external indices, the Feynman diagram contains two contributions corresponding to forward and backward time orderings
\begin{gather}
\begin{split}~\label{eq:2nd_order}
    \tilde{\Sigma}^{(2)}_{pq}(\omega) &=\frac{1}{2}\Bigg(\sum_{abi}\frac{\chi_{pi,ab}\chi_{ab,qi}}{\omega+\epsilon_i-\epsilon_a-\epsilon_b+i\eta} \\
    &\hspace{7.5mm}+ \sum_{ija}\frac{\chi_{pa,ij}\chi_{ij,qa}}{\omega+\epsilon_a-\epsilon_i-\epsilon_j-i\eta}\Bigg) \ .
\end{split}
\end{gather}
If both external indices are in the occupied space, we have
\begin{gather}
\begin{split}
    \tilde{\Sigma}^{(2)}_{ij}(\omega) &= \frac{1}{2}\Bigg(\sum_{kab}\frac{\chi_{ik,ab}\chi_{ab,jk}}{\omega+\epsilon_k-\epsilon_a-\epsilon_b+i\eta} \\
    &\hspace{7.5mm}+ \sum_{kla}\frac{\chi_{ia,kl}\chi_{kl,ja}}{\omega+\epsilon_a-\epsilon_k-\epsilon_l-i\eta}\Bigg) \ .
\end{split}
\end{gather}
However, since the matrix element $\chi_{ab,jk} = \braket{\Phi^{ab}_{jk}|\Bar{H}|\Phi_0} = 0$ via the amplitude equations, the forward-time contribution of the second-order self-energy within the occupied-occupied subspace vanishes. Therefore, we are left with 
\begin{equation}
    \tilde{\Sigma}^{(2)}_{ij}(\omega) = \frac{1}{2}\sum_{kla}\frac{\chi_{ia,kl}\chi_{kl,ja}}{\omega+\epsilon_a-\epsilon_k-\epsilon_l-i\eta} \ .
\end{equation}
This expression is exactly of the spectral form of the exact self-energy given in Eq.~\ref{eq:spec_occ}. Likewise, the second-order contribution to the virtual-virtual block is
\begin{equation}
     \tilde{\Sigma}^{(2)}_{ab}(\omega) = \frac{1}{2}\sum_{cdk} \frac{\chi_{ak,cd}\chi_{cd,bk}}{\omega+\epsilon_{k}-\epsilon_{c}-\epsilon_{d}+i\eta} \ .
\end{equation}
In this case, the backward time-contribution vanishes due to the doubles amplitude equations $\chi_{ac,ij} = 0$. Again, this contribution is exactly of the form given in Eq.~\ref{eq:spec_vir}. Turning to the second-order contribution corresponding to hole to particle propagation we have $\tilde{\Sigma}^{(2)}_{ai} = 0 $. This is because both $\chi_{bc,ij} = \chi_{ab,jk} = 0$. However, both time-orderings remain in the particle to hole block
\begin{gather}
\begin{split}
    \tilde{\Sigma}^{(2)}_{ia}(\omega) &= \frac{1}{2}\Bigg(\sum_{bcj} \frac{\chi_{ij,bc}\chi_{bc,aj}}{\omega+\epsilon_{j}-\epsilon_{b}-\epsilon_{c}+i\eta} \\
    &\hspace{7.5mm}+ \sum_{jkb}\frac{\chi_{ib,jk}\chi_{jk,ab}}{\omega+\epsilon_{b}-\epsilon_{j}-\epsilon_{k}-i\eta}\Bigg) \ .
\end{split}
\end{gather}
Therefore, at second-order, the perturbative expansion of the coupled-cluster self-energy is exactly that of the spectral form outlined in Subsection~\ref{sub:spec_se}. 

The forward-time dynamical contribution to the coupled-cluster self-energy in the occupied-occupied space vanishes as a result of the similarity transformation. These contributions appear instead in the first-order static term, $\tilde{\Sigma}^{\infty(0)}_{ij}$. This is because the forward-time contributions that would appear in the second-order electronic self-energy are instead contained in the doubles amplitudes $t^{ab}_{ij}$ that enter the similarity transformation. Therefore, if these forward-time contributions were also included in the second-order dynamical coupled-cluster self-energy, they would be double-counted. The same structure is observed for the backward time contribution of the coupled-cluster self-energy in the virtual-virtual space which are now contained in the static contribution, $\tilde{\Sigma}^{\infty(0)}_{ab}$. The occupied-virtual block of the coupled-cluster self-energy must vanish in order to consistently determine the doubles amplitudes that contain these contributions. When $T=0$, the second-order expression for the coupled-cluster self-energy in Eq.~\ref{eq:2nd_order} reduces the MP2 self-energy, \emph{i.e.} $\tilde{\Sigma}^{T=0(2)}_{pq}(\omega) = \Sigma^{\MP2}_{pq}(\omega)$ as $\chi_{pq,rs}\to\braket{pq||rs}$. 

Other second-order diagrams such as 
\begin{gather}
\begin{split}~\label{eq:3+4_II_pt}
\begin{gathered}
    \begin{fmfgraph*}(70,50)
    \fmfcurved
    \fmfset{arrow_len}{3mm}
    \fmfleft{i1,i2}
    \fmflabel{}{i1}
    \fmflabel{}{i2}
    \fmfright{o1,o2}
    \fmflabel{}{o1}
    \fmflabel{}{o2}
    \fmf{fermion}{i1,i2}
    \fmf{dashes}{v1,i1}
    \fmf{dashes}{o1,v1}
    \fmf{fermion,left=0.3,tension=0}{o1,o2}
    \fmf{fermion,left=0.3,tension=0}{v1,v2}
    \fmf{fermion,left=0.3,tension=0}{v2,v1}
    \fmf{dashes}{v2,i2}
    \fmf{dashes}{o2,v2}
    \fmf{fermion,left=0.3,tension=0}{o2,o1}
    \fmfdot{o1,o2,i1,i2,v1,v2}
\end{fmfgraph*}
\end{gathered}
\hspace{5mm}+\hspace{5mm}
\begin{gathered}
    \begin{fmfgraph*}(70,50)
    \fmfcurved
    \fmfset{arrow_len}{3mm}
    \fmfleft{i1,i2}
    \fmflabel{}{i1}
    \fmflabel{}{i2}
    \fmfright{o1,o2}
    \fmflabel{}{o1}
    \fmflabel{}{o2}
    \fmf{fermion}{i1,i2}
    \fmf{dbl_dashes}{v1,i1}
    \fmf{dbl_dashes}{o1,v1}
    \fmf{dbl_dashes}{v1,v3}
    \fmf{dbl_dashes}{v3,o1}
    \fmf{dbl_dashes}{v2,v4}
    \fmf{dbl_dashes}{v4,o2}
    \fmf{fermion,left=0.3,tension=0}{o1,o2}
    \fmf{fermion,left=0.3,tension=0}{v1,v2}
    \fmf{fermion,left=0.3,tension=0}{v2,v1}
    \fmf{fermion,left=0.3,tension=0}{v3,v4}
    \fmf{fermion,left=0.3,tension=0}{v4,v3}
    \fmf{dbl_dashes}{v2,i2}
    \fmf{dbl_dashes}{o2,v2}
    \fmf{fermion,left=0.3,tension=0}{o2,o1}
    \fmfdot{o1,o2,i1,i2,v1,v2,v3,v4}
    \fmfforce{(0.0w,0.0h)}{i1}
    \fmfforce{(1.0w,0.0h)}{o1}
    \fmfforce{(0.25w,1h)}{v2}
    \fmfforce{(0.25w,0.0h)}{v1}
    \fmfforce{(0.625w,0.0h)}{v3}
    \fmfforce{(0.625w,1.0h)}{v4}
    \fmfforce{(0.0w,1.0h)}{i2}
    \fmfforce{(1.0w,1.0h)}{o2}
\end{fmfgraph*}
\end{gathered}
\hspace{5mm}+\hspace{5mm} \cdots
\end{split}
\end{gather}
and so on, containing elements up to $N$-body from the similarity transformed Hamiltonian, vanish as a result of having four or more lines below any interaction vertex. The coupled-cluster self-energy diagram of Eq.~\ref{eq:cc_II_pt} represents an excitation processes whereby the intermediate scattered states consist of excited states containing two-particle-one-hole (2p1h) and two-hole-one-particle (2h1p) configurations. Likewise, the diagrams of Eq.~\ref{eq:3+4_II_pt} contain intermediate scattered states of 3p2h/3h2p and 4p3h/4h3p character, which vanish within coupled-cluster theory, but will contribute in the case of a generic $N$-body interaction Hamiltonian. 

The final diagrams that appear at second order are the static terms where the two-body effective interaction is contracted with the static self-energy and where the three-body effective interaction is contracted with the two-body interaction:
   \begin{gather} 
    \begin{split}
    \tilde{\Sigma}^{\infty(2)}_{pq}[G_0] = \hspace{2.5mm}
\begin{gathered}
    \begin{fmfgraph*}(40,40)
    \fmfcurved
    \fmfset{arrow_len}{3mm}
    \fmfleft{i1,i2}
    \fmflabel{}{i1}
    \fmflabel{}{i2}
    \fmfright{o1,o2}
    \fmflabel{}{o1}
    \fmflabel{}{o2}
    \fmf{dbl_wiggly}{i1,o1}
    \fmf{fermion,left=0.3,tension=0}{o1,o2}
    \fmf{dbl_dashes}{o2,i2}
    \fmf{fermion,left=0.3,tension=0}{o2,o1}
    \fmfdot{o1,o2,i1}
    \fmfv{decor.shape=cross,decor.size=1.0thic, decor.filled=full}{i2}
\end{fmfgraph*}
\end{gathered}\hspace{5mm}+\hspace{1.0mm}
        \begin{gathered}
    \begin{fmfgraph*}(70,50)
    \fmfcurved
    \fmfset{arrow_len}{3mm}
    \fmfleft{i1,i2}
    \fmflabel{}{i1}
    \fmflabel{}{i2}
    \fmfright{o1,o2}
    \fmflabel{}{o1}
    \fmflabel{}{o2}
    \fmf{dashes}{o1,v1}
    \fmf{dashes}{i1,v1}
    \fmf{fermion,left=0.3,tension=0}{o1,o2}
    \fmf{fermion,left=0.3,tension=0}{v1,v2}
    \fmf{fermion,left=0.3,tension=0}{v2,v1}
    \fmf{phantom}{v2,i2}
    \fmf{dbl_wiggly}{o2,v2}
    \fmf{fermion,left=0.3,tension=0}{o2,o1}
    \fmfdot{o1,o2,i1,v1,v2}
\end{fmfgraph*}
\end{gathered} 
    \end{split}
\end{gather}
These static contributions evaluate to give
\begin{gather}
    \begin{split}~\label{eq:se_pt_2}
        \tilde{\Sigma}^{\infty(2)}_{pq} &= \sum_{kc}\frac{\chi_{pc,qk}\tilde{\Sigma}^{\infty(0)}_{kc}}{(\epsilon_k-\epsilon_c+i\eta)}\\
        &+\frac{1}{(2!)^2}\sum_{abij} \frac{\chi_{pab,qij}\chi_{ij,ab}}{(\epsilon_{i}+\epsilon_{j}-\epsilon_{a}-\epsilon_{b}+i\eta)} \ .
    \end{split}
\end{gather}
The contributions resulting from the contraction of the four-body with the three-body effective interaction and so on vanish as a result of containing four or more lines below the interaction vertex. In Brueckner theory, the first term vanishes because $\tilde{\Sigma}^{\infty(0)}_{kc}=0$~\cite{coveney2023coupled}. However, it should be noted that these terms are manifestly interaction-reducible as they result from the perturbative expansion of the exact static self-energy, $\tilde{\Sigma}^{\infty}_{pq}$.

\subsection{Third-order diagrams}

In order to reduce the number of diagrams depicted at third-order, we present only the 1PI, skeleton and interaction-irreducible coupled-cluster self-energy diagrams. This restriction requires the self-energy to be written in terms of fully dressed Green's functions because the self-energy insertions are already included in the propagator renormalization. They are depicted below:
\begin{widetext}
    \begin{gather}
    \begin{split}~\label{eq:3rd_order}
    \tilde{\Sigma}^{(3)}[G_0] &= \hspace{2.5mm} 
    \begin{gathered}
    \begin{fmfgraph*}(50,70)
    \fmfcurved
    \fmfset{arrow_len}{3mm}
    \fmfleft{i1,i2,i3}
    \fmflabel{}{i1}
    \fmflabel{}{i2}
    \fmfright{o1,o2,o3}
    \fmflabel{}{o1}
    \fmflabel{}{o2}
    \fmf{dbl_wiggly}{i2,o2}
    \fmf{dbl_wiggly}{i3,o3}
    \fmf{dbl_wiggly}{i1,o1}
    \fmf{fermion}{i1,i2}
    \fmf{fermion}{i2,i3}
    \fmf{fermion}{o1,o2}
    \fmf{fermion}{o2,o3}
    \fmf{fermion,left=0.3}{o3,o1}
    \fmfforce{(0.0w,0.0h)}{i1}
    \fmfforce{(1.0w,0.0h)}{o1}
    \fmfforce{(0.0w,1.0h)}{i3}
    \fmfforce{(1.0w,1.0h)}{o3}
    \fmfdot{i1,i2,i3}
    \fmfdot{o1,o2,o3}
\end{fmfgraph*}
\end{gathered}
\hspace{10mm}+\hspace{5mm}
\begin{gathered}
\begin{fmfgraph*}(50,70)
    \fmfcurved
    \fmfset{arrow_len}{3mm}
    \fmfleft{i1,i2,i3}
    \fmflabel{}{i1}
    \fmflabel{}{i2}
    \fmfright{o1,o2,o3}
    \fmflabel{}{o1}
    \fmflabel{}{o2}
    \fmf{dbl_wiggly}{i1,v1}
    \fmf{dbl_wiggly}{v1,o1}
    \fmf{dbl_wiggly}{v2,o2}
    \fmf{dbl_wiggly}{i3,v3}
    \fmf{fermion}{i1,i3}
    \fmf{fermion,left=0.3}{o1,o2}
    \fmf{fermion,left=0.3}{o2,o1}
    \fmf{fermion,left=0.3}{v2,v3}
    \fmf{fermion,left=0.3}{v3,v2}
    \fmfforce{(0.0w,0.0h)}{i1}
    \fmfforce{(1.0w,0.0h)}{o1}
    \fmfforce{(0.5w,0.5h)}{v2}
    \fmfforce{(0.5w,0.0h)}{v1}
    \fmfforce{(0.5w,1.0h)}{v3}
    \fmfforce{(0.0w,1.0h)}{i3}
    \fmfforce{(1.0w,1.0h)}{o3}
    \fmfdot{v2,v3}
    \fmfdot{i1,i3}
    \fmfdot{o1,o2}
\end{fmfgraph*}
\end{gathered}
\hspace{5mm}+\hspace{5mm}
\begin{gathered}
    \begin{fmfgraph*}(50,70)
    \fmfcurved
    \fmfset{arrow_len}{3mm}
    \fmfleft{i1,i2,i3}
    \fmflabel{}{i1}
    \fmflabel{}{i2}
    \fmfright{o1,o2,o3}
    \fmflabel{}{o1}
    \fmflabel{}{o2}
    \fmf{dbl_wiggly}{i1,v1}
    \fmf{dashes}{i2,v2}
    \fmf{dashes}{v2,o2}
    \fmf{dbl_wiggly}{i3,v3}
    \fmf{fermion}{i1,i2}
    \fmf{fermion}{v1,v2}
    \fmf{fermion}{o2,v1}
    \fmf{fermion}{v2,v3}
    \fmf{fermion}{v3,o2}
    \fmf{fermion}{i2,i3}
    \fmfforce{(0.0w,0.0h)}{i1}
    \fmfforce{(1.0w,0.0h)}{o1}
    \fmfforce{(0.5w,0.5h)}{v2}
    \fmfforce{(0.5w,0.0h)}{v1}
    \fmfforce{(0.5w,1.0h)}{v3}
    \fmfforce{(0.0w,1.0h)}{i3}
    \fmfforce{(1.0w,1.0h)}{o3}
    \fmfdotn{v}{3}
    \fmfdot{i1,i2,i3}
    \fmfdot{o2}
\end{fmfgraph*}
\end{gathered}\hspace{5mm}+\hspace{5mm}
\begin{gathered}
    \begin{fmfgraph*}(50,70)
    \fmfcurved
    \fmfset{arrow_len}{3mm}
    \fmfleft{i1,i2,i3}
    \fmflabel{}{i1}
    \fmflabel{}{i2}
    \fmfright{o1,o2,o3}
    \fmflabel{}{o1}
    \fmflabel{}{o2}
    \fmf{dashes}{i1,v1}
    \fmf{dashes}{v1,o1}
    \fmf{dbl_wiggly}{v2,o2}
    \fmf{dbl_wiggly}{i3,v3}
    \fmf{fermion}{i1,i3}
    \fmf{fermion}{v1,v2}
    \fmf{fermion,right=0.3}{v3,v1}
    \fmf{fermion,left=0.3}{o1,o2}
    \fmf{fermion,left=0.3}{o2,o1}
    \fmf{fermion}{v2,v3}
    \fmfforce{(0.0w,0.0h)}{i1}
    \fmfforce{(1.0w,0.0h)}{o1}
    \fmfforce{(0.5w,0.5h)}{v2}
    \fmfforce{(0.5w,0.0h)}{v1}
    \fmfforce{(0.5w,1.0h)}{v3}
    \fmfforce{(0.0w,1.0h)}{i3}
    \fmfforce{(1.0w,1.0h)}{o3}
    \fmfdotn{v}{3}
    \fmfdot{i1,i3}
    \fmfdot{o1,o2}
\end{fmfgraph*}
\end{gathered}\\
\\
&+\hspace{5mm}
\begin{gathered}
    \begin{fmfgraph*}(50,70)
    \fmfcurved
    \fmfset{arrow_len}{3mm}
    \fmfleft{i1,i2,i3}
    \fmflabel{}{i1}
    \fmflabel{}{i2}
    \fmfright{o1,o2,o3}
    \fmflabel{}{o1}
    \fmflabel{}{o2}
    \fmf{dashes}{i1,v1}
    \fmf{dashes}{v1,o1}
    \fmf{dbl_wiggly}{v2,o2}
    \fmf{dbl_wiggly}{i3,v3}
    \fmf{fermion}{i1,i3}
    \fmf{fermion}{v2,v1}
    \fmf{fermion,left=0.3}{v1,v3}
    \fmf{fermion,left=0.3}{o1,o2}
    \fmf{fermion,left=0.3}{o2,o1}
    \fmf{fermion}{v3,v2}
    \fmfforce{(0.0w,0.0h)}{i1}
    \fmfforce{(1.0w,0.0h)}{o1}
    \fmfforce{(0.5w,0.5h)}{v2}
    \fmfforce{(0.5w,0.0h)}{v1}
    \fmfforce{(0.5w,1.0h)}{v3}
    \fmfforce{(0.0w,1.0h)}{i3}
    \fmfforce{(1.0w,1.0h)}{o3}
    \fmfdotn{v}{3}
    \fmfdot{i1,i3}
    \fmfdot{o1,o2}
\end{fmfgraph*}
\end{gathered}\hspace{5mm}+\hspace{5mm}
\begin{gathered}
    \begin{fmfgraph*}(50,70)
    \fmfcurved
    \fmfset{arrow_len}{3mm}
    \fmfleft{i1,i2,i3}
    \fmflabel{}{i1}
    \fmflabel{}{i2}
    \fmfright{o1,o2,o3}
    \fmflabel{}{o1}
    \fmflabel{}{o2}
    \fmf{dbl_wiggly}{i1,v1}
    \fmf{dbl_wiggly}{v2,o2}
    \fmf{dashes}{i3,v3}
    \fmf{dashes}{v3,o3}
    \fmf{fermion}{i1,i3}
    \fmf{fermion}{v2,v1}
    \fmf{fermion,left=0.3}{v1,v3}
    \fmf{fermion,left=0.3}{o3,o2}
    \fmf{fermion,left=0.3}{o2,o3}
    \fmf{fermion}{v3,v2}
    \fmfforce{(0.0w,0.0h)}{i1}
    \fmfforce{(1.0w,0.0h)}{o1}
    \fmfforce{(0.5w,0.5h)}{v2}
    \fmfforce{(0.5w,0.0h)}{v1}
    \fmfforce{(0.5w,1.0h)}{v3}
    \fmfforce{(0.0w,1.0h)}{i3}
    \fmfforce{(1.0w,1.0h)}{o3}
    \fmfdotn{v}{3}
    \fmfdot{i1,i3}
    \fmfdot{o2,o3}
\end{fmfgraph*}
\end{gathered}\hspace{5mm}+\hspace{5mm}
\begin{gathered}
    \begin{fmfgraph*}(50,70)
    \fmfcurved
    \fmfset{arrow_len}{3mm}
    \fmfleft{i1,i2,i3}
    \fmflabel{}{i1}
    \fmflabel{}{i2}
    \fmfright{o1,o2,o3}
    \fmflabel{}{o1}
    \fmflabel{}{o2}
    \fmf{dbl_wiggly}{i1,v1}
    \fmf{dbl_wiggly}{v2,o2}
    \fmf{dashes}{i3,v3}
    \fmf{dashes}{v3,o3}
    \fmf{fermion}{i1,i3}
    \fmf{fermion}{v1,v2}
    \fmf{fermion,right=0.3}{v3,v1}
    \fmf{fermion,left=0.3}{o3,o2}
    \fmf{fermion,left=0.3}{o2,o3}
    \fmf{fermion}{v2,v3}
    \fmfforce{(0.0w,0.0h)}{i1}
    \fmfforce{(1.0w,0.0h)}{o1}
    \fmfforce{(0.5w,0.5h)}{v2}
    \fmfforce{(0.5w,0.0h)}{v1}
    \fmfforce{(0.5w,1.0h)}{v3}
    \fmfforce{(0.0w,1.0h)}{i3}
    \fmfforce{(1.0w,1.0h)}{o3}
    \fmfdotn{v}{3}
    \fmfdot{i1,i3}
    \fmfdot{o2,o3}
\end{fmfgraph*}
\end{gathered}\hspace{5mm}+\hspace{5mm}
\begin{gathered}
    \begin{fmfgraph*}(50,70)
    \fmfcurved
    \fmfset{arrow_len}{3mm}
    \fmfleft{i1,i2,i3}
    \fmflabel{}{i1}
    \fmflabel{}{i2}
    \fmfright{o1,o2,o3}
    \fmflabel{}{o1}
    \fmflabel{}{o2}
    \fmf{dashes}{i1,v1}
    \fmf{dashes}{v1,o1}
    \fmf{dbl_wiggly}{v2,o2}
    \fmf{dashes}{i3,v3}
    \fmf{dashes}{v3,o3}
    \fmf{fermion}{i1,i3}
    \fmf{fermion}{v2,v1}
    \fmf{fermion,left=0.3}{v1,v3}
    \fmf{fermion}{o3,o2}
    \fmf{fermion}{o2,o1}
    \fmf{fermion,right=0.3}{o1,o3}
    \fmf{fermion}{v3,v2}
    \fmfforce{(0.0w,0.0h)}{i1}
    \fmfforce{(1.0w,0.0h)}{o1}
    \fmfforce{(0.5w,0.5h)}{v2}
    \fmfforce{(0.5w,0.0h)}{v1}
    \fmfforce{(0.5w,1.0h)}{v3}
    \fmfforce{(0.0w,1.0h)}{i3}
    \fmfforce{(1.0w,1.0h)}{o3}
    \fmfdotn{v}{3}
    \fmfdot{i1,i3}
    \fmfdot{o1,o2,o3}
\end{fmfgraph*}
\end{gathered}\\
\\
&+\hspace{5mm}
\begin{gathered}
    \begin{fmfgraph*}(50,70)
    \fmfcurved
    \fmfset{arrow_len}{3mm}
    \fmfleft{i1,i2,i3}
    \fmflabel{}{i1}
    \fmflabel{}{i2}
    \fmfright{o1,o2,o3}
    \fmflabel{}{o1}
    \fmflabel{}{o2}
    \fmf{dashes}{i1,v1}
    \fmf{dashes}{v1,o1}
    \fmf{dbl_wiggly}{v2,i2}
    \fmf{dashes}{i3,v3}
    \fmf{dashes}{v3,o3}
    \fmf{fermion}{i1,i2}
    \fmf{fermion}{i2,i3}
    \fmf{fermion,left=0.3}{v3,v1}
    \fmf{fermion,right=0.2}{o1,o3}
    \fmf{fermion,right=0.2}{o3,o1}
    \fmf{fermion}{v2,v3}
    \fmf{fermion}{v1,v2}
    \fmfforce{(0.0w,0.0h)}{i1}
    \fmfforce{(1.0w,0.0h)}{o1}
    \fmfforce{(0.5w,0.5h)}{v2}
    \fmfforce{(0.5w,0.0h)}{v1}
    \fmfforce{(0.5w,1.0h)}{v3}
    \fmfforce{(0.0w,1.0h)}{i3}
    \fmfforce{(1.0w,1.0h)}{o3}
    \fmfdotn{v}{3}
    \fmfdot{i1,i3,i2}
    \fmfdot{o1,o3}
\end{fmfgraph*}
\end{gathered}\hspace{5mm}+\hspace{5mm}
\begin{gathered}
    \begin{fmfgraph*}(50,70)
    \fmfcurved
    \fmfset{arrow_len}{3mm}
    \fmfleft{i1,i2,i3}
    \fmflabel{}{i1}
    \fmflabel{}{i2}
    \fmfright{o1,o2,o3}
    \fmflabel{}{o1}
    \fmflabel{}{o2}
    \fmf{dashes}{i1,v1}
    \fmf{dashes}{v1,o1}
    \fmf{dbl_wiggly}{v2,i2}
    \fmf{dashes}{i3,v3}
    \fmf{dashes}{v3,o3}
    \fmf{fermion}{i1,i2}
    \fmf{fermion}{i2,i3}
    \fmf{fermion,right=0.3}{v1,v3}
    \fmf{fermion,right=0.2}{o1,o3}
    \fmf{fermion,right=0.2}{o3,o1}
    \fmf{fermion}{v3,v2}
    \fmf{fermion}{v2,v1}
    \fmfforce{(0.0w,0.0h)}{i1}
    \fmfforce{(1.0w,0.0h)}{o1}
    \fmfforce{(0.5w,0.5h)}{v2}
    \fmfforce{(0.5w,0.0h)}{v1}
    \fmfforce{(0.5w,1.0h)}{v3}
    \fmfforce{(0.0w,1.0h)}{i3}
    \fmfforce{(1.0w,1.0h)}{o3}
    \fmfdotn{v}{3}
    \fmfdot{i1,i3,i2}
    \fmfdot{o1,o3}
\end{fmfgraph*}
\end{gathered}
\end{split}
\end{gather}
\end{widetext}
\begin{figure*}[ht]
    \centering
    \begin{gather*}
\begin{split}
\tilde{\Sigma}^{2\p1\h/2\h1\p(0)}_{pq} &= \hspace{5mm} \begin{gathered}
\begin{fmfgraph*}(30,30)
    \fmfset{arrow_len}{3mm}
    \fmfleft{i1,i2,i3}
    \fmfright{o1,o2,o3}
    \fmf{fermion}{i1,i2}
    \fmf{fermion}{i2,i3}
    \fmf{dbl_dashes}{i2,o2}
    \fmfforce{(0.0w,0.h)}{i1}
    \fmfforce{(0.0w,0.5h)}{i2}
    \fmfforce{(0.0w,1.0h)}{i3}
    \fmfdot{i2}
    \fmfv{decor.shape=cross,decor.filled=full, decor.size=1.5thic}{o2}
\end{fmfgraph*}
\end{gathered}\hspace{2.5mm}+\hspace{5mm}\begin{gathered}
    \begin{fmfgraph*}(40,40)
    \fmfcurved
    \fmfset{arrow_len}{3mm}
    \fmfleft{i1,i2}
    \fmflabel{}{i1}
    \fmflabel{}{i2}
    \fmfright{o1,o2}
    \fmflabel{}{o1}
    \fmflabel{}{o2}
    \fmf{dbl_wiggly}{i1,o1}
    \fmf{fermion,left=0.3,tension=0}{o1,o2}
    \fmf{dbl_dashes}{o2,i2}
    \fmf{fermion,left=0.3,tension=0}{o2,o1}
    \fmfdot{o1,o2,i1}
    \fmfv{decor.shape=cross,decor.size=1.0thic, decor.filled=full}{i2}
\end{fmfgraph*}
\end{gathered}\hspace{5mm}+\hspace{1.0mm}
        \begin{gathered}
    \begin{fmfgraph*}(60,40)
    \fmfcurved
    \fmfset{arrow_len}{3mm}
    \fmfleft{i1,i2}
    \fmflabel{}{i1}
    \fmflabel{}{i2}
    \fmfright{o1,o2}
    \fmflabel{}{o1}
    \fmflabel{}{o2}
    \fmf{dashes}{o1,v1}
    \fmf{dashes}{i1,v1}
    \fmf{fermion,left=0.3,tension=0}{o1,o2}
    \fmf{fermion,left=0.3,tension=0}{v1,v2}
    \fmf{fermion,left=0.3,tension=0}{v2,v1}
    \fmf{phantom}{v2,i2}
    \fmf{dbl_wiggly}{o2,v2}
    \fmf{fermion,left=0.3,tension=0}{o2,o1}
    \fmfdot{o1,o2,i1,v1,v2}
\end{fmfgraph*}
\end{gathered}\hspace{2.5mm}+\hspace{5mm}\begin{gathered}
    \begin{fmfgraph*}(40,40)
    \fmfcurved
    \fmfset{arrow_len}{3mm}
    \fmfleft{i1,i2}
    \fmflabel{}{i1}
    \fmflabel{}{i2}
    \fmfright{o1,o2}
    \fmflabel{}{o1}
    \fmflabel{}{o2}
    \fmf{fermion}{i1,i2}
    \fmf{dbl_wiggly}{o1,i1}
    \fmf{fermion,left=0.3,tension=0}{o1,o2}
    \fmf{dbl_wiggly}{o2,i2}
    \fmf{fermion,left=0.3,tension=0}{o2,o1}
    \fmfdot{o1,o2,i1,i2}
\end{fmfgraph*}
\end{gathered}\hspace{2.5mm}+\hspace{5mm}\begin{gathered}
    \begin{fmfgraph*}(40,40)
    \fmfcurved
    \fmfset{arrow_len}{3mm}
    \fmfleft{i1,i2}
    \fmflabel{}{i1}
    \fmflabel{}{i2}
    \fmfright{o1,o2}
    \fmflabel{}{o1}
    \fmflabel{}{o2}
    \fmf{fermion}{i1,v1}
    \fmf{fermion}{v1,i2}
    \fmf{dbl_wiggly}{o1,i1}
    \fmf{fermion,left=0.3,tension=0}{o1,o2}
    \fmf{dbl_wiggly}{o2,i2}
    \fmf{fermion,left=0.3,tension=0}{o2,o1}
    \fmf{dbl_dashes}{v1,v2}
    \fmfdot{o1,o2,i1,i2,v1}
    \fmfforce{(0.0w,0.0h)}{i1}
    \fmfforce{(0.0w,1.0h)}{i2}
    \fmfforce{(0.0w,0.5h)}{v1}
    \fmfforce{(0.5w,0.5h)}{v2}
     \fmfv{decor.shape=cross,decor.filled=full, decor.size=1.5thic}{v2}
\end{fmfgraph*}
\end{gathered}\\
\\
&+\hspace{2.5mm}
\begin{gathered}
    \begin{fmfgraph*}(40,40)
    \fmfcurved
    \fmfset{arrow_len}{3mm}
    \fmfleft{i1,i2}
    \fmflabel{}{i1}
    \fmflabel{}{i2}
    \fmfright{o1,o2}
    \fmflabel{}{o1}
    \fmflabel{}{o2}
    \fmf{fermion}{i1,i2}
    \fmf{dbl_wiggly}{o1,i1}
    \fmf{fermion,left=0.3,tension=0}{o1,o2}
    \fmf{dbl_wiggly}{o2,i2}
    \fmf{fermion,left=0.3,tension=0}{o2,v1}
     \fmf{fermion,left=0.3,tension=0}{v1,o1}
    \fmf{dbl_dashes}{v1,v2}
    \fmfdot{o1,o2,i1,i2,v1}
    \fmfforce{(1.0w,0.0h)}{o1}
    \fmfforce{(1.0w,1.0h)}{o2}
    \fmfforce{(1.25w,0.5h)}{v1}
    \fmfforce{(1.75w,0.5h)}{v2}
     \fmfv{decor.shape=cross,decor.filled=full, decor.size=1.5thic}{v2}
\end{fmfgraph*}
\end{gathered}
\hspace{15mm}+\hspace{2.5mm}\begin{gathered}
    \begin{fmfgraph*}(40,60)
    \fmfcurved
    \fmfset{arrow_len}{3mm}
    \fmfleft{i1,i2,i3}
    \fmflabel{}{i1}
    \fmflabel{}{i2}
    \fmfright{o1,o2,o3}
    \fmflabel{}{o1}
    \fmflabel{}{o2}
    \fmf{dbl_wiggly}{i2,o2}
    \fmf{dbl_wiggly}{i3,o3}
    \fmf{dbl_wiggly}{i1,o1}
    \fmf{fermion}{i1,i2}
    \fmf{fermion}{i2,i3}
    \fmf{fermion}{o1,o2}
    \fmf{fermion}{o2,o3}
    \fmf{fermion,left=0.3}{o3,o1}
    \fmfforce{(0.0w,0.0h)}{i1}
    \fmfforce{(1.0w,0.0h)}{o1}
    \fmfforce{(0.0w,1.0h)}{i3}
    \fmfforce{(1.0w,1.0h)}{o3}
    \fmfdot{i1,i2,i3}
    \fmfdot{o1,o2,o3}
\end{fmfgraph*}
\end{gathered}
\hspace{7.5mm}+\hspace{5mm}
\begin{gathered}
\begin{fmfgraph*}(40,60)
    \fmfcurved
    \fmfset{arrow_len}{3mm}
    \fmfleft{i1,i2,i3}
    \fmflabel{}{i1}
    \fmflabel{}{i2}
    \fmfright{o1,o2,o3}
    \fmflabel{}{o1}
    \fmflabel{}{o2}
    \fmf{dbl_wiggly}{i1,v1}
    \fmf{dbl_wiggly}{v1,o1}
    \fmf{dbl_wiggly}{v2,o2}
    \fmf{dbl_wiggly}{i3,v3}
    \fmf{fermion}{i1,i3}
    \fmf{fermion,left=0.3}{o1,o2}
    \fmf{fermion,left=0.3}{o2,o1}
    \fmf{fermion,left=0.3}{v2,v3}
    \fmf{fermion,left=0.3}{v3,v2}
    \fmfforce{(0.0w,0.0h)}{i1}
    \fmfforce{(1.0w,0.0h)}{o1}
    \fmfforce{(0.5w,0.5h)}{v2}
    \fmfforce{(0.5w,0.0h)}{v1}
    \fmfforce{(0.5w,1.0h)}{v3}
    \fmfforce{(0.0w,1.0h)}{i3}
    \fmfforce{(1.0w,1.0h)}{o3}
    \fmfdot{v2,v3}
    \fmfdot{i1,i3}
    \fmfdot{o1,o2}
\end{fmfgraph*}
\end{gathered}\hspace{2.5mm}+\hspace{5mm}
\begin{gathered}
    \begin{fmfgraph*}(40,60)
    \fmfcurved
    \fmfset{arrow_len}{3mm}
    \fmfleft{i1,i2,i3}
    \fmflabel{}{i1}
    \fmflabel{}{i2}
    \fmfright{o1,o2,o3}
    \fmflabel{}{o1}
    \fmflabel{}{o2}
    \fmf{dbl_wiggly}{i1,v1}
    \fmf{dashes}{i2,v2}
    \fmf{dashes}{v2,o2}
    \fmf{dbl_wiggly}{i3,v3}
    \fmf{fermion}{i1,i2}
    \fmf{fermion}{v1,v2}
    \fmf{fermion}{o2,v1}
    \fmf{fermion}{v2,v3}
    \fmf{fermion}{v3,o2}
    \fmf{fermion}{i2,i3}
    \fmfforce{(0.0w,0.0h)}{i1}
    \fmfforce{(1.0w,0.0h)}{o1}
    \fmfforce{(0.5w,0.5h)}{v2}
    \fmfforce{(0.5w,0.0h)}{v1}
    \fmfforce{(0.5w,1.0h)}{v3}
    \fmfforce{(0.0w,1.0h)}{i3}
    \fmfforce{(1.0w,1.0h)}{o3}
    \fmfdotn{v}{3}
    \fmfdot{i1,i2,i3}
    \fmfdot{o2}
\end{fmfgraph*}
\end{gathered}
\end{split}
\end{gather*}
    \caption{The 2p1h/2h1p excitation character restricted third-order one-particle irreducible coupled-cluster self-energy diagrams expressed with respect to the reference Green's function, $G_0$.}
    \label{fig:pert_cc_se}
\end{figure*}
Typical third-order diagrams of the form:
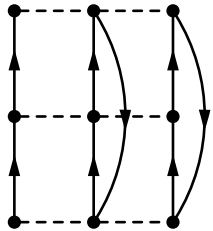
\begin{gather*}
\begin{split}
\begin{gathered}
    \begin{fmfgraph*}(50,70)
    \fmfcurved
    \fmfset{arrow_len}{3mm}
    \fmfleft{i1,i2,i3}
    \fmflabel{}{i1}
    \fmflabel{}{i2}
    \fmfright{o1,o2,o3}
    \fmflabel{}{o1}
    \fmflabel{}{o2}
    \fmf{dashes}{i1,v1}
    \fmf{dashes}{v1,o1}
    \fmf{dashes}{i2,v2}
    \fmf{dashes}{v2,o2}
    \fmf{dashes}{i3,v3}
    \fmf{dashes}{v3,o3}
    \fmf{fermion}{i1,i2}
    \fmf{fermion}{v1,v2}
    \fmf{fermion,left=0.3}{v3,v1}
    \fmf{fermion}{o1,o2}
    \fmf{fermion,left=0.3}{o3,o1}
    \fmf{fermion}{v2,v3}
    \fmf{fermion}{o2,o3}
    \fmf{fermion}{i2,i3}
    \fmfforce{(0.0w,0.0h)}{i1}
    \fmfforce{(1.0w,0.0h)}{o1}
    \fmfforce{(0.5w,0.5h)}{v2}
    \fmfforce{(0.5w,0.0h)}{v1}
    \fmfforce{(0.5w,1.0h)}{v3}
    \fmfforce{(0.0w,1.0h)}{i3}
    \fmfforce{(1.0w,1.0h)}{o3}
    \fmfdotn{v}{3}
    \fmfdot{i1,i2,i3}
    \fmfdot{o1,o2,o3}
\end{fmfgraph*}
\end{gathered}\hspace{25mm}\begin{gathered}
    \begin{fmfgraph*}(50,70)
    \fmfcurved
    \fmfset{arrow_len}{3mm}
    \fmfleft{i1,i2,i3}
    \fmflabel{}{i1}
    \fmflabel{}{i2}
    \fmfright{o1,o2,o3}
    \fmflabel{}{o1}
    \fmflabel{}{o2}
    \fmf{dbl_dashes}{i1,v1}
    \fmf{dbl_dashes}{v1,v4}
    \fmf{dbl_dashes}{v4,o1}
    \fmf{dbl_wiggly}{v2,v5}
    \fmf{dashes}{i3,v3}
    \fmf{dashes}{v3,o3}
    \fmf{fermion}{i1,i3}
    \fmf{fermion}{v1,v2}
    \fmf{fermion,right=0.3}{v3,v1}
    \fmf{fermion,left=0.3}{o1,o3}
    \fmf{fermion,left=0.3}{o3,o1}
    \fmf{fermion,left=0.3}{v5,v4}
    \fmf{fermion,left=0.3}{v4,v5}
    \fmf{fermion,left=0.3}{o3,o1}
    \fmf{fermion}{v2,v3}
    \fmfforce{(-0.1w,0.0h)}{i1}
    \fmfforce{(1.0w,0.0h)}{o1}
    \fmfforce{(0.33w,0.5h)}{v2}
    \fmfforce{(0.33w,0.0h)}{v1}
    \fmfforce{(0.33w,1.0h)}{v3}
    \fmfforce{(0.66w,0.0h)}{v4}
    \fmfforce{(0.66w,0.5h)}{v5}
    \fmfforce{(-0.1w,1.0h)}{i3}
    \fmfforce{(1.0w,1.0h)}{o3}
    \fmfdotn{v}{5}
    \fmfdot{i1,i3}
    \fmfdot{o1,o3}
\end{fmfgraph*}
\end{gathered}
\end{split}
\end{gather*}
vanish due to the fact that matrix elements of the similarity transformed Hamiltonian with four or more lines below a vertex are zero. 

Other dynamical third-order contributions such as the diagrams 
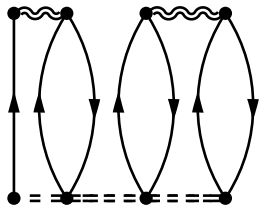
\begin{gather*}
    \begin{split}
     \begin{gathered}\begin{fmfgraph*}(70,60)
    \fmfcurved
    \fmfset{arrow_len}{3mm}
    \fmfleft{i1,i2}
    \fmflabel{}{i1}
    \fmflabel{}{i2}
    \fmfright{o1,o2}
    \fmflabel{}{o1}
    \fmflabel{}{o2}
    \fmf{dashes}{i1,v1}
    \fmf{dashes}{v1,o1}
    \fmf{dbl_wiggly}{i2,v2}
    \fmf{fermion}{i1,i2}
    \fmf{fermion,left=0.3}{v1,v2}
    \fmf{fermion,left=0.3}{v2,v1}
    \fmf{fermion,left=0.3}{o1,o2}
    \fmf{fermion,left=0.3}{o2,o1}
    \fmf{dbl_dashes}{o2,v5}
    \fmfforce{(0.0w,0.0h)}{i1}
    \fmfforce{(0.0w,1.0h)}{i2}
    \fmfforce{(1.0w,0.0h)}{o1}
    \fmfforce{(1.0w,0.5h)}{o2}
    \fmfforce{(0.5w,1.0h)}{v2}
    \fmfforce{(0.5w,0.0h)}{v1}
    \fmfforce{(1.3w,0.5h)}{v5}
    \fmfv{decor.shape=cross,decor.filled=full, decor.size=1.5thic}{v5}
    \fmfdot{v1,v2}
    \fmfdot{i1,i2}
    \fmfdot{o1,o2}
\end{fmfgraph*}
\end{gathered}\hspace{20mm}
        \begin{gathered}
    \begin{fmfgraph*}(70,60)
    \fmfcurved
    \fmfset{arrow_len}{3mm}
    \fmfleft{i1,i2}
    \fmflabel{}{i1}
    \fmflabel{}{i2}
    \fmfright{o1,o2}
    \fmflabel{}{o1}
    \fmflabel{}{o2}
    \fmf{dbl_dashes}{v1,i1}
    \fmf{dbl_dashes}{o1,v1}
    \fmf{dbl_dashes}{v1,v3}
    \fmf{dbl_dashes}{v3,o1}
    \fmf{dbl_wiggly}{v4,o2}
    \fmf{dbl_wiggly}{i2,v5}
    \fmf{fermion}{i1,i2}
    \fmf{fermion,left=0.3,tension=0}{o1,o2}
     \fmf{fermion,left=0.3,tension=0}{v1,v5}
      \fmf{fermion,left=0.3,tension=0}{v5,v1}
    \fmf{fermion,left=0.3,tension=0}{v3,v4}
    \fmf{fermion,left=0.3,tension=0}{v4,v3}
    \fmf{fermion,left=0.3,tension=0}{o2,o1}
    \fmfdot{o1,i1,v1,v3,v4,o2,i2,v5}
    \fmfforce{(0.0w,0.0h)}{i1}
    \fmfforce{(1.0w,0.0h)}{o1}
    \fmfforce{(0.25w,1h)}{v2}
    \fmfforce{(0.25w,0.0h)}{v1}
    \fmfforce{(0.625w,0.0h)}{v3}
    \fmfforce{(0.625w,1.0h)}{v4}
    \fmfforce{(0.0w,1.0h)}{i2}
    \fmfforce{(1.0w,1.0h)}{o2}
    \fmfforce{(0.25w,1.0h)}{v5}
\end{fmfgraph*}
\end{gathered}
    \end{split}
\end{gather*}
are in fact interaction-reducible as they are obtained by contracting the three-body effective interaction with a single-particle Green's function and the four-body effective interaction with a two-particle Green's function. Therefore, these diagrams will appear in the perturbative expansion of the self-energy with respect to $G_0$, but they do not contribute to the series of third-order 1PI, skeleton and interaction-irreducible diagrams that enter $\tilde{\Sigma}[\tilde{G}]$. These interaction-reducible diagrams are actually contained in the self-consistent renormalized coupled-cluster self-energy expansion at second-order to be presented in Section~\ref{sec:cc_se}.

When $T=0$, the third-order coupled-cluster self-energy diagrams exactly reduce to the two skeleton diagrams of the electronic self-energy as
\begin{gather*}
    \begin{split}
        \Sigma^{(3)}[G_0] &= \hspace{2.5mm} 
    \begin{gathered}
    \begin{fmfgraph*}(50,70)
    \fmfcurved
    \fmfset{arrow_len}{3mm}
    \fmfleft{i1,i2,i3}
    \fmflabel{}{i1}
    \fmflabel{}{i2}
    \fmfright{o1,o2,o3}
    \fmflabel{}{o1}
    \fmflabel{}{o2}
    \fmf{wiggly}{i2,o2}
    \fmf{wiggly}{i3,o3}
    \fmf{wiggly}{i1,o1}
    \fmf{fermion}{i1,i2}
    \fmf{fermion}{i2,i3}
    \fmf{fermion}{o1,o2}
    \fmf{fermion}{o2,o3}
    \fmf{fermion,left=0.3}{o3,o1}
    \fmfforce{(0.0w,0.0h)}{i1}
    \fmfforce{(1.0w,0.0h)}{o1}
    \fmfforce{(0.0w,1.0h)}{i3}
    \fmfforce{(1.0w,1.0h)}{o3}
    \fmfdot{i1,i2,i3}
    \fmfdot{o1,o2,o3}
\end{fmfgraph*}
\end{gathered}
\hspace{10mm}+\hspace{5mm}
\begin{gathered}
\begin{fmfgraph*}(50,70)
    \fmfcurved
    \fmfset{arrow_len}{3mm}
    \fmfleft{i1,i2,i3}
    \fmflabel{}{i1}
    \fmflabel{}{i2}
    \fmfright{o1,o2,o3}
    \fmflabel{}{o1}
    \fmflabel{}{o2}
    \fmf{wiggly}{i1,v1}
    \fmf{wiggly}{v1,o1}
    \fmf{wiggly}{v2,o2}
    \fmf{wiggly}{i3,v3}
    \fmf{fermion}{i1,i3}
    \fmf{fermion,left=0.3}{o1,o2}
    \fmf{fermion,left=0.3}{o2,o1}
    \fmf{fermion,left=0.3}{v2,v3}
    \fmf{fermion,left=0.3}{v3,v2}
    \fmfforce{(0.0w,0.0h)}{i1}
    \fmfforce{(1.0w,0.0h)}{o1}
    \fmfforce{(0.5w,0.5h)}{v2}
    \fmfforce{(0.5w,0.0h)}{v1}
    \fmfforce{(0.5w,1.0h)}{v3}
    \fmfforce{(0.0w,1.0h)}{i3}
    \fmfforce{(1.0w,1.0h)}{o3}
    \fmfdot{v2,v3}
    \fmfdot{i1,i3}
    \fmfdot{o1,o2}
\end{fmfgraph*}
\end{gathered} 
    \end{split}
\end{gather*}
where the interaction line is now the antisymmetrized bare Coulomb interaction.
Therefore, we have $\tilde{\Sigma}^{T=0(3)}_{pq}(\omega) = \Sigma^{(3)}_{pq}(\omega)$ as expected. 

To demonstrate the evaluation of a typical third-order contribution to the coupled-cluster self-energy, we focus on the third diagram of Eq.~\ref{eq:3rd_order}. In this diagram, from the Feynman rules~\cite{Quantum,mahan2000many,stefanucci2013nonequilibrium,carbone2013self,raimondi2018algebraic}, we have two sets of two equally oriented Green's function lines which give rise to a total symmetry factor of $\frac{1}{(2!)^2}$. There is also one closed fermion loop giving rise to an overall phase factor of $-1$. Therefore, the diagram evaluates to 
\begin{widetext}
    \begin{gather}
        \begin{split}
            \begin{gathered}
    \begin{fmfgraph*}(50,70)
    \fmfcurved
    \fmfset{arrow_len}{3mm}
    \fmfleft{i1,i2,i3}
    \fmflabel{}{i1}
    \fmflabel{}{i2}
    \fmfright{o1,o2,o3}
    \fmflabel{}{o1}
    \fmflabel{}{o2}
    \fmf{dbl_wiggly}{i1,v1}
    \fmf{dashes}{i2,v2}
    \fmf{dashes}{v2,o2}
    \fmf{dbl_wiggly}{i3,v3}
    \fmf{fermion}{i1,i2}
    \fmf{fermion}{v1,v2}
    \fmf{fermion}{o2,v1}
    \fmf{fermion}{v2,v3}
    \fmf{fermion}{v3,o2}
    \fmf{fermion}{i2,i3}
    \fmfforce{(0.0w,0.0h)}{i1}
    \fmfforce{(1.0w,0.0h)}{o1}
    \fmfforce{(0.5w,0.5h)}{v2}
    \fmfforce{(0.5w,0.0h)}{v1}
    \fmfforce{(0.5w,1.0h)}{v3}
    \fmfforce{(0.0w,1.0h)}{i3}
    \fmfforce{(1.0w,1.0h)}{o3}
    \fmfdotn{v}{3}
    \fmfdot{i1,i2,i3}
    \fmfdot{o2}
\end{fmfgraph*}
\end{gathered}\hspace{5mm}= -\frac{(i)^4}{(2!)^2}\sum_{\substack{rstuvx\\
yzwnol}}\int&\frac{d\omega_1d\omega_2d\omega_3d\omega_4}{(2\pi)^4}\chi_{pr,st}G^{0}_{su}(\omega_{1})G^{0}_{tv}(\omega_{2})G^{0}_{wr}(\omega_{1}+\omega_2-\omega)\\
&\times\chi_{uvx,yzw}G^{0}_{yn}(\omega_3)G^{0}_{zo}(\omega_4)\chi_{no,ql}G^{0}_{lx}(\omega_3+\omega_4-\omega) \ .
        \end{split}
    \end{gather}
\end{widetext}
Evaluation of all the different time-orderings associated with this diagram gives rise to six contributions. These are partitioned into three forward-time diagrams and three backward-time diagrams. This is a reflection of the general relationship: an $n$th-order self-energy Feynman diagram gives rise to $n!$ time-ordered Goldstone diagrams~\cite{Quantum,schirmer2018many}. However, it is important to note that specific time-orderings of a coupled-cluster self-energy diagram can evaluate to zero as a result of the CC amplitude equations.  

The first three diagrams of Eq.~\ref{eq:3rd_order} encode excitations that contain mixtures of 2p1h/2h1p states. In the case of three-body interactions, it was proposed that these diagrams make up the dominant contribution to the self-energy at third-order~\cite{carbone2013self,raimondi2018algebraic}. The rest of the diagrams all contain combinations of 3p2h and 2p1h excitations. Higher-body excitations are possible in the case of a generic $N$-body interaction, but vanish within coupled-cluster theory due to the structure of the similarity transformed Hamiltonian~\cite{shavitt2009many}.   

To conclude the perturbative analysis presented in this section, we depict the diagrammatic content of the set of perturbative third-order 1PI coupled-cluster self-energy diagrams restricted to the space of 2p1h/2h1p intermediate excitation states in Figure~\ref{fig:pert_cc_se}. The fifth and sixth diagrams of Figure~\ref{fig:pert_cc_se} arise by insertion of the static coupled-cluster self-energy diagram (the first diagram) into the second-order dynamical term (the fourth diagram). As a result, they are non-skeleton contributions. The second and third diagrams are static contributions that are interaction-reducible.

\section{Self-consistent renormalization of the coupled-cluster self-energy}~\label{sec:cc_se}

In this section, we construct the renormalized coupled-cluster self-energy from the exact equation-of-motion for the single-particle coupled-cluster Green's function. Our analysis demonstrates the formally consistent functional-diagrammatic formulation of the non-hermitian CC self-energy. As a result, profound connections are uncovered between the effective interactions generated by the self-consistent Green's function theory and those that arise from the functional derivatives of the BCC Lagrangian presented in Ref.~\cite{coveney2023coupled}. From this analysis, we also demonstrate how the perturbative series for the self-energy derived in Section~\ref{sec:pert} emerges from the self-consistent formalism.

\subsection{Exact equation-of-motion and the renormalized coupled-cluster self-energy}~\label{subsec:eom_gf}

Through its exact equation-of-motion, the single-particle coupled-cluster Green's function is coupled to the 4-point, 6-point, 8-point and so on Green's functions. In the spin-orbital basis, these higher-point Green's functions are defined as 
\begin{widetext}
\begin{subequations}
    \begin{align}
    \begin{split}
        &i\Tilde{G}^{4\pt}_{pq,rs}(t_1,t_2;t_3,t_4) = \braket{\Tilde{\Psi}_0|\mathcal{T}\left\{a_{q}(t_2) a_p(t_1) a^\dag_{r}(t_3)a^\dag_{s}(t_4)\right\}|\Phi_0} ,
    \end{split}\\
    \begin{split}
        &i\Tilde{G}^{6\pt}_{pqr,stu}(t_1,t_2,t_3;t_4,t_5,t_6) =\braket{\Tilde{\Psi}_0|\mathcal{T}\left\{a_{r}(t_3) a_{q}(t_2) a_p(t_1) a^\dag_{s}(t_4)a^\dag_{t}(t_5)a^\dag_{u}(t_6)\right\}|\Phi_0} \ ,
    \end{split}\\
    \begin{split}
        &i\Tilde{G}^{8\pt}_{pqrs,tuvw}(t_1,t_2,t_3,t_4;t_5,t_6,t_7,t_8) =\braket{\Tilde{\Psi}_0|\mathcal{T}\left\{a_{s}(t_4) a_{r}(t_3) a_{q}(t_2) a_p(t_1) a^\dag_{t}(t_5)a^\dag_{u}(t_6)a^\dag_{v}(t_7)a^\dag_{w}(t_8)\right\}|\Phi_0} \ .
    \end{split}
\end{align}
\end{subequations}
\end{widetext}
The 10-point and higher Green's functions are defined analogously. The relationship between the 4-point Green's function, the two-particle-hole Green's function and the two-particle reduced density matrix is given by 
\begin{subequations}
    \begin{gather}
    \begin{split}
        \tilde{G}^{4\pt}_{pq,rs}(t^+,t;t',t'^{+}) &= \tilde{G}^{2\p\h}_{pq,rs}(t-t')
    \end{split}
\end{gather}
with
\begin{gather}
    \begin{split}
        \tilde{\Gamma}_{pq,rs} &= i\tilde{G}^{2\p\h}_{pq,rs}(t-t^{+}) = \braket{\tilde{\Psi}_0|a^\dag_{r}a^\dag_s a_q a_p|\Phi_0} \ .
    \end{split}
\end{gather}
\end{subequations}
Here, the notation $t-t^{+}$ denotes evaluation of the second time argument of the Green's function with a positive infinitesimal, \emph{i.e.} $t^{+} = t+\eta$, where $\eta\to0$ from above. Analogous relationships hold between the higher-body Green's functions and corresponding reduced density matrices. 

Taking the time-derivative of the first argument of the coupled-cluster Green's function (Eq.~\ref{eq:sp_cc_gf}), using Eq.~\ref{eq:bio_heisenberg} and the definitions of the higher-order Green's functions, we have 
\begin{widetext}
\begin{gather}
    \begin{split}~\label{eq:eom_sp_ccgf}
        \sum_{r}\left(i\frac{\partial}{\partial t_1}\delta_{pr}-f_{pr}\right)\Tilde{G}_{rq}(t_1-t_2) &= \delta_{pq}\delta(t_1-t_2)
        +\sum_{r}(\bar{h}_{pr}-f_{pr})\Tilde{G}_{rq}(t_1-t_2)+\frac{1}{2!}\sum_{rst} \bar{h}_{pr,st}\tilde{G}^{4
        \pt}_{st,rq}(t_1,t_1;t_1^+,t_2)\\
        &+\frac{1}{3!\cdot2!}\sum_{\substack{rs\\tuv}} \bar{h}_{prs,tuv}\tilde{G}^{6
        \pt}_{tuv,rsq}(t_1,t_1,t_1;t_1^{++},t_1^{+},t_2) \\
        &+ \frac{1}{4!\cdot3!}\sum_{\substack{rst\\uvol}} \bar{h}_{prst,uvol}\tilde{G}^{8
        \pt}_{uvol,rstq}(t_1,t_1,t_1,t_1;t^{+++}_1,t_1^{++},t_1^{+},t_2)\\
        &+\cdots
    \end{split}
\end{gather}
In contrast to the electronic Green's function where the interaction Hamiltonian is a two-body operator, since $\bar{H}$ is an $N$-body operator, the single-particle coupled-cluster Green's function is coupled all the way up to the $N$-particle Green's function. Using the functional inverse for the `non-interacting' Green's function and taking the Fourier transformation of this expression, we rewrite the equation-of-motion as the Dyson series
    \begin{gather}
    \begin{split}~\label{eq:eom_sc}
        &\tilde{G}_{pq}(\omega) = G^{0}_{pq}(\omega) + \sum_{rs} G^{0}_{pr}(\omega)(\bar{h}_{rs}-f_{rs})\tilde{G}_{sq}(\omega)\\
        &-\frac{1}{2!}\sum_{rstu} G^{0}_{pr}(\omega)\bar{h}_{rs,tu} \int^{\infty}_{-\infty}\frac{d\omega_1d\omega_2}{(2\pi)^2}\ \tilde{G}^{4\pt}_{tu,qs}(\omega_1,\omega_2;\omega,\omega_1+\omega_2-\omega)\\
        &+\frac{1}{3!\cdot2!}\sum_{\substack{rst\\uvw}}G^{0}_{ps}(\omega)\bar{h}_{rst,uvw}\int^{\infty}_{-\infty}\frac{d\omega_1d\omega_2d\omega_3d\omega_4}{(2\pi)^4} \ \tilde{G}^{6\pt}_{uvw,rqt}(\omega_1,\omega_2,\omega_3;\omega_4,\omega,\omega_1+\omega_2+\omega_3-\omega_4-\omega) \\
        &-\frac{1}{4!\cdot3!}\sum_{\substack{rst\\uvol}} G^{0}_{pn}(\omega)\bar{h}_{nrst,uvol}\int^{\infty}_{-\infty}\frac{d\omega_1d\omega_2d\omega_3d\omega_4d\omega_5d\omega_6}{(2\pi)^6}\tilde{G}^{8
        \pt}_{uvol,qrst}(\omega_1,\omega_2,\omega_3,\omega_4;\omega,\omega_5,\omega_6,\omega_1+\omega_2+\omega_3+\omega_4-\omega_{5}-\omega_6-\omega)\\
        &+\cdots
    \end{split}
\end{gather}
\end{widetext}
where the continuing series implies coupling of the five-body interaction to the 10-point Green's function and so on. Therefore, knowledge of the single-particle Green's function requires the knowledge of all $N$-point Green's functions. In order to derive the self-consistent expression for the coupled-cluster self-energy, we first introduce expressions for the 4-point, 6-point, 8-point vertices and so on. These interaction vertices contain only 1PI diagrams and are related to the self-consistent equations of motion for the 4-point, 6-point, 8-point Green's functions and so on, respectively. 

In the frequency domain, the 4-point vertex function is defined as~\cite{mattuck1971expressing,stefanucci2013nonequilibrium} 
\begin{widetext}
\begin{gather}
\begin{split}~\label{eq:4point_vert}
        \tilde{G}^{4\pt}_{pq,rs}(\omega_1,\omega_2;\omega_3,\omega_4) &= 2\pi i\Big(\delta(\omega_1-\omega_3)\delta(\omega_2-\omega_4)\tilde{G}_{pr}(\omega_1)\tilde{G}_{qs}(\omega_2)-\delta(\omega_1-\omega_4)\delta(\omega_2-\omega_3)\tilde{G}_{ps}(\omega_1)\tilde{G}_{qr}(\omega_2)\Big)\\
        &-\sum_{tuvw} \tilde{G}_{pt}(\omega_1)\tilde{G}_{qu}(\omega_2)\tilde{\Lambda}^{4\pt}_{tu,vw}(\omega_1,\omega_2;\omega_3,\omega_4)\tilde{G}_{vr}(\omega_3)\tilde{G}_{ws}(\omega_4) \ .
\end{split}
\end{gather}
\end{widetext}
This equation presents the 4-point Green's function as the sum of the antisymmetrized propagation of two independent particles, with $\tilde{\Lambda}^{4\pt}$ representing their effective interaction. The 4-point vertex, $\tilde{\Lambda}^{4\pt}$, is very closely related to the Bethe-Salpeter kernel and `vertex' of Hedin's equations (see Section~\ref{sec:BSE} and Appendix~\ref{app:4_vertex}). However, it is constructed from 1PI diagrams as opposed to the two-particle irreducible diagrams (2PI) which enter the BSE kernel.
It is important to note that Eq.~\ref{eq:4point_vert} is formally exact and does not constitute an approximation.

Similarly, the exact equation-of-motion for the 6-point Green's function can be written as~\cite{mattuck1971expressing,stefanucci2013nonequilibrium,carbone2013self} 
\begin{widetext}
\begin{gather}
    \begin{split}~\label{eq:6point}
        &\tilde{G}^{6\pt}_{pqr,stu}(\omega_1,\omega_2,\omega_3;\omega_4,\omega_5,\omega_6) = -(2\pi)^2\mathcal{A}_{[\{p\omega_1,q\omega_2,r\omega_3\}]}\Big[\delta(\omega_1-\omega_4)\delta(\omega_2-\omega_5)\delta(\omega_3-\omega_6)\tilde{G}_{ps}(\omega_1)\tilde{G}_{qt}(\omega_2)\tilde{G}_{ru}(\omega_3)\Big]\\
        &-2\pi i \mathcal{P}_{[\{p\omega_1,q\omega_2,r\omega_3\}]}\mathcal{P}_{[\{s\omega_4,t\omega_5,u\omega_6\}]}\left(\delta(\omega_1-\omega_4)\tilde{G}_{ps}(\omega_1)\sum_{vwlo}\tilde{G}_{qv}(\omega_2)\tilde{G}_{rw}(\omega_3)\tilde{\Lambda}^{4\pt}_{vw,lo}(\omega_2,\omega_3;\omega_5,\omega_6)\tilde{G}_{lt}(\omega_5)\tilde{G}_{ou}(\omega_6)\right)\\
        &+\sum_{\substack{vwl\\
        omn}} \tilde{G}_{pv}(\omega_1)\tilde{G}_{qw}(\omega_2)\tilde{G}_{rl}(\omega_3)\tilde{\Lambda}^{6\pt}_{vwl,omn}(\omega_1,\omega_2,\omega_3;\omega_4,\omega_5,\omega_6)\tilde{G}_{os}(\omega_4)\tilde{G}_{mt}(\omega_5)\tilde{G}_{nu}(\omega_6)
    \end{split}
\end{gather} 
\end{widetext}
Here, we employ the notation of Ref.~\cite{carbone2013self} where $\mathcal{A}_{[\{p\omega_1,q\omega_2,r\omega_3\}]}$ represents the antisymmetric sum of all possible pairs of permutations of the index-frequency pairs and $\mathcal{P}_{[\{p\omega_1,q\omega_2,r\omega_3\}]}$ sums all cyclic permutations of each index-frequency pair. The antisymmetrized and cyclic permutation summations are required to ensure that the fermionic exchange symmetry is obeyed during multiparticle propagation. In general, there are $n!$ terms that contribute to the independent-particle propagation of $n$ fermions. For a single fermion propagating independently with respect to the coupled propagating of $(n-1)$ interacting fermions gives rise to $n^2$ equivalent combinations. For the case of the 6-point vertex equation, we have $3!=6$ independent-particle propagation terms, $3^2=9$ terms corresponding to the propagation of a single fermion with two interacting fermions~\cite{mattuck1971expressing}. 

From this equation, we immediately see the emergence of the coupling of the 6-point vertex function to the 4-point vertex. This is a consequence of the fact that the 6-point vertex is related to the functional derivative of the 4-point vertex with respect to the single-particle Green's function. Continuing this process up to the $N$-body Green's function and $2N$-point vertex function, yields expressions that couple to all propagators and contain the correct permutation symmetry with respect to the Fermi statistics. Again, it must be emphasised that all equations involving the vertex functions are exact by definition and inserting these expressions into Eq.~\ref{eq:eom_sc}, using the definition of the self-energy through the Dyson equation, we obtain the following expression for the irreducible coupled-cluster self-energy
\vspace{-2.5mm}
\begin{widetext}
    \begin{gather}
        \begin{split}
            &\Tilde{\Sigma}_{pq}(\omega) = \left(\bar{h}_{pq} -i \sum_{rs} \bar{h}_{pr,qs}\tilde{G}_{sr}(t-t^+) +\frac{i}{(2!)^2}\sum_{rstu} \bar{h}_{prs,qtu}\tilde{G}^{2\p\h}_{tu,rs}(t-t^+) + \cdots\right)-f_{pq}\\
            &+\frac{1}{2!}\sum_{\substack{stu\\vwl}}\left(\bar{h}_{ps,tu}-i\sum_{on}\bar{h}_{pso,tun}\tilde{G}_{no}(t-t^+)+\frac{i}{(2!)^2}\sum_{on\epsilon\kappa} \bar{h}_{pson,tu\epsilon\kappa}\tilde{G}^{2\p\h}_{\epsilon\kappa,on}(t-t^+)+\cdots\right)\\
            &\times\int\frac{d\omega_1d\omega_2}{(2\pi)^2}\tilde{G}_{tv}(\omega_1)\tilde{G}_{uw}(\omega_2)\Tilde{\Lambda}^{4\pt}_{vw,ql}(\omega_1,\omega_2;\omega,\omega_1+\omega_2-\omega) \tilde{G}_{ls}(\omega_1+\omega_2-\omega)\\
            &+\frac{1}{3!\cdot2!}\sum_{\substack{st\\uvw}}\sum_{\substack{l\sigma\\\alpha\gamma\kappa}}\left(\bar{h}_{spt,uvw} -i\sum_{\beta\epsilon}\bar{h}_{spt\beta,uvw\epsilon}\tilde{G}_{\epsilon\beta}(t-t^+)+\frac{i}{(2!)^2}\sum_{\beta\epsilon\delta\rho} \bar{h}_{spt\beta\epsilon,uvw\delta\rho}\tilde{G}^{2\p\h}_{\delta\rho,\beta\epsilon}(t-t^+)+\cdots\right)\\
            &\times \int\frac{d\omega_1d\omega_2d\omega_3d\omega_4}{(2\pi)^4}\tilde{G}_{ul}(\omega_1)\tilde{G}_{v\sigma}(\omega_2)\tilde{G}_{w\alpha}(\omega_3)\Tilde{\Lambda}^{6\pt}_{l\sigma\alpha,\gamma q \kappa}(\omega_1,\omega_2,\omega_3;\omega_4,\omega,\omega_1+\omega_2+\omega_3-\omega_4-\omega)  \\
            &\hspace{30mm}\times\tilde{G}_{\gamma s}(\omega_4)\tilde{G}_{\sigma t}(\omega_1+\omega_2+\omega_3-\omega_4-\omega)
            \\
            &+\frac{1}{4!\cdot3!}\sum_{\substack{stu\\vwol}}\sum_{\substack{\sigma g z\\y\gamma
            \epsilon\delta}}\Bigg(\bar{h}_{pstu,vwol}-i\sum_{\alpha\beta}\bar{h}_{pstu\alpha,vwol\beta}\tilde{G}_{\beta\alpha}(t-t^{+})+\frac{i}{(2!)^2}\sum_{\alpha\beta\gamma\phi}\bar{h}_{pstu\alpha\beta,vwol\gamma\phi}\tilde{G}^{2\p\h}_{\gamma\phi,\alpha\beta}(t-t^+)+\cdots\Bigg) \\
            &\times\int\frac{d\omega_1d\omega_2d\omega_3d\omega_4d\omega_5d\omega_6}{(2\pi)^6}\tilde{G}_{v\sigma}(\omega_1)\tilde{G}_{wg}(\omega_2)\tilde{G}_{oz}(\omega_3)\tilde{G}_{ly}(\omega_4)\\
            &\hspace{30mm}\times\Tilde{\Lambda}^{8\pt}_{\sigma g z y, q\gamma\epsilon\delta}(\omega_1,\omega_2,\omega_3,\omega_4;\omega,\omega_5,\omega_6,\omega_1+\omega_2+\omega_3+\omega_4-\omega_{5}-\omega_6-\omega)\\
            &\hspace{30mm}\times\tilde{G}_{\gamma s}(\omega_5)\tilde{G}_{\epsilon t}(\omega_6)\tilde{G}_{\delta u}(\omega_1+\omega_2+\omega_3+\omega_4-\omega_{5}-\omega_6-\omega)
            \\
            &+ \cdots
        \end{split}
    \end{gather}
From the equation-of-motion for the Green's function this expression continues up to the $2N$-point vertex. Remarkably, as a result of the exact vertex equations, the effective interaction elements appearing here correspond exactly to those of the similarity transformed Hamiltonian normal-ordered with respect to the exact ground state biorthogonal expectation value (Eq.~\ref{eq:norm_tot}). Therefore, we may simplify the above expression to obtain the fully renormalized coupled-cluster self-energy as 
\begin{gather}
        \begin{split}
            \Tilde{\Sigma}_{pq}(\omega) &= (\tilde{F}_{pq}-f_{pq})\\
            &+\frac{1}{2!}\sum_{\substack{stu\\vwl}}\tilde{\Xi}_{ps,tu}\int\frac{d\omega_1d\omega_2}{(2\pi)^2}\tilde{G}_{tv}(\omega_1)\tilde{G}_{uw}(\omega_2)\Tilde{\Lambda}^{4\pt}_{vw,ql}(\omega_1,\omega_2;\omega,\omega_1+\omega_2-\omega)\tilde{G}_{ls}(\omega_1+\omega_2-\omega)\\
            &+\frac{1}{3!\cdot2!}\sum_{\substack{st\\uvw}}\sum_{\substack{l\sigma\\\alpha\gamma\kappa}}\tilde{\chi}_{spt,uvw}\int\frac{d\omega_1d\omega_2d\omega_3d\omega_4}{(2\pi)^4}\tilde{G}_{ul}(\omega_1)\tilde{G}_{v\sigma}(\omega_2)\tilde{G}_{w\alpha}(\omega_3)\\
            &\hspace{15mm}\times\Tilde{\Lambda}^{6\pt}_{l\sigma\alpha,\gamma q \kappa}(\omega_1,\omega_2,\omega_3;\omega_4,\omega,\omega_1+\omega_2+\omega_3-\omega_4-\omega) \tilde{G}_{\gamma s}(\omega_4)\tilde{G}_{\sigma t}(\omega_1+\omega_2+\omega_3-\omega_4-\omega)\\
            &+\frac{1}{4!\cdot3!}\sum_{\substack{stu\\vwol}}\sum_{\substack{\sigma g z\\y\gamma
            \epsilon\delta}}\tilde{\chi}_{pstu,vwol} \int\frac{d\omega_1d\omega_2d\omega_3d\omega_4d\omega_5d\omega_6}{(2\pi)^6}\tilde{G}_{v\sigma}(\omega_1)\tilde{G}_{wg}(\omega_2)\tilde{G}_{oz}(\omega_3)\tilde{G}_{ly}(\omega_4)\\
            &\hspace{30mm}\times\Tilde{\Lambda}^{8\pt}_{\sigma g z y, q\gamma\epsilon\delta}(\omega_1,\omega_2,\omega_3,\omega_4;\omega,\omega_5,\omega_6,\omega_1+\omega_2+\omega_3+\omega_4-\omega_{5}-\omega_6-\omega)\\
            &\hspace{30mm}\times\tilde{G}_{\gamma s}(\omega_5)\tilde{G}_{\epsilon t}(\omega_6)\tilde{G}_{\delta u}(\omega_1+\omega_2+\omega_3+\omega_4-\omega_{5}-\omega_6-\omega)
            \\
            &+ \cdots
        \end{split}
\end{gather}
The Feynman diagrammatic representation of this equation is given by
\begin{gather}
    \begin{split}~\label{eq:exact_se}
        \tilde{\Sigma}[\tilde{G}] &=\hspace{7.5mm} 
\begin{gathered}
\begin{fmfgraph*}(40,40)
    \fmfset{arrow_len}{3mm}
    \fmfleft{i1}
    \fmfright{o1}
    \fmf{zigzag}{o1,i1}
    \fmfv{decor.shape=cross,decor.filled=full, decor.size=1.5thic}{o1}
    \fmfdot{i1}
\end{fmfgraph*}
\end{gathered}\hspace{5mm}+\hspace{5mm}
\begin{gathered}
    \begin{fmfgraph*}(50,50)
    \fmfset{arrow_len}{3mm}
    \fmfleft{i1,i2,i3}
    \fmfright{o1,o2,o3}
    \fmfrpolyn{label=$\Tilde{\Lambda}^{4\pt}$}{G}{4}
    \fmfforce{(0.0w,0.5h)}{G2}
    \fmfforce{(0.0w,0.0h)}{G1}
    \fmfforce{(1.0w,0.5h)}{G3}
    \fmfforce{(1.0w,0.0h)}{G4}
    \fmfforce{(0.0w,1.0h)}{i3}
    \fmfforce{(1.0w,1.0h)}{o3}
    \fmfdot{G1,G2,G3,G4,i3,o3}
    \fmf{double_arrow}{G2,i3}
    \fmf{double_arrow}{G3,o3}
    \fmf{double_arrow,left=0.4}{o3,G4}
    \fmf{dbl_zigzag}{i3,o3}
    \end{fmfgraph*}
\end{gathered}\hspace{7.5mm}+\hspace{7.5mm}
\begin{gathered}
    \begin{fmfgraph*}(50,50)
    \fmfset{arrow_len}{3mm}
    \fmfleft{i1,i2,i3}
    \fmfright{o1,o2,o3}
    \fmfrpolyn{label=$\Tilde{\Lambda}^{6\pt}$}{G}{6}
    \fmfforce{(0.0w,0.5h)}{G2}
    \fmfforce{(0.0w,0.0h)}{G1}
    \fmfforce{(0.5w,0.5h)}{G3}
    \fmfforce{(1.0w,0.5h)}{G4}
    \fmfforce{(1.0w,0.0h)}{G5}
    \fmfforce{(0.5w,0.0h)}{G6}
    \fmfforce{(0.0w,1.0h)}{i3}
    \fmfforce{(1.0w,1.0h)}{o3}
    \fmfforce{(0.5w,1.0h)}{v1}
    \fmfdot{G1,G2,G3,G4,G6,G5,i3,o3,v1}
    \fmf{double_arrow}{G2,i3}
    \fmf{double_arrow}{G3,v1}
    \fmf{double_arrow,right=0.4}{i3,G1}
    \fmf{double_arrow,left=0.4}{o3,G5}
    \fmf{double_arrow}{G4,o3}
    \fmf{dbl_dashes}{i3,v1}
    \fmf{dbl_dashes}{v1,o3}
    \end{fmfgraph*}
\end{gathered}\hspace{7.5mm}+\hspace{7.5mm}
\begin{gathered}
    \begin{fmfgraph*}(50,50)
    \fmfset{arrow_len}{3mm}
    \fmfleft{i1,i2,i3}
    \fmfright{o1,o2,o3}
    \fmfrpolyn{label=$\Tilde{\Lambda}^{8\pt}$}{G}{8}
    \fmfforce{(0.0w,0.5h)}{G2}
    \fmfforce{(0.0w,0.0h)}{G1}
    \fmfforce{(0.33w,0.5h)}{G3}
    \fmfforce{(0.66w,0.5h)}{G4}
    \fmfforce{(1.0w,0.5h)}{G5}
    \fmfforce{(1.0w,0.0h)}{G6}
    \fmfforce{(0.66w,0.0h)}{G7}
    \fmfforce{(0.33w,0.0h)}{G8}
    \fmfforce{(0.0w,1.0h)}{i3}
    \fmfforce{(1.0w,1.0h)}{o3}
    \fmfforce{(0.33w,1.0h)}{v1}
    \fmfforce{(0.66w,1.0h)}{v4}
    \fmfdot{G1,G2,G3,G4,G6,G5,G7,G8,i3,o3,v1,v4}
    \fmf{double_arrow}{G2,i3}
    \fmf{double_arrow}{G3,v1}
    \fmf{double_arrow,left=0.4}{o3,G6}
    \fmf{double_arrow}{G5,o3}
    \fmf{double_arrow,left=0.4}{v4,G7}
    \fmf{double_arrow,right=0.4}{v1,G8}
    \fmf{double_arrow}{G4,v4}
    \fmf{curly}{i3,v1}
    \fmf{curly}{v1,v4}
    \fmf{curly}{v4,o3}
    \end{fmfgraph*}
\end{gathered}\\
&+\cdots
    \end{split}
\end{gather}
where the higher-order diagrams are represented by propagators coupled to the five-body interaction and the 10-point vertex function and so on. 
Here, we have introduced the diagrammatic notation for the `vertex' functions, $\Tilde{\Lambda}^{4/6/8\pt}$. 
Diagrammatically, through the vertices, the renormalized self-energy is expressed as the sum of all 1PI, interaction-irreducible skeleton diagrams. From the Feynman rules, we see that the correct pre-factors of the self-energy are reproduced. The first vertex diagram has one set of two equally oriented Green's function lines therefore giving a total symmetry factor of $\frac{1}{2!}$. The second vertex diagram contains one set of three and one set of two equally oriented Green's function lines giving rise to the total symmetry factor of $\frac{1}{3!2!}$. The third vertex diagram contains one set of four and one set of three equally oriented Green's function lines giving rise to the total symmetry factor of $\frac{1}{4!3!}$, and so on. This results in the self-consistently renormalized series for the self-energy functional (up to third order)
\begin{gather}
\begin{split}
\tilde{\Sigma}[\tilde{G}] &=\hspace{7.5mm} 
\begin{gathered}
\begin{fmfgraph*}(40,40)
    \fmfset{arrow_len}{3mm}
    \fmfleft{i1}
    \fmfright{o1}
    \fmf{zigzag}{o1,i1}
    \fmfv{decor.shape=cross,decor.filled=full, decor.size=1.5thic}{o1}
    \fmfdot{i1}
\end{fmfgraph*}
\end{gathered}\hspace{5mm}+\hspace{5mm} 
\begin{gathered}
\begin{fmfgraph*}(50,50)
    \fmfcurved
    \fmfset{arrow_len}{3mm}
    \fmfleft{i1,i2}
    \fmflabel{}{i1}
    \fmflabel{}{i2}
    \fmfright{o1,o2}
    \fmflabel{}{o1}
    \fmflabel{}{o2}
    \fmf{double_arrow}{i1,i2}
    \fmf{dbl_zigzag}{o1,i1}
    \fmf{double_arrow,left=0.2,tension=0}{o1,o2}
    \fmf{dbl_zigzag}{o2,i2}
    \fmf{double_arrow,left=0.2,tension=0}{o2,o1}
    \fmfdot{o1,o2,i1,i2}
\end{fmfgraph*}
\end{gathered} \hspace{5mm}+\hspace{7.5mm} 
\begin{gathered}
    \begin{fmfgraph*}(50,70)
    \fmfcurved
    \fmfset{arrow_len}{3mm}
    \fmfleft{i1,i2,i3}
    \fmflabel{}{i1}
    \fmflabel{}{i2}
    \fmfright{o1,o2,o3}
    \fmflabel{}{o1}
    \fmflabel{}{o2}
    \fmf{dbl_zigzag}{i2,o2}
    \fmf{dbl_zigzag}{i3,o3}
    \fmf{dbl_zigzag}{i1,o1}
    \fmf{double_arrow}{i1,i2}
    \fmf{double_arrow}{i2,i3}
    \fmf{double_arrow}{o1,o2}
    \fmf{double_arrow}{o2,o3}
    \fmf{double_arrow,left=0.3}{o3,o1}
    \fmfforce{(0.0w,0.0h)}{i1}
    \fmfforce{(1.0w,0.0h)}{o1}
    \fmfforce{(0.0w,1.0h)}{i3}
    \fmfforce{(1.0w,1.0h)}{o3}
    \fmfdot{i1,i2,i3}
    \fmfdot{o1,o2,o3}
\end{fmfgraph*}
\end{gathered}
\hspace{7.5mm}+\hspace{7.5mm}
\begin{gathered}
\begin{fmfgraph*}(50,70)
    \fmfcurved
    \fmfset{arrow_len}{3mm}
    \fmfleft{i1,i2,i3}
    \fmflabel{}{i1}
    \fmflabel{}{i2}
    \fmfright{o1,o2,o3}
    \fmflabel{}{o1}
    \fmflabel{}{o2}
    \fmf{dbl_zigzag}{i1,v1}
    \fmf{dbl_zigzag}{v1,o1}
    \fmf{dbl_zigzag}{v2,o2}
    \fmf{dbl_zigzag}{i3,v3}
    \fmf{double_arrow}{i1,i3}
    \fmf{double_arrow,left=0.3}{o1,o2}
    \fmf{double_arrow,left=0.3}{o2,o1}
    \fmf{double_arrow,left=0.3}{v2,v3}
    \fmf{double_arrow,left=0.3}{v3,v2}
    \fmfforce{(0.0w,0.0h)}{i1}
    \fmfforce{(1.0w,0.0h)}{o1}
    \fmfforce{(0.5w,0.5h)}{v2}
    \fmfforce{(0.5w,0.0h)}{v1}
    \fmfforce{(0.5w,1.0h)}{v3}
    \fmfforce{(0.0w,1.0h)}{i3}
    \fmfforce{(1.0w,1.0h)}{o3}
    \fmfdot{v2,v3}
    \fmfdot{i1,i3}
    \fmfdot{o1,o2}
\end{fmfgraph*}
\end{gathered}
\\
\\
&+\hspace{7.5mm}
\begin{gathered}
    \begin{fmfgraph*}(50,70)
    \fmfcurved
    \fmfset{arrow_len}{3mm}
    \fmfleft{i1,i2,i3}
    \fmflabel{}{i1}
    \fmflabel{}{i2}
    \fmfright{o1,o2,o3}
    \fmflabel{}{o1}
    \fmflabel{}{o2}
    \fmf{dbl_zigzag}{i1,v1}
    \fmf{dbl_dashes}{i2,v2}
    \fmf{dbl_dashes}{v2,o2}
    \fmf{dbl_zigzag}{i3,v3}
    \fmf{double_arrow}{i1,i2}
    \fmf{double_arrow}{v1,v2}
    \fmf{double_arrow}{o2,v1}
    \fmf{double_arrow}{v2,v3}
    \fmf{double_arrow}{v3,o2}
    \fmf{double_arrow}{i2,i3}
    \fmfforce{(0.0w,0.0h)}{i1}
    \fmfforce{(1.0w,0.0h)}{o1}
    \fmfforce{(0.5w,0.5h)}{v2}
    \fmfforce{(0.5w,0.0h)}{v1}
    \fmfforce{(0.5w,1.0h)}{v3}
    \fmfforce{(0.0w,1.0h)}{i3}
    \fmfforce{(1.0w,1.0h)}{o3}
    \fmfdotn{v}{3}
    \fmfdot{i1,i2,i3}
    \fmfdot{o2}
\end{fmfgraph*}
\end{gathered}
\hspace{5mm}+\hspace{7.5mm}
\begin{gathered}\begin{fmfgraph*}(50,70)
    \fmfcurved
    \fmfset{arrow_len}{3mm}
    \fmfleft{i1,i2,i3}
    \fmflabel{}{i1}
    \fmflabel{}{i2}
    \fmfright{o1,o2,o3}
    \fmflabel{}{o1}
    \fmflabel{}{o2}
    \fmf{dbl_dashes}{i1,v1}
    \fmf{dbl_dashes}{v1,o1}
    \fmf{dbl_zigzag}{v2,o2}
    \fmf{dbl_zigzag}{i3,v3}
    \fmf{double_arrow}{i1,i3}
    \fmf{double_arrow}{v1,v2}
    \fmf{double_arrow,right=0.3}{v3,v1}
    \fmf{double_arrow,left=0.3}{o1,o2}
    \fmf{double_arrow,left=0.3}{o2,o1}
    \fmf{double_arrow}{v2,v3}
    \fmfforce{(0.0w,0.0h)}{i1}
    \fmfforce{(1.0w,0.0h)}{o1}
    \fmfforce{(0.5w,0.5h)}{v2}
    \fmfforce{(0.5w,0.0h)}{v1}
    \fmfforce{(0.5w,1.0h)}{v3}
    \fmfforce{(0.0w,1.0h)}{i3}
    \fmfforce{(1.0w,1.0h)}{o3}
    \fmfdotn{v}{3}
    \fmfdot{i1,i3}
    \fmfdot{o1,o2}
\end{fmfgraph*}
\end{gathered}\hspace{5mm} + \hspace{5mm}
\begin{gathered}
    \begin{fmfgraph*}(50,70)
    \fmfcurved
    \fmfset{arrow_len}{3mm}
    \fmfleft{i1,i2,i3}
    \fmflabel{}{i1}
    \fmflabel{}{i2}
    \fmfright{o1,o2,o3}
    \fmflabel{}{o1}
    \fmflabel{}{o2}
    \fmf{dbl_dashes}{i1,v1}
    \fmf{dbl_dashes}{v1,o1}
    \fmf{dbl_zigzag}{v2,o2}
    \fmf{dbl_zigzag}{i3,v3}
    \fmf{double_arrow}{i1,i3}
    \fmf{double_arrow}{v2,v1}
    \fmf{double_arrow,left=0.3}{v1,v3}
    \fmf{double_arrow,left=0.3}{o1,o2}
    \fmf{double_arrow,left=0.3}{o2,o1}
    \fmf{double_arrow}{v3,v2}
    \fmfforce{(0.0w,0.0h)}{i1}
    \fmfforce{(1.0w,0.0h)}{o1}
    \fmfforce{(0.5w,0.5h)}{v2}
    \fmfforce{(0.5w,0.0h)}{v1}
    \fmfforce{(0.5w,1.0h)}{v3}
    \fmfforce{(0.0w,1.0h)}{i3}
    \fmfforce{(1.0w,1.0h)}{o3}
    \fmfdotn{v}{3}
    \fmfdot{i1,i3}
    \fmfdot{o1,o2}
\end{fmfgraph*}
\end{gathered}\hspace{5mm}+\hspace{5mm}
\begin{gathered}
    \begin{fmfgraph*}(50,70)
    \fmfcurved
    \fmfset{arrow_len}{3mm}
    \fmfleft{i1,i2,i3}
    \fmflabel{}{i1}
    \fmflabel{}{i2}
    \fmfright{o1,o2,o3}
    \fmflabel{}{o1}
    \fmflabel{}{o2}
    \fmf{dbl_zigzag}{i1,v1}
    \fmf{dbl_zigzag}{v2,o2}
    \fmf{dbl_dashes}{i3,v3}
    \fmf{dbl_dashes}{v3,o3}
    \fmf{double_arrow}{i1,i3}
    \fmf{double_arrow}{v2,v1}
    \fmf{double_arrow,left=0.3}{v1,v3}
    \fmf{double_arrow,left=0.3}{o3,o2}
    \fmf{double_arrow,left=0.3}{o2,o3}
    \fmf{double_arrow}{v3,v2}
    \fmfforce{(0.0w,0.0h)}{i1}
    \fmfforce{(1.0w,0.0h)}{o1}
    \fmfforce{(0.5w,0.5h)}{v2}
    \fmfforce{(0.5w,0.0h)}{v1}
    \fmfforce{(0.5w,1.0h)}{v3}
    \fmfforce{(0.0w,1.0h)}{i3}
    \fmfforce{(1.0w,1.0h)}{o3}
    \fmfdotn{v}{3}
    \fmfdot{i1,i3}
    \fmfdot{o2,o3}
\end{fmfgraph*}
\end{gathered}\\
\\
&+\hspace{5mm}
\begin{gathered}
    \begin{fmfgraph*}(50,70)
    \fmfcurved
    \fmfset{arrow_len}{3mm}
    \fmfleft{i1,i2,i3}
    \fmflabel{}{i1}
    \fmflabel{}{i2}
    \fmfright{o1,o2,o3}
    \fmflabel{}{o1}
    \fmflabel{}{o2}
    \fmf{dbl_zigzag}{i1,v1}
    \fmf{dbl_zigzag}{v2,o2}
    \fmf{dbl_dashes}{i3,v3}
    \fmf{dbl_dashes}{v3,o3}
    \fmf{double_arrow}{i1,i3}
    \fmf{double_arrow}{v1,v2}
    \fmf{double_arrow,right=0.3}{v3,v1}
    \fmf{double_arrow,left=0.3}{o3,o2}
    \fmf{double_arrow,left=0.3}{o2,o3}
    \fmf{double_arrow}{v2,v3}
    \fmfforce{(0.0w,0.0h)}{i1}
    \fmfforce{(1.0w,0.0h)}{o1}
    \fmfforce{(0.5w,0.5h)}{v2}
    \fmfforce{(0.5w,0.0h)}{v1}
    \fmfforce{(0.5w,1.0h)}{v3}
    \fmfforce{(0.0w,1.0h)}{i3}
    \fmfforce{(1.0w,1.0h)}{o3}
    \fmfdotn{v}{3}
    \fmfdot{i1,i3}
    \fmfdot{o2,o3}
\end{fmfgraph*}
\end{gathered}\hspace{5mm}+\hspace{5mm}
\begin{gathered}
    \begin{fmfgraph*}(50,70)
    \fmfcurved
    \fmfset{arrow_len}{3mm}
    \fmfleft{i1,i2,i3}
    \fmflabel{}{i1}
    \fmflabel{}{i2}
    \fmfright{o1,o2,o3}
    \fmflabel{}{o1}
    \fmflabel{}{o2}
    \fmf{dbl_dashes}{i1,v1}
    \fmf{dbl_dashes}{v1,o1}
    \fmf{dbl_zigzag}{v2,o2}
    \fmf{dbl_dashes}{i3,v3}
    \fmf{dbl_dashes}{v3,o3}
    \fmf{double_arrow}{i1,i3}
    \fmf{double_arrow}{v2,v1}
    \fmf{double_arrow,left=0.3}{v1,v3}
    \fmf{double_arrow}{o3,o2}
    \fmf{double_arrow}{o2,o1}
    \fmf{double_arrow,right=0.3}{o1,o3}
    \fmf{double_arrow}{v3,v2}
    \fmfforce{(0.0w,0.0h)}{i1}
    \fmfforce{(1.0w,0.0h)}{o1}
    \fmfforce{(0.5w,0.5h)}{v2}
    \fmfforce{(0.5w,0.0h)}{v1}
    \fmfforce{(0.5w,1.0h)}{v3}
    \fmfforce{(0.0w,1.0h)}{i3}
    \fmfforce{(1.0w,1.0h)}{o3}
    \fmfdotn{v}{3}
    \fmfdot{i1,i3}
    \fmfdot{o1,o2,o3}
\end{fmfgraph*}
\end{gathered}\hspace{5mm}+\hspace{5mm}
\begin{gathered}
    \begin{fmfgraph*}(50,70)
    \fmfcurved
    \fmfset{arrow_len}{3mm}
    \fmfleft{i1,i2,i3}
    \fmflabel{}{i1}
    \fmflabel{}{i2}
    \fmfright{o1,o2,o3}
    \fmflabel{}{o1}
    \fmflabel{}{o2}
    \fmf{dbl_dashes}{i1,v1}
    \fmf{dbl_dashes}{v1,o1}
    \fmf{dbl_zigzag}{v2,i2}
    \fmf{dbl_dashes}{i3,v3}
    \fmf{dbl_dashes}{v3,o3}
    \fmf{double_arrow}{i1,i2}
    \fmf{double_arrow}{i2,i3}
    \fmf{double_arrow,left=0.3}{v3,v1}
    \fmf{double_arrow,right=0.2}{o1,o3}
    \fmf{double_arrow,right=0.2}{o3,o1}
    \fmf{double_arrow}{v2,v3}
    \fmf{double_arrow}{v1,v2}
    \fmfforce{(0.0w,0.0h)}{i1}
    \fmfforce{(1.0w,0.0h)}{o1}
    \fmfforce{(0.5w,0.5h)}{v2}
    \fmfforce{(0.5w,0.0h)}{v1}
    \fmfforce{(0.5w,1.0h)}{v3}
    \fmfforce{(0.0w,1.0h)}{i3}
    \fmfforce{(1.0w,1.0h)}{o3}
    \fmfdotn{v}{3}
    \fmfdot{i1,i3,i2}
    \fmfdot{o1,o3}
\end{fmfgraph*}
\end{gathered}\hspace{5mm}+\hspace{5mm}
\begin{gathered}
    \begin{fmfgraph*}(50,70)
    \fmfcurved
    \fmfset{arrow_len}{3mm}
    \fmfleft{i1,i2,i3}
    \fmflabel{}{i1}
    \fmflabel{}{i2}
    \fmfright{o1,o2,o3}
    \fmflabel{}{o1}
    \fmflabel{}{o2}
    \fmf{dbl_dashes}{i1,v1}
    \fmf{dbl_dashes}{v1,o1}
    \fmf{dbl_zigzag}{v2,i2}
    \fmf{dbl_dashes}{i3,v3}
    \fmf{dbl_dashes}{v3,o3}
    \fmf{double_arrow}{i1,i2}
    \fmf{double_arrow}{i2,i3}
    \fmf{double_arrow,right=0.3}{v1,v3}
    \fmf{double_arrow,right=0.2}{o1,o3}
    \fmf{double_arrow,right=0.2}{o3,o1}
    \fmf{double_arrow}{v3,v2}
    \fmf{double_arrow}{v2,v1}
    \fmfforce{(0.0w,0.0h)}{i1}
    \fmfforce{(1.0w,0.0h)}{o1}
    \fmfforce{(0.5w,0.5h)}{v2}
    \fmfforce{(0.5w,0.0h)}{v1}
    \fmfforce{(0.5w,1.0h)}{v3}
    \fmfforce{(0.0w,1.0h)}{i3}
    \fmfforce{(1.0w,1.0h)}{o3}
    \fmfdotn{v}{3}
    \fmfdot{i1,i3,i2}
    \fmfdot{o1,o3}
\end{fmfgraph*}
\end{gathered}\hspace{5mm}+\hspace{5mm}\cdots
\end{split}
\end{gather}
\end{widetext}
Subsequently we observe the renormalization of both propagators and interaction vertices as required by the presence of the interaction-irreducible diagrams. This is the result of the implicit normal-ordering of the Hamiltonian generated by the Green's function formalism. Third-order diagrams such as 
\begin{gather*}
\begin{split}
\begin{gathered}
    \begin{fmfgraph*}(50,70)
    \fmfcurved
    \fmfset{arrow_len}{3mm}
    \fmfleft{i1,i2,i3}
    \fmflabel{}{i1}
    \fmflabel{}{i2}
    \fmfright{o1,o2,o3}
    \fmflabel{}{o1}
    \fmflabel{}{o2}
    \fmf{dbl_dashes}{i1,v1}
    \fmf{dbl_dashes}{v1,o1}
    \fmf{dbl_dashes}{i2,v2}
    \fmf{dbl_dashes}{v2,o2}
    \fmf{dbl_dashes}{i3,v3}
    \fmf{dbl_dashes}{v3,o3}
    \fmf{double_arrow}{i1,i2}
    \fmf{double_arrow}{v1,v2}
    \fmf{double_arrow,left=0.3}{v3,v1}
    \fmf{double_arrow}{o1,o2}
    \fmf{double_arrow,left=0.3}{o3,o1}
    \fmf{double_arrow}{v2,v3}
    \fmf{double_arrow}{o2,o3}
    \fmf{double_arrow}{i2,i3}
    \fmfforce{(0.0w,0.0h)}{i1}
    \fmfforce{(1.0w,0.0h)}{o1}
    \fmfforce{(0.5w,0.5h)}{v2}
    \fmfforce{(0.5w,0.0h)}{v1}
    \fmfforce{(0.5w,1.0h)}{v3}
    \fmfforce{(0.0w,1.0h)}{i3}
    \fmfforce{(1.0w,1.0h)}{o3}
    \fmfdotn{v}{3}
    \fmfdot{i1,i2,i3}
    \fmfdot{o1,o2,o3}
\end{fmfgraph*}
\end{gathered}\hspace{25mm}\begin{gathered}
    \begin{fmfgraph*}(50,70)
    \fmfcurved
    \fmfset{arrow_len}{3mm}
    \fmfleft{i1,i2,i3}
    \fmflabel{}{i1}
    \fmflabel{}{i2}
    \fmfright{o1,o2,o3}
    \fmflabel{}{o1}
    \fmflabel{}{o2}
    \fmf{curly}{i1,v1}
    \fmf{curly}{v1,v4}
    \fmf{curly}{v4,o1}
    \fmf{dbl_zigzag}{v2,v5}
    \fmf{dbl_dashes}{i3,v3}
    \fmf{dbl_dashes}{v3,o3}
    \fmf{double_arrow}{i1,i3}
    \fmf{double_arrow}{v1,v2}
    \fmf{double_arrow,right=0.3}{v3,v1}
    \fmf{double_arrow,left=0.3}{o1,o3}
    \fmf{double_arrow,left=0.3}{o3,o1}
    \fmf{double_arrow,left=0.3}{v5,v4}
    \fmf{double_arrow,left=0.3}{v4,v5}
    \fmf{double_arrow,left=0.3}{o3,o1}
    \fmf{double_arrow}{v2,v3}
    \fmfforce{(-0.1w,0.0h)}{i1}
    \fmfforce{(1.1w,0.0h)}{o1}
    \fmfforce{(0.33w,0.5h)}{v2}
    \fmfforce{(0.33w,0.0h)}{v1}
    \fmfforce{(0.33w,1.0h)}{v3}
    \fmfforce{(0.66w,0.0h)}{v4}
    \fmfforce{(0.66w,0.5h)}{v5}
    \fmfforce{(-0.1w,1.0h)}{i3}
    \fmfforce{(1.1w,1.0h)}{o3}
    \fmfdotn{v}{5}
    \fmfdot{i1,i3}
    \fmfdot{o1,o3}
\end{fmfgraph*}
\end{gathered}
\end{split}
\end{gather*}
vanish due to the fact that the renormalized three-body, four-body and so on matrix elements of the similarity transformed Hamiltonian that contain four or more lines below a vertex are zero. This is analogous to the corresponding perturbative analysis presented in Section~\ref{sec:pert}.
The explicitly static term is given by 
\begin{gather}
    \begin{split}~\label{eq:exact_stat}
    \Tilde{\Sigma}^{\infty}_{pq} &= \hspace{5mm}
        \begin{gathered}
\begin{fmfgraph*}(40,40)
    \fmfset{arrow_len}{3mm}
    \fmfleft{i1}
    \fmfright{o1}
    \fmf{zigzag}{o1,i1}
    \fmfv{decor.shape=cross,decor.filled=full, decor.size=1.5thic}{o1}
    \fmfdot{i1}
     \fmflabel{$\substack{p\\q}$}{i1}
\end{fmfgraph*}
\end{gathered}\hspace{2.5mm}= \tilde{F}_{pq} - f_{pq}\\
&=\hspace{5mm}
\begin{gathered}
\begin{fmfgraph*}(40,40)
    \fmfset{arrow_len}{3mm}
    \fmfleft{i1,i2,i3}
    \fmfright{o1,o2,o3}
    \fmf{fermion}{i1,i2}
    \fmf{fermion}{i2,i3}
    \fmf{dbl_dashes}{i2,o2}
    \fmfforce{(0.0w,0.h)}{i1}
    \fmfforce{(0.0w,0.5h)}{i2}
    \fmfforce{(0.0w,1.0h)}{i3}
    \fmfdot{i2}
    \fmfv{decor.shape=cross,decor.filled=full, decor.size=1.5thic}{o2}
\end{fmfgraph*}
\end{gathered}\hspace{5mm}+\hspace{7.5mm}
\begin{gathered}
    \begin{fmfgraph*}(40,40)
    \fmfcurved
    \fmfset{arrow_len}{3mm}
    \fmfleft{i1,i2}
    \fmflabel{}{i1}
    \fmflabel{}{i2}
    \fmfright{o1,o2}
    \fmflabel{}{o1}
    \fmflabel{}{o2}
    \fmf{dbl_wiggly}{i1,o1}
    \fmf{fermion,left=0.3,tension=0}{o1,o2}
    \fmf{fermion,left=0.3,tension=0}{o2,o1}
    \fmf{dbl_plain}{v1,o2}
    \fmf{dbl_plain}{o2,v2}
    \fmf{fermion}{v3,i1}
    \fmf{fermion}{i1,v4}
    \fmfforce{(1.05w,1.0h)}{v2}
    \fmfforce{(0.75w,1.0h)}{v1}
    \fmfforce{(0.0w,-0.5h)}{v3}
    \fmfforce{(0.0w,0.5h)}{v4}
    \fmfforce{(0.0w,0.0h)}{i1}
    \fmfdot{o1,i1}
\end{fmfgraph*}
\end{gathered}\\
\\
\\&+\hspace{7.5mm}
\begin{gathered}
    \begin{fmfgraph*}(60,40)
    \fmfcurved
    \fmfset{arrow_len}{3mm}
    \fmfleft{i1,i2}
    \fmflabel{}{i1}
    \fmflabel{}{i2}
    \fmfright{o1,o2}
    \fmflabel{}{o1}
    \fmflabel{}{o2}
    \fmf{dashes}{o1,v1}
    \fmf{dashes}{i1,v1}
    \fmf{fermion,left=0.3,tension=0}{o1,o2}
    \fmf{fermion,left=0.3,tension=0}{v1,v2}
    \fmf{fermion,left=0.3,tension=0}{v2,v1}
    \fmf{phantom}{v2,i2}
    \fmf{dbl_plain}{o2,v2}
    \fmf{fermion,left=0.3,tension=0}{o2,o1}
    \fmf{fermion}{v3,i1}
    \fmf{fermion}{i1,v4}
    \fmfdot{o1,i1,v1}
    \fmfforce{(0.0w,-0.5h)}{v3}
    \fmfforce{(0.0w,0.5h)}{v4}
    \fmfforce{(0.0w,0.0h)}{i1}
    \fmfforce{(0.5w,1.0h)}{v2}
    \fmfforce{(0.5w,0.0h)}{v1}
\end{fmfgraph*}
\end{gathered}\hspace{5mm}+\hspace{2.5mm}\cdots\\
\\
\\
&=\tilde{\Sigma}^{\infty(0)}_{pq} + \sum_{ia}\lambda^{i}_{a}\chi_{pa,qi} + \frac{1}{(2!)^2}\sum_{ijab}\lambda^{ij}_{ab}\chi_{pab,qij} + \cdots
    \end{split}
\end{gather}
Remarkably, this exactly corresponds to the static term first derived in Ref~\cite{coveney2023coupled} from the functional derivative of the BCC Lagrangian with respect to the `non-interacting' single-particle Green's function. This profound result provides the rigorous proof of the fully diagrammatic approach taken in Ref.~\cite{coveney2023coupled}.

When $T=0$ (meaning that $\tilde{G}$ is also replaced by $G$), we find that 
\begin{gather}
    \begin{split}
        \lim_{\substack{T\to0}}\tilde{\Sigma}^{\infty}_{pq} &= \lim_{\substack{T\to0}}\tilde{F}_{pq} -f_{pq}\\
        &= -i\sum_{rs}\braket{pr||qs}\Big(G_{sr}(t-t^+) - G^{0}_{rs}(t-t^+)\Big) \\
        &= \sum_{rs}\braket{pr||qs}\Big(\gamma_{rs} - \gamma^{\rf}_{rs}\Big)\\
        &=\Sigma^{\infty}_{pq} \ ,
        \end{split}
\end{gather}
which is the exact static part of the two-body electronic self-energy (see Eq.~\ref{eq:stat_se}). It is important to note that the property $\tilde{F}_{ai}=F_{ai} = 0$ holds for the interaction-irreducible effective interactions. However, $\tilde{F}_{ai}$ is no longer simply the singles amplitude equations but contains all amplitude equations. This can clearly be seen by inspection of the structure of Eq.~\ref{eq:exact_stat}. 

The effective two-body interaction is given by the following Goldstone diagram series
\begin{gather}
        \begin{split}~\label{eq:2body}
            \tilde{\Xi}_{pq,rs} &=\hspace{5mm}\begin{gathered}
    \begin{fmfgraph*}(40,40)
    \fmfset{arrow_len}{3mm}
    \fmfleft{i1}
    \fmfright{o1}
    \fmf{dbl_zigzag}{o1,i1}
    \fmfdot{i1,o1}
    \fmflabel{$\substack{p\\r}$}{i1}
    \fmflabel{$\substack{q\\s}$}{o1}
\end{fmfgraph*}
\end{gathered}\\
&=\hspace{5mm}
\begin{gathered}
    \begin{fmfgraph*}(40,40)
    \fmfset{arrow_len}{3mm}
    \fmfleft{i1}
    \fmfright{o1}
    \fmf{dbl_wiggly}{o1,i1}
    \fmfdot{i1,o1}
\end{fmfgraph*}
\end{gathered}\hspace{5mm}+\hspace{5mm}
\begin{gathered}
    \begin{fmfgraph*}(60,40)
    \fmfcurved
    \fmfset{arrow_len}{3mm}
    \fmfleft{i1,i2}
    \fmflabel{}{i1}
    \fmflabel{}{i2}
    \fmfright{o1,o2}
    \fmflabel{}{o1}
    \fmflabel{}{o2}
    \fmf{dashes}{o1,v1}
    \fmf{dashes}{i1,v1}
    \fmf{fermion,left=0.3,tension=0}{o1,o2}
    \fmf{phantom}{v2,i2}
    \fmf{fermion,left=0.3,tension=0}{o1,o2}
    \fmf{dbl_plain}{v3,o2}
    \fmf{dbl_plain}{o2,v4}
    \fmf{fermion,left=0.3,tension=0}{o2,o1}
    \fmf{fermion,left=0.3,tension=0}{o2,o1}
    \fmfdot{o1,i1,v1}
    \fmfforce{(1.05w,1.0h)}{v3}
    \fmfforce{(0.75w,1.0h)}{v4}
\end{fmfgraph*}
\end{gathered}\\
&+\hspace{5mm}
\begin{gathered}
    \begin{fmfgraph*}(60,50)
    \fmfcurved
    \fmfset{arrow_len}{3mm}
    \fmfleft{i1,i2}
    \fmflabel{}{i1}
    \fmflabel{}{i2}
    \fmfright{o1,o2}
    \fmflabel{}{o1}
    \fmflabel{}{o2}
    \fmf{dbl_dashes}{v1,i1}
    \fmf{dbl_dashes}{o1,v1}
    \fmf{dbl_dashes}{v1,v3}
    \fmf{dbl_dashes}{v3,o1}
    \fmf{dbl_plain}{v4,o2}
    \fmf{fermion,left=0.3,tension=0}{o1,o2}
    \fmf{fermion,left=0.3,tension=0}{v3,v4}
    \fmf{fermion,left=0.3,tension=0}{v4,v3}
    \fmf{fermion,left=0.3,tension=0}{o2,o1}
    \fmfdot{o1,i1,v1,v3}
    \fmfforce{(0.0w,0.0h)}{i1}
    \fmfforce{(1.0w,0.0h)}{o1}
    \fmfforce{(0.25w,1h)}{v2}
    \fmfforce{(0.25w,0.0h)}{v1}
    \fmfforce{(0.625w,0.0h)}{v3}
    \fmfforce{(0.625w,1.0h)}{v4}
    \fmfforce{(0.0w,1.0h)}{i2}
    \fmfforce{(1.0w,1.0h)}{o2}
\end{fmfgraph*}
\end{gathered}\hspace{5mm}+\hspace{2.5mm} \cdots\\
\\
&= \chi_{pq,rs} + \sum_{ia}\lambda^{i}_{a}\chi_{pqa,rsi} + \frac{1}{(2!)^2}\sum_{ijab}\lambda^{ij}_{ab}\chi_{pqab,rsij} \\
&+ \cdots
        \end{split}
    \end{gather}
Once more, we uncover the connection between the self-consistent Green's function formalism and the CC Lagrangian. This effective interaction was also derived in Ref.~\cite{coveney2023coupled} by taking the second derivative of the BCC Lagrangian with respect to the non-interacting Green's function and corresponds to the exact static component of the BSE kernel (see Section~\ref{sec:BSE}). We also have $\tilde{\Xi}_{ab,ij}=\chi_{ab,ij}=0$ due to the coupled-cluster amplitude equations. 

Using the expression for the effective two-body interaction, the corresponding second-order self-consistent coupled-cluster self-energy takes the form
\vspace{10mm}
\begin{gather}
    \begin{split}~\label{eq:2nd}
    \tilde{\Sigma}^{(2)}[\tilde{G}] &= 
\begin{gathered}\begin{fmfgraph*}(50,50)
    \fmfcurved
    \fmfset{arrow_len}{3mm}
    \fmfleft{i1,i2}
    \fmflabel{}{i1}
    \fmflabel{}{i2}
    \fmfright{o1,o2}
    \fmflabel{}{o1}
    \fmflabel{}{o2}
    \fmf{double_arrow}{i1,i2}
    \fmf{dbl_zigzag}{o1,i1}
    \fmf{double_arrow,left=0.2,tension=0}{o1,o2}
    \fmf{dbl_zigzag}{o2,i2}
    \fmf{double_arrow,left=0.2,tension=0}{o2,o1}
    \fmfdot{o1,o2,i1,i2}
\end{fmfgraph*}
\end{gathered}\\
\\
\\
&=\hspace{3mm}
\begin{gathered}\begin{fmfgraph*}(50,50)
    \fmfcurved
    \fmfset{arrow_len}{3mm}
    \fmfleft{i1,i2}
    \fmflabel{}{i1}
    \fmflabel{}{i2}
    \fmfright{o1,o2}
    \fmflabel{}{o1}
    \fmflabel{}{o2}
    \fmf{double_arrow}{i1,i2}
    \fmf{dbl_wiggly}{o1,i1}
    \fmf{double_arrow,left=0.2,tension=0}{o1,o2}
    \fmf{dbl_wiggly}{o2,i2}
    \fmf{double_arrow,left=0.2,tension=0}{o2,o1}
    \fmfdot{o1,o2,i1,i2}
\end{fmfgraph*}
\end{gathered}\hspace{3mm}+\hspace{3mm}
        \begin{gathered}\begin{fmfgraph*}(50,50)
    \fmfcurved
    \fmfset{arrow_len}{3mm}
    \fmfleft{i1,i2}
    \fmflabel{}{i1}
    \fmflabel{}{i2}
    \fmfright{o1,o2}
    \fmflabel{}{o1}
    \fmflabel{}{o2}
    \fmf{dashes}{i1,v1}
    \fmf{dashes}{v1,o1}
    \fmf{dbl_wiggly}{i2,v2}
    \fmf{double_arrow}{i1,i2}
    \fmf{double_arrow,left=0.3}{v1,v2}
    \fmf{double_arrow,left=0.3}{v2,v1}
    \fmf{fermion,left=0.3}{o1,o2}
    \fmf{fermion,left=0.3}{o2,o1}
    \fmf{dbl_plain}{v4,o2}
    \fmf{dbl_plain}{o2,v5}
    \fmfforce{(0.0w,0.0h)}{i1}
    \fmfforce{(0.0w,1.0h)}{i2}
    \fmfforce{(1.0w,0.0h)}{o1}
    \fmfforce{(1.0w,0.5h)}{o2}
    \fmfforce{(0.5w,1.0h)}{v2}
    \fmfforce{(0.5w,0.0h)}{v1}
    \fmfforce{(1.15w,0.5h)}{v4}
    \fmfforce{(0.85w,0.5h)}{v5}
    \fmfdot{v1,v2}
    \fmfdot{i1,i2}
    \fmfdot{o1}
\end{fmfgraph*}
\end{gathered}\\
\\
\\
&+\hspace{5mm}
\begin{gathered}
    \begin{fmfgraph*}(50,50)
    \fmfcurved
    \fmfset{arrow_len}{3mm}
    \fmfleft{i1,i2}
    \fmflabel{}{i1}
    \fmflabel{}{i2}
    \fmfright{o1,o2,o3}
    \fmflabel{}{o1}
    \fmflabel{}{o2}
    \fmf{dashes}{i2,v2}
    \fmf{dashes}{v2,o2}
    \fmf{dbl_wiggly}{i1,v1}
    \fmf{double_arrow}{i1,i2}
    \fmf{double_arrow,left=0.3}{v1,v2}
    \fmf{double_arrow,left=0.3}{v2,v1}
     \fmf{fermion,left=0.3}{o2,o3}
    \fmf{fermion,left=0.3}{o3,o2}
    \fmf{dbl_plain}{v4,o3}
    \fmf{dbl_plain}{o3,v5}
    \fmfforce{(0.0w,0.0h)}{i1}
    \fmfforce{(0.0w,1.0h)}{i2}
    \fmfforce{(1.0w,0.0h)}{o1}
    \fmfforce{(1.0w,1.0h)}{o2}
    \fmfforce{(1.0w,1.5h)}{o3}
    \fmfforce{(0.5w,1.0h)}{v2}
    \fmfforce{(0.5w,1.5h)}{v3}
    \fmfforce{(0.5w,0.0h)}{v1}
        \fmfforce{(1.15w,1.5h)}{v4}
    \fmfforce{(0.85w,1.5h)}{v5}
    \fmfdot{v1,v2}
    \fmfdot{i1,i2}
    \fmfdot{o2}
    \end{fmfgraph*}
\end{gathered}\hspace{5mm}+\hspace{3mm}
\begin{gathered}
    \begin{fmfgraph*}(50,50)
    \fmfcurved
    \fmfset{arrow_len}{3mm}
    \fmfleft{i1,i2}
    \fmflabel{}{i1}
    \fmflabel{}{i2}
    \fmfright{o1,o2,o3}
    \fmflabel{}{o1}
    \fmflabel{}{o2}
    \fmf{dashes}{i2,v2}
    \fmf{dashes}{v2,o2}
    \fmf{dashes}{i1,v1}
    \fmf{dashes}{v1,o1}
    \fmf{double_arrow}{i1,i2}
    \fmf{double_arrow,left=0.3}{v1,v2}
    \fmf{double_arrow,left=0.3}{v2,v1}
     \fmf{fermion,left=0.3}{o2,o3}
    \fmf{fermion,left=0.3}{o3,o2}
    \fmf{dbl_plain}{v4,o3}
    \fmf{dbl_plain}{o3,v5}
    \fmf{fermion,left=0.3}{o1,v6}
    \fmf{fermion,left=0.3}{v6,o1}
    \fmf{dbl_plain}{v7,v6}
    \fmf{dbl_plain}{v6,v8}
    \fmfforce{(0.0w,0.0h)}{i1}
    \fmfforce{(0.0w,1.0h)}{i2}
    \fmfforce{(1.0w,0.0h)}{o1}
    \fmfforce{(1.0w,1.0h)}{o2}
    \fmfforce{(1.0w,1.5h)}{o3}
    \fmfforce{(0.5w,1.0h)}{v2}
    \fmfforce{(0.5w,1.5h)}{v3}
    \fmfforce{(0.5w,0.0h)}{v1}
        \fmfforce{(1.15w,1.5h)}{v4}
    \fmfforce{(0.85w,1.5h)}{v5}
    \fmfforce{(1.0w,0.5h)}{v6}
    \fmfforce{(1.15w,0.5h)}{v7}
    \fmfforce{(0.85w,0.5h)}{v8}
    \fmfdot{v1,v2}
    \fmfdot{i1,i2}
    \fmfdot{o1,o2}
    \end{fmfgraph*}
\end{gathered}
\hspace{5mm}+\hspace{5mm} \cdots
\end{split}
\end{gather}
and so on for the terms containing the higher-body de-excitation amplitudes. Here, we have introduced the mixed Feynman-Goldstone diagram notation to represent the Green's functions and interaction-reducible matrix elements. The dressed propagators are to be identified as Green's functions, whilst undressed fermion lines contracting with the $\lambda$-matrices are to be interpreted as Goldstone diagrams that contribute to the effective interaction, $\tilde{\chi}_{pq,rs}$. From the diagrammatic expansion it is straightforward to see that the interaction-reducible diagrams are contained in the interaction-irreducible effective interactions. 

The three-body effective interaction is similarly given by  
\begin{gather}
        \begin{split}
            \tilde{\chi}_{pqr,stu} &=\hspace{5mm}\begin{gathered}
    \begin{fmfgraph*}(40,40)
    \fmfset{arrow_len}{3mm}
    \fmfleft{i1}
    \fmfright{o1}
    \fmf{dbl_dashes}{i1,v1}
    \fmf{dbl_dashes}{v1,o1}
    \fmfdot{i1,o1,v1}
    \fmflabel{$\substack{p\\s}$}{i1}
    \fmflabel{$\substack{q\\ \\ t}$}{v1}
    \fmflabel{$\substack{r\\u}$}{o1}
\end{fmfgraph*}
\end{gathered}\\
&=\hspace{5mm}\begin{gathered}
    \begin{fmfgraph*}(40,40)
    \fmfset{arrow_len}{3mm}
    \fmfleft{i1}
    \fmfright{o1}
    \fmf{dashes}{i1,v1}
    \fmf{dashes}{v1,o1}
    \fmfdot{i1,o1,v1}
\end{fmfgraph*}
\end{gathered}\hspace{5mm}+\hspace{5mm}
\begin{gathered}
    \begin{fmfgraph*}(60,40)
    \fmfcurved
    \fmfset{arrow_len}{3mm}
    \fmfleft{i1,i2}
    \fmflabel{}{i1}
    \fmflabel{}{i2}
    \fmfright{o1,o2}
    \fmflabel{}{o1}
    \fmflabel{}{o2}
    \fmf{dbl_dashes}{i1,v1}
    \fmf{dbl_dashes}{v1,v5}
    \fmf{dbl_dashes}{v5,o1}
    \fmf{fermion,left=0.3,tension=0}{o1,o2}
    \fmf{phantom}{v2,i2}
    \fmf{fermion,left=0.3,tension=0}{o1,o2}
    \fmf{dbl_plain}{v3,o2}
    \fmf{dbl_plain}{o2,v4}
    \fmf{fermion,left=0.3,tension=0}{o2,o1}
    \fmf{fermion,left=0.3,tension=0}{o2,o1}
    \fmfdot{o1,i1,v1,v5}
    \fmfforce{(1.05w,1.0h)}{v3}
    \fmfforce{(0.75w,1.0h)}{v4}
    \fmfforce{(0.67w,0.0h)}{v5}
    \fmfforce{(0.33w,0.0h)}{v1}
\end{fmfgraph*}
\end{gathered}\\
\\
&+\hspace{5mm}
\begin{gathered}
    \begin{fmfgraph*}(60,50)
    \fmfcurved
    \fmfset{arrow_len}{3mm}
    \fmfleft{i1,i2}
    \fmflabel{}{i1}
    \fmflabel{}{i2}
    \fmfright{o1,o2}
    \fmflabel{}{o1}
    \fmflabel{}{o2}
    \fmf{dbl_dashes}{i1,v1}
    \fmf{dbl_dashes}{v1,v5}
    \fmf{dbl_dashes}{v5,v3}
    \fmf{dbl_dashes}{v3,o1}
    \fmf{dbl_plain}{v4,o2}
    \fmf{fermion,left=0.3,tension=0}{o1,o2}
    \fmf{fermion,left=0.3,tension=0}{v3,v4}
    \fmf{fermion,left=0.3,tension=0}{v4,v3}
    \fmf{fermion,left=0.3,tension=0}{o2,o1}
    \fmfdot{o1,i1,v1,v3,v5}
    \fmfforce{(0.0w,0.0h)}{i1}
    \fmfforce{(1.0w,0.0h)}{o1}
    \fmfforce{(0.25w,1h)}{v2}
    \fmfforce{(0.25w,0.0h)}{v1}
    \fmfforce{(0.625w,0.0h)}{v3}
    \fmfforce{(0.625w,1.0h)}{v4}
    \fmfforce{(0.0w,1.0h)}{i2}
    \fmfforce{(1.0w,1.0h)}{o2}
    \fmfforce{(0.425w,0.0h)}{v5}
\end{fmfgraph*}
\end{gathered}\hspace{5mm}+\hspace{2.5mm} \cdots
        \end{split}
    \end{gather}
\begin{figure*}[ht]
    \centering
    \begin{gather*}
\begin{split}~\label{eq:sc_cc}
\tilde{\Sigma}^{2\p1\h/2\h1\p}_{pq} &=\hspace{5mm}\begin{gathered}
\begin{fmfgraph*}(30,30)
    \fmfset{arrow_len}{3mm}
    \fmfleft{i1}
    \fmfright{o1}
    \fmf{zigzag}{o1,i1}
    \fmfv{decor.shape=cross,decor.filled=full, decor.size=1.5thic}{o1}
    \fmfdot{i1}
\end{fmfgraph*}
\end{gathered}\hspace{2.5mm}+\hspace{2.5mm}  \begin{gathered}\begin{fmfgraph*}(40,40)
    \fmfcurved
    \fmfset{arrow_len}{3mm}
    \fmfleft{i1,i2}
    \fmflabel{}{i1}
    \fmflabel{}{i2}
    \fmfright{o1,o2}
    \fmflabel{}{o1}
    \fmflabel{}{o2}
    \fmf{double_arrow}{i1,i2}
    \fmf{dbl_zigzag}{o1,i1}
    \fmf{double_arrow,left=0.2,tension=0}{o1,o2}
    \fmf{dbl_zigzag}{o2,i2}
    \fmf{double_arrow,left=0.2,tension=0}{o2,o1}
    \fmfdot{o1,o2,i1,i2}
\end{fmfgraph*}
\end{gathered}\hspace{2.5mm}+\hspace{2.5mm}
\begin{gathered}
    \begin{fmfgraph*}(40,60)
    \fmfcurved
    \fmfset{arrow_len}{3mm}
    \fmfleft{i1,i2,i3}
    \fmflabel{}{i1}
    \fmflabel{}{i2}
    \fmfright{o1,o2,o3}
    \fmflabel{}{o1}
    \fmflabel{}{o2}
    \fmf{dbl_zigzag}{i2,o2}
    \fmf{dbl_zigzag}{i3,o3}
    \fmf{dbl_zigzag}{i1,o1}
    \fmf{double_arrow}{i1,i2}
    \fmf{double_arrow}{i2,i3}
    \fmf{double_arrow}{o1,o2}
    \fmf{double_arrow}{o2,o3}
    \fmf{double_arrow,left=0.3}{o3,o1}
    \fmfforce{(0.0w,0.0h)}{i1}
    \fmfforce{(1.0w,0.0h)}{o1}
    \fmfforce{(0.0w,1.0h)}{i3}
    \fmfforce{(1.0w,1.0h)}{o3}
    \fmfdot{i1,i2,i3}
    \fmfdot{o1,o2,o3}
\end{fmfgraph*}
\end{gathered}
\hspace{7.5mm}+\hspace{5mm}
\begin{gathered}
\begin{fmfgraph*}(40,60)
    \fmfcurved
    \fmfset{arrow_len}{3mm}
    \fmfleft{i1,i2,i3}
    \fmflabel{}{i1}
    \fmflabel{}{i2}
    \fmfright{o1,o2,o3}
    \fmflabel{}{o1}
    \fmflabel{}{o2}
    \fmf{dbl_zigzag}{i1,v1}
    \fmf{dbl_zigzag}{v1,o1}
    \fmf{dbl_zigzag}{v2,o2}
    \fmf{dbl_zigzag}{i3,v3}
    \fmf{double_arrow}{i1,i3}
    \fmf{double_arrow,left=0.3}{o1,o2}
    \fmf{double_arrow,left=0.3}{o2,o1}
    \fmf{double_arrow,left=0.3}{v2,v3}
    \fmf{double_arrow,left=0.3}{v3,v2}
    \fmfforce{(0.0w,0.0h)}{i1}
    \fmfforce{(1.0w,0.0h)}{o1}
    \fmfforce{(0.5w,0.5h)}{v2}
    \fmfforce{(0.5w,0.0h)}{v1}
    \fmfforce{(0.5w,1.0h)}{v3}
    \fmfforce{(0.0w,1.0h)}{i3}
    \fmfforce{(1.0w,1.0h)}{o3}
    \fmfdot{v2,v3}
    \fmfdot{i1,i3}
    \fmfdot{o1,o2}
\end{fmfgraph*}
\end{gathered}\hspace{5mm}+\hspace{5mm}
\begin{gathered}
    \begin{fmfgraph*}(40,60)
    \fmfcurved
    \fmfset{arrow_len}{3mm}
    \fmfleft{i1,i2,i3}
    \fmflabel{}{i1}
    \fmflabel{}{i2}
    \fmfright{o1,o2,o3}
    \fmflabel{}{o1}
    \fmflabel{}{o2}
    \fmf{dbl_zigzag}{i1,v1}
    \fmf{dbl_dashes}{i2,v2}
    \fmf{dbl_dashes}{v2,o2}
    \fmf{dbl_zigzag}{i3,v3}
    \fmf{double_arrow}{i1,i2}
    \fmf{double_arrow}{v1,v2}
    \fmf{double_arrow}{o2,v1}
    \fmf{double_arrow}{v2,v3}
    \fmf{double_arrow}{v3,o2}
    \fmf{double_arrow}{i2,i3}
    \fmfforce{(0.0w,0.0h)}{i1}
    \fmfforce{(1.0w,0.0h)}{o1}
    \fmfforce{(0.5w,0.5h)}{v2}
    \fmfforce{(0.5w,0.0h)}{v1}
    \fmfforce{(0.5w,1.0h)}{v3}
    \fmfforce{(0.0w,1.0h)}{i3}
    \fmfforce{(1.0w,1.0h)}{o3}
    \fmfdotn{v}{3}
    \fmfdot{i1,i2,i3}
    \fmfdot{o2}
\end{fmfgraph*}
\end{gathered}
\end{split}
\end{gather*}
    \caption{The 2p1h/2h1p excitation character restricted third-order one-particle irreducible skeleton coupled-cluster self-energy diagrams expressed with respect to the exact single-particle coupled-cluster Green's function, $\tilde{G}$.}
    \label{fig:sc_cc_se}
\end{figure*}
and so on. The higher-body terms are constructed analogously by contracting the higher-order effective interaction matrix elements $\{\chi\}$ with the corresponding $\lambda$-matrices. These relationships demonstrate that the effective interactions generated by the self-consistent Green's function theory correspond to the higher-order derivatives of the CC Lagrangian with respect to the non-interacting Green's function. Within the CCSD approximation, all dressed effective interactions contributions are truncated to contain contractions of terms up to the doubles de-excitation amplitudes, $\{\lambda^{ij}_{ab}\}$, only.

Using the Green's function residues and poles defined in Section~\ref{sec:ex_sp_ccgf}, as an example, we evaluate the second-order self-consistent contribution to the renormalized coupled-cluster self-energy to give
\begin{gather}
    \begin{split}~\label{eq:sc_2nd_se}
        \tilde{\Sigma}^{(2)}_{pq}[\tilde{G}](\omega) = \frac{1}{2}\sum_{\substack{rst\\uvw}}\tilde{\Xi}_{pr,tu}\tilde{G}^{2\p1\h/2\h1\p}_{tur,svw}(\omega)\tilde{\Xi}_{sv,qw}    .
    \end{split}
\end{gather}
Here, we have identified the resolvents generated in Eq.~\ref{eq:sc_2nd_se} in terms of the propagator
\begin{gather}
    \begin{split}~\label{eq:resolv}
    \tilde{G}^{2\p1\h/2\h1\p}_{tur,svw}(\omega) &= \sum_{\mu_1\mu_2\nu_1}\frac{(\Tilde{Y}^{\mu_1}_{t}\Tilde{Y}^{\mu_2}_{u}\Tilde{X}^{\nu_1}_{r})\bar{Y}^{\mu_1}_s\bar{Y}^{\mu_2}_v\bar{X}^{\nu_1}_w}{\omega+\varepsilon^{N-1}_{\nu_1}-\varepsilon^{N+1}_{\mu_1}-\varepsilon^{N+1}_{\mu_2}+i\eta} \\
    &+ \sum_{\nu_1 \nu_2 \mu_1}\frac{\bar{X}^{\nu_1}_{t}\bar{X}^{\nu_2}_{u}\bar{Y}^{\mu_1}_{r}(\tilde{X}^{\nu_1}_s\tilde{X}^{\nu_2}_v\tilde{Y}^{\mu_1}_w)}{\omega+\varepsilon^{N+1}_{\mu_1}-\varepsilon^{N-1}_{\nu_1}-\varepsilon^{N-1}_{\nu_2}-i\eta} ,
    \end{split}
\end{gather}
representing correlated simultaneous charged excitation processes involving three-particles. The space spanned by the indices $\mu$ and $\nu$ is much larger than the spin-orbital space as they index the removal/addition poles of the Green's function and are directly connected to charged excitation processes.

To conclude this subsection, in Figure~\ref{fig:sc_cc_se} we depict the diagrammatic content of the set of renormalized third-order 1PI skeleton coupled-cluster self-energy diagrams restricted to the space of 2p1h/2h1p intermediate excitation states. These diagrams contain excitations involving renormalized intermediate states of 2p1h/2h1p excitation character and will therefore constitute the most important contributions to the coupled-cluster self-energy at third-order. All diagrams depicted in  Figure~\ref{fig:sc_cc_se} are 1PI, skeleton and interaction-irreducible.

\subsection{Aside on the vertex functions}

The vertex functions introduced in Section~\ref{subsec:eom_gf} provide us with a formally exact, closed-form equation for the coupled-cluster self-energy. However, they are clearly related to the equation-of-motion of their corresponding higher-body Green's function, thereby constituting a self-consistent self of equations. 

For example, by taking the time-derivative of the first argument of the 4-point coupled-cluster Green's function, using the biorthogonal Heisenberg equation-of-motion and taking the Fourier transform, we have
\begin{widetext}
\begin{gather}
\begin{split}~\label{eq:eom_g4pt}
    &\tilde{G}^{4\pt}_{pq,rs}(\omega_1,\omega_2;\omega_3,\omega_4) = \\
    &2\pi i\left(\delta(\omega_1-\omega_3)\delta(\omega_2-\omega_4)G^{0}_{pr}(\omega_1)\tilde{G}_{qs}(\omega_2)-\delta(\omega_1-\omega_4)\delta(\omega_2-\omega_3)G^{0}_{ps}(\omega_1)\tilde{G}_{qr}(\omega_2)\right)\\
    &+\sum_{ut} G^{0}_{pu}(\omega_1)\left(\bar{h}_{ut}-f_{ut}\right)\tilde{G}^{4\pt}_{ut,rs}(\omega_1,\omega_2;\omega_3,\omega_4) \\
    &- \frac{1}{2!}\sum_{utvw} G^{0}_{pu}(\omega_1)\bar{h}_{ut,vw} \int\frac{d\omega d\omega'}{(2\pi)^2} \tilde{G}^{6\pt}_{vwq,rts}(\omega,\omega',\omega_2;\omega_3,\omega+\omega'-\omega_1,\omega_4)\\
    &+\frac{1}{3!\cdot2!}\sum_{\substack{utv\\
    wl\sigma}} G^{0}_{pu}(\omega_1) \bar{h}_{utv,wl\sigma}\int\frac{d\omega d\omega' d\omega'' d\omega'''}{(2\pi)^4} \tilde{G}^{8\pt}_{wl\sigma q,rtvs}(\omega,\omega',\omega'',\omega_2;\omega_3,\omega'',\omega+\omega'+\omega''-\omega_1-\omega''',\omega_4) \hspace{3mm} \\
    &+\hspace{3mm} \cdots
\end{split}
\end{gather}
\end{widetext}
In Eq.~\ref{eq:4point_vert}, all single-particle Green's functions appearing are exact, whereas in Eq.~\ref{eq:eom_g4pt} we see that one of the propagators is undressed in the first line. However, inserting the vertex equations that define $\tilde{\Lambda}^{4\pt}$, $\tilde{\Lambda}^{6\pt}$, $\tilde{\Lambda}^{8\pt}$ and so on into Eq.~\ref{eq:eom_g4pt} results in order-by-order self-consistent dressing of the non-interacting Green's function. As a result, we are left with a self-consistent equation for the 4-point vertex function $\tilde{\Lambda}^{4\pt}$ that is coupled to all higher-order vertices $\tilde{\Lambda}^{6\pt}$, $\tilde{\Lambda}^{8\pt}$ and so on. In addition to the dressing of the single-particle Green's function arising from the vertex equations, we also find that the interaction vertices are also renormalized and dressed. This procedure is analogous to that observed in the case of the EOM for the single-particle Green's function as discussed in Subsection~\ref{subsec:eom_gf}. The same analysis also applies to the 6-point, 8-point and so on vertex functions. The relationship between the 4-point vertex and the BSE kernel is discussed at length in Appendix~\ref{app:4_vertex}. To lowest-order, the $n$-point vertices are given by the corresponding $(\frac{n}{2})$-body anti-symmetrized effective interactions. The $n$-point vertex is generated by the functional derivative of the $(n-2)$-point vertex with respect to the single-particle Green's function, with the 4-point vertex generated by the functional derivative of the self-energy with respect to the Green's function.

\subsection{Recovering the perturbative expansion}

To recover the perturbative expansion for the coupled-cluster self-energy requires careful consideration. This is due to the fact that the interaction-irreducible matrix elements of the similarity transformed Hamiltonian also depend on the exact Green's function. The self-consistent self-energy at second-order is composed of two diagrams 
\begin{gather}
\begin{split}~\label{eq:sc_2}
\tilde{\Sigma}^{\sco(2)}[\tilde{G}]= \hspace{2.5mm}
\begin{gathered}
\begin{fmfgraph*}(40,40)
    \fmfset{arrow_len}{3mm}
    \fmfleft{i1}
    \fmfright{o1}
    \fmf{zigzag}{o1,i1}
    \fmfv{decor.shape=cross,decor.filled=full, decor.size=1.5thic}{o1}
    \fmfdot{i1}
\end{fmfgraph*}
\end{gathered}\hspace{5mm}+\hspace{3mm}
\begin{gathered}
    \begin{fmfgraph*}(50,50)
    \fmfcurved
    \fmfset{arrow_len}{3mm}
    \fmfleft{i1,i2}
    \fmflabel{}{i1}
    \fmflabel{}{i2}
    \fmfright{o1,o2}
    \fmflabel{}{o1}
    \fmflabel{}{o2}
    \fmf{double_arrow}{i1,i2}
    \fmf{dbl_zigzag}{o1,i1}
    \fmf{double_arrow,left=0.3,tension=0}{o1,o2}
    \fmf{dbl_zigzag}{o2,i2}
    \fmf{double_arrow,left=0.3,tension=0}{o2,o1}
    \fmfdot{o1,o2,i1,i2}
\end{fmfgraph*}
\end{gathered}
\end{split}
\end{gather}
However, the effective interactions as well as Green's function lines have corresponding Dyson expansions. Diagrammatically, these take the form (see Appendix~\ref{app_pert_int})
\begin{gather}
    \begin{split}~\label{eq:eff-body}
    \begin{gathered}
        \begin{fmfgraph*}(40,40)
    \fmfset{arrow_len}{3mm}
    \fmfleft{i1}
    \fmfright{o1}
    \fmf{zigzag}{o1,i1}
    \fmfv{decor.shape=cross,decor.filled=full, decor.size=1.5thic}{o1}
    \fmfdot{i1}
\end{fmfgraph*}
\end{gathered}\hspace{2.5mm} &= \hspace{2.5mm} 
\begin{gathered}
\begin{fmfgraph*}(40,40)
    \fmfset{arrow_len}{3mm}
    \fmfleft{i1}
    \fmfright{o1}
    \fmf{dbl_dashes}{o1,i1}
    \fmfv{decor.shape=cross,decor.filled=full, decor.size=1.5thic}{o1}
    \fmfdot{i1}
\end{fmfgraph*}
\end{gathered}\hspace{2.5mm}+\hspace{2.5mm}
\begin{gathered}
    \begin{fmfgraph*}(40,40)
    \fmfcurved
    \fmfset{arrow_len}{3mm}
    \fmfleft{i1,i2}
    \fmflabel{}{i1}
    \fmflabel{}{i2}
    \fmfright{o1,o2}
    \fmflabel{}{o1}
    \fmflabel{}{o2}
    \fmf{dbl_wiggly}{i1,o1}
    \fmf{fermion,left=0.3,tension=0}{o1,o2}
    \fmf{dbl_dashes}{o2,i2}
    \fmf{fermion,left=0.3,tension=0}{o2,o1}
    \fmfdot{o1,o2,i1}
    \fmfv{decor.shape=cross,decor.size=1thic, decor.filled=full}{i2}
\end{fmfgraph*}
\end{gathered}\\
&+\hspace{2.5mm}
\begin{gathered}
    \begin{fmfgraph*}(65,50)
    \fmfcurved
    \fmfset{arrow_len}{3mm}
    \fmfleft{i1,i2}
    \fmflabel{}{i1}
    \fmflabel{}{i2}
    \fmfright{o1,o2}
    \fmflabel{}{o1}
    \fmflabel{}{o2}
    \fmf{dashes}{o1,v1}
    \fmf{dashes}{i1,v1}
    \fmf{fermion,left=0.3,tension=0}{o1,o2}
    \fmf{fermion,left=0.3,tension=0}{v1,v2}
    \fmf{fermion,left=0.3,tension=0}{v2,v1}
    \fmf{phantom}{v2,i2}
    \fmf{dbl_wiggly}{o2,v2}
    \fmf{fermion,left=0.3,tension=0}{o2,o1}
    \fmfdot{o1,o2,i1,v1,v2}
\end{fmfgraph*}
\end{gathered}\hspace{2.5mm}+\hspace{2.5mm} \cdots
    \end{split}
\end{gather}
with 
\begin{gather}
    \begin{split}~\label{eq:2-body}
        \begin{gathered}
        \begin{fmfgraph*}(40,40)
    \fmfset{arrow_len}{3mm}
    \fmfleft{i1}
    \fmfright{o1}
    \fmf{dbl_zigzag}{o1,i1}
    \fmfdot{i1,o1}
\end{fmfgraph*}
\end{gathered}\hspace{2.5mm} &= \hspace{2.5mm} 
\begin{gathered}
        \begin{fmfgraph*}(40,40)
    \fmfset{arrow_len}{3mm}
    \fmfleft{i1}
    \fmfright{o1}
    \fmf{dbl_wiggly}{o1,i1}
    \fmfdot{i1,o1}
\end{fmfgraph*}
\end{gathered}\hspace{2.5mm}+\hspace{2.5mm}
\begin{gathered}
    \begin{fmfgraph*}(60,40)
    \fmfcurved
    \fmfset{arrow_len}{3mm}
    \fmfleft{i1,i2}
    \fmflabel{}{i1}
    \fmflabel{}{i2}
    \fmfright{o1,o2}
    \fmflabel{}{o1}
    \fmflabel{}{o2}
    \fmf{dashes}{o1,v1}
    \fmf{dashes}{i1,v1}
    \fmf{fermion,left=0.3,tension=0}{o1,o2}
    \fmf{phantom}{v2,i2}
    \fmf{fermion,left=0.3,tension=0}{o1,o2}
    \fmf{dbl_dashes}{o2,v2}
    \fmf{fermion,left=0.3,tension=0}{o2,o1}
    \fmfv{decor.shape=cross,decor.size=1thic, decor.filled=full}{v2}
    \fmf{fermion,left=0.3,tension=0}{o2,o1}
    \fmfdot{o1,o2,i1,v1}
\end{fmfgraph*}
\end{gathered}\\
&+\hspace{2.5mm}
\begin{gathered}
    \begin{fmfgraph*}(60,50)
    \fmfcurved
    \fmfset{arrow_len}{3mm}
    \fmfleft{i1,i2}
    \fmflabel{}{i1}
    \fmflabel{}{i2}
    \fmfright{o1,o2}
    \fmflabel{}{o1}
    \fmflabel{}{o2}
    \fmf{dbl_dashes}{v1,i1}
    \fmf{dbl_dashes}{o1,v1}
    \fmf{dbl_dashes}{v1,v3}
    \fmf{dbl_dashes}{v3,o1}
    \fmf{dbl_wiggly}{v4,o2}
    \fmf{fermion,left=0.3,tension=0}{o1,o2}
    \fmf{fermion,left=0.3,tension=0}{v3,v4}
    \fmf{fermion,left=0.3,tension=0}{v4,v3}
    \fmf{fermion,left=0.3,tension=0}{o2,o1}
    \fmfdot{o1,o2,i1,v1,v3,v4}
    \fmfforce{(0.0w,0.0h)}{i1}
    \fmfforce{(1.0w,0.0h)}{o1}
    \fmfforce{(0.25w,1h)}{v2}
    \fmfforce{(0.25w,0.0h)}{v1}
    \fmfforce{(0.625w,0.0h)}{v3}
    \fmfforce{(0.625w,1.0h)}{v4}
    \fmfforce{(0.0w,1.0h)}{i2}
    \fmfforce{(1.0w,1.0h)}{o2}
\end{fmfgraph*}
\end{gathered}\hspace{5mm}+\hspace{2.5mm} \cdots
    \end{split}
\end{gather}
Now, retaining only terms up to second-order (diagrams containing up to two interaction lines), we insert Eqs~\ref{eq:eff-body} and~\ref{eq:2-body} into Eq.~\ref{eq:sc_2} to obtain the second-order perturbative expansion of the self-energy as 
\begin{gather}
\begin{split}~\label{eq:sc_2_pt}
\tilde{\Sigma}^{(2)}[G_0] &= \hspace{2.5mm}
\begin{gathered}
\begin{fmfgraph*}(40,40)
    \fmfset{arrow_len}{3mm}
    \fmfleft{i1}
    \fmfright{o1}
    \fmf{dbl_dashes}{o1,i1}
    \fmfv{decor.shape=cross,decor.filled=full, decor.size=1.5thic}{o1}
    \fmfdot{i1}
\end{fmfgraph*}
\end{gathered}\hspace{5mm}+\hspace{3mm}
\begin{gathered}
    \begin{fmfgraph*}(50,50)
    \fmfcurved
    \fmfset{arrow_len}{3mm}
    \fmfleft{i1,i2}
    \fmflabel{}{i1}
    \fmflabel{}{i2}
    \fmfright{o1,o2}
    \fmflabel{}{o1}
    \fmflabel{}{o2}
    \fmf{fermion}{i1,i2}
    \fmf{dbl_wiggly}{o1,i1}
    \fmf{fermion,left=0.3,tension=0}{o1,o2}
    \fmf{dbl_wiggly}{o2,i2}
    \fmf{fermion,left=0.3,tension=0}{o2,o1}
    \fmfdot{o1,o2,i1,i2}
\end{fmfgraph*}
\end{gathered}\hspace{2.5mm}+\hspace{2.5mm}\begin{gathered}
    \begin{fmfgraph*}(40,40)
    \fmfcurved
    \fmfset{arrow_len}{3mm}
    \fmfleft{i1,i2}
    \fmflabel{}{i1}
    \fmflabel{}{i2}
    \fmfright{o1,o2}
    \fmflabel{}{o1}
    \fmflabel{}{o2}
    \fmf{dbl_wiggly}{i1,o1}
    \fmf{fermion,left=0.3,tension=0}{o1,o2}
    \fmf{dbl_dashes}{o2,i2}
    \fmf{fermion,left=0.3,tension=0}{o2,o1}
    \fmfdot{o1,o2,i1}
    \fmfv{decor.shape=cross,decor.size=1thic, decor.filled=full}{i2}
\end{fmfgraph*}
\end{gathered}\\
&+\hspace{2.5mm}
\begin{gathered}
    \begin{fmfgraph*}(65,50)
    \fmfcurved
    \fmfset{arrow_len}{3mm}
    \fmfleft{i1,i2}
    \fmflabel{}{i1}
    \fmflabel{}{i2}
    \fmfright{o1,o2}
    \fmflabel{}{o1}
    \fmflabel{}{o2}
    \fmf{dashes}{o1,v1}
    \fmf{dashes}{i1,v1}
    \fmf{fermion,left=0.3,tension=0}{o1,o2}
    \fmf{fermion,left=0.3,tension=0}{v1,v2}
    \fmf{fermion,left=0.3,tension=0}{v2,v1}
    \fmf{phantom}{v2,i2}
    \fmf{dbl_wiggly}{o2,v2}
    \fmf{fermion,left=0.3,tension=0}{o2,o1}
    \fmfdot{o1,o2,i1,v1,v2}
\end{fmfgraph*}
\end{gathered}
\end{split}
\end{gather}
This is exactly the expression obtained at second order in the perturbative expansion as presented in Section~\ref{sec:pert}. All higher-order perturbative diagrams are obtained from the self-consistent expression by successive diagrammatic expansion of exact propagators and interaction vertices.

\section{The Coupled-cluster Ground State Energy}~\label{sec:gs_energy}

In this section, we demonstrate how the coupled-cluster correlation energy can be obtained from the coupled-cluster self-energy through the equation-of-motion for the SP-CCGF. The coupled-cluster ground state energy is given by the expectation value 
\begin{gather}
    \begin{split}~\label{eq:gs_energy}
        E^{\CC}_0 &= \braket{\Phi_0|\bar{H}|\Phi_0}\\
        &=\sum_{pq}\bar{h}_{pq}\gamma^{\rf}_{pq} + \frac{1}{(2!)^2}\sum_{pq,rs} \bar{h}_{pq,rs}\Gamma^{\rf}_{pq,rs}\\
        &+\frac{1}{(3!)^2}\sum_{pqr,stu}\bar{h}_{pqr,stu} \Gamma^{\rf}_{pqr,stu} + \cdots \ ,
    \end{split}
\end{gather}
From Eq.~\ref{eq:gs_energy}, we have 
\begin{gather}
    \begin{split}~\label{eq:energy_ref}
        E^{\CC}_0 &= \sum_{i}\bar{h}_{ii} + \frac{1}{2}\sum_{ij} \bar{h}_{ij,ij} \\
        &= E^{\rf}_0+\frac{1}{4}\sum_{ijab}\braket{ij||ab}(t^{ab}_{ij}+2t^{a}_it^b_j) + \sum_{ia}f_{ia}t^{a}_{i}\\
        &= E^{\rf}_0 + E^{\CC}_{c}
        \ .
    \end{split}
\end{gather}
All additional higher-body terms evaluate to zero due to the structure of the coupled-cluster similarity transformed Hamiltonian as any matrix elements with four or more lines below the interaction vertex vanish~\cite{shavitt2009many}. The ground state energy is also given by the coupled-cluster Lagrangian via 
\begin{widetext}
\begin{gather}
\begin{split}~\label{eq:energy_lagrange}
    \mathcal{L}^{\CC}_0 &= \braket{\tilde{\Psi}_0|\bar{H}|\Phi_0}=\braket{\tilde{\Psi}_0|\bar{h}_1|\Phi_0}+\braket{\tilde{\Psi}_0|\bar{V}|\Phi_0}+\braket{\tilde{\Psi}_0|\bar{W}|\Phi_0}+\cdots\\
    &= \sum_{pq}\bar{h}_{pq}\gamma^{\rf}_{pq} + \frac{1}{(2!)^2}\sum_{pq,rs} \bar{h}_{pq,rs}\Gamma^{\rf}_{pq,rs}+\frac{1}{(3!)^2}\sum_{pqr,stu}\bar{h}_{pqr,stu} \Gamma^{\rf}_{pqr,stu} + \cdots \\
    &+\sum_{pq}\bar{h}_{pq}\braket{\Lambda|a^\dag_pa_q|\Phi_0}+\frac{1}{(2!)^2}\sum_{pq,rs} \bar{h}_{pq,rs}\braket{\Lambda|a^\dag_pa^\dag_qa_sa_r|\Phi_0}+\frac{1}{(3!)^2}\sum_{pqr,stu}\bar{h}_{pqr,stu}\braket{\Lambda|a^\dag_pa^\dag_qa^\dag_ra_ua_ta_s|\Phi_0}
    +  \cdots
\end{split}
\end{gather}
\end{widetext}
where $\bra{\Lambda} = \bra{\Phi_0}\Lambda$ is the state obtained from application of the de-excitation operator onto the reference. From the equation-of-motion for the single-particle CC Green's function (Eq.~\ref{eq:eom_sp_ccgf}), it appears that the total ground-state energy cannot be obtained from the self-energy because
\begin{gather}
    \begin{split}~\label{eq:cc_gm}
        &\frac{1}{2}\lim_{\eta\to0^+}\sum_{pq}\int^{\infty}_{-\infty}\frac{d\omega}{2\pi i}e^{i\eta\omega}\tilde{\Sigma}_{pq}(\omega)\tilde{G}_{qp}(\omega) = \\
        &\frac{1}{2}\sum_{pq}(\bar{h}_{pq}-f_{pq})\braket{\Phi_0|(\mathbbm{1}+\Lambda)a^\dag_p a_q|\Phi_0}\\
        &+ \frac{1}{(2!)^2}\sum_{pq,rs} \bar{h}_{pq,rs}\braket{\Phi_0|(\mathbbm{1}+\Lambda)a^\dag_pa^\dag_qa_sa_r|\Phi_0}\\
        &+\frac{1}{3!\cdot(2!)^2} \sum_{pqr,stu}\bar{h}_{pqr,stu}\braket{\Phi_0|(\mathbbm{1}+\Lambda)a^\dag_pa^\dag_qa^\dag_ra_ua_ta_s|\Phi_0} \\
        &+  \cdots
    \end{split}
\end{gather}
which equates to 
\begin{gather}
    \begin{split}~\label{eq:cc_gm_1}
        &\frac{1}{2}\lim_{\eta\to0^+}\sum_{pq}\int^{\infty}_{-\infty}\frac{d\omega}{2\pi i}e^{i\eta\omega}\tilde{\Sigma}_{pq}(\omega)\tilde{G}_{qp}(\omega) =\\
        &\frac{1}{2}\braket{\Phi_0|(\mathbbm{1}+\Lambda)(\bar{h}_1-F)|\Phi_0}+ \braket{\Phi_0|(\mathbbm{1}+\Lambda)\bar{V}|\Phi_0}\\
        &+\frac{3}{2}\braket{\Phi_0|(\mathbbm{1}+\Lambda)\bar{W}|\Phi_0} +  \cdots
    \end{split}
\end{gather}
Hence, the correct prefactors required to give the coupled-cluster Lagrangian in Eq.~\ref{eq:energy_lagrange} are not generated by the equation-of-motion for the single-particle Green's function as demonstrated by Eq.~\ref{eq:cc_gm_1}. Therefore the amplitude equations are not correctly reproduced from the equation-of-motion. This is analogous to the fact that the ground state energy of a three-body hermitian interaction Hamiltonian is not obtainable from the equation-of-motion for the single-particle Green's function~\cite{carbone2013self}. However, the coupled-cluster ground state energy is still accessible from the coupled-cluster self-energy. 

To demonstrate how this is the case, we evaluate the frequency integral in Eq.~\ref{eq:cc_gm_1} analytically by making use of the pole structure of the SP-CCGF and associated coupled-cluster self-energy given in Section~\ref{sec:ex_sp_ccgf}. As the coupled-cluster self-energy consists of a static and dynamical part, $\tilde{\Sigma}(\omega) = \tilde{\Sigma}^{\infty}+\tilde{\Sigma}^{\text{D}}(\omega)$, we have
\begin{gather}
    \begin{split}
        &-\frac{i}{2}\lim_{\eta\to0^+}\sum_{pq} \int^{\infty}_{-\infty}\frac{d\omega}{2\pi}e^{i\eta\omega} \tilde{\Sigma}_{pq}(\omega)\tilde{G}_{qp}(\omega) = \frac{1}{2}\sum_{pq}\tilde{\Sigma}^{\infty}_{pq}\tilde{\gamma}_{pq}\\
        &-\frac{i}{2}\lim_{\eta\to0^+}\sum_{ij} \int^{\infty}_{-\infty}\frac{d\omega}{2\pi}e^{i\eta\omega} \tilde{\Sigma}^{\text{D}}_{ij}(\omega)\tilde{G}_{ji}(\omega)\\
        &-\frac{i}{2}\lim_{\eta\to0^+}\sum_{ab} \int^{\infty}_{-\infty}\frac{d\omega}{2\pi}e^{i\eta\omega} \tilde{\Sigma}^{\text{D}}_{ab}(\omega)\tilde{G}_{ba}(\omega) ,
    \end{split}
\end{gather}
where we have used $\tilde{G}_{ai}(\omega)=\tilde{\Sigma}^{\text{D}}_{ai}(\omega)=0$. The dynamical part of the self-energy is given by the sum of the forward- and backward-time contribuitons: $\tilde{\Sigma}^{\text{D}}=\tilde{\Sigma}^{\text{F}}+\tilde{\Sigma}^{\text{B}}$. The one-particle reduced density matrix is given by $\tilde{\gamma}_{pq}=\braket{\tilde{\Psi}_0|a^\dag_p a_q|\Phi_0}$.  The poles of the occupied-occupied block of the self-energy and Green's function lie above the real axis and the poles of the virtual-virtual block of the self-energy and Green's function lie below the real axis. Therefore, from the residue theorem, the integrals over frequency vanish and we are left only with the static contribution
\begin{gather}
    \begin{split}~\label{eq:gm_cc}
        -\frac{i}{2}\lim_{\eta\to0^+}\sum_{pq} \int^{\infty}_{-\infty}\frac{d\omega}{2\pi}e^{i\eta\omega} \tilde{\Sigma}_{pq}(\omega)\tilde{G}_{qp}(\omega) = \frac{1}{2}\sum_{pq}\tilde{\Sigma}^{\infty}_{pq}\tilde{\gamma}_{pq} \ .
    \end{split}
\end{gather}
From the definition of the one-particle reduced density matrix and the static component of the coupled-cluster self-energy, Eq.~\ref{eq:gm_cc} evaluates to  
\begin{gather}
    \begin{split}~\label{eq:gm_energy_cc}
        &-\frac{i}{2}\lim_{\eta\to0^+}\sum_{pq} \int^{\infty}_{-\infty}\frac{d\omega}{2\pi}e^{i\eta\omega} \tilde{\Sigma}_{pq}(\omega)\tilde{G}_{qp}(\omega) \\
        &=\frac{1}{2}\sum_{i}\tilde{\Sigma}^{\infty}_{ii}+\frac{1}{2}\sum_{ai}\lambda^{i}_{a}\tilde{\Sigma}^{\infty(0)}_{ai} \\
        &= \frac{1}{2}\sum_{ii}  \tilde{\Sigma}^{\infty(0)}_{ii} + \frac{1}{2}\sum_{ika}\lambda^{k}_{a}\chi_{ia,ik} + \frac{1}{(2!)^3}\sum_{iklab}\lambda^{kl}_{ab}\chi_{iab,ikl} + \cdots\\
        &+\frac{1}{2}\sum_{ai}\lambda^{i}_{a}(F_{ai}-f_{ai}) \ ,
    \end{split}
\end{gather}
where we have used Eq.~\ref{eq:exact_stat} to identify $\tilde{\Sigma}^{\infty}_{ai}=\tilde{\Sigma}^{\infty(0)}_{ai}$. We have also used Eq.~\ref{eq:exact_stat} to write the second equality of Eq~\ref{eq:gm_energy_cc}, alongside the definition $\tilde{\Sigma}^{\infty(0)}_{pq}=F_{pq}-f_{pq}$, given in Section~\ref{sec:pert}. We have derived these equations using a general reference state $\ket{\Phi_0}$, such that $\Sigma^{\infty(0)}_{ai}\neq f_{ai}\neq 0$. 
Using Eqs~\ref{eq:energy_lagrange} and~\ref{eq:cc_gm_1}, we can write the CC ground state energy as 
 \begin{gather}
     \begin{split}~\label{eq:energy_full}
         E^{\CC}_0 &= \frac{1}{2}\sum_{i}\tilde{\Sigma}^{\infty}_{ii}+\frac{1}{2}\sum_{ia}\lambda^{i}_{a}(F_{ai}-f_{ai}) \\
         &+ \frac{1}{2}\braket{\tilde{\Psi}_0|(\bar{h}_1+F)|\Phi_0}
         - \frac{1}{2}\braket{\tilde{\Psi}_0|\bar{W}|\Phi_0} + \cdots
     \end{split}
 \end{gather}
The second and third terms evaluate to give $\frac{1}{2}\braket{\tilde{\Psi}_0|(\bar{h}_1+F)|\Phi_0}=\frac{1}{2}\sum_{i}(\bar{h}_{ii}+f_{ii})+\sum_{ai}\lambda^{i}_{a}(\bar{h}_{ai}+f_{ai})$. The additional three-body component results in the expression
\begin{gather}
    \begin{split}
        -\frac{1}{2}\braket{\tilde{\Psi}_0|\bar{W}|\Phi_0} &= -\frac{1}{(2!)^2}\sum_{ijka}\lambda^{k}_{a}\bar{h}_{aij,kij}\\
        &-\frac{1}{(2!)^3}\sum_{kijab}\lambda^{ij}_{ab}\bar{h}_{abk,ijk}\\
        &-\frac{1}{2\cdot(3!)^2}\sum_{ijkabc}\lambda^{ijk}_{abc}\bar{h}_{abc,ijk} \ .
    \end{split}
\end{gather}
The higher-body terms in Eq.~\ref{eq:energy_full} give rise to a similar structure of terms whereby the higher-body interactions are contracted with different de-excitation amplitudes. Using these identities we can simplify Eq.~\ref{eq:energy_full} to give
\begin{gather}
    \begin{split}~\label{eq:proof_energy}
        E^{\CC}_0 &= \frac{1}{2}\sum_{i}\Big(h_{ii}+f_{ii}+\tilde{\Sigma}^{\infty(0)}_{ii}+\sum_{a}h_{ia}t^{a}_{i}\Big) \\
        &+ \frac{1}{2}\sum_{ia}\lambda^{i}_{a}(F_{ai}+\bar{h}_{ai})\\
        &+ \frac{1}{2}\sum_{ika}\lambda^{k}_{a}\chi_{ai,ki} + \frac{1}{(2!)^3}\sum_{iklab}\lambda^{kl}_{ab}\chi_{abi,kli} + \cdots\\
        &-\frac{1}{(2!)^2}\sum_{ijka}\lambda^{k}_{a}\bar{h}_{aij,kij}-\frac{1}{(2!)^3}\sum_{kijab}\lambda^{ij}_{ab}\bar{h}_{abk,ijk}\\
        &-\frac{1}{2\cdot(3!)^2}\sum_{ijkabc}\lambda^{ijk}_{abc}\bar{h}_{abc,ijk}+\cdots
    \end{split}
\end{gather}
where we have used $\bar{h}_{ii} = h_{ii}+\sum_{a}h_{ia}t^{a}_{i}$. We have also used the antisymmetry of the matrix elements to re-write terms originating from the trace of the static component of the self-energy ($\tilde{\Sigma}^{\infty}_{pq}$) such as: $\chi_{ia,ik}=\chi_{ai,ki}$ and so on. Using the definition of $\tilde{\Sigma}^{\infty(0)}_{ii}$ from Eq.~\ref{eq:cc_stat_se}, we find 
\begin{gather}
    \begin{split}
         \frac{1}{2}\sum_{i}\tilde{\Sigma}^{\infty(0)}_{ii} &=\frac{1}{4}\sum_{ijab}\braket{ij||ab}(t^{ab}_{ij}+2t^{a}_it^b_j) + \frac{1}{2}\sum_{ia}f_{ia}t^{a}_{i}\\
         &+\frac{1}{2}\sum_{iaj}\braket{ij||ai}t^{a}_{j} \ ,
    \end{split}
\end{gather}
remembering that $f_{ia}\neq 0$ for a general reference state. 
Using this result, we identify the first term of Eq.~\ref{eq:proof_energy} as the coupled-cluster ground state energy obtained from the ground state coupled-cluster equations as 
\begin{gather}
    \begin{split}
        E^{\CC}_0 &= \frac{1}{2}\sum_{i}\Big(h_{ii}+f_{ii}+\tilde{\Sigma}^{\infty(0)}_{ii}+\sum_{a}h_{ia}t^{a}_{i}\Big)\\
        &= E^{\rf}_0 + \frac{1}{4}\sum_{ijab}\braket{ij||ab}(t^{ab}_{ij}+2t^{a}_it^b_j) +\sum_{ia}f_{ia}t^{a}_{i} \ ,
    \end{split}
\end{gather}
where we have used the definition of the reference energy as $E^{\rf}_0 = \frac{1}{2}\sum_{i}(h_{ii}+f_{ii})$. This result is clearly identical to Eq.~\ref{eq:energy_ref}. Therefore, we have 
\begin{gather}
    \begin{split}~\label{eq:proof_energy2}
        E^{\CC}_0 &= E^{\CC}_0 + \frac{1}{2}\sum_{ia}\lambda^{i}_{a}(F_{ai}+\bar{h}_{ai})+ \frac{1}{2}\sum_{ika}\lambda^{k}_{a}\chi_{ai,ki} \\
        &+ \frac{1}{(2!)^3}\sum_{iklab}\lambda^{kl}_{ab}\chi_{abi,kli} + \cdots\\
        &-\frac{1}{(2!)^2}\sum_{ijka}\lambda^{k}_{a}\bar{h}_{aij,kij}-\frac{1}{(2!)^3}\sum_{kijab}\lambda^{ij}_{ab}\bar{h}_{abk,ijk}\\
        &-\frac{1}{2\cdot(3!)^2}\sum_{ijkabc}\lambda^{ijk}_{abc}\bar{h}_{abc,ijk}+\cdots 
    \end{split}
\end{gather}
In order for this equation to be correct, we must demonstrate that the remaining terms of Eq.~\ref{eq:proof_energy2} vanish as a result of the CC amplitude equations. To do so, we use the relationship between the effective interactions given in Appendix~\ref{appendix_2}. The additional terms generated from the expectation values such as $-\frac{1}{2}\braket{\tilde{\Psi}_0|\bar{W}|\Phi_0}$ and so on give rise to the correct prefactors required to generate the CC amplitude equations. This can be seen by re-arranging the terms in Eq.~\ref{eq:proof_energy2} as 
\begin{widetext}
\begin{gather}
    \begin{split}~\label{eq:proof_energy3}
        E^{\CC}_0 &= E^{\CC}_0 + \frac{1}{2}\sum_{ia}\lambda^{i}_{a}F_{ai}+ \frac{1}{2}\sum_{ka}\lambda^{k}_{a}\Bigg(\bar{h}_{ak}+\sum_{i}\chi_{ai,ki}-\frac{1}{2!}\sum_{ij}\bar{h}_{aij,kij}+\cdots\Bigg) \\
        &+ \frac{1}{(2!)^3}\sum_{iklab}\lambda^{kl}_{ab}\Bigg(\chi_{abi,kli} -\bar{h}_{abi,kli}-\sum_{m}\bar{h}_{abim,klim}-\cdots\Bigg)-\frac{1}{2\cdot(3!)^2}\sum_{ijkabc}\lambda^{ijk}_{abc}\Bigg(\bar{h}_{abc,ijk}+\sum_{m}\bar{h}_{abim,klim}+\cdots\Bigg) +\cdots 
    \end{split}
\end{gather}
\end{widetext}
where the additional terms outside the brackets contain the four-body interaction matrices contracted with the corresponding four-body de-excitation amplitudes and so on.  
We can identify the terms in brackets as the effective interactions defined in Appendix~\ref{appendix_2},
$\{F_{pq},\chi_{pq,rs},\chi_{pqr,stu},\cdots\}$, that arise when normal-ordering the similarity transformed Hamiltonian with respect to the reference state. To make this explicit, for the first bracketed term of Eq.~\ref{eq:proof_energy3}, we have terms of the form
\begin{gather}
    \begin{split}
        &\bar{h}_{ak}+\sum_{i}\chi_{ai,ki}-\frac{1}{2!}\sum_{ij}\bar{h}_{aij,kij}+\cdots \\
        &=\bar{h}_{ak}+\sum_{i}\bar{h}_{ai,ki}+\sum_{ij}\bar{h}_{aij,kij}+\cdots-\frac{1}{2!}\sum_{ij}\bar{h}_{aij,kij}+\cdots\\
        &= \bar{h}_{ak}+\sum_{i}\bar{h}_{ai,ki}+\frac{1}{2!}\sum_{ij}\bar{h}_{aij,kij}+\cdots\\
        &=F_{ai} \ ,
    \end{split}
\end{gather}
where we have used the expansion for $\chi_{ai,ki}$ given in Appendix~\ref{appendix_2}.
From this analysis it is clear that the correct pre-factors required to form the CC amplitude equations are generated by subtraction of the missing higher-body expectation values that are not included in the coupled-cluster Galitskii-Migdal formula (Eq.~\ref{eq:cc_gm_1}). The first term of the second line of Eq.~\ref{eq:proof_energy3} vanishes as the series in the brackets cancels the effective interaction, $\chi_{abi,kli}$. Combining these results together, Eq.~\ref{eq:proof_energy3} can be re-written as
\begin{gather}
    \begin{split}~\label{eq:proof_energy4}
        E^{\CC}_0 &= E^{\CC}_0 + \sum_{ia}\lambda^{i}_{a}F_{ai}-\frac{1}{2\cdot(3!)^2}\sum_{ijkabc}\lambda^{ijk}_{abc}\chi_{abc,ijk}+\cdots
    \end{split}
\end{gather}
Written is this form, it is clear that the additional terms in Eq.~\ref{eq:proof_energy4} vanish as they contain exactly the CC amplitude equations, $F_{ai}=\chi_{ab,ij}=\chi_{abc,ijk} = 0$ and so on. Therefore, within the Green's function formalism, the coupled-cluster correlation energy is generally given by the expression
\begin{gather}
    \begin{split}~\label{eq:cc_en}
        E^{\CC}_c &= \frac{1}{2}\sum_{i} \left(\tilde{\Sigma}^{\infty(0)}_{ii}+\sum_{a}h_{ia}t^{a}_{i}\right) . 
    \end{split}
\end{gather}
In the Breuckner basis, the second term vanishes and the correlation energy is simply given by the first term of the static contribution to the coupled-cluster self-energy~\cite{coveney2023coupled}. Hence, the coupled-cluster ground state correlation energy is completely determined by the coupled-cluster self-energy. From Eq.~\ref{eq:cc_en}, we find that the coupled-cluster ground state energy obtained at any
level of approximation of the CC self-energy gives the same ground state energy as that obtained at the same
level of approximation of the ground state coupled-cluster
equations. As explicitly mentioned in Section~\ref{subsec:cc_rep}, this is \emph{not} the case for the coupled-cluster representation of the electronic Green's function.

\section{The Coupled-cluster Dyson supermatrix and self-energy approximations}~\label{sec:adc}

In this section, we provide an overview of the coupled-cluster quasiparticle equation and Dyson supermatrix. Standard Green's function theory is centered around the frequency-dependent quasiparticle equation, which gives access to the poles and residues of the Green's function. This allows us to directly combine coupled-cluster theory with the methodology employed in the Green's function formalism, leading us to introduce several approximations for the coupled-cluster self-energy that preserve its correct spectral form. 

The Dyson equation for the SP-CCGF can be used to derive the coupled-cluster quasiparticle Hamiltonian using the identity
\begin{gather}
    \begin{split}
        \tilde{H}_{pq}(\omega) &= \Big(\omega\delta_{pq}-\tilde{G}^{-1}_{pq}(\omega)\Big)= f_{pq} + \tilde{\Sigma}_{pq}(\omega) \ , 
    \end{split}
\end{gather}
where: $\tilde{G}^{-1}_{pq}(\omega)=[G_{0}]^{-1}_{pq}(\omega)-\tilde{\Sigma}_{pq}(\omega)$, with $[G_{0}]^{-1}_{pq}(\omega)=(\omega\delta_{pq}-f_{pq})$. The poles and residues of the SP-CCGF are then given by the eigenvalues and eigenvectors of the self-consistent frequency-dependent coupled-cluster quasiparticle equation: 
\begin{gather}
    \begin{split}~\label{eq:cc_qp}
        \sum_{q}\tilde{H}_{pq}(\varepsilon_{\gamma})\bar{A}^{\gamma}_{q} = \bar{A}^{\gamma}_{p}\varepsilon_{\gamma} \ . 
    \end{split}
\end{gather}
Here, $\gamma$ is a composite index that spans both forward- and backward-time poles and residues of the SP-CCGF: $\gamma=\mu,\nu$. When the full set of 1PI diagrams are included in the coupled-cluster self-energy, the eigenvalues of Eq.~\ref{eq:cc_qp} are the exact addition and removal energies of the system. As discussed in Section~\ref{sec:ex_sp_ccgf}, the coupled-cluster self-energy is inherently non-hermitian. Therefore, the coupled-cluster quasiparticle Hamiltonian possesses distinct left eigenstates given by 
\begin{gather}
    \begin{split}~\label{eq:cc_qp_left}
        \sum_{q}\tilde{A}^{\gamma}_{q}\tilde{H}_{qp}(\varepsilon_{\gamma}) = \varepsilon_{\gamma}\tilde{A}^{\gamma}_{p} \ .  
    \end{split}
\end{gather}
The right and left eigenvectors are related to the SP-CCGF residues defined in Section~\ref{sec:ex_sp_ccgf} through 
\begin{subequations}~\label{eq:res}
    \begin{align}
        \begin{split}
            \bar{A}^{\gamma=\mu,\nu}_{p} &\equiv \Bigg\{\begin{array}{c}
                 \tilde{Y}^{\mu}_{p}=\braket{\tilde{\Psi}_0|a_{p}|\Psi^{N+1}_{\mu}}  \\
                \bar{X}^{\nu}_{p} = \braket{\tilde{\Psi}^{N-1}_{\nu}|a_{p}|\Phi_0} 
             \end{array}
        \end{split}\\
    \begin{split}
        \tilde{A}^{\gamma=\mu,\nu}_{p} &\equiv \Bigg\{\begin{array}{c}
        \bar{Y}^{\mu}_{p}=\braket{\tilde{\Psi}^{N+1}_{\mu}|a^\dag_{p}|\Phi_0} \\
                 \tilde{X}^{\nu}_{p}=\braket{\tilde{\Psi}_{0}|a^\dag_{p}|\Psi^{N-1}_{\nu}}  \ .
                 \end{array}
    \end{split}
    \end{align}
\end{subequations}
Similarly, the eigenvalues are related to the poles of the SP-CCGF as 
\begin{gather}
        \begin{split}~\label{eq:poles}
            \varepsilon_{\gamma=\mu,\nu} &\equiv \Bigg\{\begin{array}{c}
                 \varepsilon_{\mu}^{N+1}  \\
                \varepsilon_{\nu}^{N-1} \ ,
             \end{array}
        \end{split}
\end{gather}
which are at the formally exact removal and addition energies of the system.

Using the spectral form of the CC self-energy, given in Eq.~\ref{eq:spec_se}, the frequency-dependent coupled-cluster quasiparticle equation (Eq.~\ref{eq:cc_qp}) can be written in terms of the frequency-independent `upfolded' coupled-cluster Dyson supermatrix eigenvalue problem
\begin{gather}
    \begin{split}~\label{eq:super}
        \left(\begin{array}{ccc}
            \mathbf{\bar{H}} & \mathbf{\tilde{U}} & \mathbf{\bar{V}} \\
            \mathbf{\bar{U}} & \mathbf{\bar{K}}^{>}+\mathbf{\bar{C}}^{>} & \mathbf{0} \\
             \mathbf{\tilde{V}} & \mathbf{0} & \mathbf{\bar{K}}^{<}+\mathbf{\bar{C}}^{<}
        \end{array}\right)&\left(\begin{array}{c}
             \mathbf{\bar{A}}  \\
              \mathbf{\bar{W}}^{+}\\
              \mathbf{\bar{W}}^{-}
        \end{array}\right) =\varepsilon \left(\begin{array}{c}
             \mathbf{\bar{A}}  \\
              \mathbf{\bar{W}}^{+}\\
              \mathbf{\bar{W}}^{-}
        \end{array}\right),
    \end{split}
\end{gather}
where $\mathbf{\bar{H}}=\mathbf{f}+\mathbf{\tilde{\Sigma}}^{\infty}$ is the sum of the Fock matrix and the exact static component of the CC self-energy. Here we have employed matrix notation for the coupling and interactions between the forward and backward time multi-particle-hole ISCs. The eigenvalues, $\varepsilon$, are the exact addition and removal energies of the system. Diagonalization of the coupled-cluster Dyson supermatrix yields all the forward- and backward-time poles and residues of the SP-CCGF at once. The CC Dyson supermatrix was first arrived at in Ref.~\cite{coveney2023coupled}. However, the explicit analysis of its relationship to the SP-CCGF as well as its eigenvector components was not presented. Throughout the rest of this paper we denote the CC Dyson supermatrix as $\mathbf{\tilde{D}}^{\text{CC}}$.

Since the coupled-cluster self-energy is non-hermitian, the coupled-cluster Dyson supermatrix also has distinct left and right supereigenvectors. However, the exact CC Dyson supermatrix is pseudo-hermitian, due to the similarity transformed Hamiltonian $\bar{H}$, as it possesses a real spectrum. In Appendix~\ref{app:psd} we show that the exact coupled-cluster self-energy is \emph{biorthogonally} positive semi-definite. This property follows from the pseudo-hermiticity of the CC Dyson supermatrix and CC self-energy. It should be noted that the exact electronic self-energy of the hermitian electronic structure Hamiltonian is positive semi-definite, which is an important property that ensures the correct causal structure of the theory~\cite{winter1972study,schirmer1983new,bruneval2025gw+}. 

The components of the right and left eigenvectors in the space of excited multi-particle-hole ISCs are found as 
\begin{subequations}
    \begin{align}
    \begin{split}
        \bar{W}^{+,\gamma}_{J} &= \sum_{J''p} \Big(\varepsilon_{\gamma}\mathbbm{1}-(\mathbf{\bar{K}}^{>}+\mathbf{\bar{C}}^{>})\Big)^{-1}_{JJ''}\bar{U}_{J'',p}\bar{A}^{\gamma}_{p}
    \end{split}\\
    \begin{split}
        \bar{W}^{-,\gamma}_{A} &= \sum_{A''p} \Big(\varepsilon_{\gamma}\mathbbm{1}-(\mathbf{\bar{K}}^{<}+\mathbf{\bar{C}}^{<})\Big)^{-1}_{AA''}\tilde{V}_{A'',p}\bar{A}^{\gamma}_{p}
    \end{split}\\
\begin{split}
        \tilde{W}^{+,\gamma}_{J} &= \sum_{pJ''} \tilde{A}^{\gamma}_{p}\tilde{U}_{p,J''}\Big(\varepsilon_{\gamma}\mathbbm{1}-(\mathbf{\bar{K}}^{>}+\mathbf{\bar{C}}^{>})\Big)^{-1}_{J''J}
    \end{split}\\
    \begin{split}
        \tilde{W}^{+,\gamma}_{A} &= \sum_{pA''} \tilde{A}^{\gamma}_{p}\bar{V}_{p,A''}\Big(\varepsilon_{\gamma}\mathbbm{1}-(\mathbf{\bar{K}}^{<}+\mathbf{\bar{C}}^{<})\Big)^{-1}_{A''A} \ ,
    \end{split}
\end{align}
\end{subequations}
where $\left(\begin{array}{ccc}
             \mathbf{\tilde{A}}  & \mathbf{\tilde{W}}^{+} & \mathbf{\tilde{W}}^{-}
             \end{array}\right)$ is the left supereigenvector solution. From these formal solutions, it should be noted that the $\bar{W}^{\pm,\gamma}/\tilde{W}^{\pm,\gamma}$ components of the eigenvectors are explicit functions of the corresponding eigenvalue, $\varepsilon_{\gamma}$.

             If we impose the normalization condition: $\sum_{p}\tilde{A}^{\gamma}_{p}\bar{A}^{\gamma}_{p}=1$,
             the quasiparticle renormalization factor can be found from the inverse of the overlap of the supereigenvectors as
\begin{gather}
    \begin{split}~\label{eq:qp_factor}
       Z_{\gamma} &= \Bigg(\sum_{p} \tilde{A}^{\gamma}_{p}\bar{A}^{\gamma}_{p} + \sum_{J} \tilde{W}^{+,\gamma}_{J}\bar{W}^{+,\gamma}_{J}+\sum_{A} \tilde{W}^{-,\gamma}_{A}
        \bar{W}^{-,\gamma}_{A}\Bigg)^{-1} \\
        &= \Bigg(1-\frac{\partial\tilde{\Sigma}_{\gamma\gamma}(\omega)}{\partial\omega}\Bigg|_{\omega=\varepsilon_{\gamma}}\Bigg)^{-1} \ .
    \end{split}
\end{gather}
However, if we instead require that full supereigenvectors be normalized under the biorthogonal inner product then the renormalization factor is simply given by: $Z_{\gamma} = \sum_{p}\tilde{A}^{\gamma}_{p}\bar{A}^{\gamma}_{p}$. 

From the right and left supereigenvector solutions of the CC Dyson supermatrix (Eq.~\ref{eq:super}), it is possible to reconstruct the exact SP-CCGF by taking the resolvent of the Dyson supermatrix in the single-particle spin-orbital subspace. For the retarded SP-CCGF, the result is given by
\begin{gather}
    \begin{split}
        \tilde{G}^{R}_{pq}(\omega) &= \Big((\omega+i\eta)\mathbbm{1}-\mathbf{\tilde{D}}^{\text{CC}}\Big)^{-1}_{pq}\\
        &= \sum_{\gamma=\mu,\nu}\frac{\bar{A}^{\gamma}_{p}\tilde{A}^{\gamma}_{q}}{\omega-\varepsilon_{\gamma}+i\eta} \\
        &=\sum_{\mu}\frac{\tilde{Y}^{\mu}_{p}\bar{Y}^{\mu}_{q}}{\omega-\varepsilon^{N+1}_{\mu}+i\eta} + \sum_{\nu}\frac{\tilde{X}^{\nu}_{q}\bar{X}^{\nu}_{p}}{\omega-\varepsilon^{N-1}_{\nu}+i\eta}\ ,
    \end{split}
\end{gather}
where we have used the relationships given in Eqs~\ref{eq:res} and~\ref{eq:poles} to go from the second equality to the third equality. As can be seen, this expression agrees with the exact Lehmann representation given in Eq.~\ref{eq:exact_leh} (remembering that the difference between the time-ordered and retarded single-particle Green's function is that the retarded propagator simply places all the poles below the real axis~\cite{Quantum,mahan2000many,stefanucci2013nonequilibrium}).

By introducing the coupled-cluster Dyson supermatrix, we preserve the correct analytical structure of the self-energy required to obtain the correct convergence of the Feynman-Dyson diagrammatic perturbation expansion~\cite{hirata2024nonconvergence}. This representation forms the basis of the Algebraic Diagrammatic Construction method~\cite{schirmer1983new,schirmer2018many,raimondi2018algebraic}. In the following subsections, we construct the upfolded Dyson supermatrices for different 2p1h/2h1p CC self-energy approximations that preserve the correct analytic structure. To be consistent with the treatment of 2p1h/2h1p excitations, all coupling and effective interaction matrices appearing in these subsections will be implicitly taken within the CCSD approximation ($T=T_1+T_2$). These CC self-energy approximations reveal the connection to IP/EA-EOM-CCSD theory and allow us to leverage the connections between Green's function and coupled-cluster theory.

The equations derived in this work are no more complex than those that have been derived in the context of EOM-CC theory~\cite{nooijen1995equation,hirata2000high,hirata2000high1}, the Algebraic Diagrammatic Construction method~\cite{schirmer1983new,raimondi2018algebraic} and the Gorkov-Green's function formalism~\cite{soma2011ab,barbieri2022gorkov}. Therefore, similar hierarchies of approximations and powerful numerical techniques can be ported over for the efficient solution of the coupled-cluster Dyson supermatrix~\cite{weikert1996block,marino2024diagrammatic,soma2014ab,raimondi2018algebraic,bintrim2021full,bintrim2022full,nooijen1997similarity,stanton1999simple,musial2003equation,backhouse2022constructing,opoku2023new,opoku2023new_prop,opoku2024new_prop,marino2026gorkov}.

In the following subsections, we will use the fermionic antisymmetry of the effective interaction matrix elements to restrict our sums to be over ordered sets of single-particle Green's function indices: $(i<j;a)$, $(i<j<k;a<b)$,  $(\nu_1<\nu_2;\mu_1)$, $(\nu_1<\nu_2<\nu_3;\mu_1<\mu_2)$ and so on. We can recover the general case using unrestricted summations by introduction of the relevant symmetry factors of the Feynman diagrams presented in Sections~\ref{sec:pert} and~\ref{sec:cc_se}.

\subsection{Second-order perturbative and self-consistent approximations}~\label{sub:2nd}
From the second-order perturbative coupled-cluster self-energy (Eq.~\ref{eq:2nd_order}), we see that $J\equiv(i;a<b)$/$A\equiv(i<j;a)$ are restricted to the space of 2p1h/2h1p excitations such that:
\begin{subequations}
        \begin{align}
            \begin{split}~\label{eq:couplings}
                \tilde{U}^{(0)}_{p,iab} &= \chi_{pi,ab}
                \end{split}\\
                \begin{split}
                \bar{U}^{(0)}_{iab,q} &= \chi_{ab,qi}
                \end{split}\\
                \begin{split}
                    \tilde{V}^{(0)}_{ija,q} &= \chi_{ij,qa}
                \end{split}\\
                \begin{split}
                \bar{V}^{(0)}_{p,ija} &= \chi_{pa,ij}
            \end{split}
        \end{align}
\end{subequations}
The ISC interaction matrices are given by $\mathbf{\bar{K}}^{>}_{iab,jcd} = (\epsilon_{a}+\epsilon_{b}-\epsilon_{i})(\delta_{ac}\delta_{bd}-\delta_{ad}\delta_{bc})\delta_{ij}$ and $\mathbf{\bar{K}}^{<}_{ija,klb} = (\epsilon_{i}+\epsilon_{j}-\epsilon_{a})(\delta_{ik}\delta_{jl}-\delta_{il}\delta_{jk})\delta_{ab}$, respectively. This leads to the following upfolded CC Dyson supermatrix
\begin{gather}
    \begin{split}~\label{eq:Dyson_2nd_pt}
        \mathbf{\tilde{D}}^{\CC}_{(2)} = \left(\begin{array}{ccc}
            f_{pq}+\tilde{\Sigma}^{\infty(0)}_{pq}+\tilde{\Sigma}^{\infty(2)}_{pq} & \tilde{U}^{(0)}_{p,iab} &\bar{V}^{(0)}_{p,ija} \\
            \bar{U}^{(0)}_{iab,q} & \mathbf{\bar{K}}^{>}_{iab,jcd} & \mathbf{0} \\
             \tilde{V}^{(0)}_{ija,q} & \mathbf{0} & \mathbf{\bar{K}}^{<}_{ija,klb}
        \end{array}\right) .
    \end{split}
\end{gather}
Alternatively, we also carry out this procedure for the second-order self-consistent, renormalized coupled-cluster self-energy given in Eq.~\ref{eq:sc_2nd_se}. This self-energy also consists of 2p1h/2h1p excitation character ($J\equiv(\nu_1;\mu_1<\mu_2)$/$A\equiv(\nu_1<\nu_2;\mu_1)$) and gives rise to the following coupling matrices: 
\begin{subequations}
\begin{align}
    \begin{split}
    \tilde{U}^{\sco(2)}_{p,\nu_1\mu_1\mu_2} &= \sum_{rtu}\tilde{\Xi}_{pr,tu}\Tilde{Y}^{\mu_1}_{t}\Tilde{Y}^{\mu_2}_{u}\Tilde{X}^{\nu_1}_{r}
    \end{split}\\
    \begin{split}
    \bar{U}^{\sco(2)}_{\nu_1\mu_1\mu_2,q} &= \sum_{svw}\tilde{\Xi}_{sv,qw}\bar{Y}^{\mu_1}_s\bar{Y}^{\mu_2}_v\bar{X}^{\nu_1}_w
    \end{split}\\
    \begin{split}
    \bar{V}^{\sco(2)}_{p,\nu_1\nu_2\mu_1} &= \sum_{rtu}\tilde{\Xi}_{pr,tu}\bar{X}^{\nu_1}_{t}\bar{X}^{\nu_2}_{u}\bar{Y}^{\mu_1}_{r}
    \end{split}\\ 
    \begin{split}
        \tilde{V}^{\sco(2)}_{\nu_1\nu_2\mu_1,q} &= \sum_{svw}\tilde{\Xi}_{sv,qw}\tilde{X}^{\nu_1}_s\tilde{X}^{\nu_2}_v\tilde{Y}^{\mu_1}_w \ .
    \end{split}
\end{align}
\end{subequations}
 The interaction matrices become $\mathbf{\bar{K}}^{>}_{\nu_1\mu_1\mu_2,\nu_2\mu_3\mu_4}=(\varepsilon^{N+1}_{\mu_1}+\varepsilon^{N+1}_{\mu_2}-\varepsilon^{N-1}_{\nu_1})(\delta_{\mu_1\mu_3}\delta_{\mu_2\mu_4}-\delta_{\mu_1\mu_4}\delta_{\mu_2\mu_3})\delta_{\nu_1\nu_2}$ and $\mathbf{\bar{K}}^{<}_{\nu_1\nu_2\mu_1,\nu_3\nu_4\mu_2}=(\varepsilon^{N-1}_{\nu_1}+\varepsilon^{N-1}_{\nu_2}-\varepsilon^{N+1}_{\nu_1})(\delta_{\nu_1\nu_3}\delta_{\nu_2\nu_4}-\delta_{\nu_1\nu_4}\delta_{\nu_2\nu_3})\delta_{\mu_1\mu_2}$. Within this approximation, the resulting coupled-cluster Dyson supermatrix is given by
\begin{gather}
    \begin{split}~\label{eq:Dyson_2nd}
        \mathbf{\tilde{D}}^{\CC}_{\sco(2)} = \left(\begin{array}{ccc}
            f_{pq}+\tilde{\Sigma}^{\infty\CCSD}_{pq}& \tilde{U}^{\sco(2)}_{p,\nu_1\mu_1\mu_2} & \bar{V}^{\sco(2)}_{p,\nu_1\nu_2\mu_1} \\
            \bar{U}^{\sco(2)}_{\nu_1\mu_1\mu_2,q} & \mathbf{\bar{K}}^{>}_{\nu_1\mu_1\mu_2,\nu_2\mu_3\mu_4} & \mathbf{0} \\
             \tilde{V}^{\sco(2)}_{\nu_1\nu_2\mu_1,q}  & \mathbf{0} & \mathbf{\bar{K}}^{<}_{\nu_1\nu_2\mu_1,\nu_3\nu_4\mu_2}
        \end{array}\right) .
    \end{split}
\end{gather}
The static contribution, $\tilde{\Sigma}^{\infty\CCSD}_{pq}$, is taken within the CCSD approximation to be consistent with the treatment of 2p1h/2h1p excitations contained in the second-order self-energy:
\begin{gather}
    \begin{split}
        \tilde{\Sigma}^{\infty\CCSD}_{pq} = \tilde{\Sigma}^{\infty(0)}_{pq} + \sum_{ia}\lambda^{i}_{a}\chi_{pa,qi} + \frac{1}{(2!)^2}\sum_{ijab}\lambda^{ij}_{ab}\chi_{pab,qij} \ .
    \end{split}
\end{gather}
The solution of Eq.~\ref{eq:Dyson_2nd} requires iterative diagonalization, whereby the coupling and interaction matrices are updated at each step. The zeroth-order iteration is taken as the second-order perturbative supermatrix defined in Eq.~\ref{eq:Dyson_2nd_pt} and the resulting eigenvectors and eigenvalues are subsequently fed into the definition of the coupling and interaction matrices. This processes is repeated until convergence of the spectrum of $\mathbf{\tilde{D}}^{\CC}_{\sco(2)}$ is achieved.

\subsection{Coupled-cluster self-energy approximations including the complete set of 2p1h/2h1p interactions}

The second-order coupled-cluster self-energy contains interactions between ISCs that are fundamentally of 2p1h/2h1p excitation character. This is demonstrated by the fact that the ISCs are parametrized by the ordered sets of single-particle Green's function indices: $J\equiv(i;a<b)$/$A\equiv(i<j;a)$. We can include the complete set of 2p1h/2h1p interactions between these ISCs by defining the interaction matrices as
\begin{subequations}~\label{eq:eom_ints}
    \begin{align}
    \begin{split}
    \mathbf{\bar{K}}^{>}_{iab,jcd}+\mathbf{\bar{C}}^{>(0)}_{iab,jcd} &=  \braket{\Phi^{ab}_{i}|\bar{H}_N|\Phi^{cd}_{j}}
    \end{split}\\
    \begin{split}
 \mathbf{\bar{K}}^{<}_{ija,klb}+\mathbf{\bar{C}}^{<(0)}_{ija,klb} &=-\braket{\Phi^{b}_{kl}|\bar{H}_N|\Phi^{a}_{ij}} \ , 
    \end{split}
    \end{align}
\end{subequations}
where the matrices $\mathbf{\bar{K}}^{>}_{iab,jcd}/\mathbf{\bar{K}}^{<}_{ija,klb}$ are defined in Subsection~\ref{sub:2nd} and are block-diagonal by definition (see Section~\ref{sec:ex_sp_ccgf}). This approximation corresponds to allowing all particles to propagate independently, to interact pairwise via the static component of the two-body interaction as well as the combined three-body interaction of the similarity transformed Hamiltonian normal-ordered with respect to the reference determinant~\cite{coveney2023coupled,riva2023multichannel,riva2024derivation}. The explicit expression for the interaction matrices are therefore given by 
\begin{subequations}
\begin{align}
    \begin{split}
        \mathbf{\bar{C}}^{>(0)}_{iab,jcd} &= \tilde{\Sigma}^{\infty(0)}_{ac}\delta_{ij}\delta_{bd}-\tilde{\Sigma}^{\infty(0)}_{ad}\delta_{ij}\delta_{bc} + \tilde{\Sigma}^{\infty(0)}_{bd}\delta_{ac}\delta_{ij}\\
        &-\tilde{\Sigma}^{\infty(0)}_{bc}\delta_{ad}\delta_{ij}- \tilde{\Sigma}^{\infty(0)}_{ji}\delta_{ac}\delta_{bd}+\tilde{\Sigma}^{\infty(0)}_{ji}\delta_{ad}\delta_{bc}\\
        &+\chi_{ja,ci}\delta_{bd}+\chi_{jb,di}\delta_{ac}+ \chi_{ab,cd}\delta_{ij}\\
        &- \chi_{ja,di}\delta_{bc}- \chi_{jb,ci}\delta_{ad} + \chi_{jab,cid} 
    \end{split}\\
    \begin{split}
        \mathbf{\bar{C}}^{<(0)}_{ija,klb} &= \tilde{\Sigma}^{\infty(0)}_{ik}\delta_{ab}\delta_{jl}-\tilde{\Sigma}^{\infty(0)}_{il}\delta_{ab}\delta_{jk}+ \tilde{\Sigma}^{\infty(0)}_{jl}\delta_{ab}\delta_{ik}\\
        &-\tilde{\Sigma}^{\infty(0)}_{jk}\delta_{ab}\delta_{il}+\tilde{\Sigma}^{\infty(0)}_{ba}\delta_{il}\delta_{jk} -\tilde{\Sigma}^{\infty(0)}_{ba}\delta_{ik}\delta_{jl}\\
        &+\chi_{ib,al}\delta_{jk}+\chi_{jb,ak}\delta_{il}- \chi_{ij,kl}\delta_{ab} \\
        &- \chi_{ib,ak}\delta_{jl}- \chi_{jb,al}\delta_{ik} + \chi_{ijb,kal} \ .
    \end{split}
\end{align}
\end{subequations}
To include the full set of time-orderings of the coupled-cluster self-energy diagrams corresponding to the interaction matrices defined in Eqs~\ref{eq:eom_ints} requires the modification of the 2p1h and 2h1p coupling matrices given in Subsection~\ref{sub:2nd} as
\begin{subequations}
    \begin{align}
        \begin{split}~\label{eq:mod_coupling1}
            \tilde{U}^{2\p1\h(0)}_{p,iab} &= \chi_{pi,ab}+\frac{1}{2}\sum_{kl}\chi_{pi,kl}\frac{\chi_{kl,ab}}{\epsilon_{k}+\epsilon_{l}-\epsilon_{a}-\epsilon_{b}}\\
            &+\sum_{kc}\chi_{pc,ak}\frac{\chi_{ik,bc}}{\epsilon_{i}+\epsilon_{k}-\epsilon_{b}-\epsilon_{c}}\\
            &-\sum_{kc}\chi_{pc,bk}\frac{\chi_{ik,ac}}{\epsilon_{i}+\epsilon_{k}-\epsilon_{a}-\epsilon_{c}}\\
            &-\sum_{k}\frac{\chi_{pi,kb}\tilde{\Sigma}^{\infty(0)}_{ka}}{\epsilon_{k}-\epsilon_{a}} + \sum_{k}\frac{\chi_{pi,ka}\tilde{\Sigma}^{\infty(0)}_{kb}}{\epsilon_{k}-\epsilon_{b}}\\
            &+\sum_{c}\frac{\chi_{pc,ab}\tilde{\Sigma}^{\infty(0)}_{ic}}{\epsilon_{i}-\epsilon_{c}} \ ,
        \end{split}\\
        \begin{split}~\label{eq:mod_coupling2}
            \tilde{V}^{2\h1\p(0)}_{ija,q} &= \chi_{ij,qa}+\frac{1}{2}\sum_{cd}\frac{\chi_{ij,cd}}{\epsilon_{i}+\epsilon_{j}-\epsilon_{c}-\epsilon_{d}}\chi_{cd,qa}\\
            &+ \sum_{kc}\frac{\chi_{kj,ca}}{\epsilon_{k}+\epsilon_{j}-\epsilon_{c}-\epsilon_{a}}\chi_{ic,qk}\\
            &-\sum_{kc}\frac{\chi_{ki,ca}}{\epsilon_{k}+\epsilon_{i}-\epsilon_{c}-\epsilon_{a}}\chi_{jc,qk}\\
            &+\sum_{c}\frac{\tilde{\Sigma}^{\infty(0)}_{ic}\chi_{cj,qa}}{\epsilon_{i}-\epsilon_{c}}-\sum_{c}\frac{\tilde{\Sigma}^{\infty(0)}_{jc}\chi_{ci,qa}}{\epsilon_{j}-\epsilon_{c}}\\
            &-\sum_{k}\frac{\tilde{\Sigma}^{\infty(0)}_{ka}\chi_{ij,qk}}{\epsilon_{k}-\epsilon_{a}} \ .
        \end{split}
    \end{align}
\end{subequations}
In the Breuckner basis, the final three terms of Eqs~\ref{eq:mod_coupling1} and~\ref{eq:mod_coupling2} vanish. The coupling matrices $\bar{U}^{(0)}_{iab,q}/\bar{V}^{(0)}_{p,ija}$ remain the same as those defined in Subsection~\ref{sub:2nd}. This reflects the inherent non-hermitian structure of the coupled-cluster self-energy. There are no terms in the coupling matrices containing the three-body effective interaction as these terms vanish as a result of the CC amplitude equations.

Using the definition of the interaction and coupling matrices defined above leads to the upfolded CC Dyson supermatrix
\begin{gather}
    \begin{split}~\label{eq:cc_eom_se}
        &\mathbf{\tilde{D}}^{\CC}_{\text{ADC(2p1h(0))}} =\\
        &\left(\begin{array}{ccc}
            f_{pq}+\tilde{\Sigma}^{\infty\text{pt2}}_{pq} & \tilde{U}^{2\p1\h(0)}_{p,abi} &\bar{V}^{(0)}_{p,ija} \\
           \bar{U}^{(0)}_{cdj,q} & \mathbf{\bar{K}}^{>}_{iab,jcd}+\mathbf{\bar{C}}^{>(0)}_{iab,jcd} & \mathbf{0} \\
             \tilde{V}^{2\h1\p(0)}_{klb,q} & \mathbf{0} &  \mathbf{\bar{K}}^{<}_{ija,klb}+\mathbf{\bar{C}}^{<(0)}_{ija,klb}
        \end{array}\right) ,
    \end{split}
\end{gather}
where $\tilde{\Sigma}^{\infty\text{pt2}}_{pq}=\tilde{\Sigma}^{\infty(0)}_{pq}+\tilde{\Sigma}^{\infty(2)}_{pq}$.
This approximation corresponds to the infinite-order summation of the coupled-cluster self-energy diagrams depicted in Figure~\ref{fig:pert_cc_se} and is closely related to IP/EA-EOM-CC theory truncated to the space of 2p1h/2h1p excitations. However, the upper left block of the coupled-cluster Dyson supermatrix is defined over the combined set of occupied and virtual spin-orbitals. This definition of the interaction and coupling matrices results in the 2p1h(0)/2h1p(0) coupled-cluster self-energy approximation
\begin{gather}
    \begin{split}~\label{eq:occ_cc_se}
        &\tilde{\Sigma}^{\text{ADC(2p1h(0))}}_{pq}(\omega) = \tilde{\Sigma}^{\infty(0)}_{pq} + \tilde{\Sigma}^{\infty(2)}_{pq}\\
        &+ \frac{1}{(2!)^2} \sum_{\substack{iab\\jcd}}\tilde{U}^{2\p1\h(0)}_{p,iab}\Big((\omega+i\eta)\mathbbm{1}-(\mathbf{\bar{K}}^{>}+\mathbf{\bar{C}}^{>(0)})\Big)^{-1}_{iab,jcd}\bar{U}^{(0)}_{jcd,q}\\
        &+\frac{1}{(2!)^2} \sum_{\substack{kla\\mnb}}\bar{V}^{(0)}_{p,kla}\Big((\omega-i\eta)\mathbbm{1}-(\mathbf{\bar{K}}^{<}+\mathbf{\bar{C}}^{<(0)})\Big)^{-1}_{kla,mnb}\tilde{V}^{2\h1\p(0)}_{mnb,q} .
    \end{split}
\end{gather}

Similarly, we can further generalize the 2p1h/2h1p coupled-cluster self-energy approximation by allowing the interaction matrices to be composed of the full effective interactions arising when normal-ordering $\bar{H}$ with respect to the biorthogonal ground state expectation value. This results in the following interaction matrices
\begin{figure*}[ht]
    \centering
    \begin{gather*}
\begin{split}
\tilde{\Sigma}^{2\p1\h/2\h1\p}_{pq} &= \hspace{2.5mm}\begin{gathered}
\begin{fmfgraph*}(30,30)
    \fmfset{arrow_len}{3mm}
    \fmfleft{i1}
    \fmfright{o1}
    \fmf{zigzag}{o1,i1}
    \fmfv{decor.shape=cross,decor.filled=full, decor.size=1.5thic}{o1}
    \fmfdot{i1}
\end{fmfgraph*}
\end{gathered}\hspace{2.5mm}+\hspace{2.5mm}
\begin{gathered}
    \begin{fmfgraph*}(40,40)
    \fmfcurved
    \fmfset{arrow_len}{3mm}
    \fmfleft{i1,i2}
    \fmflabel{}{i1}
    \fmflabel{}{i2}
    \fmfright{o1,o2}
    \fmflabel{}{o1}
    \fmflabel{}{o2}
    \fmf{fermion}{i1,i2}
    \fmf{dbl_zigzag}{o1,i1}
    \fmf{fermion,left=0.3,tension=0}{o1,o2}
    \fmf{dbl_zigzag}{o2,i2}
    \fmf{fermion,left=0.3,tension=0}{o2,o1}
    \fmfdot{o1,o2,i1,i2}
\end{fmfgraph*}
\end{gathered} \hspace{2.5mm}+\hspace{5mm} 
    \begin{gathered}
    \begin{fmfgraph*}(40,60)
    \fmfcurved
    \fmfset{arrow_len}{3mm}
    \fmfleft{i1,i2,i3}
    \fmflabel{}{i1}
    \fmflabel{}{i2}
    \fmfright{o1,o2,o3}
    \fmflabel{}{o1}
    \fmflabel{}{o2}
    \fmf{dbl_zigzag}{i2,o2}
    \fmf{dbl_zigzag}{i3,o3}
    \fmf{dbl_zigzag}{i1,o1}
    \fmf{fermion}{i1,i2}
    \fmf{fermion}{i2,i3}
    \fmf{fermion}{o1,o2}
    \fmf{fermion}{o2,o3}
    \fmf{fermion,left=0.3}{o3,o1}
    \fmfforce{(0.0w,0.0h)}{i1}
    \fmfforce{(1.0w,0.0h)}{o1}
    \fmfforce{(0.0w,1.0h)}{i3}
    \fmfforce{(1.0w,1.0h)}{o3}
    \fmfdot{i1,i2,i3}
    \fmfdot{o1,o2,o3}
\end{fmfgraph*}
\end{gathered}
\hspace{7.5mm}+\hspace{5mm}
\begin{gathered}
\begin{fmfgraph*}(40,60)
    \fmfcurved
    \fmfset{arrow_len}{3mm}
    \fmfleft{i1,i2,i3}
    \fmflabel{}{i1}
    \fmflabel{}{i2}
    \fmfright{o1,o2,o3}
    \fmflabel{}{o1}
    \fmflabel{}{o2}
    \fmf{dbl_zigzag}{i1,v1}
    \fmf{dbl_zigzag}{v1,o1}
    \fmf{dbl_zigzag}{v2,o2}
    \fmf{dbl_zigzag}{i3,v3}
    \fmf{fermion}{i1,i3}
    \fmf{fermion,left=0.3}{o1,o2}
    \fmf{fermion,left=0.3}{o2,o1}
    \fmf{fermion,left=0.3}{v2,v3}
    \fmf{fermion,left=0.3}{v3,v2}
    \fmfforce{(0.0w,0.0h)}{i1}
    \fmfforce{(1.0w,0.0h)}{o1}
    \fmfforce{(0.5w,0.5h)}{v2}
    \fmfforce{(0.5w,0.0h)}{v1}
    \fmfforce{(0.5w,1.0h)}{v3}
    \fmfforce{(0.0w,1.0h)}{i3}
    \fmfforce{(1.0w,1.0h)}{o3}
    \fmfdot{v2,v3}
    \fmfdot{i1,i3}
    \fmfdot{o1,o2}
\end{fmfgraph*}
\end{gathered}\hspace{2.5mm}+\hspace{5mm}
\begin{gathered}
    \begin{fmfgraph*}(40,60)
    \fmfcurved
    \fmfset{arrow_len}{3mm}
    \fmfleft{i1,i2,i3}
    \fmflabel{}{i1}
    \fmflabel{}{i2}
    \fmfright{o1,o2,o3}
    \fmflabel{}{o1}
    \fmflabel{}{o2}
    \fmf{dbl_zigzag}{i1,v1}
    \fmf{dbl_dashes}{i2,v2}
    \fmf{dbl_dashes}{v2,o2}
    \fmf{dbl_zigzag}{i3,v3}
    \fmf{fermion}{i1,i2}
    \fmf{fermion}{v1,v2}
    \fmf{fermion}{o2,v1}
    \fmf{fermion}{v2,v3}
    \fmf{fermion}{v3,o2}
    \fmf{fermion}{i2,i3}
    \fmfforce{(0.0w,0.0h)}{i1}
    \fmfforce{(1.0w,0.0h)}{o1}
    \fmfforce{(0.5w,0.5h)}{v2}
    \fmfforce{(0.5w,0.0h)}{v1}
    \fmfforce{(0.5w,1.0h)}{v3}
    \fmfforce{(0.0w,1.0h)}{i3}
    \fmfforce{(1.0w,1.0h)}{o3}
    \fmfdotn{v}{3}
    \fmfdot{i1,i2,i3}
    \fmfdot{o2}
\end{fmfgraph*}
\end{gathered}\\
\\
&+\hspace{5mm}\begin{gathered}
    \begin{fmfgraph*}(40,40)
    \fmfcurved
    \fmfset{arrow_len}{3mm}
    \fmfleft{i1,i2}
    \fmflabel{}{i1}
    \fmflabel{}{i2}
    \fmfright{o1,o2}
    \fmflabel{}{o1}
    \fmflabel{}{o2}
    \fmf{fermion}{i1,v1}
    \fmf{fermion}{v1,i2}
    \fmf{dbl_zigzag}{o1,i1}
    \fmf{fermion,left=0.3,tension=0}{o1,o2}
    \fmf{dbl_zigzag}{o2,i2}
    \fmf{fermion,left=0.3,tension=0}{o2,o1}
    \fmf{zigzag}{v1,v2}
    \fmfdot{o1,o2,i1,i2,v1}
    \fmfforce{(0.0w,0.0h)}{i1}
    \fmfforce{(0.0w,1.0h)}{i2}
    \fmfforce{(0.0w,0.5h)}{v1}
    \fmfforce{(0.5w,0.5h)}{v2}
     \fmfv{decor.shape=cross,decor.filled=full, decor.size=1.5thic}{v2}
\end{fmfgraph*}
\end{gathered}\hspace{5mm}+\hspace{5mm}
\begin{gathered}
    \begin{fmfgraph*}(40,40)
    \fmfcurved
    \fmfset{arrow_len}{3mm}
    \fmfleft{i1,i2}
    \fmflabel{}{i1}
    \fmflabel{}{i2}
    \fmfright{o1,o2}
    \fmflabel{}{o1}
    \fmflabel{}{o2}
    \fmf{fermion}{i1,i2}
    \fmf{dbl_zigzag}{o1,i1}
    \fmf{fermion,left=0.3,tension=0}{o1,o2}
    \fmf{dbl_zigzag}{o2,i2}
    \fmf{fermion,left=0.3,tension=0}{o2,v1}
     \fmf{fermion,left=0.3,tension=0}{v1,o1}
    \fmf{zigzag}{v1,v2}
    \fmfdot{o1,o2,i1,i2,v1}
    \fmfforce{(1.0w,0.0h)}{o1}
    \fmfforce{(1.0w,1.0h)}{o2}
    \fmfforce{(1.25w,0.5h)}{v1}
    \fmfforce{(1.75w,0.5h)}{v2}
     \fmfv{decor.shape=cross,decor.filled=full, decor.size=1.5thic}{v2}
\end{fmfgraph*}
\end{gathered}
\end{split}
\end{gather*}
    \caption{The set of third-order 2p1h/2h1p excitation character restricted 1PI coupled-cluster self-energy diagrams that are obtained from the diagrams in Figure~\ref{fig:sc_cc_se} by replacing the self-consistent propagators with their lowest-order linearized equivalents: $\tilde{G}\rightarrow G_0 + G_0\tilde{\Sigma}^{\infty}G_0$.}
    \label{fig:adc(3)}
\end{figure*}

\begin{subequations}
\begin{align}
    \begin{split}~\label{eq:forward}
        \mathbf{\bar{C}}^{>}_{iab,jcd} &= \tilde{\Sigma}^{\infty}_{ac}\delta_{ij}\delta_{bd}-\tilde{\Sigma}^{\infty}_{ad}\delta_{ij}\delta_{bc} + \tilde{\Sigma}^{\infty}_{bd}\delta_{ac}\delta_{ij}\\
        &-\tilde{\Sigma}^{\infty}_{bc}\delta_{ad}\delta_{ij}- \tilde{\Sigma}^{\infty}_{ji}\delta_{ac}\delta_{bd}+\tilde{\Sigma}^{\infty}_{ji}\delta_{ad}\delta_{bc}
        \\
        &+\tilde{\Xi}_{ja,ci}\delta_{bd}+\tilde{\Xi}_{jb,di}\delta_{ac}+ \tilde{\Xi}_{ab,cd}\delta_{ij}\\
        &- \tilde{\Xi}_{ja,di}\delta_{bc}- \tilde{\Xi}_{jb,ci}\delta_{ad} + \tilde{\chi}_{jab,cid} 
    \end{split}\\
    \begin{split}~\label{eq:backward}
        \mathbf{\bar{C}}^{<}_{ija,klb} &= \tilde{\Sigma}^{\infty}_{ik}\delta_{ab}\delta_{jl}-\tilde{\Sigma}^{\infty}_{il}\delta_{ab}\delta_{jk}+ \tilde{\Sigma}^{\infty}_{jl}\delta_{ab}\delta_{ik}\\
        &-\tilde{\Sigma}^{\infty}_{jk}\delta_{ab}\delta_{il}+\tilde{\Sigma}^{\infty}_{ba}\delta_{il}\delta_{jk} -\tilde{\Sigma}^{\infty}_{ba}\delta_{ik}\delta_{jl}\\
        &+\tilde{\Xi}_{ib,al}\delta_{jk}+\tilde{\Xi}_{jb,ak}\delta_{il}- \tilde{\Xi}_{ij,kl}\delta_{ab} \\
        &- \tilde{\Xi}_{ib,ak}\delta_{jl}- \tilde{\Xi}_{jb,al}\delta_{ik} + \tilde{\chi}_{ijb,kal} \ .
    \end{split}
\end{align}
\end{subequations}
To reiterate, in order to be consistent with the treatment of the 2p1h/2h1p excitation character of the intermediate state configurations, the static coupled-cluster self-energy contribution and the effective interactions $\tilde{\Xi}_{pq,rs}$ and $\tilde{\chi}_{pqr,stu}$ appearing in Eqs~\ref{eq:forward} and~\ref{eq:backward} are taken in the CCSD approximation. Then, we define the coupling matrices to be composed of the renormalized interactions as follows:
\begin{subequations}
    \begin{align}
        \begin{split}~\label{eq:mod_coupling1sc}
            \tilde{U}^{2\p1\h}_{p,iab} &= \tilde{\Xi}_{pi,ab}+\frac{1}{2}\sum_{kl}\tilde{\Xi}_{pi,kl}\frac{\tilde{\Xi}_{kl,ab}}{\epsilon_{k}+\epsilon_{l}-\epsilon_{a}-\epsilon_{b}}\\
            &+\sum_{kc}\tilde{\Xi}_{pc,ak}\frac{\tilde{\Xi}_{ik,bc}}{\epsilon_{i}+\epsilon_{k}-\epsilon_{b}-\epsilon_{c}}\\
            &-\sum_{kc}\tilde{\Xi}_{pc,bk}\frac{\tilde{\Xi}_{ik,ac}}{\epsilon_{i}+\epsilon_{k}-\epsilon_{a}-\epsilon_{c}}\\
            &-\sum_{k}\frac{\tilde{\Xi}_{pi,kb}\tilde{\Sigma}^{\infty}_{ka}}{\epsilon_{k}-\epsilon_{a}} + \sum_{k}\frac{\tilde{\Xi}_{pi,ka}\tilde{\Sigma}^{\infty}_{kb}}{\epsilon_{k}-\epsilon_{b}}\\
            &+\sum_{c}\tilde{\Xi}_{pc,ab}\frac{\tilde{\Sigma}^{\infty}_{ic}}{\epsilon_{i}-\epsilon_{c}} \ ,
        \end{split}\\
        \begin{split}~\label{eq:mod_coupling}
            \bar{U}^{2\p1\h}_{iab,q} &= \tilde{\Xi}_{ab,qi} \ ,
        \end{split}\\
        \begin{split}~\label{eq:mod_coupling2sc}
            \tilde{V}^{2\h1\p}_{ija,q} &= \tilde{\Xi}_{ij,qa}+\frac{1}{2}\sum_{cd}\frac{\tilde{\Xi}_{ij,cd}}{\epsilon_{i}+\epsilon_{j}-\epsilon_{c}-\epsilon_{d}}\tilde{\Xi}_{cd,qa}\\
            &+ \sum_{kc}\frac{\tilde{\Xi}_{kj,ca}}{\epsilon_{k}+\epsilon_{j}-\epsilon_{c}-\epsilon_{a}}\tilde{\Xi}_{ic,qk}\\
            &-\sum_{kc}\frac{\tilde{\Xi}_{ki,ca}}{\epsilon_{k}+\epsilon_{i}-\epsilon_{c}-\epsilon_{a}}\tilde{\Xi}_{jc,qk}\\
            &+\sum_{c}\frac{\tilde{\Sigma}^{\infty}_{ic}\tilde{\Xi}_{cj,qa}}{\epsilon_{i}-\epsilon_{c}}-\sum_{c}\frac{\tilde{\Sigma}^{\infty}_{jc}\tilde{\Xi}_{ci,qa}}{\epsilon_{j}-\epsilon_{c}}\\
            &-\sum_{k}\frac{\tilde{\Sigma}^{\infty}_{ka}\tilde{\Xi}_{ij,qk}}{\epsilon_{k}-\epsilon_{a}} \ ,
        \end{split}\\
        \begin{split}
            \bar{V}^{2\h1\p}_{p,ija} &= \tilde{\Xi}_{pa,ij} \ .
        \end{split}
    \end{align}
\end{subequations}
This definition of the coupling matrices ensures that all time-orderings of the 2p1h/2h1p CC self-energy diagrams are included in the CC Dyson supermatrix, thereby going beyond the analysis presented in Ref.~\cite{coveney2023coupled}. The CC Dyson supermatrix within this approximation is given by

\begin{gather}
    \begin{split}~\label{eq:cc_eom_se_sc}
        &\mathbf{\tilde{D}}^{\CC}_{\text{ADC(2p1h)}} = \\
        &\left(\begin{array}{ccc}
            f_{pq}+\tilde{\Sigma}^{\infty\CCSD}_{pq} & \tilde{U}^{2\p1\h}_{p,iab} & \bar{V}^{2\h1\p}_{p,ija}  \\
            \bar{U}^{2\p1\h}_{jcd,q} & \mathbf{\bar{K}}^{>}_{iab,jcd}+\mathbf{\bar{C}}^{>}_{iab,jcd} & \mathbf{0} \\
             \tilde{V}^{2\h1\p}_{klb,q} & \mathbf{0} &  \mathbf{\bar{K}}^{<}_{ija,klb}+\mathbf{\bar{C}}^{<}_{ija,klb}
        \end{array}\right) .
    \end{split}
\end{gather}

The CC Dyson supermatrix approximations, defined in Eqs.~\ref{eq:cc_eom_se} and~\ref{eq:cc_eom_se_sc}, can be viewed as infinite-order summations of coupled-cluster self-energy diagrams. Eq.~\ref{eq:cc_eom_se} is constructed from the infinite-order summation of the perturbative coupled-cluster self-energy diagrams depicted in Figure~\ref{fig:pert_cc_se}, whereas Eq.~\ref{eq:cc_eom_se_sc} results from the infinite-order summation of the diagrammatic series depicted in Figure~\ref{fig:adc(3)}. 

The coupled-cluster self-energy diagrams depicted in Figure~\ref{fig:adc(3)} are obtained from the self-consistent diagrams presented in Figure~\ref{fig:sc_cc_se} by retaining the third-order 2p1h/2h1p self-energy diagrams that result upon replacement of the self-consistent propagators with their lowest-order linearized equivalents: $\tilde{G}\rightarrow G_0 + G_0\tilde{\Sigma}^{\infty}G_0$. This replacement gives rise to the sixth and seventh diagrams of Figure~\ref{fig:adc(3)} which are one-particle irreducible, non-skeleton contributions.

Downfolding the CC Dyson supermatrix in Eq.~\ref{eq:cc_eom_se_sc} into the single-particle spin-orbital subspace spanned by the upper left block results in the following coupled-cluster self-energy approximation
\begin{gather}
    \begin{split}~\label{eq:se_cc}
        &\tilde{\Sigma}^{\text{ADC(2p1h)}}_{pq}(\omega) = \tilde{\Sigma}^{\infty\CCSD}_{pq} \\
        &+ \frac{1}{(2!)^2} \sum_{\substack{iab\\jcd}}\tilde{U}^{2\p1\h}_{p,iab}\Big((\omega+i\eta)\mathbbm{1}-(\mathbf{\bar{K}}^{>}+\mathbf{\bar{C}}^{>})\Big)^{-1}_{iab,jcd}\bar{U}^{2\p1\h}_{jcd,q}\\
        &+ \frac{1}{(2!)^2} \sum_{\substack{kla\\mnb}}\bar{V}^{2\h1\p}_{p,kla}\Big((\omega-i\eta)\mathbbm{1}-(\mathbf{\bar{K}}^{<}+\mathbf{\bar{C}}^{<})\Big)^{-1}_{kla,mnb}\tilde{V}^{2\h1\p}_{mnb,q} .
    \end{split}
\end{gather}

As demonstrated in Appendix~\ref{app:ADC}, the coupled-cluster Dyson supermatrices of Eqs~\ref{eq:cc_eom_se} and~\ref{eq:cc_eom_se_sc}  are identical to the third-order Algebraic Diagrammatic Construction method (ADC(3)) applied to the 2p1h/2h1p coupled-cluster self-energy diagrams depicted in Figures~\ref{fig:pert_cc_se} and~\ref{fig:adc(3)}. However, it should be noted that the vanishing different Goldstone time-orderings of the third-order CC self-energy diagrams, as a result of the CC amplitude equations, is in contradistinction to the hermitian ADC(3) electronic Dyson supermatrix construction~\cite{schirmer1983new,schirmer2018many}. In the case of the hermitian electronic self-energy, these different time-orderings are non-zero and give rise to additional contributions to the coupling matrices. The reader is referred to Refs~\cite{schirmer1983new,schirmer2018many,raimondi2018algebraic} for a comprehensive review of the Algebraic Diagrammatic Construction method.

\section{Relationship to IP/EA-EOM-CC and $G_0W_0$ approximations}~\label{sec:comp}

In this section, we provide an overview of the connections between approximations for the coupled-cluster self-energy, IP/EA-EOM-CC and $G_0W_0$ theory. Subsequently, we arrive at CC-$G_0W_0$ theory.

\subsection{Equation-of-motion coupled-cluster theory}

The IP-EOM-CC method was presented in Section~\ref{sec:cc_overview}. The electron removal energies are obtained from the IP-EOM-CC eigenvalue problem given in  
Eq.~\ref{eq:eom_cc}. Restricting the IP-EOM-CC supermatrix to contain only 2h1p interactions gives the IP-EOM-CCSD approximation:
\begin{gather}
    \begin{split}
       \mathbf{\bar{H}}^{\text{IP-EOM-CCSD}} = -\left(\begin{array}{cc}
           \braket{\Phi_{i}|\bar{H}_N|\Phi_{j}}  & \braket{\Phi_{i}|\bar{H}_N|\Phi^{a}_{kl}} \\
            \braket{\Phi^{b}_{mn}|\bar{H}_N|\Phi_{j}} & \braket{\Phi^{b}_{mn}|\bar{H}_N|\Phi^{a}_{kl}} 
        \end{array}\right) \ .
    \end{split}
\end{gather}
By identifying that the upper left block is $-\braket{\Phi_{i}|\bar{H}_N|\Phi_{j}}=f_{ji}+\tilde{\Sigma}^{\infty(0)}_{ji}$, that in the Brueckner basis, $\braket{\Phi_{i}|\bar{H}_N|\Phi^{a}_{kl}} =-\chi_{kl,ia}$ and $\braket{\Phi^{b}_{mn}|\bar{H}_N|\Phi_{j}}=-\chi_{jb,mn}$, we re-write the IP-EOM-CCSD supermatrix as
\begin{gather}
    \begin{split}
       \mathbf{\bar{H}}^{\text{IP-EOM-CCSD}} = \left(\begin{array}{cc}
            f_{ji}+\tilde{\Sigma}^{\infty(0)}_{ji} & \chi_{kl,ia} \\
            \chi_{jb,mn} &  \mathbf{\bar{K}}^{<}_{kla,mnb}+\mathbf{\bar{C}}^{<(0)}_{kla,mnb}
        \end{array}\right) \ ,
    \end{split}
\end{gather}
where we have used Eqs~\ref{eq:eom_ints} to identify the interaction matrices as the lower right block of the IP-EOM-CCSD supermatrix. 
Downfolding this supermatrix into the single-particle occupied-occupied subspace, we have~\cite{coveney2023coupled,coveney2025uncovering} 
\begin{gather}
    \begin{split}~\label{eq:ip_eom_se}
        &\bar{H}^{\text{EOM-CCSD}}_{ij}(\omega) =  f_{ji}+\tilde{\Sigma}^{\infty(0)}_{ji} \\
        &+ \frac{1}{4} \sum_{\substack{kla\\mnb}}\chi_{ja,kl}\Big(\omega\mathbbm{1}-(\mathbf{\bar{K}}^{<}+\mathbf{\bar{C}}^{<(0)})\Big)^{-1}_{kla,mnb}\chi_{mn,ib} .
    \end{split}
\end{gather}
Transposing Eq.~\ref{eq:ip_eom_se}, we identify the effective EOM-CCSD `non-Dyson self-energy' as:
\begin{gather}
    \begin{split}~\label{eq:ip_eom_eff}
    &\bar{\Sigma}^{\text{EOM-CCSD}}_{ij}(\omega) = \bar{H}^{\text{EOM-CCSD}}_{ji}(\omega)-f_{ij}\\
        &=  \tilde{\Sigma}^{\infty(0)}_{ij} + \frac{1}{4} \sum_{\substack{kla\\mnb}}\chi_{ia,kl}\Big(\omega\mathbbm{1}-(\mathbf{\bar{K}}^{<}+\mathbf{\bar{C}}^{<(0)})\Big)^{-1}_{kla,mnb}\chi_{mn,jb} .
    \end{split}
\end{gather}
By comparison with Eq.~\ref{eq:occ_cc_se}, we see that the IP-EOM-CCSD effective self-energy is related to the transpose of the $2\p1\h(0)$ coupled-cluster self-energy restricted to the occupied-occupied block. However, the contribution $\tilde{\Sigma}^{\infty(2)}_{ij}$ does not appear in IP-EOM-CCSD theory. It arises from the correct treatment of the static component in the perturbative expansion of the coupled-cluster self-energy resulting from the Green's function formalism. Additionally, the coupling matrices of the ADC(2p1h(0)) CC self-energy (Eq.~\ref{eq:occ_cc_se}) contain different time-orderings that are neglected in the IP-EOM-CCSD approximation which is based on a Configuration Interaction expansion. As a result, ADC(2p1h(0)) CC self-energy diagrams such as 
\begin{gather*}
        \begin{split}
            \begin{gathered}
    \begin{fmfgraph*}(50,70)
    \fmfcurved
    \fmfset{arrow_len}{3mm}
    \fmfleft{i1,i2,i3}
    \fmflabel{}{i1}
    \fmflabel{}{i2}
    \fmfright{o1,o2,o3}
    \fmflabel{}{o1}
    \fmflabel{}{o2}
    \fmf{dbl_wiggly}{o2,i2}
    \fmf{dbl_wiggly}{i1,o1}
    \fmf{dbl_wiggly}{i3,o3}
    \fmf{fermion}{i3,i1}
    \fmf{fermion}{i2,i3}
    \fmf{fermion}{o1,o2}
    \fmf{fermion}{o2,o3}
    \fmf{fermion}{o3,o1}
    \fmfforce{(0.0w,0.0h)}{i1}
    \fmfforce{(0.66w,0.0h)}{o1}
     \fmfforce{(0.75w,0.66h)}{i2}
    \fmfforce{(1.25w,0.66h)}{o2}
    \fmfforce{(0.0w,1.0h)}{i3}
    \fmfforce{(0.66w,1.0h)}{o3}
    \fmfdot{i1,i2,i3}
    \fmfdot{o2,o1,o3}
\end{fmfgraph*}
\end{gathered}
\end{split}
\end{gather*}
are not included in IP-EOM-CCSD theory. However, it should be noted that the untruncated algebraic-wavefuction based IP/EA-EOM-CC theory and the full set of 1PI coupled-cluster self-energy diagrams are both formally exact approaches for obtaining the charged-excitation spectrum of the similarity-transformed Hamiltonian despite their differing mathematical formulations~\cite{coveney2023coupled}.

\subsection{The $G_0W_0$ self-energy and \textbf{CC}-$G_0W_0$ theory}

The dynamical $G_0W_0$ self-energy is given by~\cite{hedin1965new,berkelbach2018communication,bintrim2021full,tolle2023exact}
\begin{equation}~\label{eq:gw_se}
    \Sigma^{G_0W_0}_{pq}(\omega) =  \sum_{a\nu}\frac{W_{p,a\nu}W_{a\nu,q}}{\omega-\Omega_{\nu}-\epsilon_{a}+i\eta}+\sum_{i\nu}\frac{W_{p,i\nu}W_{i\nu,q}}{\omega+\Omega_{\nu}-\epsilon_{i}-i\eta} \ ,
\end{equation}
where the coupling and interaction matrix elements $W_{p,q\nu}$ and $\Omega_{\nu}$ are usually found by solution of the RPA equations 
\begin{gather}
    \begin{split}~\label{eq:rpa}
        \left(\begin{array}{cc}
            \mathbf{A} & \mathbf{B} \\
            -\mathbf{B}^* & -\mathbf{A}^*
        \end{array}\right)\left(\begin{array}{c}
             \mathbf{X}  \\
              \mathbf{Y}
        \end{array}\right) = \mathbf{\Omega}\left(\begin{array}{c}
             \mathbf{X}  \\
              \mathbf{Y}
        \end{array}\right)
    \end{split}
\end{gather}
where
\begin{subequations}
    \begin{align}
        \begin{split}
        A_{ia,jb} &= \left(\epsilon_a-\epsilon_i\right)\delta_{ab}\delta_{ij} + \braket{ib||aj} \ , 
    \end{split}\\
    \begin{split}
        B_{ia,jb} &=  \braket{ij||ab} \ ,
    \end{split}\\
    \begin{split}~\label{eq:W}
        W_{p,q\nu} &= \sum_{ia} \Big(\braket{pa||qi}X^{\nu}_{ia}+\braket{pi||qa}Y^{\nu}_{ia}\Big) \ .
    \end{split}
    \end{align}
\end{subequations}
\begin{figure*}[ht]
    \centering
     \begin{gather*}
\begin{split}
\tilde{\Sigma}^{\text{CC-$G_0W_0$}}_{pq} &= \hspace{5mm} \begin{gathered}
\begin{fmfgraph*}(30,30)
    \fmfset{arrow_len}{3mm}
    \fmfleft{i1,i2,i3}
    \fmfright{o1,o2,o3}
    \fmf{fermion}{i1,i2}
    \fmf{fermion}{i2,i3}
    \fmf{dbl_dashes}{i2,o2}
    \fmfforce{(0.0w,0.h)}{i1}
    \fmfforce{(0.0w,0.5h)}{i2}
    \fmfforce{(0.0w,1.0h)}{i3}
    \fmfdot{i2}
    \fmfv{decor.shape=cross,decor.filled=full, decor.size=1.5thic}{o2}
\end{fmfgraph*}
\end{gathered}\hspace{2.5mm}+\hspace{2.5mm}\begin{gathered}
    \begin{fmfgraph*}(60,40)
    \fmfcurved
    \fmfset{arrow_len}{3mm}
    \fmfleft{i1,i2}
    \fmflabel{}{i1}
    \fmflabel{}{i2}
    \fmfright{o1,o2}
    \fmflabel{}{o1}
    \fmflabel{}{o2}
    \fmf{dashes}{o1,v1}
    \fmf{dashes}{i1,v1}
    \fmf{fermion,left=0.3,tension=0}{o1,o2}
    \fmf{fermion,left=0.3,tension=0}{v1,v2}
    \fmf{fermion,left=0.3,tension=0}{v2,v1}
    \fmf{phantom}{v2,i2}
    \fmf{dbl_plain}{o2,v2}
    \fmf{fermion,left=0.3,tension=0}{o2,o1}
    \fmfdot{o1,i1,v1}
    \fmfforce{(0.0w,-0.5h)}{v3}
    \fmfforce{(0.0w,0.5h)}{v4}
    \fmfforce{(0.0w,0.0h)}{i1}
    \fmfforce{(0.5w,1.0h)}{v2}
    \fmfforce{(0.5w,0.0h)}{v1}
\end{fmfgraph*}
\end{gathered}\hspace{2.5mm}+\hspace{5mm}\begin{gathered}
    \begin{fmfgraph*}(40,40)
    \fmfcurved
    \fmfset{arrow_len}{3mm}
    \fmfleft{i1,i2}
    \fmflabel{}{i1}
    \fmflabel{}{i2}
    \fmfright{o1,o2}
    \fmflabel{}{o1}
    \fmflabel{}{o2}
    \fmf{fermion}{i1,i2}
    \fmf{dbl_wiggly}{o1,i1}
    \fmf{fermion,left=0.3,tension=0}{o1,o2}
    \fmf{dbl_wiggly}{o2,i2}
    \fmf{fermion,left=0.3,tension=0}{o2,o1}
    \fmfdot{o1,o2,i1,i2}
\end{fmfgraph*}
\end{gathered}\hspace{2.5mm}+\hspace{5mm}
\begin{gathered}
\begin{fmfgraph*}(40,60)
    \fmfcurved
    \fmfset{arrow_len}{3mm}
    \fmfleft{i1,i2,i3}
    \fmflabel{}{i1}
    \fmflabel{}{i2}
    \fmfright{o1,o2,o3}
    \fmflabel{}{o1}
    \fmflabel{}{o2}
    \fmf{dbl_wiggly}{i1,v1}
    \fmf{dbl_wiggly}{v1,o1}
    \fmf{dbl_wiggly}{v2,o2}
    \fmf{dbl_wiggly}{i3,v3}
    \fmf{fermion}{i1,i3}
    \fmf{fermion,left=0.3}{o1,o2}
    \fmf{fermion,left=0.3}{o2,o1}
    \fmf{fermion,left=0.3}{v2,v3}
    \fmf{fermion,left=0.3}{v3,v2}
    \fmfforce{(0.0w,0.0h)}{i1}
    \fmfforce{(1.0w,0.0h)}{o1}
    \fmfforce{(0.5w,0.5h)}{v2}
    \fmfforce{(0.5w,0.0h)}{v1}
    \fmfforce{(0.5w,1.0h)}{v3}
    \fmfforce{(0.0w,1.0h)}{i3}
    \fmfforce{(1.0w,1.0h)}{o3}
    \fmfdot{v2,v3}
    \fmfdot{i1,i3}
    \fmfdot{o1,o2}
\end{fmfgraph*}
\end{gathered}\hspace{2.5mm}+\hspace{2.5mm}\begin{gathered}
\begin{fmfgraph*}(40,60)
    \fmfcurved
    \fmfset{arrow_len}{3mm}
    \fmfleft{i1,i2,i3,i4}
    \fmflabel{}{i1}
    \fmflabel{}{i2}
    \fmfright{o1,o2,o3,o4}
    \fmflabel{}{o1}
    \fmflabel{}{o2}
    \fmf{dbl_wiggly}{i1,o1}
    \fmf{dbl_wiggly}{o2,v2}
    \fmf{dbl_wiggly}{v3,v4}
    \fmf{dbl_wiggly}{i4,o4}
    \fmf{fermion}{i1,i4}
    \fmf{fermion,left=0.3}{o1,o2}
    \fmf{fermion,left=0.3}{o2,o1}
    \fmf{fermion,left=0.3}{v4,o4}
    \fmf{fermion,left=0.3}{o4,v4}
    \fmf{fermion,left=0.3}{v2,v3}
    \fmf{fermion,left=0.3}{v3,v2}
    \fmfforce{(0.0w,0.0h)}{i1}
    \fmfforce{(1.0w,0.0h)}{o1}
    \fmfforce{(0.66w,0.33h)}{v2}
    \fmfforce{(0.33w,0.0h)}{v1}
    \fmfforce{(0.66w,0.66h)}{v3}
    \fmfforce{(0.33w,0.66h)}{v4}
    \fmfforce{(0.0w,1.0h)}{i4}
    \fmfforce{(0.33w,1.0h)}{o4}
    \fmfdot{v2,v3,v4}
    \fmfdot{i1,i4}
    \fmfdot{o1,o2,o4}
\end{fmfgraph*}
\end{gathered}\hspace{2.5mm}+\hspace{2.5mm}\cdots
\end{split}
\end{gather*}
    \caption{Diagrammatic representation of the CC-$G_0W_0$ self-energy. The bubble diagrams are summed to infinite-order through the $\mathbf{\tilde{D}}^{\text{CC-$G_0W_0$}}$ supermatrix representation (Eq.~\ref{eq:cc_gowo_super}) and the two-body antisymmetrized effective interactions are evaluated within the rCCD approximation. The non-interacting Green's function lines appearing in the dynamical self-energy diagrams are not fully antisymmetrized with respect to fermionic exchange.}
    \label{fig:CC_G0W0}
\end{figure*}
In the $G_0W_0$ approximation, the exchange interaction is usually separated from the direct interaction and one is left with only the direct RPA (dRPA) equations. Here, we instead choose to work with the antisymmetrized RPA equations to more easily exploit the connections to coupled-cluster theory. The exchange contributions may be trivially removed from our equations to recover the $G_0W_0$ approximation using dRPA~\cite{lange2018relation,tolle2023exact}. 
In terms of the upfolded Dyson supermatrix representation, the $G_0W_0$ eigenvalue problem takes the form 
\begin{gather}
    \begin{split}
        \mathbf{D}^{G_0W_0} = \left(\begin{array}{ccc}
            f_{pq} & W_{p,a\nu} &W_{p,i\nu} \\
            W_{a\nu,q} & \mathbf{\Lambda}^{G_0W_0>}_{a\nu,b\nu'} & \mathbf{0} \\
             W_{i\nu,q}& \mathbf{0} & \mathbf{\Lambda}^{G_0W_0<}_{i\nu,j\nu'}
        \end{array}\right)
    \end{split}
\end{gather}
with 
\begin{subequations}
    \begin{align}
    \begin{split}
        \mathbf{\Lambda}^{G_0W_0>}_{a\nu,b\nu'} &= (\epsilon_{a}+\Omega_{\nu})\delta_{ab}\delta_{\nu\nu'}
        \end{split}\\
        \begin{split}
        \mathbf{\Lambda}^{G_0W_0<}_{i\nu,j\nu'} &= (\epsilon_{i}-\Omega_{\nu})\delta_{ij}\delta_{\nu\nu'} \ .
    \end{split}
\end{align}
\end{subequations}
where the index $\nu$ is a quasi-boson index that spans a set of values much larger than the spin-orbital index. The RPA eigenvalue problem can be re-written in terms of effective coupled-cluster doubles amplitudes by identifying $\mathbf{T}=\mathbf{YX}^{-1}$. Using this expression in the RPA equations (Eq.~\ref{eq:rpa}) leads to the ring CCD (rCCD) approximation~\cite{scuseria2008ground,scuseria2013particle,coveney2023coupled}
\begin{gather}
    \begin{split}~\label{eq:rCCD_eqns}
        \mathbf{B}^* + \mathbf{A^*T} + \mathbf{TA} + \mathbf{TBT} = \mathbf{0} \ ,
    \end{split}
\end{gather}
with the RPA Hamiltonian written as 
\begin{gather}
    \begin{split}
        H^{\RPA}_{ia,jb} = (\epsilon_{a}-\epsilon_{i})\delta_{ab}\delta_{ij} + \braket{ib||aj}+\sum_{kc}\braket{ik||ac}(t^{cb}_{kj})_{\rCCD} .
    \end{split}
\end{gather}
As a result, the RPA eigenvalue problem becomes
\begin{gather}
    \begin{split}
        \sum_{jb}H^{\RPA}_{ia,jb}X^{\nu}_{jb} = \Omega_{\nu}X^{\nu}_{ia} \ .
    \end{split}
\end{gather}
where the rCCD doubles amplitudes do not retain fermionic anti-symmetry~\cite{scuseria2008ground}.

To demonstrate the connection between the $G_0W_0$ approximation and the coupled-cluster self-energy, we restrict our analysis of the coupled-cluster self-energy to the space of 2h1p/2p1h excitations. In doing so, we introduce the CC-$G_0W_0$ Dyson supermatrix as 
    \begin{gather}
    \begin{split}~\label{eq:cc_gowo_super}
        &\mathbf{\tilde{D}}^{\text{CC-$G_0W_0$}} = \left(\begin{array}{ccc}
            f_{pq} + \tilde{\Sigma}^{\infty{\rCCD}}_{pq} & \chi^{\rCCD}_{pi,ab} &\chi^{\rCCD}_{pa,ij} \\
            \chi^{\rCCD}_{ab,qi} & \mathbf{\bar{\Lambda}}^{G_0W_0>}_{iab,jcd} & \mathbf{0} \\
             \chi^{\rCCD}_{ij,qa}& \mathbf{0} & \mathbf{\bar{\Lambda}}^{G_0W_0<}_{ija,klb}
        \end{array}\right)
    \end{split}
\end{gather}
where the effective interactions $\{\chi^{\rCCD}_{pq,rs}\}$ are found from the two-body matrix elements of $\bar{H}$, but are formed from only the rCCD diagrams and amplitudes. For example, the element $\chi^{\rCCD}_{ia,kl}$ is given by
\begin{gather}
    \begin{split}
        \chi^{\rCCD}_{ia,kl} = \braket{ia||kl} + \sum_{cm}\braket{im||kc}(t^{ca}_{ml})_{\rCCD}  \ .
    \end{split}
\end{gather} 
The interaction matrices are defined as 
\begin{subequations}
\begin{align}
    \begin{split}
        \mathbf{\bar{\Lambda}}^{G_0W_0>}_{iab,jcd} &= (\epsilon_{a}\delta_{bd}\delta_{ij}+H^{\dRPA}_{ib,jd})\delta_{ac} 
    \end{split}\\
    \begin{split}
    \mathbf{\bar{\Lambda}}^{G_0W_0<}_{ija,klb} &= (\epsilon_{i}\delta_{jl}\delta_{ab}-H^{\dRPA}_{ja,lb})\delta_{ik} \ .
    \end{split}
\end{align}
\end{subequations}
The static contribution, $\tilde{\Sigma}^{\infty{\rCCD}}_{pq}$, can be evaluated exactly within the rCCD approximation and is given by 
\begin{gather}
    \begin{split}
        \tilde{\Sigma}^{\infty{\rCCD}}_{pq}=\tilde{\Sigma}^{\infty(0)\rCCD}_{pq} + \frac{1}{(2!)^2}\sum_{ijab}\chi^{\rCCD}_{pab,qij}(\lambda^{ij}_{ab})_{\rCCD} \ ,
    \end{split}
\end{gather}
where $(\lambda^{ij}_{ab})_{\rCCD}$ are the rCCD de-excitation amplitudes obtained from the rCCD Lagrangian~\cite{rishi2020route}. There are no singles amplitudes in the coupled-cluster doubles approximation. The three-body interaction elements can be found in Ref.~\cite{shavitt2009many}, an example of which is the element
\begin{gather}
    \begin{split}
        \chi^{\rCCD}_{iab,jkl} = \sum_{n}\chi^{\rCCD}_{in,jk}(t^{ab}_{nl})_{\rCCD} + \sum_{c}\chi^{\rCCD}_{ia,ck}(t^{cb}_{jl})_{\rCCD} \ .
    \end{split}
\end{gather}
Using these coupling and interaction matrix elements, the diagrammatic representation of the CC-$G_0W_0$ self-energy diagrams contained in the $\mathbf{\tilde{D}}^{\text{CC-$G_0W_0$}}$ supermatrix is depicted in Figure~\ref{fig:CC_G0W0}. We clearly see that this approximation results in the infinite-order summation of the coupled-cluster bubble self-energy diagrams, analogous to the $G_0W_0$ approximation which performs an infinite-order summation of electronic bubble self-energy diagrams. 

Finally, from Eq.~\ref{eq:cc_en}, the ground state correlation energy in CC-$G_0W_0$ theory is the RPA correlation energy~\cite{scuseria2008ground,coveney2023coupled} 
\begin{gather}
\begin{split}
    E^{\text{CC-$G_0W_0$}}_{c} = \frac{1}{2}\sum_{i}\tilde{\Sigma}^{\infty(0)\rCCD}_{ii} = \frac{1}{4}\sum_{ij,ab}\braket{ij||ab}(t^{ab}_{ij})_{\rCCD} \ .
\end{split}
\end{gather}
Through this reformulation of the $G_0W_0$ approximation in the framework of coupled-cluster theory, we may now include `vertex' corrections that go beyond the $G_0W_0$ approximation, while maintaining the correct spectral form of the self-energy. This is achieved by simply modifying the doubles excitation and de-excitation amplitudes that enter the approximation. For example, using Eq.~\ref{eq:cc_gowo_super}, one may use the full CCSD singles and doubles excitation and de-excitation amplitudes instead of the rCCD amplitudes. This would provide a simple way to extend the $G_0W_0$ approximation to include effects of singles excitations and go beyond RPA screening by including the complete set of interactions, $\braket{\Phi^{b}_{j}|\bar{H}_N|\Phi^{a}_{i}}$, between singly excited configurations in the definition of the interaction matrices defined in $\mathbf{\bar{\Lambda}}^{G_0W_0>}_{iab,jcd}/\mathbf{\bar{\Lambda}}^{G_0W_0<}_{ija,klb}$~\cite{coveney2023coupled}.

\section{The Coupled-cluster Bethe-Salpeter Equation}~\label{sec:BSE}

Within the Green's function formalism, the Bethe-Salpeter equation (BSE) describes two-particle correlation processes and gives access to neutral and two-particle excitation energies. The coupled-cluster Bethe-Salpeter equation is written as 
\begin{widetext}
\begin{gather}
    \begin{split}~\label{eq:BSE}
        \tilde{L}_{pq,rs}(t_1,t_2;t_3,t_4) = \tilde{L}^{0}_{pq,rs}(t_1,t_2;t_3,t_4) +\sum_{tuvw}\int dtdt'dt''dt''' \tilde{L}^{0}_{pt,ru}(t_1,t;t_3,t')\tilde{\Xi}_{uv,tw}(t',t'';t,t''')\tilde{L}_{wq,vs}(t''',t_2;t'',t_4)
    \end{split}
\end{gather}
\end{widetext}
where $\tilde{L}^{0}_{pq,rs}(t_1,t_2;t_3,t_4) = -i\tilde{G}_{ps}(t_1,t_4)\tilde{G}_{qr}(t_2,t_3)$ corresponds to the propagation of two-independent particles~\cite{stefanucci2013nonequilibrium}, and $\tilde{\Xi}$ is the coupled-cluster Bethe-Salpeter kernel. The correlated four-point linear response function $\tilde{L}$ is defined as 
\begin{gather}
    \begin{split}
        \tilde{L}_{pq,rs}(t_1,t_2;t_3,t_4)&=\tilde{G}^{4\pt}_{pq,rs}(t_1,t_2;t_3,t_4)\\
        &-i\tilde{G}_{pr}(t_1,t_3)\tilde{G}_{qs}(t_2,t_4)\ ,
    \end{split}
\end{gather}
and can be expressed as the functional derivative of the single-particle Green's function with respect to a fictitious external potential via application of Schwinger's functional derivative technique to the biorthogonal Interaction picture (Section~\ref{sec:bio_qm}).

It is important to note that the coupled-cluster BSE is not an equation for the coupled-cluster representation of the electronic 4-point response function, which is defined as 
        \begin{gather*}
    \begin{split}
        \bar{L}_{pq,rs}(t_1,t_2;t_3,t_4) &= \bar{G}^{4\pt}_{pq,rs}(t_1,t_2;t_3,t_4)\\
        &-i\bar{G}_{pr}(t_1,t_3)\bar{G}_{qs}(t_2,t_4) \ ,
    \end{split}
\end{gather*}
where
\begin{gather*}
    \begin{split}
        &i\bar{G}^{4\pt}_{pq,rs}(t_1,t_2;t_3,t_4) = \braket{\Tilde{\Psi}_0|\mathcal{T}\left\{\bar{a}_{q}(t_2) \bar{a}_p(t_1) \bar{a}^\dag_{r}(t_3)\bar{a}^\dag_{s}(t_4)\right\}|\Phi_0} \ 
    \end{split}
\end{gather*}
is the coupled-cluster representation of the electronic 4-point Green's function. This is because the coupled-cluster representation of the electronic 4-point electronic response function can only be obtained by direct computation using quantities obtained from ground state and EE-EOM-CC theory, which remains largely unexplored. Instead, the coupled-cluster BSE naturally arises as a result of the coupling  between the SP-CCGF and the four-point response function through the diagrammatic coupled-cluster self-energy.

In the real time-domain, the kernel of the coupled-cluster BSE is given by the functional derivative of the coupled-cluster self-energy with respect to the single-paricle coupled-cluster Green's function as 
\begin{equation}
    \tilde{\Xi}_{pq,rs}(t_1,t_2;t_3,t_4) = i\frac{\delta\tilde{\Sigma}_{pr}(t_1,t_3)}{\delta\tilde{G}_{sq}(t_4,t_2)} \ .
\end{equation}
The kernel is composed of two-particle irreducible (2PI) diagrams~\cite{stefanucci2013nonequilibrium}. The BSE is fundamentally different from the corresponding Dyson equation for the single-particle Green's function because it cannot be cast as a closed form equation for the ph/hp response function. The ph/hp response function is defined as~\cite{schirmer2018many} 
\begin{equation}~\label{eq:ph_resp}
    \tilde{L}^{\p\h}_{pq,rs}(t_1,t_2) = \tilde{L}_{pq,rs}(t_1,t_2;t^{+}_1,t^{+}_2) \ .
\end{equation}
The Fourier transform of the exact coupled-cluster ph/hp response function (Eq.~\ref{eq:ph_resp}), obtained by resolving the identity of $N$-particle eigenstates of $\Bar{H}$, possess poles at the \emph{exact} neutral excitation spectrum: 
\begin{gather}
    \begin{split}
        \tilde{L}^{\text{ph}}_{pq,rs}(\omega) = \sum_{\nu\neq0} &\Bigg(\frac{\braket{\tilde{\Psi}_0|a^\dag_{s}a_q|\bar{\Psi}^{N}_{\nu}}\braket{\tilde{\Psi}^{N}_{\nu}|a^\dag_{r}a_p|\Phi_0}}{\omega-\Omega_{\nu}+i\eta}\\
        &-\frac{\braket{\tilde{\Psi}_0|a^\dag_{r}a_p|\bar{\Psi}^{N}_{\nu}}\braket{\tilde{\Psi}^{N}_{\nu}|a^\dag_{s}a_q|\Phi_0}}{\omega+\Omega_{\nu}-i\eta}\Bigg) ,
    \end{split}
\end{gather}
where $\Omega_{\nu}=E^{N}_{n}-E^{N}_0$ are the exact neutral excitation energies of the system. 

Now, restricting the BSE to that of the ph/hp sector, we have 
\begin{widetext}
\begin{gather}
    \begin{split}~\label{eq:ph}
        \tilde{L}^{\p\h}_{pq,rs}(t_1,t_2) &= \tilde{L}^{\p\h(0)}_{pq,rs}(t_1,t_2) +\sum_{tuvw}\int dtdt'dt''dt''' \tilde{L}^{0}_{pt,ru}(t_1,t;t^{+}_1,t')\tilde{\Xi}_{uv,tw}(t',t'';t,t''')\tilde{L}_{wq,vs}(t''',t_2;t'',t^{+}_2) \ .
    \end{split}
\end{gather}
\end{widetext}
It is the presence of the kernel, $\tilde{\Xi}$, that means we cannot send the time arguments in the propagators to the corresponding ones for the ph/hp propagator~\cite{strinati1988application,romaniello2009double,zhang2013dynamical,sangalli2011double,rebolini2016range,olevano2019formally}. Therefore, we do not have a closed equation for the ph/hp response function as Eq.~\ref{eq:ph} is not a functional of $\tilde{L}^{\p\h}$. This is because the BSE contains a fundamental coupling between the particle-hole, particle-particle and hole-hole response functions through the kernel~\cite{marie2025anomalous}.

\subsection{The coupled-cluster Bethe-Salpeter kernel}

By taking the functional derivative of the coupled-cluster self-energy defined in Eq.~\ref{eq:exact_se} with respect to the coupled-cluster Green's function we generate the coupled-cluster BSE kernel. Diagrammatically, this functional derivative is found by cutting the Green's function and interaction lines in the Feynman diagram series for the renormalized coupled-cluster self-energy. To do this, we first analyze the variation of the interaction lines with respect to the Green's function. In the real-time domain, the static component of the self-energy is given by 
\begin{gather}
    \begin{split}
        \tilde{\Sigma}^{\infty}_{pq}(t_1,t_3) &= \tilde{\Sigma}^{\infty}_{pq}\delta(t_1,t_3)\\
        &= \Bigg(\bar{h}_{pq} -i \sum_{rs} \bar{h}_{pr,qs}\tilde{G}_{sr}(t_1-t_1^+) \\
        &+ \frac{i}{4}\sum_{rs,tu} \bar{h}_{prs,qtu}\tilde{G}^{2\p\h}_{tu,rs}(t_1-t_1^+) + \cdots\Bigg)\delta(t_1,t_3) \ .
    \end{split}
\end{gather}
The functional derivative of the static component with respect to the exact Green's function is therefore 
\begin{widetext}
\begin{gather}
    \begin{split}
        i\frac{\delta\tilde{\Sigma}^{\infty}_{pr}(t_1,t_3)}{\delta\tilde{G}_{sq}(t_4,t_2)} &= \left( \bar{h}_{pq,rs}- i\sum_{rs,tu} \bar{h}_{pqt,rsu}\tilde{G}_{ut}(t_1-t_1^+) + \cdots\right)\delta(t_1,t_3)\delta(t_1,t_2)\delta(t^+_1,t_4)+ \tilde{\Xi}^{c}_{pq,rs}(t_1,t_2;t^+_1,t_4)\delta(t_1,t_3) \ ,
    \end{split}
\end{gather}
\end{widetext}
where $\tilde{\Xi}^{c}_{pq,rs}(t_1,t_2;t^+_1,t_4)$ represents the contribution that is dependent on functional derivatives involving the $\tilde{\Lambda}^{2n\pt}$-vertices and is explicitly related to multiparticle correlation effects. This additional term $\tilde{\Xi}^c$ is similar in origin to the functional derivative of $W$ with respect to the Green's function in the $GW$ approximation that arises when deriving the electronic $GW$-BSE kernel. The first term inside the brackets is simply the series for the two-body effective interaction (see Appendix~\ref{appendix_2}) and so we have 
\begin{gather}
    \begin{split}
        i\frac{\delta\tilde{\Sigma}^{\infty}_{pr}(t_1,t_3)}{\delta\tilde{G}_{sq}(t_4,t_2)} &= \tilde{\Xi}_{pq,rs}\delta(t_1,t_3)\delta(t_1,t_2)\delta(t^+_1,t_4) \\
        &+ \tilde{\Xi}^{c}_{pq,rs}(t_1,t_2;t^+_1,t_4)\delta(t_1,t_3) \ .
    \end{split}
\end{gather}
Therefore, at first-order, the functional derivative is the static two-body effective interaction and arises due to `independent particle' propagation. The additional term $\tilde{\Xi}^c$ must also be accounted for and is a dynamical contribution. 

To derive the additional component $\tilde{\Xi}^{c}$ we adopt a simplified notation whereby the spin-orbital indices and time/frequency arguments are removed. With this notation, we write the 4-point and 6-point vertex equations from Eqs~\ref{eq:4point_vert} and~\ref{eq:6point} as 
\begin{gather}
    \begin{split}
        \tilde{G}^{4\pt} = i\mathcal{A}\{\tilde{G}\tilde{G}\} -\tilde{G}\tilde{G}\tilde{\Lambda}^{4\pt}\tilde{G}\tilde{G}
    \end{split}
\end{gather}
and
\begin{gather}
    \begin{split}
        \tilde{G}^{6\pt} = -\mathcal{A}\{\tilde{G}\tilde{G}\tilde{G}\} -i\mathcal{P}\{\tilde{G}\tilde{G}\tilde{G}\tilde{\Lambda}^{4\pt}\tilde{G}\tilde{G}\} +\tilde{G}\tilde{G}\tilde{G}\tilde{\Lambda}^{6\pt}\tilde{G}\tilde{G}\tilde{G} \ .
    \end{split}
\end{gather}
where $\mathcal{A}$ generates the correct fermionic symmetry associated with independent-particle propagation and $\mathcal{P}$ is the cyclic permutation symmetry required to correctly describe fermionic statistics of two correlated particles propagating with respect to a third independent-particle. 
Using this simplified notation, the static component of the self-energy becomes
\begin{gather}
\begin{split}
    \tilde{\Sigma}^{\infty} = \bar{h}^{1b}-i\bar{h}^{2b}\tilde{G} +\frac{i}{4}\bar{h}^{3b}\tilde{G}^{4\pt} -\frac{i}{(3!)^2}\bar{h}^{4b}\tilde{G}^{6\pt} + \cdots
\end{split}
\end{gather}
where $\bar{h}^{nb}$ represent the $n$-body matrix elements of the similarity transformed Hamiltonian. The contribution of the functional derivative $\frac{\delta\tilde{\Sigma}^{\infty}}{\delta\tilde{G}}$ due to anti-symmetrized independent-particle propagation gives the static two-body effective interaction, $\tilde{\Xi}^{2b}$. The functional derivatives of the higher-body Green's functions with respect to the single-particle propagator (removing the anti-symmetrized independent-particle propagation) gives rise to the general structure 
\vspace{-2.5mm}
\begin{subequations}
    \begin{gather}
        \begin{split}
             \frac{\delta}{\delta\tilde{G}}\Big(\tilde{G}^{4\pt}-i\mathcal{A}\{\tilde{G}\tilde{G}\}\Big) = - 2\tilde{G}\tilde{\Lambda}^{4\pt}\tilde{G}\tilde{G} -\tilde{G}\tilde{G}\frac{\delta\tilde{\Lambda}^{4\pt}}{\delta\tilde{G}}\tilde{G}\tilde{G}
        \end{split}
    \end{gather}
    with
    \begin{gather}
        \begin{split}
             \frac{\delta}{\delta\tilde{G}}\Big(\tilde{G}^{6\pt}+\mathcal{A}\{\tilde{G}\tilde{G}\tilde{G}\}\Big) &= -(3!)^2i\tilde{G}\tilde{G}\tilde{\Lambda}^{4\pt}\tilde{G}\tilde{G}\\
             &-3^2i\tilde{G}\tilde{G}\tilde{G}\frac{\delta\tilde{\Lambda}^{4\pt}}{\delta\tilde{G}}\tilde{G}\tilde{G}\\
             &+  3\tilde{G}\tilde{G}\tilde{\Lambda}^{6\pt}\tilde{G}\tilde{G}\tilde{G} \\
             &+\tilde{G}\tilde{G}\tilde{G}\frac{\delta\tilde{\Lambda}^{6\pt}}{\delta\tilde{G}}\tilde{G}\tilde{G}\tilde{G}
        \end{split}
    \end{gather}
\end{subequations}
and so on for the higher derivatives. In these equations, the pre-factors represent the number of equivalent terms generated by the different functional derivatives with respect to permutation of the fermionic states. Therefore, the functional derivative of the static component of the coupled-cluster self-energy gives rise to 
\begin{widetext}
\begin{gather}
    \begin{split}
        i\frac{\delta\tilde{\Sigma}^{\infty}}{\delta\tilde{G}} &= \tilde{\Xi}^{2b} +\frac{1}{(2!)^2}\bar{h}^{3b}\Big(2!\tilde{G}\tilde{\Lambda}^{4\pt}\tilde{G}\tilde{G}+\tilde{G}\tilde{G}\frac{\delta\tilde{\Lambda}^{4\pt}}{\delta\tilde{G}}\tilde{G}\tilde{G}\Big) \\
        &+\frac{1}{(3!)^2} \bar{h}^{4b}\Big(-(3!)^2i\tilde{G}\tilde{G}\tilde{\Lambda}^{4\pt}\tilde{G}\tilde{G}-3^2i\tilde{G}\tilde{G}\tilde{G}\frac{\delta\tilde{\Lambda}^{4\pt}}{\delta\tilde{G}}\tilde{G}\tilde{G}+3\tilde{G}\tilde{G}\tilde{\Lambda}^{6\pt}\tilde{G}\tilde{G}\tilde{G}+\tilde{G}\tilde{G}\tilde{G}\frac
        {\delta\tilde{\Lambda}^{6\pt}}{\delta\tilde{G}}\tilde{G}\tilde{G}\tilde{G}\Big)\\
        &+ \cdots
    \end{split}
\end{gather}
Collecting like terms together along with the higher-order contributions arising from functional derivatives of the higher-order Green's functions we have
\begin{gather}
    \begin{split}
        i\frac{\delta\tilde{\Sigma}^{\infty}}{\delta\tilde{G}} &= \tilde{\Xi}^{2b} +\frac{1}{2!}\Big(\bar{h}^{3b}-i\bar{h}^{4b}\tilde{G}+\cdots\Big)\Big(\tilde{G}\tilde{\Lambda}^{4\pt}\tilde{G}\tilde{G}\Big)
        +\frac{1}{(2!)^2}\Big(\bar{h}^{3b}-i\bar{h}^{4b}\tilde{G}+\cdots\Big)\Big(\tilde{G}\tilde{G}\frac{\delta\tilde{\Lambda}^{4\pt}}{\delta\tilde{G}}\tilde{G}\tilde{G}\Big) \\
        &+\frac{1}{(3!)^2} \Big(\bar{h}^{4b}-i\bar{h}^{5b}\tilde{G}+\cdots\Big)\Big(3\tilde{G}\tilde{G}\tilde{\Lambda}^{6\pt}\tilde{G}\tilde{G}\tilde{G}+\tilde{G}\tilde{G}\tilde{G}\frac
        {\delta\tilde{\Lambda}^{6\pt}}{\delta\tilde{G}}\tilde{G}\tilde{G}\tilde{G}\Big) + \cdots
    \end{split}
\end{gather}
\end{widetext}
Identifying the expressions in brackets as the effective interactions that arise when normal-ordering the similarity transformed Hamiltonian with respect to the biorthogonal ground state expectation value, we find 
\begin{gather}
    \begin{split}~\label{eq:outline}
        i\frac{\delta\tilde{\Sigma}^{\infty}}{\delta\tilde{G}} &= \tilde{\Xi}^{2b} +\frac{1}{2!}\tilde{\chi}^{3b}\Big(\tilde{G}\tilde{\Lambda}^{4\pt}\tilde{G}\tilde{G}\Big)
        +\frac{1}{(2!)^2}\tilde{\chi}^{3b}\Big(\tilde{G}\tilde{G}\frac{\delta\tilde{\Lambda}^{4\pt}}{\delta\tilde{G}}\tilde{G}\tilde{G}\Big)\\
        &+\frac{1}{3!\cdot 2!} \tilde{\chi}^{4b}\tilde{G}\tilde{G}\Lambda^{6\pt}\tilde{G}\tilde{G}\tilde{G}\\
        &+\frac{1}{(3!)^2}\tilde{\chi}^{4b}\tilde{G}\tilde{G}\tilde{G}\frac
        {\delta\Lambda^{6\pt}}{\delta\tilde{G}}\tilde{G}\tilde{G}\tilde{G} + \cdots
    \end{split}
\end{gather}
where the series continues with the higher-order effective interactions contracting with higher-order vertices in this schematic derivation. At first-order, the explicitly frequency-dependent contribution gives rise to the following diagrams, which correspond to the second term of the right hand side of Eq.~\ref{eq:outline}
\vspace{5mm}
\begin{gather}
    \begin{split}~\label{eq:add_cont}
        \tilde{\Xi}^{c(1)}_{pq,rs}\hspace{2.5mm}=\hspace{5mm}\begin{gathered}\begin{fmfgraph*}(50,50)
    \fmfcurved
    \fmfset{arrow_len}{3mm}
    \fmfleft{i1,i2}
    \fmflabel{}{i1}
    \fmflabel{}{i2}
    \fmfright{o1,o2}
    \fmflabel{}{o1}
    \fmflabel{}{o2}
    \fmf{dbl_dashes}{i1,v1}
    \fmf{dbl_dashes}{v1,o1}
    \fmf{dbl_zigzag}{o2,v2}
    \fmf{double_arrow}{o1,o2}
    \fmf{double_arrow,left=0.3}{v1,v2}
    \fmf{double_arrow,left=0.3}{v2,v1}
    \fmfforce{(0.0w,0.0h)}{i1}
    \fmfforce{(0.0w,1.0h)}{i2}
    \fmfforce{(1.0w,0.0h)}{o1}
    \fmfforce{(1.0w,1.0h)}{o2}
    \fmfforce{(0.5w,1.0h)}{v2}
    \fmfforce{(0.5w,0.0h)}{v1}
    \fmfdotn{v}{2}
    \fmfdot{i1}
    \fmfdot{o2,o1}
     \fmflabel{$\substack{p\\r}$}{i1}
     \fmflabel{$s$}{o1}
     \fmflabel{$q$}{o2}
\end{fmfgraph*}
\end{gathered}\hspace{5mm} + \hspace{5mm}\begin{gathered}\begin{fmfgraph*}(50,50)
    \fmfcurved
    \fmfset{arrow_len}{3mm}
    \fmfleft{i1,i2}
    \fmflabel{}{i1}
    \fmflabel{}{i2}
    \fmfright{o1,o2}
    \fmflabel{}{o1}
    \fmflabel{}{o2}
    \fmf{dbl_dashes}{i1,v1}
    \fmf{dbl_dashes}{v1,o1}
    \fmf{dbl_zigzag}{o2,v2}
    \fmf{double_arrow}{v1,v2}
    \fmf{double_arrow,left=0.3}{o1,o2}
    \fmf{double_arrow,left=0.3}{o2,o1}
    \fmfforce{(0.0w,0.0h)}{i1}
    \fmfforce{(0.0w,1.0h)}{i2}
    \fmfforce{(1.0w,0.0h)}{o1}
    \fmfforce{(1.0w,1.0h)}{o2}
    \fmfforce{(0.5w,1.0h)}{v2}
    \fmfforce{(0.5w,0.0h)}{v1}
    \fmfdotn{v}{2}
    \fmfdot{i1}
    \fmfdot{o2,o1}
     \fmflabel{$\substack{p\\r}$}{i1}
     \fmflabel{$s$}{v1}
     \fmflabel{$q$}{v2}
\end{fmfgraph*}
\end{gathered}
    \end{split}
\end{gather}
Here, we have re-inserted the spin-orbital indices and accounted for the different external index permutations. Diagrammatically, this term arises from cutting the Green's function lines in the following diagram contributing to the static component of the self-energy:
\begin{gather}
\begin{split}~\label{eq:corr_cont}
    \tilde{\Sigma}^{\infty(c,1)}_{pr} = \hspace{2.5mm}\begin{gathered}\begin{fmfgraph*}(50,50)
    \fmfcurved
    \fmfset{arrow_len}{3mm}
    \fmfleft{i1,i2}
    \fmflabel{}{i1}
    \fmflabel{}{i2}
    \fmfright{o1,o2}
    \fmflabel{}{o1}
    \fmflabel{}{o2}
    \fmf{dbl_dashes}{i1,v1}
    \fmf{dbl_dashes}{v1,o1}
    \fmf{dbl_zigzag}{o2,v2}
    \fmf{double_arrow,left=0.3}{v1,v2}
    \fmf{double_arrow,left=0.3}{v2,v1}
    \fmf{double_arrow,left=0.3}{o1,o2}
    \fmf{double_arrow,left=0.3}{o2,o1}
    \fmfforce{(0.0w,0.0h)}{i1}
    \fmfforce{(0.0w,1.0h)}{i2}
    \fmfforce{(1.0w,0.0h)}{o1}
    \fmfforce{(1.0w,1.0h)}{o2}
    \fmfforce{(0.5w,1.0h)}{v2}
    \fmfforce{(0.5w,0.0h)}{v1}
    \fmfdotn{v}{2}
    \fmfdot{i1}
    \fmfdot{o2,o1}
     \fmflabel{$\substack{p\\r}$}{i1}
\end{fmfgraph*}
\end{gathered}
\end{split}
\end{gather}
This diagram appears when expanding the 4-point Green's function in the expression for $\tilde{\Sigma}^{\infty}$ through the four-point vertex equation (Eq.~\ref{eq:4point_vert}). The resulting contribution is explicitly dynamical. A similar expression holds for the two-body effective interaction elements 
\begin{gather}
    \begin{split}~\label{eq:1st_2body}
        i\frac{\delta\tilde{\Xi}_{pq,rs}}{\delta\tilde{G}_{ut}(t_4,t_2)}\delta(t_1,t_3) &= \tilde{\chi}_{pqt,rsu}\delta(t_1,t_3)\delta(t_1,t_2)\delta(t^+_1,t_4) \\
        &+ \tilde{\chi}^{c}_{pqt,rsu}(t_1,t_2;t^+_1t_4)\delta(t_1,t_3) \ .
    \end{split}
\end{gather}
Using our simplified notation, we find 
\begin{gather}
    \begin{split}~\label{eq:outline1}
        i\frac{\delta\tilde{\Xi}^{2b}}{\delta\tilde{G}} &= \tilde{\chi}^{3b} +\frac{1}{2!}\tilde{\chi}^{4b}\Big(\tilde{G}\tilde{\Lambda}^{4\pt}\tilde{G}\tilde{G}\Big)
        +\frac{1}{(2!)^2}\tilde{\chi}^{4b}\Big(\tilde{G}\tilde{G}\frac{\delta\tilde{\Lambda}^{4\pt}}{\delta\tilde{G}}\tilde{G}\tilde{G}\Big)\\
        &+\frac{1}{3!\cdot2!} \tilde{\chi}^{5b}\tilde{G}\tilde{G}\Lambda^{6\pt}\tilde{G}\tilde{G}\tilde{G}\\
        &+\frac{1}{(3!)^2}\tilde{\chi}^{5b}\tilde{G}\tilde{G}\tilde{G}\frac
        {\delta\Lambda^{6\pt}}{\delta\tilde{G}}\tilde{G}\tilde{G}\tilde{G} + \cdots
    \end{split}
\end{gather}
and so on for the derivatives of higher-body interactions. In the case of a three-body Hamiltonian, this derivative would only generate the bare three-body interaction, $i\frac{\delta\tilde{\Xi}^{2b}}{\delta\tilde{G}}=\tilde{\chi}^{3b} =\bar{h}^{3b}$, as there are no higher-body interactions present. Using these results, the corresponding expansion for the coupled-cluster BSE kernel is given by the following series of Feynman diagrams (up to second-order)
\begin{widetext}
\begin{gather}
\begin{split}~\label{eq:bse_kernel}
\tilde{\Xi}_{pq,rs}[\tilde{G}] &=\hspace{7.5mm} 
\begin{gathered}
\begin{fmfgraph*}(40,40)
    \fmfset{arrow_len}{3mm}
    \fmfleft{i1}
    \fmfright{o1}
    \fmf{dbl_zigzag}{o1,i1}
    \fmflabel{$\substack{p\\r}$}{i1}
    \fmflabel{$\substack{q\\s}$}{o1}
    \fmfdot{i1,o1}
\end{fmfgraph*}
\end{gathered}\hspace{5mm}+\hspace{5mm} 
\begin{gathered}
\begin{fmfgraph*}(50,50)
    \fmfcurved
    \fmfset{arrow_len}{3mm}
    \fmfleft{i1,i2}
    \fmflabel{}{i1}
    \fmflabel{}{i2}
    \fmfright{o1,o2}
    \fmflabel{}{o1}
    \fmflabel{}{o2}
    \fmf{double_arrow}{i1,i2}
    \fmf{dbl_zigzag}{o1,i1}
    \fmf{dbl_zigzag}{o2,i2}
    \fmf{double_arrow,tension=0}{o2,o1}
    \fmfdot{o1,o2,i1,i2}
    \fmflabel{$r$}{i1}
    \fmflabel{$q$}{o1}
    \fmflabel{$p$}{i2}
    \fmflabel{$s$}{o2}
\end{fmfgraph*}
\end{gathered} \hspace{5mm}+\hspace{7.5mm} 
\begin{gathered}
\begin{fmfgraph*}(50,50)
    \fmfcurved
    \fmfset{arrow_len}{3mm}
    \fmfleft{i1,i2}
    \fmflabel{}{i1}
    \fmflabel{}{i2}
    \fmfright{o1,o2}
    \fmflabel{}{o1}
    \fmflabel{}{o2}
    \fmf{double_arrow}{i1,i2}
    \fmf{dbl_zigzag}{o1,i1}
    \fmf{dbl_zigzag}{o2,i2}
    \fmf{double_arrow,tension=0}{o1,o2}
    \fmfdot{o1,o2,i1,i2}
    \fmflabel{$r$}{i1}
    \fmflabel{$s$}{o1}
    \fmflabel{$p$}{i2}
    \fmflabel{$q$}{o2}
\end{fmfgraph*}
\end{gathered} \hspace{5mm}+\hspace{7.5mm}
\begin{gathered}
\begin{fmfgraph*}(50,50)
    \fmfcurved
    \fmfset{arrow_len}{3mm}
    \fmfleft{i1,i2}
    \fmflabel{}{i1}
    \fmflabel{}{i2}
    \fmfright{o1,o2}
    \fmflabel{}{o1}
    \fmflabel{}{o2}
    \fmf{dbl_zigzag}{o1,i1}
    \fmf{double_arrow,left=0.2,tension=0}{o1,o2}
    \fmf{dbl_zigzag}{o2,i2}
    \fmf{double_arrow,left=0.2,tension=0}{o2,o1}
    \fmfdot{o1,o2,i1,i2}
    \fmflabel{$\substack{q\\r}$}{i1}
    \fmflabel{$\substack{p\\s}$}{i2}
\end{fmfgraph*}
\end{gathered}
\\ 
\\
\\ &+\hspace{7.5mm}
\begin{gathered}\begin{fmfgraph*}(50,50)
    \fmfcurved
    \fmfset{arrow_len}{3mm}
    \fmfleft{i1,i2}
    \fmflabel{}{i1}
    \fmflabel{}{i2}
    \fmfright{o1,o2}
    \fmflabel{}{o1}
    \fmflabel{}{o2}
    \fmf{dbl_dashes}{i1,v1}
    \fmf{dbl_dashes}{v1,o1}
    \fmf{dbl_zigzag}{i2,v2}
    \fmf{double_arrow}{i1,i2}
    \fmf{double_arrow,left=0.3}{v1,v2}
    \fmf{double_arrow,left=0.3}{v2,v1}
    \fmfforce{(0.0w,0.0h)}{i1}
    \fmfforce{(0.0w,1.0h)}{i2}
    \fmfforce{(1.0w,0.0h)}{o1}
    \fmfforce{(0.5w,1.0h)}{v2}
    \fmfforce{(0.5w,0.0h)}{v1}
    \fmfdotn{v}{2}
    \fmfdot{i1,i2}
    \fmfdot{o1}
     \fmflabel{$r$}{i1}
     \fmflabel{$p$}{i2}
    \fmflabel{$\substack{q\\s}$}{o1}
\end{fmfgraph*}
\end{gathered}
\hspace{5mm}+\hspace{7.5mm}
\begin{gathered}\begin{fmfgraph*}(50,50)
    \fmfcurved
    \fmfset{arrow_len}{3mm}
    \fmfleft{i1,i2}
    \fmflabel{}{i1}
    \fmflabel{}{i2}
    \fmfright{o1,o2}
    \fmflabel{}{o1}
    \fmflabel{}{o2}
    \fmf{dbl_dashes}{i2,v2}
    \fmf{dbl_dashes}{v2,o2}
    \fmf{dbl_zigzag}{i1,v1}
    \fmf{double_arrow}{i1,i2}
    \fmf{double_arrow,left=0.3}{v1,v2}
    \fmf{double_arrow,left=0.3}{v2,v1}
    \fmfforce{(0.0w,0.0h)}{i1}
    \fmfforce{(0.0w,1.0h)}{i2}
    \fmfforce{(1.0w,0.0h)}{o1}
    \fmfforce{(0.5w,1.0h)}{v2}
    \fmfforce{(0.5w,0.0h)}{v1}
    \fmfdotn{v}{2}
    \fmfdot{i1,i2}
    \fmfdot{o2}
    \fmflabel{$r$}{i1}
     \fmflabel{$p$}{i2}
    \fmflabel{$\substack{q\\s}$}{o2}
\end{fmfgraph*}
\end{gathered}\hspace{7.5mm} + \hspace{5mm}\begin{gathered}\begin{fmfgraph*}(50,50)
    \fmfcurved
    \fmfset{arrow_len}{3mm}
    \fmfleft{i1,i2}
    \fmflabel{}{i1}
    \fmflabel{}{i2}
    \fmfright{o1,o2}
    \fmflabel{}{o1}
    \fmflabel{}{o2}
    \fmf{dbl_dashes}{i1,v1}
    \fmf{dbl_dashes}{v1,o1}
    \fmf{dbl_zigzag}{o2,v2}
    \fmf{double_arrow}{o1,o2}
    \fmf{double_arrow,left=0.3}{v1,v2}
    \fmf{double_arrow,left=0.3}{v2,v1}
    \fmfforce{(0.0w,0.0h)}{i1}
    \fmfforce{(0.0w,1.0h)}{i2}
    \fmfforce{(1.0w,0.0h)}{o1}
    \fmfforce{(1.0w,1.0h)}{o2}
    \fmfforce{(0.5w,1.0h)}{v2}
    \fmfforce{(0.5w,0.0h)}{v1}
    \fmfdotn{v}{2}
    \fmfdot{i1}
    \fmfdot{o2,o1}
     \fmflabel{$\substack{p\\r}$}{i1}
     \fmflabel{$s$}{o1}
     \fmflabel{$q$}{o2}
\end{fmfgraph*}
\end{gathered}\hspace{7.5mm} + \hspace{5mm}\begin{gathered}\begin{fmfgraph*}(50,50)
    \fmfcurved
    \fmfset{arrow_len}{3mm}
    \fmfleft{i1,i2}
    \fmflabel{}{i1}
    \fmflabel{}{i2}
    \fmfright{o1,o2}
    \fmflabel{}{o1}
    \fmflabel{}{o2}
    \fmf{dbl_dashes}{i1,v1}
    \fmf{dbl_dashes}{v1,o1}
    \fmf{dbl_zigzag}{o2,v2}
    \fmf{double_arrow}{v1,v2}
    \fmf{double_arrow,left=0.3}{o1,o2}
    \fmf{double_arrow,left=0.3}{o2,o1}
    \fmfforce{(0.0w,0.0h)}{i1}
    \fmfforce{(0.0w,1.0h)}{i2}
    \fmfforce{(1.0w,0.0h)}{o1}
    \fmfforce{(1.0w,1.0h)}{o2}
    \fmfforce{(0.5w,1.0h)}{v2}
    \fmfforce{(0.5w,0.0h)}{v1}
    \fmfdotn{v}{2}
    \fmfdot{i1}
    \fmfdot{o2,o1}
     \fmflabel{$\substack{p\\r}$}{i1}
     \fmflabel{$s$}{v1}
     \fmflabel{$q$}{v2}
\end{fmfgraph*}
\end{gathered}\\
\\
\\
&+\hspace{2.5mm}\cdots
\end{split}
\end{gather}
\end{widetext}
We obtain the perturbative expansion, $\tilde{\Xi}[G_0]$, to second-order by inserting Eq.~\ref{eq:2-body} into Eq.~\ref{eq:bse_kernel} while replacing all effective interactions and Green's functions by their reference counterparts: $\tilde{\chi}\to\chi$, $\tilde{G}\to G_0$.
As can be seen from the general diagrammatic series, the CC-BSE kernel reduces to the electronic BSE kernel when $T=0$ and the effects of the similarity transformation are removed. The second, third and fourth diagrams of the first line of Eq.~\ref{eq:bse_kernel} arise by cutting the Green's function lines of the second-order dynamical self-energy diagram of Eq.~\ref{eq:2nd}. The fifth and sixth diagrams of Eq.~\ref{eq:bse_kernel} appear when taking the functional derivative of the two-body effective interaction lines appearing in the second-order self-energy diagram. They are the first-order contributions according to Eq.~\ref{eq:1st_2body}. The final two diagrams of Eq.~\ref{eq:bse_kernel} are those that arise due to derivatives of the static component of the self-energy as outlined in Eq.~\ref{eq:add_cont}.  

We can also take the functional derivative of the vanishing second-order self-energy diagram (see Section~\ref{sec:pert})
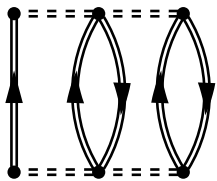
\begin{gather*}
\begin{split}
\begin{gathered}
    \begin{fmfgraph*}(70,50)
    \fmfcurved
    \fmfset{arrow_len}{3mm}
    \fmfleft{i1,i2}
    \fmflabel{}{i1}
    \fmflabel{}{i2}
    \fmfright{o1,o2}
    \fmflabel{}{o1}
    \fmflabel{}{o2}
    \fmf{double_arrow}{i1,i2}
    \fmf{dbl_dashes}{v1,i1}
    \fmf{dbl_dashes}{o1,v1}
    \fmf{double_arrow,left=0.3,tension=0}{o1,o2}
    \fmf{double_arrow,left=0.3,tension=0}{v1,v2}
    \fmf{double_arrow,left=0.3,tension=0}{v2,v1}
    \fmf{dbl_dashes}{v2,i2}
    \fmf{dbl_dashes}{o2,v2}
    \fmf{double_arrow,left=0.3,tension=0}{o2,o1}
    \fmfdot{o1,o2,i1,i2,v1,v2}
\end{fmfgraph*}
\end{gathered}
\end{split}
\end{gather*}
and find that this gives rise to the BSE kernel diagrams
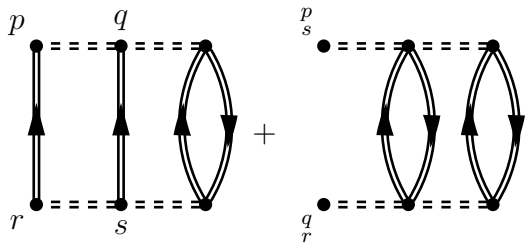
\vspace{2.5mm}
\begin{gather*}
\begin{split}
\begin{gathered}
    \begin{fmfgraph*}(70,50)
    \fmfcurved
    \fmfset{arrow_len}{3mm}
    \fmfleft{i1,i2}
    \fmflabel{}{i1}
    \fmflabel{}{i2}
    \fmfright{o1,o2}
    \fmflabel{}{o1}
    \fmflabel{}{o2}
    \fmf{double_arrow}{i1,i2}
    \fmf{dbl_dashes}{v1,i1}
    \fmf{dbl_dashes}{o1,v1}
    \fmf{double_arrow,left=0.3,tension=0}{o1,o2}
    \fmf{double_arrow}{v1,v2}
    \fmf{dbl_dashes}{v2,i2}
    \fmf{dbl_dashes}{o2,v2}
    \fmf{double_arrow,left=0.3,tension=0}{o2,o1}
    \fmfdot{o1,o2,i1,i2,v1,v2}
    \fmfforce{(0.5w,1.0h)}{v2}
    \fmfforce{(0.5w,0.0h)}{v1}
    \fmflabel{$r$}{i1}
    \fmflabel{$p$}{i2}
     \fmflabel{$s$}{v1}
     \fmflabel{$q$}{v2}
\end{fmfgraph*}
\end{gathered}\hspace{2.5mm}+\hspace{2.5mm}
\begin{gathered}
    \begin{fmfgraph*}(70,50)
    \fmfcurved
    \fmfset{arrow_len}{3mm}
    \fmfleft{i1,i2}
    \fmflabel{}{i1}
    \fmflabel{}{i2}
    \fmfright{o1,o2}
    \fmflabel{}{o1}
    \fmflabel{}{o2}
    \fmf{dbl_dashes}{v1,i1}
    \fmf{dbl_dashes}{o1,v1}
    \fmf{double_arrow,left=0.3,tension=0}{o1,o2}
    \fmf{double_arrow,left=0.3,tension=0}{v1,v2}
    \fmf{double_arrow,left=0.3,tension=0}{v2,v1}
    \fmf{dbl_dashes}{v2,i2}
    \fmf{dbl_dashes}{o2,v2}
    \fmf{double_arrow,left=0.3,tension=0}{o2,o1}
    \fmfdot{o1,o2,i1,i2,v1,v2}
    \fmflabel{$\substack{p\\s}$}{i2}
    \fmflabel{$\substack{q\\r}$}{i1}
\end{fmfgraph*}
\end{gathered}
\end{split}
\end{gather*}
However, as these diagrams have four Green's function lines below the three-body vertex, they also vanish due to the coupled-cluster similarity transformation. 
If we take the functional derivative of the two-body interaction in Eq.~\ref{eq:corr_cont}, at first-order we generate the complementary exchange diagram 
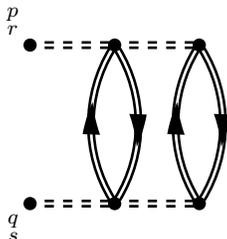
\vspace{1.0mm}
\begin{gather*}
\begin{split}
\begin{gathered}
    \begin{fmfgraph*}(70,50)
    \fmfcurved
    \fmfset{arrow_len}{3mm}
    \fmfleft{i1,i2}
    \fmflabel{}{i1}
    \fmflabel{}{i2}
    \fmfright{o1,o2}
    \fmflabel{}{o1}
    \fmflabel{}{o2}
    \fmf{dbl_dashes}{v1,i1}
    \fmf{dbl_dashes}{o1,v1}
    \fmf{double_arrow,left=0.3,tension=0}{o1,o2}
    \fmf{double_arrow,left=0.3,tension=0}{v1,v2}
    \fmf{double_arrow,left=0.3,tension=0}{v2,v1}
    \fmf{dbl_dashes}{v2,i2}
    \fmf{dbl_dashes}{o2,v2}
    \fmf{double_arrow,left=0.3,tension=0}{o2,o1}
    \fmfdot{o1,o2,i1,i2,v1,v2}
    \fmflabel{$\substack{p\\r}$}{i2}
    \fmflabel{$\substack{q\\s}$}{i1}
\end{fmfgraph*}
\end{gathered}
\end{split}
\end{gather*}
which also vanishes. At first-order, the functional derivative of the three-body effective interaction in Eq.~\ref{eq:corr_cont} gives rise to the diagram
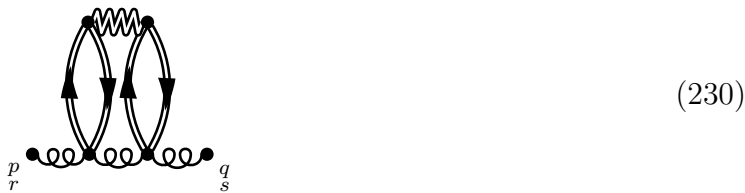
\begin{gather*}
    \begin{split}
        \begin{gathered}
    \begin{fmfgraph*}(50,40)
    \fmfcurved
    \fmfset{arrow_len}{3mm}
    \fmfleft{i1,i2}
    \fmflabel{}{i1}
    \fmflabel{}{i2}
    \fmfright{o1,o2}
    \fmflabel{}{o1}
    \fmflabel{}{o2}
    \fmf{curly}{i1,v1}
    \fmf{curly}{v1,v3}
    \fmf{curly}{v3,o1}
    \fmf{dbl_zigzag}{v4,v5}
    \fmf{double_arrow,left=0.3,tension=0}{v1,v5}
    \fmf{double_arrow,left=0.3,tension=0}{v3,v4}
    \fmf{double_arrow,left=0.3,tension=0}{v4,v3}
    \fmf{double_arrow,left=0.3,tension=0}{v5,v1}
    \fmfdot{o1,i1,v1,v3,v5,v4}
    \fmfforce{(-0.1w,0.0h)}{i1}
    \fmfforce{(1.0w,0.0h)}{o1}
    \fmfforce{(0.258w,1h)}{v2}
    \fmfforce{(0.258w,0.0h)}{v1}
    \fmfforce{(0.625w,0.0h)}{v3}
    \fmfforce{(0.625w,1.0h)}{v4}
    \fmfforce{(0.0w,1.0h)}{i2}
    \fmfforce{(1.0w,1.0h)}{o2}
    \fmfforce{(0.25w,1.0h)}{v5}
    \fmflabel{$\substack{p\\r}$}{i1}
    \fmflabel{$\substack{q\\s}$}{o1}
\end{fmfgraph*}
\end{gathered}
    \end{split}
\end{gather*}
which explicitly contains the 4-body effective interaction. Depending on the ordering of the external indices, this diagram may not vanish and will give a non-zero contribution to the CC-BSE kernel. 

From our analysis, we see that the diagrams of the Bethe-Salpeter kernel generated by an $N$-body interaction result from functional derivatives of different components of the self-energy. We note that the corresponding expression for the 6-point vertex function is obtained by the functional derivative of the BSE kernel with respect to the single-particle Green's function, $\tilde{\Lambda}^{6\pt}\sim i\frac{\delta\tilde{\Xi}}{\delta\tilde{G}}$, and so on for the higher-order vertices. These relationships arise as the 6-point Green's function is generated from the functional derivative of the 4-point Green's function and so on. 

Using the general analysis presented in this work, we can generate the diagrammatic series for the coupled-cluster BSE kernel order-by-order. When restricted to the Tamm-Dancoff approximation, the perturbative expansion of the coupled-cluster BSE kernel derived here is closely related to the EE-EOM-CC eigenvalue problem. To conclude this section, we write the terms appearing in the second-order perturbative expansion of the CC-BSE kernel as
\begin{widetext}
    \begin{gather}
    \begin{split}~\label{eq:bse_2nd}
        \tilde{\Xi}^{(2)}_{pq,rs}(\omega) &= \chi_{pq,rs} + \sum_{kc}\frac{\chi_{pqc,rsk}\tilde{\Sigma}^{\infty(0)}_{kc}}{\epsilon_{k}-\epsilon_{c}} + \frac{1}{(2!)^2}\sum_{abij}\frac{\chi_{pqab,rsij}\chi_{ij,ab}}{\epsilon_{i}-\epsilon_{j}-\epsilon_{a}-\epsilon_{b}}+\sum_{kc}\frac{\chi_{pk,sc}\chi_{cq,rk}}{\omega-(\epsilon_{c}-\epsilon_{k})+i\eta}\\
        &+\sum_{kc}\frac{\chi_{pc,ks}\chi_{kq,rc}}{\omega-(\epsilon_{k}-\epsilon_{c})-i\eta}-\frac{1}{2}\sum_{kl}\frac{\chi_{pq,kl}\chi_{kl,rs}}{\omega-(\epsilon_{k}+\epsilon_{l})-i\eta}+\frac{1}{2}\sum_{cd}\frac{\chi_{pq,cd}\chi_{cd,rs}}{\omega-(\epsilon_{c}+\epsilon_{d})+i\eta}\\
        &+\frac{1}{2}\sum_{abk}\frac{(\chi_{abq,rks}\chi_{pk,ab}+\chi_{pkq,abs}\chi_{ab,rk}+\chi_{pab,rks}\chi_{kq,ab}+\chi_{pab,rsk}\chi_{qk,ab})}{\omega+\epsilon_{k}-\epsilon_{a}-\epsilon_{b}+i\eta}\\
        &+\frac{1}{2}\sum_{kla}\frac{(\chi_{klq,ras}\chi_{pa,kl}+\chi_{paq,kls}\chi_{kl,ra}+\chi_{pkl,ras}\chi_{aq,kl}+\chi_{pkl,rsa}\chi_{qa,kl})}{\omega+\epsilon_{a}-\epsilon_{k}-\epsilon_{l}-i\eta} \ .
        \end{split}
\end{gather}
\end{widetext}
Here, the first three terms arise due to the perturbative expansion of the two-body effective interaction (see Eq.~\ref{eq:2-body}). The remaining terms correspond to the remaining diagrams of Eq.~\ref{eq:bse_kernel} where the Green's function and interaction lines are replaced by $G_0$ and $\chi$ equivalents. We clearly see the similarity between the fourth, fifth, sixth and seventh terms of Eq.~\ref{eq:bse_2nd} and those obtained within the second-order perturbative approximation for the electronic BSE kernel~\cite{zhang2013dynamical,sangalli2011double,rebolini2016range,monino2023connections}. Using Eq.~\ref{eq:bse_2nd} for the CC-BSE kernel, we can formulate closed equations for the particle-hole response function which give access to neutral excitation energies.

\section{Conclusions and Outlook}~\label{sec:conclusions}

In summary, we have presented the diagrammatic theory of the irreducible self-energy and Bethe-Salpeter kernel that is generated by a general $N$-body hermitian or non-hermitian interaction Hamiltonian. We have centered our analysis around the non-hermitian coupled-cluster similarity transformed Hamiltonian.

To generate a consistent functional-diagrammatic theory of the coupled-cluster self-energy, we introduce the single-particle Green's function  $\tilde{G}$, of a general $N$-body non-hermitian interaction within the biorthogonal quantum theory. We extend the GML theorem to include adiabatic generation of the right and left ground states of a psuedo-hermitian Hamiltonian, which leads to the perturbative expansion of the non-hermitian single-particle Green's function. Using our extended GML theorem, we define the non-hermitian Dyson equation and generate the perturbative expansion for the coupled-cluster self-energy, $\tilde{\Sigma}[G_0]$. From the exact equation-of-motion for the single-particle CC Green's function we reveal the structure of the self-consistent renormalized coupled-cluster self-energy, $\tilde{\Sigma}[\tilde{G}]$ and demonstrate how the perturbative self-energy is generated from this functional. The analysis presented in this work clearly demonstrates the emergence of different effective interactions that appear when expanding the self-energy with respect to either the non-interacting or the exact Green's function. These interactions correspond to those appearing when normal-ordering the Hamiltonian with respect to the reference or exact ground state. Further, we show how the electronic self-energy emerges when the similarity transformation of the electronic structure Hamiltonian is removed. 

We subsequently derive the coupled-cluster Dyson supermatrix from the exact, analytic spectral representation of the coupled-cluster self-energy and present several approximate expressions for the coupled-cluster self-energy that are restricted to the space of 2p1h/2h1p excitations. In the process, we have firmly demonstrated the relationship and connections between the coupled-cluster self-energy, IP/EA-EOM-CC theory and the $G_0W_0$ approximation. This analysis lead us to introduce CC-$G_0W_0$ theory which combines aspects of the $G_0W_0$ approximation with the coupled-cluster self-energy presented in Section~\ref{sec:cc_se}. This is a novel approximation that will be explored in future work. Our formulation also allows us to explore approximations beyond $G_0W_0$ theory by using the full CCSD singles and doubles excitation and de-excitation amplitudes in place of the rCCD amplitudes. Use of the full CCSD amplitudes goes beyond RPA screening by including the complete set of interactions, $\braket{\Phi^{b}_{j}|\bar{H}_N|\Phi^{a}_{i}}$, between singly excited configurations~\cite{coveney2023coupled}. 

Finally, we derive the coupled-cluster BSE kernel in terms of the functional derivative of the coupled-cluster self-energy with respect to the single-particle coupled-cluster Green's function. We demonstrate that diagrams of the CC-BSE kernel are generated both by cutting Green's function lines in self-energy diagrams whilst also taking derivatives of the interaction vertices. These diagrams are closely related to the terms that arise within EE-EOM-CC theory once further approximations are made in defining an effective kernel that depends only on a single frequency~\cite{rebolini2016range,sangalli2011double,strinati1988application}.

Of particular interest in the context of non-hermitian quantum systems are $\mathcal{PT}$-symmetric Hamiltonians. However, in certain parameter regimes, not every eigenstate is also $\mathcal{PT}$-symmetric with a corresponding real eigenvalue solution~\cite{meden2023mathcal,zhang2025nonadiabatic}. It has been well established that a $\mathcal{PT}$-broken phase can exist where eigenstates take on complex eigenvalues~\cite{bender1998real,bender2002complex,mostafazadeh2002pseudo,brody2013biorthogonal}. The Green's function formalism presented here is rooted in the biorthogonal theory whereby the left and right eigenstates of the Hamiltonian are dealt with explicitly. As a result, our approach is not concerned with the construction of a $\mathcal{PT}$-symmetric metric for the Hilbert space. Therefore, our formalism allows for the calculation of the full spectra of a generic non-hermitian Hamiltonian, including complex eigenvalues corresponding to a $\mathcal{PT}$-broken phase. This is also supported by the fact that the CC similarity transformed Hamiltonian does not possess $\mathcal{PT}$-symmetry. In fact, the presence of complex solutions can occur in truncated CC theories~\cite{kohn2007can,kjonstad2017resolving}, although their interpretation in the context of CC theory indicates a deficiency in the similarity-transformation rather than the presence a $\mathcal{PT}$-broken phase.

Our formalism provides the rigorous theoretical foundation for many possible implementations as a result of different approximations of the coupled-cluster amplitude equations. A potential further application of the formalism presented in this work would be to generate the self-energy and BSE kernel for the transcorrelated Hamiltonian~\cite{handy1969energies,tsuneyuki2008transcorrelated,mcardle2020improving,lee2023studies}. The theory presented in this work provides a unified and complete description of the seemingly disconnected approaches of coupled-cluster theory and the Green’s function formalism. Our findings should form the basis for improved formulations of Green’s function theory for systems of many interacting particles.


\appendix

\section{Proof of extended Gell-Mann and Low theorem}~\label{app_gm}

From the `adiabatic switiching on' procedure, we have the time-evolution operator in the interaction picture as 
\begin{gather}
    \begin{split}
        \bar{U}_{\eta}(0,-\infty) = \mathcal{T}\left\{\exp\left(-i\int^{0}_{-\infty}dt\ \Bar{H}_{1}(t)e^{-\eta|t|}\right)\right\}
    \end{split}
\end{gather}
If the limit
\begin{gather}
    \begin{split}~\label{eq:cov_tew_theorem}
        \lim_{\eta\to0} \frac{\bar{U}_{\eta}(0,-\infty)\ket{\Phi_0}}{\braket{\Tilde{\Phi}_0|\bar{U}_{\eta}(0,-\infty)|\Phi_0}} \equiv \frac{\ket{\Psi_0}}{\braket{\Tilde{\Phi}_0|\Psi_0}}
    \end{split}
\end{gather}
exits then:
\begin{gather}
    \begin{split}
        \Bar{H}\frac{\ket{\Psi_0}}{\braket{\Tilde{\Phi}_0|\Psi_0}} = E\frac{\ket{\Psi_0}}{\braket{\Tilde{\Phi}_0|\Psi_0}} \ .
    \end{split}
\end{gather}
Therefore, the quantity $\frac{\ket{\Psi_0}}{\braket{\Tilde{\Phi}_0|\Psi_0}}$ is an eigenstate of the full Hamiltonian, $\bar{H}$. 

\textbf{Proof:} Consider the quantity 
\begin{gather}
\begin{split}~\label{eq:comm}
    (\Bar{H}_0-E^{(0)})\ket{\Psi_0(\eta)} &= (\Bar{H}_0-E^{(0)})\bar{U}_{\eta}(0,-\infty)\ket{\Phi_0}\\
    &= \left[\bar{H}_0,\bar{U}_{\eta}(0,-\infty)\right]\ket{\Phi_0} \ .
\end{split}
\end{gather}
In the biorthogonal Interaction picture, we have that 
\begin{equation}
    i\frac{\partial}{\partial t}\Bar{H}_1(t) = \left[\Bar{H}_1(t),\Bar{H}_0\right]
\end{equation}
Evaluating the commutators arising in Eq.~\ref{eq:comm}, we have 
\begin{widetext}
    \begin{gather}
    \begin{split}
        (\Bar{H}_0-E^{(0)})\ket{\Psi_0(\eta)} = -\sum_{n=1}^{\infty} \frac{(-i)^{n-1}}{n!} \int^{0}_{-\infty} dt_1 \cdots& \int^{0}_{-\infty} dt_n e^{-\eta|t_1+...+t_n|}\left(\sum^{n}_{i=1}\frac{\partial}{\partial t_i}\right)\mathcal{T}\left\{\Bar{H}_1(t_1)...\Bar{H}_1(t_n)\right\}\ket{\Phi_0}
    \end{split}
\end{gather}
\end{widetext}
Performing integration by parts with respect to $t_1$, we have 
\begin{gather}
    \begin{split}
        (\Bar{H}_0-E^{(0)})\ket{\Psi_0(\eta)} = - \Bar{H}_1\ket{\Psi_0(\eta)} +i\eta g\frac{\partial}{\partial g}\ket{\Psi_0(\eta)} 
    \end{split}
\end{gather}
where we have assumed that the interaction $\Bar{H}_1$ is proportional to an arbitrary coupling constant, $g$, which allows us to identify~\cite{gell1954quantum,Quantum} 
\begin{gather}
    \begin{split}
        \frac{(-i)^{n-1}}{(n-1)!}g^{n} = ig\frac{\partial}{\partial g}\left(\frac{(-i)^{n}}{n!}g^{n}\right) \ .
    \end{split}
\end{gather}
Therefore, we have
\begin{equation}~\label{eq:app_1}
    (\Bar{H}-E^{(0)})\ket{\Psi_0(\eta)} = i\eta g\frac{\partial}{\partial g}\ket{\Psi_0(\eta)} 
\end{equation}
and projecting on both sides with $\frac{\bra{\Tilde{\Phi}_0}}{\braket{\Tilde{\Phi}_0|\Psi_0(\eta)}}$, we get 
\begin{equation}~\label{eq:app_2}
    \frac{\braket{\Tilde{\Phi}_0|\Bar{H}_1|\Psi_0(\eta)}}{\braket{\Tilde{\Phi}_0|\Psi_0(\eta)}} = i\eta g\frac{\partial}{\partial g}\Big(\log\braket{\Tilde{\Phi}_0|\Psi_0(\eta)} \Big) = \Delta E(\eta) \ .
\end{equation}
However, we also have that 
\begin{gather}
    \begin{split}
        (\Bar{H}-E^{(0)})\frac{\ket{\Psi_0(\eta)}}{\braket{\Tilde{\Phi}_0|\Psi_0(\eta)}} = \frac{i\eta g}{\braket{\Tilde{\Phi}_0|\Psi_0(\eta)}}\frac{\partial}{\partial g}\ket{\Psi_0(\eta)}
    \end{split}
\end{gather}
by multiplying Eq~\ref{eq:app_1} by $\frac{1}{\braket{\Tilde{\Phi}_0|\Psi_0(\eta)}}$. This equation can be rearranged to give 
\begin{gather}
    \begin{split}
        &\left(\Bar{H}-E^{(0)}-i\eta g\frac{\partial}{\partial g}\right)\frac{\ket{\Psi_0(\eta)}}{\braket{\Tilde{\Phi}_0|\Psi_0(\eta)}} =\\
        &\frac{\ket{\Psi_0(\eta)}}{\braket{\Tilde{\Phi}_0|\Psi_0(\eta)}}\times\left(i\eta g\frac{\partial}{\partial g} \log\braket{\Tilde{\Phi}_0|\Psi_0(\eta)} \right)
    \end{split}
\end{gather}
However, by Eq.~\ref{eq:app_2}, we may rearrange the expression to give 
\begin{gather}
    \begin{split}
         \left (\bar{H}-E^{(0)}-\Delta E(\eta)\right)&\frac{\ket{\Psi_0(\eta)}}{\braket{\Tilde{\Phi}_0|\Psi_0(\eta)}} = i\eta g\frac{\partial}{\partial g}\frac{\ket{\Psi_0(\eta)}}{\braket{\Tilde{\Phi}_0|\Psi_0(\eta)}}
    \end{split}
\end{gather}
Now, we are in a position to take the limit as $\eta\to0$. By assumption, the right hand side is finite to all orders in perturbation theory. Therefore, the multiplication by $\eta$ as $\eta\to0$ sends the right hand side to zero. As a result, we have 
\begin{equation}
    \left(\Bar{H}-E\right)\lim_{\eta\to0}\frac{\ket{\Psi_0(\eta)}}{\braket{\Tilde{\Phi}_0|\Psi_0(\eta)}} = 0
\end{equation}
and hence we have the extended Gell-Mann and Low theorem. By analogy, we also have
\begin{gather}
    \begin{split}
        \left(\Bar{H}^\dag-E\right)\lim_{\eta\to0}\frac{\ket{\Tilde{\Psi}_0(\eta)}}{\braket{\Phi_0|\Tilde{\Psi}_0(\eta)}} = 0 
    \end{split}
\end{gather}
for the left eigenstate. As a result, the ground state energy shift is given by 
\begin{equation}
    \Delta E = \frac{\braket{\tilde{\Phi}_0|\Bar{H}_1|\Psi_0}}{\braket{\tilde{\Phi}_0|\Psi_0}} = \frac{\braket{\Phi_0|\Bar{H}^\dag_1|\Tilde{\Psi}_0}}{\braket{\Phi_0|\Tilde{\Psi}_0}} \ ,
\end{equation}
where $\Delta E = E-E^{(0)}$.

\begin{widetext}
\section{Normal-orderings of the similarity transformed Hamiltonian}~\label{appendix_2}

The similarity transformed Hamiltonian is written as 
\begin{gather}
    \begin{split}
        \Bar{H} = e^{-T}He^{T} &= \sum_{pq} \bar{h}_{pq} a^\dag_pa_q + \frac{1}{(2!)^2}\sum_{pq,rs}\bar{h}_{pq,rs} a^\dag_pa^\dag_qa_sa_r + \frac{1}{(3!)^2}\sum_{pqr,stu}\bar{h}_{pqr,stu} a^\dag_pa^\dag_qa^\dag_ra_ua_ta_s + \cdots
    \end{split}
\end{gather}
Normal-ordering this Hamiltonian with respect to the reference determinant gives the expression 
\begin{gather}
    \begin{split}
        \Bar{H} &= \braket{\Phi_0|\bar{H}|\Phi_0} + \sum_{pq}F_{pq} \left\{a^\dag_pa_q\right\}_0 + \frac{1}{(2!)^2}\sum_{pq,rs}\chi_{pq,rs} \left\{a^\dag_pa^\dag_qa_sa_r\right\}_0 + \frac{1}{(3!)^2}\sum_{pqr,stu} \chi_{pqr,stu} \left\{a^\dag_pa^\dag_qa^\dag_ra_ua_ta_s\right\}_0 + \cdots
    \end{split}
\end{gather}
where the effective interaction matrix elements are given by~\cite{coveney2023coupled} 
\begin{subequations}
    \begin{align}
        \begin{split}
            F_{pq} &= \bar{h}_{pq} -i \sum_{rs} \bar{h}_{pr,qs}G^{0}_{sr}(t-t^+) + \frac{i}{(2!)^2}\sum_{rs,tu} \bar{h}_{prs,qtu}G^{2\p\h,(0)}_{tu,rs}(t-t^+) + \cdots
        \end{split}\\
        \begin{split}
            \chi_{pq,rs} &= \bar{h}_{pq,rs} -i\sum_{tu} \bar{h}_{pqt,rsu}G^0_{ut}(t-t^+) + \frac{i}{(2!)^2}\sum_{tuwv}\bar{h}_{pqtu,rswv}G^{2\p\h,(0)}_{wv,tu}(t-t^+) + \cdots
        \end{split}
    \end{align}
\end{subequations}
and so on for the three-body and higher effective interactions. Here, $-iG^{0}_{pq}(t-t^+)=\braket{\Phi_0|a^\dag_{q}a_{p}|\Phi_0}$ and $iG^{2\p\h(0)}_{pq,rs}(t-t^{+}) =\braket{\Phi_0|a^\dag_{r}a^\dag_s a_q a_p|\Phi_0}$ are the reference one- and two-particle reduced density matrices, respectively. The higher-order contributions are given by contracting higher-body matrix elements with reference reduced density matrices. However, we can also write the similarity transformed Hamiltonian as 
\begin{gather}
    \begin{split}
        \Bar{H} &= \braket{\tilde{\Psi}_0|\bar{H}|\Phi_0} + \sum_{pq}\Tilde{F}_{pq} \left\{a^\dag_pa_q\right\} + \frac{1}{(2!)^2}\sum_{pq,rs}\Tilde{\Xi}_{pq,rs} \left\{a^\dag_pa^\dag_qa_sa_r\right\} + \frac{1}{(3!)^2}\sum_{pqr,stu} \Tilde{\chi}_{pqr,stu} \left\{a^\dag_pa^\dag_qa^\dag_ra_ua_ta_s\right\}
        + \cdots 
    \end{split}
\end{gather}
where $E^{\CC}_0=\braket{\tilde{\Psi}_0|\bar{H}|\Phi_0}=\braket{\Phi_0|\bar{H}|\Phi_0}$ from the coupled-cluster Lagrangian. The normal-ordered effective interactions become
\begin{subequations}
    \begin{align}
        \begin{split}
            \Tilde{F}_{pq} &= \bar{h}_{pq} -i \sum_{rs} \bar{h}_{pr,qs}\tilde{G}_{sr}(t-t^+) + \frac{i}{(2!)^2}\sum_{rs,tu} \bar{h}_{prs,qtu}\tilde{G}^{2\p\h}_{tu,rs}(t-t^+) + \cdots\\
            &= F_{pq} + \sum_{ai} \chi_{pa,qi}\lambda^{i}_a + \frac{1}{(2!)^2}\sum_{ijab} \chi_{pab,qij}\lambda^{ij}_{ab} + \cdots
        \end{split}\\
        \begin{split}
            \Tilde{\Xi}_{pq,rs} &= \bar{h}_{pq,rs} -i \sum_{tu} \bar{h}_{pqt,rsu}\tilde{G}_{ut}(t-t^+)+ \frac{i}{(2!)^2}\sum_{tuwv}\chi_{pqtu,rswv}\tilde{G}^{2\p\h}_{wv,tu}(t-t^+) + \cdots\\
            &=\chi_{pq,rs} + \sum_{ia} \chi_{pqa,rsi}\lambda^{i}_{a} + \frac{1}{(2!)^2}\sum_{ijab} \chi_{pqab,rsij}\lambda^{ij}_{ab} + \cdots
        \end{split}
    \end{align}
    and so on for the three-body and higher effective interactions. Here, $-i\tilde{G}_{pq}(t-t^+)=\braket{\tilde{\Psi}_0|a^\dag_{q}a_{p}|\Phi_0}$ and $i\tilde{G}^{2\p\h}_{pq,rs}(t-t^{+}) =\braket{\tilde{\Psi}_0|a^\dag_{r}a^\dag_s a_q a_p|\Phi_0}$ are the biorthogonal one- and two-particle reduced density matrices, respectively. The higher-order contributions are given by contracting higher-body matrix elements with biorthogonal reduced density matrices. 
\end{subequations}

\section{Relationship between the BSE kernel and the 4-point vertex function}~\label{app:4_vertex}

The BSE kernel $\tilde{\Xi}$, and the 4-point vertex function $\tilde{\Lambda}^{4\pt}$, are closely related. From Eq.~\ref{eq:4point_vert} for the 4-point vertex function, in the real-time domain, we have 
\begin{gather}
    \begin{split}
        \tilde{G}^{4\pt}_{pq,rs}(t_1,t_2;t_3,t_4) &= i\tilde{G}_{pr}(t_1,t_3)\tilde{G}_{qs}(t_2,t_4) -i\tilde{G}_{ps}(t_1,t_4)\tilde{G}_{qr}(t_2,t_3) \\
        &-\sum_{tuvw}\int dtdt'dt''dt''' \tilde{G}_{pt}(t_1,t')\tilde{G}_{qu}(t_2,t'')\tilde{\Lambda}^{4\pt}_{tu,vw}(t',t''';t,t'')\tilde{G}_{vr}(t,t_3)\tilde{G}_{ws}(t''',t_4) \ .
    \end{split}
\end{gather}
This equation can be rewritten using the definition of the 4-point response function $\tilde{L}$ and its non-interacting counterpart $\tilde{L}_0$, as 
\begin{gather}
    \begin{split}~\label{eq:4-point_BSE}
        \tilde{L}_{pq,rs}(t_1,t_2;t_3,t_4) &= \tilde{L}^{0}_{pq,rs}(t_1,t_2;t_3,t_4) \\
        &+\sum_{tuvw}\int dtdt'dt''dt''' \tilde{L}^0_{pv,rt}(t_1,t;t_3,t')\tilde{\Lambda}^{4\pt}_{tu,vw}(t',t'';t,t''')\tilde{L}^{0}_{wq,us}(t''',t_2;t'',t_4) \ .
    \end{split}
\end{gather}
Comparison of Eq.~\ref{eq:4-point_BSE} with Eq.~\ref{eq:BSE}, we see that the 4-point vertex function is the reducible BSE kernel. With this identification, we can immediately write the relationship between the BSE kernel and 4-point vertex as 
\begin{gather}
\begin{split}
    \tilde{\Lambda}^{4\pt}_{pq,rs}(t_1,t_2;t_3,t_4) &= \tilde{\Xi}_{pq,rs}(t_1,t_2;t_3,t_4) \\
    &+ \sum_{tuvw}\int dtdt'dt''dt'''  \tilde{\Xi}_{pv,rt}(t_1,t;t_3,t') \tilde{L}^0_{tu,vw}(t',t'';t,t''')\tilde{\Lambda}^{4\pt}_{wq,us}(t''',t_2;t'',t_4) \ .
\end{split}
\end{gather}
Using this relationship, we have 
\begin{gather}
    \begin{split}
        \tilde{L} &= \tilde{L}_0 + \tilde{L}_0\tilde{\Lambda}^{4\pt}\tilde{L}_0\\
          &= \tilde{L}_0 + \tilde{L}_0(\tilde{\Xi}+\tilde{\Xi} \tilde{L}_0\tilde{\Lambda}^{4\pt})\tilde{L}_0\\
          &= \tilde{L}_0 + \tilde{L}_0\tilde{\Xi} \tilde{L} 
    \end{split}
\end{gather}
which is the Bethe-Salpeter equation (Eq.~\ref{eq:BSE}). Since the 4-point vertex is the reducible kernel of the BSE, it consists of 1PI diagrams and hence appears in the self-consistent expansion for the self-energy. As a result, the 4-point vertex can be viewed as the functional derivative of $\tilde{\Sigma}$ with respect to $\tilde{G}$: $\tilde{\Lambda}^{4\pt}\sim i\frac{\delta\tilde{\Sigma}}{\delta\tilde{G}}$. The irreducible BSE kernel $\tilde{\Xi}$, consists only of 2PI diagrams as the BSE performs the geometric series summation of these terms.

\section{Expansion of interaction-irreducible vertices}~\label{app_pert_int}

Here, we present the diagrammatic expansion of the interaction-irreducible one-body vertex. The derivation for higher-order interaction vertices is analogous. The one-body interaction-irreducible vertex is given by 
\begin{gather}
\begin{split}~\label{eq:eff_1}
    \tilde{F}_{pq} &= \bar{h}_{pq} -i \sum_{rs} \bar{h}_{pr,qs}\tilde{G}_{sr}(t-t^+) + \frac{i}{(2!)^2} \sum_{rstu}\bar{h}_{prs,qtu}\tilde{G}^{2\p\h}_{tu,rs}(t-t^+) + \cdots
\end{split}
\end{gather}
Using the Dyson equation for the single-particle Green's function
\begin{gather}
    \begin{split}
        &\tilde{G}_{pq}(t-t') = G^{0}_{pq}(t-t') + \int dt_1dt_2\sum_{vw} G^0_{pv}(t-t_1)\tilde{\Sigma}_{vw}(t_1-t_2)\tilde{G}_{wq}(t_2-t')
    \end{split}
\end{gather}
and expanding the equal time two-particle Green's function as the antisymmetrized sum of the first two terms from independent-particle propagation
\begin{gather}
    \begin{split}~\label{eq:2p_GF}
        \Tilde{G}^{2\p\h}_{tu,rs}(t-t^+) &\approx i\tilde{G}_{tr}(t-t^+)\tilde{G}_{us}(t-t^+)-i\tilde{G}_{ts}(t-t^+)\tilde{G}_{ur}(t-t^+) , 
    \end{split}
\end{gather}
we insert the Dyson equation into Eqs~\ref{eq:eff_1} and~\ref{eq:2p_GF} to obtain 
\begin{gather}
\begin{split}~\label{eq:eff_expand}
    \tilde{F}_{pq} = F_{pq} -i \sum_{rt} &\left(\bar{h}_{pr,qt}-i\sum_{su}  \bar{h}_{prs,qtu}G^{0}_{us}(t-t^+)+ \frac{i}{(2!)^2}\sum_{suw\sigma}\bar{h}_{prsu,qtw\sigma}G^{2\p\h(0)}_{w\sigma,su}(t-t^+)+ ...\right)\\
    &\times\sum_{vw}\int dt_1 dt_2 G^0_{tv}(t-t_1)\Sigma_{vw}(t_1-t_2)\tilde{G}_{wr}(t_2-t^+)
    + \cdots
\end{split}
\end{gather}
Now, the expression in brackets is exactly that of the effective interaction $\chi_{pr,qt}$ and so this expression simplifies to 
\begin{gather}
\begin{split}
     &\tilde{F}_{pq} = F_{pq} -i \sum_{rtvw} \chi_{pr,qt}\times\int dt_1 dt_2 G^0_{tv}(t-t_1)\tilde{\Sigma}_{vw}(t_1-t_2)\tilde{G}_{wr}(t_2-t^+)+\cdots
\end{split}
\end{gather}
By further perturbative expansion of the higher-body Green's functions, we arrive at the expression
\begin{gather}
\begin{split}
     \tilde{\chi}_{pq} &= \chi_{pq} -i \sum_{rtvw} \chi_{pr,qt}\times\int dt_1 dt_2 G^0_{tv}(t-t_1)\tilde{\Sigma}_{vw}(t_1-t_2)\tilde{G}_{wr}(t_2-t^+)\\
     &+\frac{i}{(2!)^2}\sum_{\substack{rstu\\vwlo}}\chi_{prs,qtu}\times\int dt_1 dt_2\tilde{G}_{tl}(t_1-t_2)\tilde{G}_{no}(t_1-t_2)\tilde{\Lambda}^{4\pt}_{lo,vw}(t_1,t_1;t_2,t_2)\tilde{G}_{vr}(t_2-t_1)\tilde{G}_{ws}(t_2-t_1) \\
     &+ \cdots
\end{split}
\end{gather}
Perturbative expansion of the Green's function $\tilde{G}\approx G_0$, the self-energy $\tilde{\Sigma}\approx\tilde{\Sigma}^{\infty(0)}$ and the 4-point vertex function as $\tilde{\Lambda}^{4\pt(0)}_{pq,rs}\approx\chi_{pq,rs}$, allows us to write this expression diagrammatically
\begin{gather}
    \begin{split}~\label{eq:1-body}
    \begin{gathered}
        \begin{fmfgraph*}(40,40)
    \fmfset{arrow_len}{3mm}
    \fmfleft{i1}
    \fmfright{o1}
    \fmf{zigzag}{o1,i1}
    \fmfv{decor.shape=cross,decor.filled=full, decor.size=1.5thic}{o1}
    \fmfdot{i1}
\end{fmfgraph*}
\end{gathered}\hspace{2.5mm} = \hspace{2.5mm} 
\begin{gathered}
\begin{fmfgraph*}(40,40)
    \fmfset{arrow_len}{3mm}
    \fmfleft{i1}
    \fmfright{o1}
    \fmf{dbl_dashes}{o1,i1}
    \fmfv{decor.shape=cross,decor.filled=full, decor.size=1.5thic}{o1}
    \fmfdot{i1}
\end{fmfgraph*}
\end{gathered}\hspace{2.5mm}+\hspace{2.5mm}
\begin{gathered}
    \begin{fmfgraph*}(40,40)
    \fmfcurved
    \fmfset{arrow_len}{3mm}
    \fmfleft{i1,i2}
    \fmflabel{}{i1}
    \fmflabel{}{i2}
    \fmfright{o1,o2}
    \fmflabel{}{o1}
    \fmflabel{}{o2}
    \fmf{dbl_wiggly}{i1,o1}
    \fmf{fermion,left=0.3,tension=0}{o1,o2}
    \fmf{dbl_dashes}{o2,i2}
    \fmf{fermion,left=0.3,tension=0}{o2,o1}
    \fmfdot{o1,o2,i1}
    \fmfv{decor.shape=cross,decor.size=1thic, decor.filled=full}{i2}
\end{fmfgraph*}
\end{gathered}\hspace{5mm}+\hspace{2.5mm}
\begin{gathered}
    \begin{fmfgraph*}(60,40)
    \fmfcurved
    \fmfset{arrow_len}{3mm}
    \fmfleft{i1,i2}
    \fmflabel{}{i1}
    \fmflabel{}{i2}
    \fmfright{o1,o2}
    \fmflabel{}{o1}
    \fmflabel{}{o2}
    \fmf{dashes}{o1,v1}
    \fmf{dashes}{i1,v1}
    \fmf{fermion,left=0.3,tension=0}{o1,o2}
    \fmf{fermion,left=0.3,tension=0}{v1,v2}
    \fmf{fermion,left=0.3,tension=0}{v2,v1}
    \fmf{phantom}{v2,i2}
    \fmf{dbl_wiggly}{o2,v2}
    \fmf{fermion,left=0.3,tension=0}{o2,o1}
    \fmfdot{o1,o2,i1,v1,v2}
\end{fmfgraph*}
\end{gathered}\hspace{2.5mm}+\hspace{2.5mm} \cdots
    \end{split}
\end{gather}
This is exactly the diagrammatic series presented in Section~\ref{sec:cc_se} when recovering the perturbative expansion of the coupled-cluster self-energy. The derivation for the expansion of the two-body effective interaction is entirely analogous. 

\end{widetext}
\section{Formal properties of the CC Dyson supermatrix and CC self-energy}~\label{app:psd}

The CC quasiparticle Hamiltonian (Eq.~\ref{eq:cc_qp}) and CC Dyson supermatrix (Eq.~\ref{eq:super}) are non-hermitian but possess the same \emph{real} spectrum. As first demonstrated in Ref.~\cite{mostafazadeh2002pseudo}, a diagonalizable non-hermitian matrix possesses a \emph{real} spectrum \emph{if and only if} it is pseudo-hermitian with respect to a linear hermitian automorphism (metric) of the form 
\begin{gather}
    \begin{split}
        \mathbf{\tau} = \mathbf{A}^\dag \mathbf{A} \ .
    \end{split}
\end{gather}
This results in a biorthogonal formulation of quantum mechanics whereby the relationship between an arbitrary vector and its associated vector is given by~\cite{brody2013biorthogonal}
\begin{gather}
    \begin{split}
        \mathbf{\tilde{v}} = \tau\mathbf{v} \ . 
    \end{split}
\end{gather}
This definition requires the CC Dyson supermatrix to be pseudo-hermitian:
\begin{gather}
    \begin{split}
        (\mathbf{\tilde{D}}^{\text{CC}})^\dag = \tau\mathbf{\tilde{D}}^{\text{CC}}\tau^{-1} \ . 
    \end{split}
\end{gather}
It can also be shown that a diagonalizable matrix possesses a \emph{real} spectrum \emph{if and only if} there is a pseudo-canonical transformation that maps the matrix into a hermitian one~\cite{mostafazadeh2002pseudo}. This requires the CC and electronic Dyson supermatrices to be related by the similarity transformation
\begin{gather}
    \begin{split}
        \mathbf{D}^{\text{el}} = \mathbf{A}\mathbf{\tilde{D}}^{\text{CC}}\mathbf{A}^{-1} \ ,
    \end{split}
\end{gather}
such that $\mathbf{D}^{\text{el}}=(\mathbf{D}^{\text{el}})^\dag$. Therefore, we can show that for an arbitrary state $\mathbf{y}$, we have 
\begin{gather}
    \begin{split}
        \mathbf{y}^\dag\mathbf{D}^{\text{el}}\mathbf{y} = \mathbf{y}^\dag\mathbf{A}\mathbf{\tilde{D}}^{\text{CC}}\mathbf{A}^{-1}\mathbf{y} \ .
    \end{split}
\end{gather}
Now, defining the arbitrary state $\mathbf{v} = \mathbf{A}^{-1}\mathbf{y}$, we can write its associated state as  $\mathbf{\tilde{v}} = \mathbf{A}^\dag\mathbf{A}(\mathbf{A}^{-1}\mathbf{y})=\mathbf{A}^\dag\mathbf{y}$. This results in the general relationship
\begin{gather}
    \begin{split}
        \mathbf{y}^\dag\mathbf{D}^{\text{el}}\mathbf{y} = \mathbf{\tilde{v}}^\dag\mathbf{\tilde{D}}^{\text{CC}}\mathbf{v}
    \end{split}
\end{gather}
for any arbitrary vectors $\mathbf{y}$ and $\mathbf{v}$ related to each other by the invertible transformation $\mathbf{y} = \mathbf{A}\mathbf{v}$. 

We can also provide a similar analysis for the dynamical CC quasiparticle equation. By working in the canonical basis that diagonalizes the Fock operator, we can write the frequency-dependent CC quasiparticle equation (Eq.~\ref{eq:cc_qp}) as 
\begin{gather}
    \begin{split}
        \mathbf{\tilde{\Sigma}}(\varepsilon)\mathbf{C} = \Delta\varepsilon\mathbf{C}
    \end{split}
\end{gather}
where $\Delta\varepsilon=\varepsilon-\mathbf{\epsilon}$ is real and $\mathbf{\epsilon}$ is the matrix of eigenvalues of the Fock operator. Therefore, as the eigenvalues of the non-hermitian coupled-cluster self-energy are real, the coupled-cluster self-energy must also be pseudo-hermitian. The pseudo-hermiticity of the CC self-energy is expressed with respect to the metric
\begin{gather}
    \begin{split}
        \eta = \mathbf{B}^\dag\mathbf{B} \ .
    \end{split}
\end{gather}
Therefore, a general state and its associated state are related via $\mathbf{\tilde{v}}=\eta\mathbf{v}$ for some general vector $\mathbf{v}$. As this is the case, we can define the pseudo-canonical transformation that takes the coupled-cluster self-energy into the hermitian electronic self-energy via 
\begin{gather}
    \begin{split}
        \mathbf{\Sigma}(\omega) = \mathbf{B}\mathbf{\tilde{\Sigma}}(\omega)\mathbf{B}^{-1} \ , 
    \end{split}
\end{gather}
with $\mathbf{\Sigma}(\omega)=(\mathbf{\Sigma}(\omega))^\dag$. 
Therefore, for any arbitrary state $\mathbf{x}$, we have the biorthogonal relationship
\begin{gather}
    \begin{split}
        \mathbf{x}^\dag\mathbf{\Sigma}(\omega)\mathbf{x} = \mathbf{\tilde{w}}^\dag\mathbf{\tilde{\Sigma}}(\omega)\mathbf{w}
    \end{split}
\end{gather}
which holds for the general states $\mathbf{x}$ and $\mathbf{w}$ related by the invertible transformation $\mathbf{w}=\mathbf{B}^{-1}\mathbf{x}$. As a result, we immediately conclude that the coupled-cluster self-energy is biorthogonally positive semi-definite as a result of the positive semi-definite property of the electronic self-energy~\cite{winter1972study,schirmer1983new,bruneval2025gw+}. This property is summarized by the following equations for the imaginary part of the \emph{time-ordered} CC self-energy
\begin{gather}
    \begin{split}
             \begin{array}{c}
                 \text{Im} \left(\mathbf{\tilde{w}}^\dag\mathbf{\tilde{\Sigma}}(\omega)\mathbf{w}\right) \geq 0 \hspace{2.5mm}:\hspace{2.5mm}  \omega > \mu  \\
                 \vspace{1.5mm}
                \text{Im} \left(\mathbf{\tilde{w}}^\dag\mathbf{\tilde{\Sigma}}(\omega)\mathbf{w}\right) \leq 0  \hspace{2.5mm}:\hspace{2.5mm}   \omega < \mu
             \end{array}
\end{split}
\end{gather}
for an arbitrary state $\mathbf{w}$ and its associated state $\mathbf{\tilde{w}}$, where $\mu$ is the chemical potential. Therefore, the CC self-energy preserves causality in the biorthogonal formalism.  

The pseudo-hermitian property displayed in coupled-cluster theory can be further demonstrated by the relationship between the hermitian electronic structure Hamiltonian $H$, and the similarity transformed non-hermitian Hamiltonian $\Bar{H}$:
\begin{equation}
    H = e^{T}\Bar{H}e^{-T} \ .
\end{equation}
From this relationship, we find that the similarity transformed Hamiltonian is necessarily pseudo-hermitian with respect to the metric defined by 
\begin{equation}~\label{eq:cc_met}
    \tau = (e^{T})^\dag e^{T}\ .
\end{equation}
Clearly this metric is hermitian and possesses an inverse. Therefore, coupled-cluster theory can be formulated as a biorthogonal pseudo-hermitian quantum theory with a non trivial metric defined by Eq.~\ref{eq:cc_met}.

\begin{widetext}
\section{Algebraic Diagrammatic Construction of the CC self-energy}~\label{app:ADC}

To demonstrate the connection between the diagrammatic coupled-cluster self-energy and the ADC method  we use the Lippmann-Schwinger equation

\begin{gather}
\begin{split}~\label{eq:schwinger}
    &\Big((\omega\mp i\eta)\mathbbm{1}-(\mathbf{\bar{K}}^{\lessgtr}+\mathbf{\bar{C}}^{\lessgtr})\Big)^{-1}_{LL'}=\sum_{L''}\Big((\omega\mp i\eta)\mathbbm{1}-\mathbf{\bar{K}}^{\lessgtr}\Big)^{-1}_{LL''}\sum_{n=0}^{\infty}\Bigg(\mathbf{\bar{C}}^{\lessgtr}\Big((\omega\mp i\eta)\mathbbm{1}-\mathbf{\bar{K}}^{\lessgtr}\Big)^{-1}\Bigg)^{n}_{L''L'} 
\end{split}
\end{gather} 
where $L=J,A$ is a general composite index representing either forward- or backward-time ISCs. By taking the second Born approximation for this propagator~\cite{coveney2024rearrangement}, we have
\begin{gather}
\begin{split}~\label{eq:2nd_Born}
    &\Big((\omega\mp i\eta)\mathbbm{1}-(\mathbf{\bar{K}}^{\lessgtr}+\mathbf{\bar{C}}^{\lessgtr})\Big)^{-1}_{LL'}\approx\Big((\omega\mp i\eta)\mathbbm{1}-\mathbf{\bar{K}}^{\lessgtr}\Big)^{-1}_{LL'}+\sum_{\substack{L''L'''}}\Big((\omega\mp i\eta)\mathbbm{1}-\mathbf{\bar{K}}^{\lessgtr}\Big)^{-1}_{LL''}\mathbf{\bar{C}}^{\lessgtr}_{L''L'''}\Big((\omega\mp i\eta)\mathbbm{1}-\mathbf{\bar{K}}^{\lessgtr}\Big)^{-1}_{L'''L'} .
\end{split}
\end{gather} 
Inserting Eq.~\ref{eq:2nd_Born} into Eq.~\ref{eq:se_cc}, using the expression for the interaction matrices in Eqs~\ref{eq:forward} and~\ref{eq:backward}, gives the series of Feynman diagrams depicted in Figure~\ref{fig:adc(3)}. At third-order, a self-energy Feynman diagram contains $3!=6$ different time-orderings or Goldstone self-energy diagrams~\cite{Quantum,schirmer1983new,schirmer2018many}. These are distributed equally in the form of three distinct forward-time and three distinct backward-time Goldstone self-energy diagrams, $\tilde{\Sigma}^{\F}(\omega)/\tilde{\Sigma}^{\B}(\omega)$. To illustrate these results, we take the forward-time self-energy contribution of the fifth diagram of Figure~\ref{fig:adc(3)}. There are three different Goldstone diagrams corresponding to the forward-time component of this diagram. The first time-ordering gives the Goldstone diagram:
\begin{gather}
        \begin{split}
            \begin{gathered}
    \begin{fmfgraph*}(50,70)
    \fmfcurved
    \fmfset{arrow_len}{3mm}
    \fmfleft{i1,i2,i3}
    \fmflabel{}{i1}
    \fmflabel{}{i2}
    \fmfright{o1,o2,o3}
    \fmflabel{}{o1}
    \fmflabel{}{o2}
    \fmf{dbl_zigzag}{i1,v1}
    \fmf{dbl_dashes}{i2,v2}
    \fmf{dbl_dashes}{v2,o2}
    \fmf{dbl_zigzag}{i3,v3}
    \fmf{fermion}{i1,i2}
    \fmf{fermion}{v1,v2}
    \fmf{fermion}{o2,v1}
    \fmf{fermion}{v2,v3}
    \fmf{fermion}{v3,o2}
    \fmf{fermion}{i2,i3}
    \fmfforce{(0.0w,0.0h)}{i1}
    \fmfforce{(1.0w,0.0h)}{o1}
    \fmfforce{(0.5w,0.5h)}{v2}
    \fmfforce{(0.5w,0.0h)}{v1}
    \fmfforce{(0.5w,1.0h)}{v3}
    \fmfforce{(0.0w,1.0h)}{i3}
    \fmfforce{(1.0w,1.0h)}{o3}
    \fmfdotn{v}{3}
    \fmfdot{i1,i2,i3}
    \fmfdot{o2}
\end{fmfgraph*}
\end{gathered} \hspace{5mm} = &\sum_{\substack{JJ'\\J''J'''}}\tilde{U}^{2\p1\h,2\text{b}}_{p,J}\Big((\omega+i\eta)\mathbbm{1}-\mathbf{\bar{K}}^{>}\Big)^{-1}_{JJ'}\mathbf{\bar{C}}^{>,3\text{b}}_{J'J''}\Big((\omega+i\eta)\mathbbm{1}-\mathbf{\bar{K}}^{>}\Big)^{-1}_{J''J'''}\bar{U}^{2\p1\h}_{J''',q}  \ ,
\end{split}
\end{gather}
remembering that the 2p1h ISC index, $J\equiv(i;a<b)$. 
From the diagram, we immediately identify the coupling matrix elements as $\tilde{U}^{2\p1\h,2\text{b}}_{p,J}=\tilde{\Xi}_{pi,ab}$ and $\bar{U}^{2\p1\h}_{J''',q}=\tilde{\Xi}_{cd,qj}$. The three-body interaction matrix element is $\mathbf{\bar{C}}^{>,3\text{b}}_{J'J''}=-\tilde{\chi}_{abj,cdi}=\tilde{\chi}_{jab,cid}$, where we use the antisymmetry of the effective interaction matrix element. 

The second time-ordering gives rise to the following Goldstone diagram:
\begin{gather}
        \begin{split}
            \begin{gathered}
    \begin{fmfgraph*}(50,70)
    \fmfcurved
    \fmfset{arrow_len}{3mm}
    \fmfleft{i1,i2,i3}
    \fmflabel{}{i1}
    \fmflabel{}{i2}
    \fmfright{o1,o2,o3}
    \fmflabel{}{o1}
    \fmflabel{}{o2}
    \fmf{dbl_zigzag}{o2,v2}
    \fmf{dbl_dashes}{i1,v1}
    \fmf{dbl_dashes}{v1,o1}
    \fmf{dbl_zigzag}{i3,v3}
    \fmf{fermion}{i1,i3}
    \fmf{fermion}{o1,o2}
    \fmf{fermion}{v2,i1}
    \fmf{fermion}{o2,v1}
    \fmf{fermion}{v1,v3}
    \fmf{fermion,left=0.3}{v3,o1}
    \fmfforce{(0.0w,0.0h)}{i1}
    \fmfforce{(1.0w,0.0h)}{o1}
    \fmfforce{(0.35w,0.5h)}{v2}
    \fmfforce{(0.5w,0.0h)}{v1}
     \fmfforce{(0.75w,0.5h)}{o2}
    \fmfforce{(0.5w,1.0h)}{v3}
    \fmfforce{(0.0w,1.0h)}{i3}
    \fmfforce{(1.0w,1.0h)}{o3}
    \fmfdotn{v}{3}
    \fmfdot{i1,i3}
    \fmfdot{o2,o1}
\end{fmfgraph*}
\end{gathered}\hspace{5mm}&=\sum_{JJ'}\tilde{U}^{2\p1\h,2\text{b}}_{p,J}\Big((\omega+i\eta)\mathbbm{1}-\mathbf{\bar{K}}^{>}\Big)^{-1}_{JJ'}\bar{U}^{3\text{b}}_{J',q} \ , 
\end{split}
\end{gather}
from which, we can immediately identify the first coupling matrix as $\tilde{U}^{2\p1\h,2\text{b}}_{p,J}=\tilde{\Xi}_{pi,ab}$. The second coupling matrix element contains effects resulting from the three-body interaction and is given by 
\begin{gather}
    \begin{split}
        \bar{U}^{3\text{b}}_{J,q} = \frac{1}{2}\sum_{klc} \frac{\tilde{\chi}_{abc,kli}\tilde{\Xi}_{kl,qc}}{\epsilon_{k}+\epsilon_{l}+\epsilon_{i}-\epsilon_{a}-\epsilon_{b}-\epsilon_{c}} = 0  \ . 
    \end{split}
\end{gather}
This term vanishes as a result of the CC amplitude equations: $\tilde{\chi}_{abc,kli}= 0$. Therefore, the second time-ordered contribution vanishes as a result of the CC amplitude equations.  The third and final time-ordered contribution to this coupled-cluster self-energy Feynman diagram is given by the Goldstone diagram:
\begin{gather}~\label{eq:3rd_order_import}
        \begin{split}
            \begin{gathered}
    \begin{fmfgraph*}(50,70)
    \fmfcurved
    \fmfset{arrow_len}{3mm}
    \fmfleft{i1,i2,i3}
    \fmflabel{}{i1}
    \fmflabel{}{i2}
    \fmfright{o1,o2,o3}
    \fmflabel{}{o1}
    \fmflabel{}{o2}
    \fmf{dbl_zigzag}{i1,v1}
    \fmf{dbl_dashes}{i3,v3}
    \fmf{dbl_dashes}{v3,o3}
    \fmf{dbl_zigzag}{o2,v2}
    \fmf{fermion}{i1,i3}
    \fmf{fermion}{v1,v3}
    \fmf{fermion}{o2,o3}
    \fmf{fermion}{o3,v1}
    \fmf{fermion}{v3,o2}
    \fmf{fermion}{i3,v2}
    \fmfforce{(0.0w,0.0h)}{i1}
    \fmfforce{(1.0w,0.0h)}{o1}
    \fmfforce{(0.9w,0.5h)}{v2}
    \fmfforce{(0.5w,0.0h)}{v1}
    \fmfforce{(0.5w,1.0h)}{v3}
    \fmfforce{(0.0w,1.0h)}{i3}
    \fmfforce{(1.0w,1.0h)}{o3}
    \fmfforce{(1.5w,0.5h)}{o2}
    \fmfdotn{v}{3}
    \fmfdot{i1,i3}
    \fmfdot{o2,o3}
\end{fmfgraph*}
\end{gathered} \hspace{12.5mm} = \sum_{JJ'}\tilde{U}^{3\text{b}}_{p,J}\Big((\omega+i\eta)\mathbbm{1}-\mathbf{\bar{K}}^{>}\Big)^{-1}_{JJ'}\bar{U}^{2\p1\h}_{J',q} \ .
\end{split}
\end{gather}
The first coupling matrix element contains the three-body effective interaction and evaluates to give 
\begin{gather}
\begin{split}
    \tilde{U}^{3\text{b}}_{p,J} = \frac{1}{2}\sum_{klc}\frac{\tilde{\chi}_{kli,abc}\tilde{\Xi}_{pc,kl}}{\epsilon_{k}+\epsilon_{l}+\epsilon_{i}-\epsilon_{a}-\epsilon_{b}-\epsilon_{c}} = 0\ .
\end{split}
\end{gather}
This coupling term also vanishes due to the structure of the CC similarity transformed Hamiltonian as effective interactions that have four or more lines below a vertex are zero: $\tilde{\chi}_{kli,abc}=0$. The second coupling matrix element is identified as $\bar{U}^{2\p1\h}_{J',q}=\tilde{\Xi}_{ab,qi}$. Therefore, the third time-ordered contribution to the CC self-energy Feynman diagram also vanishes as a result of the CC similarity transformation.

Bringing these results together, the forward-time contribution of this third-order self-energy diagram, $\tilde{\Sigma}^{\F,3\text{b}}_{pq}(\omega)$, is given by
\begin{gather}
    \begin{split}~\label{eq:three_se}
        \tilde{\Sigma}^{\F,3\text{b}}_{pq}(\omega) = &\sum_{\substack{JJ'\\J''J'''}}\tilde{U}^{2\p1\h,2\text{b}}_{p,J}\Big((\omega+i\eta)\mathbbm{1}-\mathbf{\bar{K}}^{>}\Big)^{-1}_{JJ'}\mathbf{\bar{C}}^{>,3\text{b}}_{J'J''}\Big((\omega+i\eta)\mathbbm{1}-\mathbf{\bar{K}}^{>}\Big)^{-1}_{J''J'''}\bar{U}^{2\p1\h}_{J''',q} \ .
    \end{split}
\end{gather}
The forward-time component of the second-order diagram, the second diagram of Figure~\ref{fig:adc(3)}, evaluates to 
\begin{gather}
    \begin{split}~\label{eq:two_se}
        \tilde{\Sigma}^{\F,2\text{b}}_{pq}(\omega) = \sum_{\substack{JJ'}}\tilde{U}^{2\p1\h,2\text{b}}_{p,J}\Big((\omega+i\eta)\mathbbm{1}-\mathbf{\bar{K}}^{>}\Big)^{-1}_{JJ'}\bar{U}^{2\p1\h}_{J',q} \ .
    \end{split}
\end{gather}
Adding Eqs~\ref{eq:three_se} and~\ref{eq:two_se} together, we can use the spectral representation of the self-energy and the corresponding Lippmann-Schwinger equation (Eq.~\ref{eq:schwinger}) to perform the infinite-order summation of these two diagrams by writing
\begin{gather}
    \begin{split}
        \tilde{\Sigma}^{\F,2\p1\h(3\text{b})}_{pq}(\omega) &= \sum_{\substack{JJ'}}\tilde{U}^{2\p1\h,2\text{b}}_{p,J}\Big((\omega+i\eta)\mathbbm{1}-(\mathbf{\bar{K}}^{>}+\mathbf{\bar{C}}^{>,3\text{b}})\Big)^{-1}_{JJ'}\bar{U}^{2\p1\h}_{J',q}  \ .
    \end{split}
\end{gather}
\end{widetext}
Using the upfolded Dyson supermatrix representation, we can simply sum this self-energy contribution to infinite-order by identification of the three-body interaction matrix element: $\mathbf{\bar{C}}^{>,3\text{b}}_{JJ'}$. Performing this analysis for all the forward- and backward-time interaction matrix elements defined in Eqs~\ref{eq:forward} and~\ref{eq:backward} will give the infinite-order summation of the CC self-energy diagrams depicted in Figure~\ref{fig:adc(3)}. The same analysis presented here can also be applied to the set of diagrams depicted in Figure~\ref{fig:pert_cc_se} and yields the coupled-cluster Dyson supermatrix of Eq.~\ref{eq:cc_eom_se}.

\end{fmffile}

\bibliography{cc_se}

\end{document}